%% file: comp_sALE.tex
\documentclass[11pt]{article}
\usepackage{verbatim,amsmath,amssymb}
\usepackage{epsfig,float,color}
\usepackage[font=small,labelfont=bf]{caption}
\usepackage{geometry}
\usepackage{setspace}
\usepackage{natbib}
\usepackage{amsthm,mathrsfs,amsfonts,dsfont} 
\usepackage{hyperref}

\definecolor{darkblue}{rgb}{0,0,1}
\definecolor{col1}{rgb}{1,0,1}
\definecolor{col2}{rgb}{0,0.7,0}
\definecolor{colb}{rgb}{1,.85,.0}

\hypersetup{pdftex=true, colorlinks=true, breaklinks=true, linkcolor=darkblue, menucolor=darkblue, pagecolor=darkblue, citecolor=darkblue, urlcolor=darkblue}

\newtheoremstyle{rem}%name
{6pt}%Space above
{6pt}%Space below
{\small}%Body font
{}%Indent amount
{\bf}%Theorem head font
{:}%Punctuation after theorem head
{.5em}%Space after theorem head
{}%Theorem head spec(can be left empty, meaning ‘normal’)

\theoremstyle{rem}
\newtheorem{remark}{Remark}[section]

\newfloat{boxtable}{luc}{Box}
\newfloat{Box}{luc}{Box}

\input{neco.tex}

\newcommand{\tmu}{\mu_\mrm}
\begin{document}

\begin{center}

\Large{\bf{A curvilinear surface ALE formulation for self-evolving Navier-Stokes manifolds -- Stabilized finite element formulation}} \\[4mm]

\end{center}

\renewcommand{\thefootnote}{\fnsymbol{footnote}}

\begin{center}
\large{Roger A. Sauer$^{\mra,\mrb,\mrc,}$\footnote[1]{corresponding author, email: roger.sauer@rub.de}}
\vspace{3mm}

\small{\textit{
$^\mra$Institute for Structural Mechanics, Ruhr University Bochum, 44801 Bochum, Germany \\[1mm]
$^\mrb$Department of Structural Mechanics, Gda\'{n}sk University of Technology, 80-233 Gda\'{n}sk, Poland \\[1mm]
$^\mrc$Mechanical Engineering, Indian Institute of Technology Guwahati, Assam 781039, India}}

\vspace{3mm}

\small{Published\footnote[2]{This pdf is the personal version of an article whose journal version is available at \href{https://doi.org/10.1016/j.cma.2025.118331}{www.sciencedirect.com}} 
in \textit{Comput.~Methods Appl.~Mech.~Eng.}, \href{https://doi.org/10.1016/j.cma.2025.118331}{DOI: 10.1016/j.cma.2025.118331} \\
Submitted on 3 June 2025; Revised on 15 August 2025; Accepted on 17 August 2025} 

\end{center}

\renewcommand{\thefootnote}{\arabic{footnote}}

\vspace{-4mm}

\rule{\linewidth}{.15mm}
{\bf Abstract:}
This work presents a stabilized finite element formulation of the arbitrary Lagran-gian-Eulerian (ALE) surface theory for Navier-Stokes flow on self-evolving manifolds developed in \citet{ALEtheo}. 
The formulation is physically frame-invariant, applicable to large deformations, and relevant to fluidic surfaces such as soap films, capillary menisci and lipid membranes, which are complex and inherently unstable physical systems.
It is applied here to area-incompressible surface flows using a stabilized pressure-velocity (or surface tension-velocity) formulation based on quadratic finite elements and implicit time integration.
The unknown ALE mesh motion is determined by membrane elasticity such that the in-plane mesh motion is stabilized without affecting the physical behavior of the system. 
The resulting three-field system is monolithically coupled, and fully linearized within the Newton-Rhapson solution method.
The new formulation is demonstrated on several challenging examples including shear flow on self-evolving surfaces and inflating soap bubbles with partial inflow on evolving boundaries.
Optimal convergence rates are obtained in all cases.
Particularly advantageous are $C^1$-continuous surface discretizations, for example based on NURBS.

{\bf Keywords:} Arbitrary Lagrangian-Eulerian formulation, area-incompressibility, fluid-structure interaction, Navier-Stokes equations, nonlinear finite element methods, pressure stabilization

\vspace{-5mm}
\rule{\linewidth}{.15mm}

\section{Introduction}\label{s:intro}

Simulating fluid flow on evolving surfaces is a very challenging problem due to various physical and computational complexities.
There are the usual challenges of the Navier-Stokes equations -- their nonlinearity, time-evolution, spatial resolution and stabilization, in particular for incompressibility.
This then combines with the nonlinearities and potential instabilities of the large shape changes of an evolving manifold.
Since fluids are characterized by velocity, and shapes by deformation, the overall problem is a fluid-structure interaction (FSI) problem.
The arbitrary Lagrangian-Eulerian (ALE) formulation is a classical approach for FSI problems, and so it is a natural fit for evolving surface flows.
\citet{ALEtheo} provides a new, general ALE formulation for Navier-Stokes flow on self-evolving surfaces.
Here now, a corresponding stable finite element implementation is developed, and then tested extensively on several challenging examples.

The theory for surface flows goes back to the sixties with the seminal work of \citet{scriven60},
while computational ALE formulations for bulk flows go back almost as far, see \citet{donea04} and references therein.
However, a 3D simulation framework for Navier-Stokes flow on self-evolving surfaces, (where the surface shape is fully unknown), has appeared only ecently with the concurrent works of \citet{torres19} and \citet{ALE}.
Still, these and subsequent works have restrictions that are the motivation behind the present work.

The literature on evolving surface flows can be grouped into three categories -- known surface motion, Lagrangian surface descriptions and ALE surface descriptions. 

Many works consider known surfaces -- either fixed surfaces or prescribed surface motion.
The first general 3D finite element (FE) formulation for area-incompressible flows on fixed surfaces seems to be the work of \citet{nitschke12}.
It uses a vorticity-stream function formulation, and it has been extended to prescribed surface motions \citep{reuther15} and a discrete exterior calculus discretization \citep{nitschke17}.
Subsequent pressure-velocity formulations have appeared using Taylor-Hood FE \citep{fries18}, trace FE \citep{olshanskii18}, Chorin projection \citep{reuther18}, div-conforming FE \citep{bonito19,lederer20} and FE exterior calculus \citep{toshniwal21}.
Trace FE provide a level-set surface representation from a surrounding 3D FE mesh \citep{olshanskii17}, while in the other cases surface FE are used, e.g.~following \citet{dziuk13}.
Apart from FE methods, also spectral methods and meshless methods have been used for simulating surface flows \citep{gross18,gross20}.
In subsequent work, known surface formulations have been extended to study Kevin-Helmholtz instabilities \citep{lederer20,jankuhn20},
active flows \citep{rank21}, immersed particles \citep{rower22}, and phase separations \citep{sun22} on curved surfaces.
The mathematical analysis of known surface flows also remains an ongoing topic \citep{olshanskii23,reusken24,elliott24}.
A recent comparison of different discretization schemes for Stokes flow on fixed surfaces has been investigated in \citet{hardering24}.

Known surfaces are simpler, as the unknown fluid velocity is only characterized by two unknown tangential components.
For Newtonian fluids, these velocity components are governed by the tangential surface Navier-Stokes equations that contain surface-projection operators. For self-evolving surfaces the fluid velocity has an additional unknown out-of-plane component.
Instead of using this out-of-plane component together with the two tangential components as unknowns, one can take the three velocity components of a fixed Cartesian background coordinate system as unknowns, which actually facilitates the formulation and discretization of the governing equations (as no projection operators are needed), even when their solution is harder than the fixed surface case.
This description has a long tradition in large-deformation FE formulations for solid membranes and shells \citep{oden67,simo90}.
They are commonly described in a convected, Lagrangian surface frame, which is much more straightforward than an Eulerian frame.

Due to this, initial formulations for evolving (fluid) surfaces have neglected in-plane flow and focussed on shape changes in a Lagrangian framework.
The first 3D FE formulation of this kind is the work of \citet{brown80b} that studies the deformation of static droplets by solving the Young-Laplace equation.
Hence there is only one degree-of-freedom (dof) per surface node.
This is different in the surface evolution algorithms of \citet{dziuk91} and \citet{brakke92} that have three nodal dofs. 
But in this case the surface flow can lead to large mesh distortions.
These essentially correspond to zero in-plane shear modes, as was noted by \citet{feng06} in their study of lipid bilayer membranes.  
Subsequently, \citet{ma08} proposed an iterative mesh redistribution scheme to regularize tangential mesh motions. 
Mesh redistribution schemes were also employed by \citet{barrett08a} and \citet{dziuk08} in their study of Willmore flows.
On the other hand, the problem of zero in-plane shear modes does not arise in axisymmetric surface evolution, as are studied by \citet{arroyo09}.
Apart from mesh redistribution, also remeshing can be used, as was done by \citet{elliott13} to study evolving biphasic lipid bilayers.
Another approach to stabilize the tangential mesh motion is to add in-plane shear stiffness in such a way that the out-of-plane surface motion is not affected. Such a scheme was developed in \citet{membrane} and then adapted to deforming droplets \citep{droplet,dropslide} and lipid bilayer membranes \citep{liquidshell}.
A similar mesh stabilization scheme was later proposed by \citet{dharmavaram21}.
A further approach is to eliminate the tangential mesh motion by normal projection \citep{droplet,rangarajan15}. 
In subsequent work, Lagrangian surface descriptions have been used to simulate deforming vesicles in 3D flow \citep{barrett15,barrett16},
phase separation on deforming surfaces \citep{phaseshell,valizadeh19}, Navier-Stokes surface flows with meshless methods \citep{suchde20,bharadwaj22},
and particles or filaments embedded in fluidic membranes \citep{dharmavaram22,sharma23}.

As mentioned already, Lagrangian surface descriptions for evolving surfaces are interesting when in-plane shear flows are small or not important. 
If large in-plane shear flows occur, it is more suitable to consider in-plane Eulerian or in-plane ALE surface descriptions.
Initial works of this kind have restricted themselves to axisymmetry \citep{rahimi12} or Monge parametrizations \citep{rangamani13}. 
In the latter, the out-of-plane motion is defined through the in-plane parameterization, which does not allow for arbitrary large surface motions. 
More general are the ALE formulations of \citet{torres19}, \citet{ALE}, \citet{reuther20} and \citet{alizzi23}.
However, they restrict their simulations to Eulerian or near-Eulerian in-plane flow descriptions.
Since these can still cause large mesh distortions when the surface evolves \citep{ALEtheo}, further generalization on the in-plane mesh motion is required. This is offered by an in-plane ALE formulation.
In this way the tangential mesh velocity can be prescribed, e.g.~based on the average surface velocity \citep{kinkelder21}, or computed through suitable mesh equations.
\citet{krause23} use the curvature-dependent mesh redistribution scheme of \citet{barrett08} for this.
While the approach has been successfully used to model phase transitions \citep{bachini23} and wrinkling \citep{krause24} of fluidic membranes, the scheme can be expected to become unstable under mesh refinement, as the surface curvature is generally invariant with respect to tangential mesh motion.
This problem is absent in the ALE formulation of \citet{sahu24}, which uses in-plane viscosity to control the mesh motion. 
But also this can become problematic, as it can be expected to lose stability when the mesh velocity approaches zero.
This is avoided if the mesh motion is governed by in-plane elasticity, which is the approach taken here.
As noted above, this can be done in such a way that the governing physical equations, and hence the material motion, are not affected.

The presented formulation focuses on self-evolving, area-incompressible Navier-Stokes manifolds with arbitrary shape and arbitrary (but unchanging) topology.
The problem is governed by four coupled fields -- the unknowns being the material velocity, the surface tension (= negative in-plane fluid pressure), the surface deformation and its time derivative, the ALE frame velocity. 
Pressure stabilization is achieved through the scheme of \citet{dohrmann04}, while mesh stabilization is achieved through the in-plane elasticity scheme of \citet{droplet}.

In summary, this work contains several important computational novelties: \\[-8mm]
\begin{itemize}
\item It presents a general surface ALE implementation for transient Navier-Stokes flow on self-evolving manifolds, \\[-7mm]
\item using a stabilized pressure/velocity-based finite element formulation, \\[-7mm]
\item with unknown mesh motion governed by in-plane membrane elasticity. \\[-7mm]
\item The implementation avoids unnecessary splits into local bases and coordinate systems. \\[-7mm]
\item It uses monolithic coupling and second order time integration. \\[-7mm]
\item Its full linearization is provided, including follower loads on evolving inflow boundaries. \\[-7mm]
\item Its proper convergence behavior is demonstrated on several challenging examples, \\[-7mm]
\item that include vortices on deforming soap films and varying inflow on evolving boundaries. \\[-7mm]
\end{itemize}

The remainder of this paper is organized as follows.
Sec.~\ref{s:theo} summarizes the governing equations of the ALE flow description.
The area-compressibility constraint is stabilized with the Dohrmann-Bochev scheme as discussed in Sec.~\ref{s:stab}.
The finite element discretization is then provided in Sec.~\ref{s:FE}.
This is illustrated by several numerical examples for fixed, prescribed and self-evolving surfaces in Secs.~\ref{s:Nexf} and \ref{s:Nex}. 
The paper concludes with Sec.~\ref{s:concl}.

\section{Governing ALE equations for surface flow}\label{s:theo}

This section provides a (brief) summary of the most important equations for the kinematics, weak form and constitutive relations in the ALE frame.
Their detailed presentation and derivation can be found in \citet{ALEtheo}.

\subsection{ALE surface description}

\begin{figure}[h]
\begin{center} \unitlength1cm
\begin{picture}(0,7.2)
\put(-6.2,-1.7){\includegraphics[width=122mm]{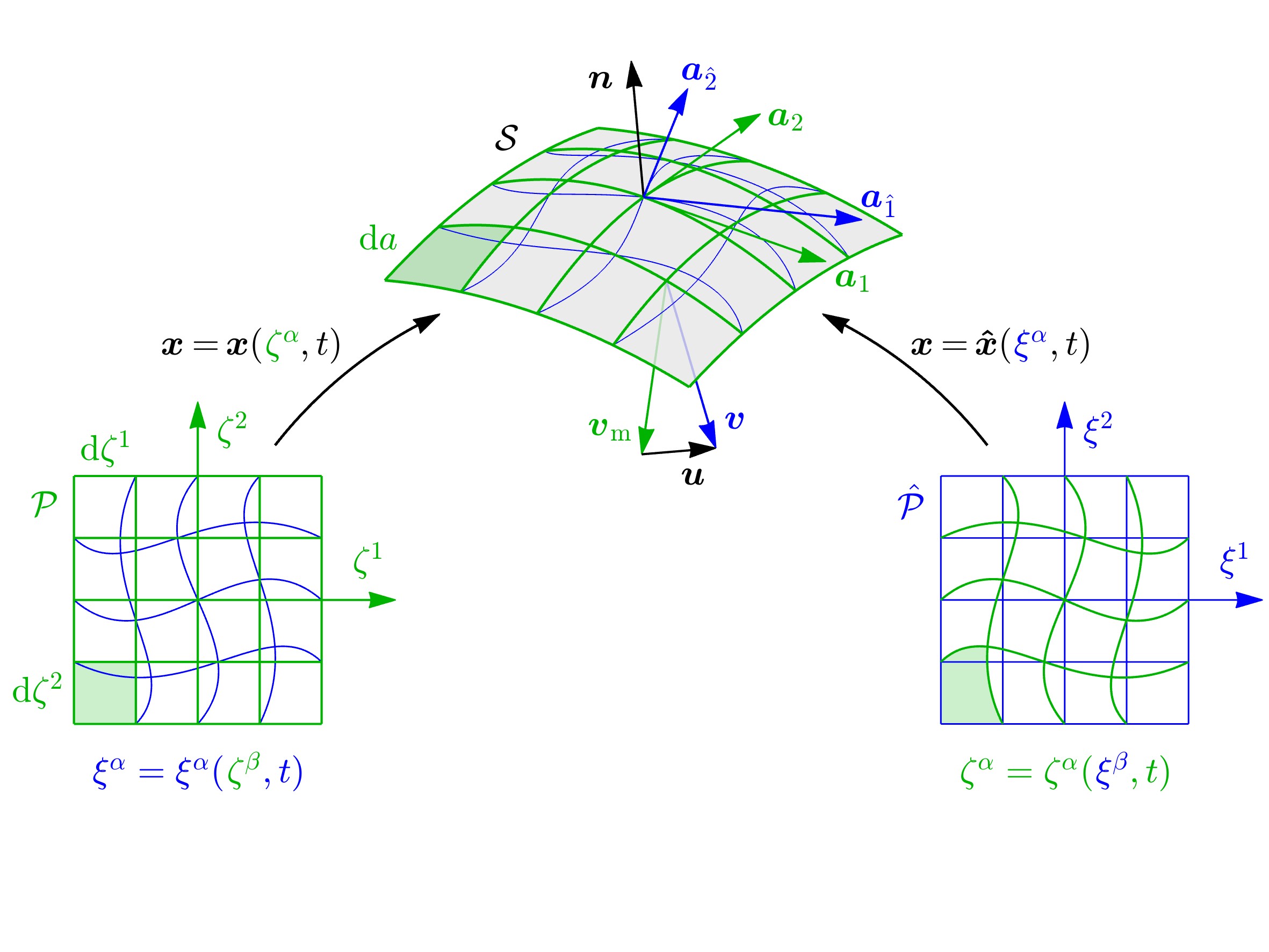}}
\end{picture}
\caption{ALE surface description:
The fluid surface $\sS$ is described by the ALE mapping $\bx=\bx(\zeta^\alpha,t)$ from the ALE parameter domain $\sP$. 
This mapping is then used for surface integration and discretization.
It is generally different from the Lagrangian mapping $\bx=\hat\bx(\xi^\alpha,t)$ from the material parameter domain $\hat\sP$.
From the two mappings follow the surface velocities $\bv_\mrm := \partial\bx/\partial t|_{\zeta^\alpha}$ and $\bv := \partial\hat\bx/\partial t|_{\xi^\alpha}$, and their in-plane difference $\bu := \bv - \bv_\mrm$.}
\label{f:ALE}
\end{center}
\end{figure}

The current fluid surface $\sS$ is described by the parameterization
\eqb{l}
\bx = \bx(\zeta^\alpha,t)\,,
\label{e:bx}\eqe
where the curvilinear coordinate $\zeta^\alpha$ describes the ALE frame.
It admits the special cases $\zeta^\alpha=\xi^\alpha$ and $\zeta^\alpha = \theta^\alpha$, that denote the Lagrangian and Eulerian surface coordinates, respectively.
In order to distinguish these special cases in the mapping, we write $\bx=\hat\bx(\xi^\alpha,t)$ and $\bx=\tilde\bx(\theta^\alpha,t)$.
Fig.~\ref{f:ALE} shows an illustration of mappings $\bx = \bx(\zeta^\alpha,t)$ and $\bx=\hat\bx(\xi^\alpha,t)$.
%-----------------------------------------------------------------
% run ALEplot4
%-----------------------------------------------------------------

Given \eqref{e:bx}, the tangent vectors in the ALE frame follow as
\eqb{l}
\ba_\alpha := \ds\pa{\bx}{\zeta^\alpha}\,.
\label{e:ba}\eqe
Together with the surface normal
\eqb{l}
\bn := \ds\frac{\ba_1\times\ba_2}{\norm{\ba_1\times\ba_2}}\,,
\label{e:bn}\eqe
they form a basis to decompose vectors and tensors into their in-plane and out-of-plane components.
It is noted that other basis vectors can be defined, importantly $\ba_{\hat\alpha} := \partial\hat\bx/\partial\xi^\alpha$ and $\ba_{\tilde\alpha} := \partial\tilde\bx/\partial\theta^\alpha$.
While those are important for the theoretical development given in \citet{ALEtheo}, they are not needed for the following computational description.
In the following all Greek indices exclusivity refer to the general ALE parameterization. 

From \eqref{e:ba}-\eqref{e:bn} follow the surface metric and its inverse,
\eqb{l}
a_{\alpha\beta} = \ba_\alpha\cdot\ba_\beta\,,
\quad \big[a^{\alpha\beta}\big] = \big[a_{\alpha\beta}\big]^{-1}\,,
\label{e:aab}\eqe
the dual basis vectors
\eqb{l}
\ba^\alpha = a^{\alpha\beta}\ba_\beta\,,
\label{e:badual}\eqe
the parametric derivative
\eqb{l}
..._{,\alpha} := \ds\pa{...}{\zeta^\alpha}\,,
\label{e:dalpha}\eqe
the surface curvature tensor components
\eqb{l}
b_{\alpha\beta} := \ba_{\alpha,\beta}\cdot\bn\,, 
\eqe
and the Christoffel symbols
\eqb{l}
\Gamma^\gamma_{\alpha\beta} := \ba_{\alpha,\beta}\cdot\ba^\gamma\,.
\label{e:Gamma}\eqe
The latter two are the out-of-plane and in-plane components of $\ba_{\alpha,\beta}$, respectively.

A reference configuration is introduced at time $t=0$, and denoted by $\sS_0$. 
Capital letters are used for the kinematical quantities from above on $\sS_0$, i.e.~$\bX=\bx\big|_{t=0}$, $\bA_\alpha=\ba_\alpha\big|_{t=0}$, $\bN=\bn\big|_{t=0}$, etc. 
Area elements in the current and reference configuration can be related to the ALE parameter domain by
\eqb{llllll}
\dif a \is J_a\,\dif\zeta^1\,\dif\zeta^2\,, & J_a \dis \sqrt{\det[a_{\alpha\beta}]}\,, \\[1.5mm]
\dif A \is J_A\,\dif\zeta^1\,\dif\zeta^2\,, & J_A \dis \sqrt{\det[A_{\alpha\beta}]} \,,
\label{e:da}\eqe
see Fig.~\ref{f:ALE}.
Area changes in the ALE frame are then given by
\eqb{l}
J_\mrm := \ds\frac{J_a}{J_A}\,.
\label{e:Jm}\eqe

The time derivative of the ALE frame is
\eqb{l}
(...)' := \ds\pa{...}{t}\Big|_{\zeta^\alpha}\,.
\label{e:bvp}\eqe
It defines the frame, or mesh, velocity
\eqb{l}
\bv_\mrm := \bx' := \ds\pa{\bx}{t}\Big|_{\zeta^\alpha}\,.
\label{e:vm}\eqe
In contrast to this, there is the material time derivative
\eqb{l}
\dot{(...)} = \ds\pa{...}{t}\Big|_{\xi^\alpha}
\label{e:dot}\eqe
that defines the fluid velocity
\eqb{l}
\bv = \dot\bx = \ds\pa{\hat\bx}{t}\Big|_{\xi^\alpha}\,.
\label{e:matvel}\eqe
The two velocities are related by 
\eqb{l}
\bv = \bv_\mrm + \dot\zeta^\alpha\, \ba_\alpha\,.
\label{e:bv}\eqe
where
\eqb{l} 
\dot\zeta^\alpha = \ba^\alpha\cdot(\bv-\bv_\mrm)\,,
\label{e:zetadot}\eqe
or, equivalently, $\dot\zeta^\alpha = v^\alpha-v_\mrm^\alpha$.
For convenience we also introduce the relative fluid velocity
\eqb{l}
\bu:=\bv-\bv_\mrm 
\label{e:bu}\eqe
that has the components $u^\alpha=\dot\zeta^\alpha$ and $u=0$ in the $\{\ba_1,\ba_2,\bn\}$ basis, see Fig.~\ref{f:ALE}.
Eq.~\eqref{e:bv} is an example of the fundamental surface ALE equation
\eqb{l} 
\boxed{\dot{(...)} = (...)' + ..._{,\alpha}\,\dot\zeta^\alpha} 
\label{e:ALE}\eqe
that is needed to translate time derivatives.
For example, in case of the acceleration it reads
\eqb{l}
\dot\bv = \bv' + \bv_{\!,\alpha}\,\big(v^\alpha-v^\alpha_\mrm\big)\,.
\label{e:vALEa}\eqe

Another important relation is,
\eqb{l}
\bv_{\!,\alpha} = \dot\ba_\alpha + \dot\zeta^\gamma_{,\alpha}\,\ba_\gamma\,,
\label{e:badot}\eqe
where $\dot\zeta^\gamma_{,\alpha}$ is the parametric derivative of $\dot\zeta^\gamma$, and not the material time derivative of $\zeta^\gamma_{,\alpha}$ (which is zero), 
as generally spatial and temporal differentiation do not commute for the ALE parameterization.
To be precise, spatial differentiation w.r.t.~$\zeta^\alpha$ does not generally commute with the material time derivative (at fixed $\xi^\alpha$).
Only spatial differentiation w.r.t.~to $\xi^\alpha$ does. 
Likewise, spatial differentiation w.r.t.~$\zeta^\alpha$ commutes with temporal differentiation at fixed $\zeta^\alpha$.

From the parametric derivative follows the surface divergence,
\eqb{l}
\divs\bv = \bv_{\!,\alpha}\cdot\ba^\alpha\,,
\label{e:sdiv}\eqe
and the surface gradient,
\eqb{l}
\nabla_{\!\mrs}\,\bv = \bv_{\!,\alpha}\otimes\ba^\alpha\,,
\label{e:sgrad}\eqe
of the fluid velocity.
The symmetric part of the latter is the rate of surface deformation tensor
\eqb{l}
\bd = (\bv_{\!,\alpha}\otimes\ba^\alpha + \ba^\alpha\otimes\bv_{\!,\alpha})/2\,.
\eqe
Like $\nabla_{\!\mrs}\,\bv$, it has both in-plane and out-of-plane components.
Its in-plane components are
\eqb{l}
d^{\alpha\beta} = \ba^\alpha\cdot\bd\,\ba^\beta = (a^{\alpha\gamma}\bv_{\!,\gamma}\cdot\ba^\beta + a^{\beta\gamma}\bv_{\!,\gamma}\cdot\ba^\alpha)/2\,.
\label{e:dab}\eqe
It is noted that $d^{\alpha\beta}a_{\alpha\beta} = \divs\bv$ and
\eqb{l}
2d^{\alpha\beta}\ba_\beta = \ba^{\alpha\beta}_\mrs\,\bv_{\!,\beta}\,,
\label{e:2dabab}\eqe
with the surface tensor
\eqb{l}
\ba^{\alpha\beta}_\mrs := a^{\alpha\beta}\,\bi + \ba^\beta\otimes\ba^\alpha\,.
\label{e:bas}\eqe
Eqs.~\eqref{e:2dabab}-\eqref{e:bas} are needed for the FE formulation later.
It is finally noted that $d^{\alpha\beta}$ is symmetric, while $\ba^{\alpha\beta}_\mrs$ is not.

\subsection{Surface constitution in ALE form}

The stress tensor of an area-incompressible Newtonian fluid surface can be written as \citep{ALEtheo}
\eqb{l}
\bsig = q\,\bi + 2\eta\,\bd_\mrs\,,
\label{e:bsig}\eqe
where $q$ is the surface tension, $\bi:=\ba_\alpha\otimes\ba^\alpha$ the surface identity tensor, $\eta$ the surface viscosity, and $\bd_\mrs = d^{\alpha\beta}\ba_\alpha\otimes\ba_\beta$ the in-plane part of $\bd$.
Generally, $q$, $\bi$ and $\bd_\mrs$ are functions of position $\bx$.
The components of $\bsig = \sig^{\alpha\beta}\ba_\alpha\otimes\ba_\beta$ in the ALE frame are
\eqb{l}
\sig^{\alpha\beta} = q\,a^{\alpha\beta} + 2\eta\,d^{\alpha\beta}\,.
\label{e:sigab}\eqe
Also the velocity gradient within $d^{\alpha\beta}$ is described with respect to the ALE frame according to Eqs.~\eqref{e:sgrad} and \eqref{e:dalpha}.
The constitutive model defined by Eqs.~\eqref{e:bsig} and \eqref{e:sigab} can be extended to surface fluids with bending resistance, see \citet{ALEtheo}, but this is not considered here.
The surface tension $q$ is defined through the area-incompressibility constraint
\eqb{l}
\divs\bv = 0\,.
\label{e:incomp}\eqe

\subsection{Governing equations in ALE form}\label{s:gov}

The governing strong form equations are derived in \citet{ALEtheo} and summarized in Box~1 there.
They consists of four coupled PDEs for the four unknown fields:
fluid velocity $\bv$, surface tension $q$, mesh velocity $\bv_\mrm$ and surface position $\bx$.
Since $\bv$, $\bv_\mrm$ and $\bx$ are vectors, there are 10 unknown components altogether.
The associated PDEs for functions $\bv(\zeta^\alpha,t)$, $q(\zeta^\alpha,t)$, $\bv_\mrm(\zeta^\alpha,t)$ and $\bx(\zeta^\alpha,t)$ are the surface Navier-Stokes equation,
\eqb{l}
\divs\bsig^\mrT + \bff = \rho\,\dot\bv\,, 
\label{e:surfNS}\eqe
(which is a vector-valued equation that has both in-plane and out-of-plane components; also for the body force $\bff$), the area-incompressibility constraint \eqref{e:incomp}, suitable mesh equations, and $\bv_\mrm=\bx'$.
The last equation implies that the problem is time-dependent, even when steady flow conditions are considered ($\rho\,\bv'\approx\mathbf{0}$).
It will be approximated by a finite difference-based time integration scheme, effectively eliminating one unknown field, e.g.~$\bx$.
The remaining three unknowns fields are solved by applying the finite element method to the remaining PDEs.
For fixed or prescribed surface motion, i.e.~known $\bx(t)$, only two unknown fields remain: $\bv$ and $q$.
This is considered in the examples of Sec.~\ref{s:Nexf}.
Sec.~\ref{s:Nex} then consider unknown $\bv$, $q$ and $\bv_\mrm$.

Two mesh equations are needed:
Firstly, the constraint
\eqb{l}
\bn\cdot\bv_\mrm = \bn\cdot\bv\,,
\eqe
which ascertains that the out-of-plane velocity components $v_\mrn :=\bv\cdot\bn$ and $v_{\mrm\mrn} :=\bv_\mrm\cdot\bn$ agree. 
The equation could be used to eliminate one of these components, but this is not done here. 
Instead the constraint will be enforced in weak form.
Secondly, an equation for the in-plane components of the mesh velocity is needed.
Two options will be used for that here: 
(I) Setting them to zero, which implies that the in-plane description is Eulerian (i.e.~$\zeta^\alpha = \theta^\alpha$), or (II) determining the in-plane mesh motion from
\eqb{l}
\sig^{\alpha\beta}_{\mrm\,;\beta} = 0\,,
\label{e:membrane}\eqe
which is the in-plane equilibrium equation of a membrane.
Here, the simple nonlinear elasticity model
\eqb{l}
\tau^{\alpha\beta}_\mrm = \tmu\,\big(A^{\alpha\beta}-a^{\alpha\beta}\big)
\label{e:taum}\eqe
is used for the membrane stress $\sig^{\alpha\beta}_\mrm = \tau^{\alpha\beta}_\mrm/J_\mrm$ in \eqref{e:membrane}.
Since this membrane stress is not entering the surface Navier-Stokes equations, it does not affect the physical behavior of the system.
It only effects the in-plane mesh motion, by effectively stabilizing it through the shear stiffness $\tmu$, which is a purely numerical parameter.
This stabilization was developed in \citet{droplet} to simulate quasi-static droplets.
\\
Both mesh options will also be enforced in weak form.

The weak form equations for $\bv$, $q$ and $\bv_\mrm$ are given by the statements
\eqb{llllll}
G \dis G_\mathrm{in} + G_\mathrm{int} - G_\mathrm{ext} \is 0 ~~ &  \forall~\delta\bx\in\sV\,, \\[1mm]
\bar G \dis \bar G_\mathrm{div} \is 0 & \forall~\delta q\in\sQ\,, \\[1mm]
\tilde G \dis \tilde G_\mri + \tilde G_\mro \is 0 & \forall~\bw\in\sW\,.
\label{e:G}\eqe
derived in \citet{ALEtheo}.
They can also be combined in the single statement
\eqb{l}
G + \bar G + \tilde G = 0\,,\quad \forall~\delta\bx\in\sV,~\delta q\in\sQ,~\bw\in\sW.
\label{e:Gc}\eqe
Here and in \eqref{e:G}, $\delta\bx$, $\delta q$ and $\bw$ denote the test functions associated with $\bv$, $q$ and $\bv_\mrm$.
The weak form is complemented by boundary and initial conditions, as well as the Dohrmann-Bochev stabilization scheme, discussed in Sec.~\ref{s:stab}. 
This section also contains a summary of all the individual contributions of Eq.~\eqref{e:G}. 

\subsection{Boundary and initial conditions in ALE form}

The governing equations of Sec.~\ref{s:gov} require boundary and initial conditions in order to be uniquely solvable.
For the fluid flow the usual velocity and traction boundary conditions,
\eqb{rlll}
\bv(\zeta^\alpha,t) \is \bar\bv(\zeta^\alpha,t) & \forall\,\bx \in \partial_v\sS~~$and$~~\forall\,t\in\sT\,, \\[1mm]
\bT(\zeta^\alpha,t) \is \bar\bT(\zeta^\alpha,t) & \forall\,\bx \in \partial_T\sS~~$and$~~\forall\,t\in\sT\,,
\label{e:BC1}\eqe
apply.
Here $\partial_v\sS$ and $\partial_T\sS$ denote the boundaries of the fluid surface
where velocities or tractions are prescribed, while $\sT$ denotes the considered time interval.
$\partial_v\sS$ contains inflow and outflow boundaries.
These do \underline{not} require to observe mass conservation, as long as the surface area can change. 
An example is provided in Sec.~\ref{s:SB}.
This example also shows that traction $\bar\bT$ can be surface deformation-dependent (a so-called follower load), which requires linearization in implicit time-stepping schemes.
There may also be no boundary ($\partial_v\sS=\partial_T\sS=\emptyset$), as in the examples of Secs.~\ref{s:Nexf}-\ref{s:octodefo}.
The traction BC (\ref{e:BC1}.2) provides a condition for surface tension $q$.
In the absence of traction BCs ($\partial_T\sS=\emptyset$), $q$ has to be specified at some location, to ascertain its uniqueness.
If mesh velocity $\bv_\mrm$ is an unknown (instead of a prescribed) field, it also requires a boundary condition (if a boundary is present). 
In the example of Sec.~\ref{s:SB} the zero Dirichlet BC,
\eqb{rlll}
\bv_\mrm(\zeta^\alpha,t) \is \mathbf{0} & \forall\,\bx \in \partial\sS~~$and$~~\forall\,t\in\sT\,,
\label{e:BC2}\eqe
is used.
In the subsequent FE formulation, BCs (\ref{e:BC1}.1) and \eqref{e:BC2} are enforced by direct elimination, while BC (\ref{e:BC1}.2) enters the external virtual work $G_\mathrm{ext}$, see Box~\ref{t:WF} and Sec.~\ref{s:f}.
For fixed surfaces, the additional velocity boundary condition 
\eqb{rlll}
v_\mrn(\zeta^\alpha,t) \is 0 & \forall\,\bx \in \sS~~$and$~~\forall\,t\in\sT\,,
\label{e:BC3}\eqe
appears. 
It is a Dirichlet BC that can be replaced by the equivalent Neumann BC on the surface pressure $p$, see Sec.~\ref{s:NBC1} for an example.

The initial conditions at $t = 0$ are
\eqb{rlll}
\bv(\zeta^\alpha,0) \is \bv_0(\zeta^\alpha) & \forall\,\bx\in\sS\,, \\[1mm]
q(\zeta^\alpha,0) \is q_0(\zeta^\alpha) & \forall\,\bx\in\sS\,, \\[1mm]
\bv_\mrm(\zeta^\alpha,0) \is \bv_\mrm^0(\zeta^\alpha) & \forall\,\bx\in\sS\,, \\[1mm]
\bx(\zeta^\alpha,0) \is \bX(\zeta^\alpha) & \forall\,\bx\in\sS\,.
\eqe
The chosen initial fluid velocity, surface tension, mesh velocity and mesh position fields $\bv_0$, $q_0$, $\bv_\mrm^0$ and $\bX$ have to be compatible, i.e.~agree with Eqs.~\eqref{e:incomp}-\eqref{e:membrane}.

\section{Dohrmann-Bochev stabilization and resulting weak form}\label{s:stab}

Incompressible flows -- such as the one described here -- are mixed problems.
Their discretization needs to satisfy the Ladyzhenskaya-Babu\v ska-Brezzi (LBB) condition \citep{ladyzhenskaya69,babuska73,brezzi74}, or use a stabilization technique.
Popular approaches are LBB-conforming velocity-pressure -- e.g.~Taylor-Hood -- discretizations, div-conforming discretizations, or stabilization methods that circumvent the LBB condition such as the pressure stabilizing Petrov Galerkin (PSPG) approach \citep{hughes86} and the polynomial pressure projection stabilization (PPPS) approach by \citet{dohrmann04}.
Most existing surface Navier-Stokes flows use Taylor-Hood \citep{fries18, krause23} or Taylor-Hood-like discretizations \citep{torres19}. 
The Dohrmann-Bochev (DB) scheme has only recently been adapted to surface flows \citep{ALE}.
The scheme is also based on velocity-pressure discretizations, but instead of picking the pressure from a lower polynomial space in order to satisfy the LBB condition directly, it can be picked from the same space, and then projected to a lower dimensional space that satisfies the LBB condition.
For instance one can use bi-quadratic velocity and pressure approximations, and then project the latter onto a linear subspace, which is the case used here.
The projection is achieved by square error minimization.
This projection is simple and can be treated purely on the element level, which makes the method comparably easy to implement.
Local, inexpensive static condensation can then be used to eliminate the auxiliary pressure/surface tension variable.

The DB scheme subtracts the stabilization term
\eqb{l}
G_\mathrm{DB} = \ds\frac{\alpha_\mathrm{DB}}{\eta}\int_{\sS}(\delta q - \delta\check q)(q-\check q)\,\dif a
\label{e:GDB}\eqe
from weak form \eqref{e:Gc}.
Here, $q(\zeta^\alpha,t)$ is the actual surface tension, while $\check q(\zeta^\alpha,t)$ is an auxiliary surface tension that is of lower interpolation order than $q$.
Further,  $\eta$ is the viscosity and $\alpha_\mathrm{DB}$ is a chosen stabilization parameter.
Eq.~\eqref{e:GDB} can be derived from the variation of the potential
\eqb{l}
\Pi_\mathrm{DB} = \ds\frac{\alpha_\mathrm{DB}}{2\eta}\int_{\sS}(q-\check q)^2\,\dif a\,.
\eqe
In the continuum limit of the discretization, $\check q$ becomes equal to $q$, and hence nothing is changed by including \eqref{e:GDB} in the weak form.
Contrary to \citet{ALE}, we choose to integrate the stabilization term over the current surface $\sS$ instead of the initial, reference surface $\sS_0$, as this leads to more accurate results for large surface stretches. 
When surface stretches are small, \eqref{e:GDB} can be integrated over $\sS_0$ without significant accuracy loss, as is seen in Sec.~\ref{s:SB}.

%-------------------------------------------------------------------------------------------------------------------------------
\begin{Box}[h]
\begin{center}
\begin{tabular}{|l|}
\hline
\\[-3mm]
Coupled weak form equations: \\[2mm]
\begin{tabular}{lllll}
$G$ & $\back\! :=\, G_\mathrm{in} + G_\mathrm{int} - G_\mathrm{ext}$ & $\back = 0~~$ &  $\forall~\delta\bx\in\sV\,,~~$ 
	& (weak form for fluid velocity $\bv$) \\[1.5mm]
$\bar G$ & $\back\! :=\, \bar G_\mathrm{div} - \bar G_\mathrm{DB}$ & $\back = 0$ & $\forall~\delta q\in\sQ\,,$ 
	& (weak form for surface tension $q$) \\[1.5mm]
$\check G$ & $\back\! :=\, \check G_\mathrm{DB}$ & $\back = 0$ & $\forall~\delta\check q\in\check\sQ\,,$ 
	& (weak form for stabilization field $\check q$) 
 \\[1mm]
$\tilde G$ & $\back\! := \Bigg\{\back\!$ \begin{tabular}{ll} $\tilde G_0\back$ & (option I) \\[1mm] 
	$\tilde G_\mathrm{el}\back$ & (option II) \end{tabular}  $\back\!\Bigg\}\!\!$ 
	& $\back = 0$ & $\forall~\bw\in\sW\,,$ 
	& (weak form for mesh velocity $\bv_\mrm$) \\[3mm]
\end{tabular} \\
with: \\
\begin{tabular}{llll}
$G_\mathrm{in}$ & $\back\!\! := \ds\int_{\sS} \delta\bx\cdot\rho\,\dot\bv\,\dif a = G_\mathrm{trans} + G_\mathrm{conv}\,,$ 
	& $G_\mathrm{trans}$ & $\back\!\! :=\ds\int_{\sS} \delta\bx\cdot\rho\,\bv'\,\dif a\,,$ \\[3.5mm]
$G_\mathrm{conv}$ & $\back\!\! := \ds\int_{\sS} \delta\bx\cdot\rho\,\bv_{,\alpha}\,\big(v^\alpha-v^\alpha_\mrm\big)\,\dif a\,,$
	& $G_\mathrm{int}$ & $\back\!\! := \ds\int_{\sS} \delta\bx_{,\alpha}\cdot\sig^{\alpha\beta}\,\ba_\beta \, \dif a \,,$ \\[3.5mm]
$G_\mathrm{ext}$ & $\back\!\! := \ds\int_{\sS}\delta\bx\cdot\bff\,\dif a + \int_{\partial_T\sS}\delta\bx\cdot\bar\bT\,\dif s\,,$ 
	& $\bar G_\mathrm{div}$ & $\back\!\! := \ds\int_{\sS}\delta q\,\divs\bv\,\dif a\,,$ \\[3.5mm]
$\tilde G_0$ & $\back\!\! := \alpha_\mrm\ds\int_{\sS_0}\bw\cdot\big(\bv_\mrm-(\bn\otimes\bn)\,\bv\big)\,\dif A\,,$
	& $\bar G_\mathrm{DB}$ & $\back\!\! := \ds\frac{\alpha_\mathrm{DB}}{\eta}\int_{\sS}\delta q\,(q-\check q)\,\dif a\,,\back$ \\[3.5mm]		
$\tilde G_\mathrm{el}$ & $\back\!\! := \ds\int_{\sS_0}w_{\alpha;\beta}\,\tau_\mrm^{\alpha\beta}\,\dif A 
	+ \alpha_\mrm \int_{\sS_0}w\,\bn\cdot\big(\bv_\mrm-\bv\big)\,\dif A\,,\,~$ 	
	& $\check G_\mathrm{DB}$ & $\back\!\! := \ds\frac{\alpha_\mathrm{DB}}{\eta}\int_{\sS}\delta\check q\,(q-\check q)\,\dif a\,.\back$ \\[3.5mm]
\end{tabular}
\\
\hline
\end{tabular}
\vspace{-5mm}
\end{center}
\caption{Coupled weak form equations for stabilized area-incompressible surface flows described in the ALE frame. 
Their finite element discretization is discussed in Sec.~\ref{s:FE}.
For this, the auxiliary variable $\check q$ of the Dohrmann-Bochev stabilization scheme can be eliminated locally, such that only three discretized weak form equations remain.
Two options are used for the in-plane components of $\bv_\mrm$: 
Setting them to zero (I), or determining them from the in-plane equations of an elastic membrane (II).}
\label{t:WF}
\end{Box}
%-------------------------------------------------------------------------------------------------------------------------------
Box~\ref{t:WF} summarizes all the weak form equations resulting from \eqref{e:G} and \eqref{e:GDB}. 
The Euler-Lagrange equations of this weak form are the equation of motion \eqref{e:surfNS}, the mesh equations ($\bv_\mrm\cdot\bn=\bv\cdot\bn$ and either \eqref{e:membrane} or $\bv_\mrm\cdot\ba_\alpha=0$)
and
\eqb{rll}
\eta\,\mathrm{div}_\mrs\,\bv - \alpha_\mathrm{DB}\big(q - \check q\big) \is 0\,, \\[1mm]
q - \check q \is 0\,,
\eqe
implying $\check q = q$ and $\mathrm{div}_\mrs\,\bv=0$.
Thus nothing changes in the continuum limit.
In the discrete setting however, different approximation spaces are chosen for $q$ and $\check q$, and so extra terms appear.
As noted above, here $q$ is chosen from the same bi-quadratic approximation space as $\bv$, while $\check q$ is picked from a linear 
subspace such that it can be eliminated efficiently on the element level.
This is presented in the following section.

\begin{remark}\label{r:WFunits}
Written separately, the weak forms $G=0$, $\bar G=0$, $\check G=0$ and $\tilde G=0$ 
do not need to have agreeing units.
Variations $\delta\bx$ and $\bw$ can thus be interpreted as displacements, and $\delta q$ and $\delta\check q$ as surface tensions, giving units
of energy for $G$ and $\tilde G$, and power for $\bar G$ and $\check G$.
But written as the sum $G + \bar G + \check G + \tilde G = 0$,
the unit of each term should agree. 
This can either be achieved by including factors in front of each term, such as $\alpha_\mathrm{DB}$ and $\alpha_\mrm$, or by reinterpreting $\delta\bx$ and $\bw$ as velocities.
\end{remark}

\begin{remark}\label{r:SUPG}
Apart from pressure stabilization, there is also convective stabilization, such as streamline upwind Petrov-Galerkin (SUPG) stabilization \citep{brooks82}. 
It is required when the convective velocity $\bu = \bv-\bv_\mrm$ is large in comparison to the computational grid.
The need for convective stabilization can be assessed from the local, mesh-specific P\'eclet number
\eqb{l}
\mathrm{Pe} := \ds\frac{h\norm{\bu}}{2\nu} = \frac{\rho h \norm{\bu}}{2\eta}\,,
\label{e:Pe}\eqe
where $h$ is the grid spacing in flow direction (half the element length for quadratic elements), $\rho$ is the surface density, and $\nu$ and $\eta$ the kinematic and dynamic (surface) viscosities. 
Instabilities can arise when Pe exceeds unity \citep{donea}.
In the subsequent examples of Sec.~\ref{s:Nexf} and \ref{s:Nex} meshes are resolved to ensure Pe $<1$.
\end{remark}

\section{Finite element discretization}\label{s:FE}

This section discussed the finite element discretization and solution of the preceding weak form.
First the spatial discretization of the different fields and their integrals is discussed in Secs.~\ref{s:FEI}-\ref{s:DWF}, 
followed by a suitable implicit time stepping scheme in Sec.~\ref{s:TI}, the resulting Newton-Raphson procedure in Sec.~\ref{s:NR}, and its normalization in Sec.~\ref{s:Norm}.

\subsection{FE interpolation}\label{s:FEI}

\subsubsection{Surface deformation and mesh motion}

The surface is discretized into $n_\mathrm{el}$ finite elements using $n_\mathrm{no}$ nodes.
Within each finite element $\Omega^e$, the surface description \eqref{e:bx} is approximated by the interpolation
\eqb{l}
\bx \approx \bx^h(\zeta^\alpha,t) = \ds\sum_{I=1}^{n_e}N_I(\zeta^\alpha)\,\bx_I(t)\,,
\label{e:bxh}\eqe
where $N_I$ denotes the shape function, $\bx_I$ denotes the position of node $I$, and $n_e$ denotes the number of nodes of element $\Omega^e$.
Eq.~\eqref{e:bxh} can be rewritten in the compact form
\eqb{l}
\bx^h = \mN_e\,\mx_e\,,
\label{e:bxha}\eqe
with $\mN_e := [N_1\,\bone,\,N_2\,\bone,\,...,\,N_{n_e}\,\bone]$ and $\mx_e := [\bx_1^\mrT,\,\bx_2^\mrT,\,...,\,\bx^\mrT_{n_e}]^\mrT$.
Here, $\bone$ denotes the identity tensor in $\bbR^3$, while
\eqb{l}
\sC_e := \{e_1,\,e_2,\,...,\,e_{n_e}\}
\label{e:con}\eqe
is the set of global node numbers of element $e$ and thus specifies the element connectivity.
Evaluating $\bx^h$ at $t=0$ lets us define a fixed reference configuration for the surface, written as
\eqb{l}
\bX^h = \mN_e\,\mX_e\,,\quad \mX_e = \mx_e(0)\,.
\eqe 
Note that, in general (as long as $\zeta^\alpha\neq\xi^\alpha$) $\mx_I(t)$ is not the current position of the fluid particle that was initially at $\mx_I(0)$.
Instead the initial position of the particle follows from the integration of $\bv$ along the path defined by fixed $\xi^\alpha$. 

According to \eqref{e:ba}, the discretized tangential basis vectors then follow from the interpolation
\eqb{l}
\ba_\alpha^h = \mN_{e,\alpha}\,\mx_e\,,
\label{e:bah}\eqe
where $\mN_{e,\alpha} := [N_{1,\alpha}\,\bone,\,N_{2,\alpha}\,\bone,\,...,\,N_{n_e,\alpha}\,\bone]$ and $N_{I,\alpha} = \partial N_I/\partial\zeta^\alpha$.
This then allows to evaluate all the surface quantities defined in Eqs.~\eqref{e:aab}-\eqref{e:Gamma} for the current surface. 
Further, evaluating \eqref{e:bah} at $t=0$ defines the initial basis
\eqb{l}
\bA_\alpha^h = \mN_{e,\alpha}\,\mX_e\,,
\eqe
which allows to evaluate the surface quantities defined in Eqs.~\eqref{e:aab}-\eqref{e:Gamma} for the initial surface. 

According to definition \eqref{e:vm}, the mesh velocity is then approximated by the interpolation
\eqb{l}
\bv_\mrm^h = \ds\sum_{I=1}^{n_e}N_I(\zeta^\alpha)\,\bx'_I(t)\,,
\label{e:bvmh}\eqe
where $\bx_I':=\partial\bx_I/\partial t =: \bv_{\mrm I}$ is the nodal mesh velocity.
In compact notation this becomes
\eqb{l}
\bv_\mrm^h = \mN_e\,\mx'_e\,,
\label{e:bvmha}\eqe
with $\mx_e':=\partial\mx_e/\partial t =: \mv_\mrm^e$.

\begin{remark}
The discretized time derivative of $\ba_\alpha$ in the ALE frame follows from \eqref{e:bah} as
\eqb{l}
\ba'^{\,h}_\alpha = \mN_{e,\alpha}\,\mx'_e\,.
\eqe
This is the same as $\bv^h_{\mrm,\alpha}$, as Eq.~\eqref{e:bvmha} indicates.
However, $\ba'_\alpha$ is generally not the same as the material time derivative of $\ba_\alpha$, which according to \eqref{e:ALE} is 
\eqb{l}
\dot\ba_\alpha = \ba'_\alpha + \dot\zeta^\beta\,\ba_{\alpha,\beta}\,.
\label{e:bap}\eqe
In the discrete setting this becomes
\eqb{l}
\dot\ba^h_\alpha = \mN_{e,\alpha}\,\mx'_e + \dot\zeta^{\beta\,h}\,\mN_{e,\alpha\beta}\,\mx_e\,,
\eqe
where $\dot\zeta^{\beta\,h}$ follows from Remark \ref{r:zetadot}.
\end{remark}

\subsubsection{Fluid velocity and its derivatives}

Analogous to interpolation \eqref{e:bxh}, the fluid velocity field is interpolated by
\eqb{l}
\bv^h = \ds\sum_{I=1}^{n_e}N_I(\zeta^\alpha)\,\bv_I(t)\,,
\label{e:bvh}\eqe
or in compact notation,
\eqb{l}
\bv^h = \mN_e\,\mv_e\,,
\label{e:bvha}\eqe
with $\mv_e := [\bv_1^\mrT,\,\bv_2^\mrT,\,...,\,\bv^\mrT_{n_e}]^\mrT$.
Note, that in general, $\bv_I \neq \partial\bx_I/\partial t = \bv_{\mrm\,I}$ (as long as $\zeta^\alpha\neq\xi^\alpha$).
Instead $\mv_I$ is an independent variable -- independent from $\mx_I$.
It characterizes the material fluid velocity, while $\mx_I$ characterizes the mesh motion.
Their relative velocity according to \eqref{e:bu} now follows from expressions \eqref{e:bvmha} and \eqref{e:bvha} as
\eqb{l}
\bu^h = \mN_e\,(\mv_e-\mx'_e)\,.
\label{e:buh}\eqe

\begin{remark}\label{r:zetadot}
Given \eqref{e:buh} one can obtain a discretized expression for its in-plane component $u^\alpha = \dot\zeta^\alpha$ according to \eqref{e:zetadot}, i.e.
\eqb{l}
\dot\zeta^{\alpha\,h} = \ba^{\alpha\,h}\cdot\mN_e\,(\mv_e-\mx'_e)\,,
\label{e:zetadoth}\eqe
where $\ba^{\alpha\,h}$ is the dual basis to
$\ba_\alpha^h$.
\end{remark}

From \eqref{e:bvp} and \eqref{e:bvha} follows 
\eqb{l}
\bv'^{\,h} = \mN_e\,\mv'_e\,,
\label{e:bvpha}\eqe
where $\mv_e':=\partial\mv_e/\partial t$.
From \eqref{e:bvha} further follows
\eqb{l}
\bv_{\!,\alpha}^h = \mN_{e,\alpha}\,\mv_e\,.
\label{e:badha}\eqe
It is reiterated that in the ALE frame, spatial differentiation (w.r.t.~$\zeta^\alpha$) does not commute with the material time derivative,
so that $\bv_{\!,\alpha}^h \neq \dot\ba^h_\alpha$. 
Instead \eqref{e:badot} applies.

\begin{remark}
If needed, $\dot\zeta^\gamma_{,\alpha}$ can now be obtained as follows.
From \eqref{e:badot} first follows $\dot\zeta^\gamma_{,\alpha} = \ba^\gamma\cdot(\bv_{,\alpha}-\dot\ba_\alpha)$.
With \eqref{e:bap} and \eqref{e:Gamma} this then yields
\eqb{l}
\dot\zeta^\gamma_{,\alpha} = \ba^\gamma\cdot(\bv_{\!,\alpha}-\ba'_\alpha)-\dot\zeta^\beta\,\Gamma^\gamma_{\alpha\beta}\,.
\eqe
In the discrete setting this becomes
\eqb{l}
\dot\zeta^{\gamma\,h}_{,\alpha} 
= \big[\ba^{\gamma\,h}\cdot\mN_{e,\alpha} - \Gamma^{\gamma\,h}_{\alpha\beta}\,\ba^{\beta\,h}\cdot\mN_e\big](\mv_e-\mx'_e)\,, 
\eqe
due to \eqref{e:zetadoth}.
\end{remark}

The surface velocity gradient and divergence can be evaluated from \eqref{e:sgrad}, \eqref{e:sdiv}, \eqref{e:bah}, \eqref{e:badual} and \eqref{e:badha}.
The discretized surface divergence, in particular, can conveniently be written as 
\eqb{l}
\divs^{\!\!\!\!h}\,\bv^h = \mD_e\,\mv_e\,,
\label{e:sdivh}\eqe
where the surface divergence operator
\eqb{l}
\mD_e := \ba^{\alpha\,h}\cdot\mN_{e,\alpha}
\label{e:De}\eqe
depends on the surface and needs to be linearized if the surface is changing.\\
Further, the discretization of Eq.~\eqref{e:2dabab} is
\eqb{l}
2d^{\alpha\beta}_h\ba_\beta^h = \ba_\mrs^{\alpha\beta}\,\mN_{e,\beta}\,\mv_e\,.
\label{e:2dababa}\eqe
It is needed for the internal FE force vector in Sec.~\ref{s:DWF}.

\subsubsection{Surface tension and its stabilization}

The Lagrange multiplier $q$ is interpolated similarly to $\bx$ and $\bv$, i.e.
\eqb{l}
q^h = \ds\sum_{I=1}^{m_e}L_I(\zeta^\alpha)\,q_I(t)\,,
\label{e:qh}\eqe
where $L_I$, $I=1,\,...,\,m_e$ denote the nodal shape functions for $q$ that can be generally different from $N_I$.
Also, $m_e$, the number of corresponding nodes can be generally different from $n_e$.
In compact notation, \eqref{e:qh} becomes
\eqb{l}
q^h = \mL_e\,\mq_e\,,
\label{e:qha}\eqe
where $\mL_e := [L_1,\,L_2,\,...,\,L_{m_e}]$ and $\mq_e := [q_1,\,q_2,\,...,\,q_{m_e}]^\mrT$.
A common approach is to pick $L_I$ and $m_e$ such that the LBB condition is satisfied. 
An alternative is to pick $L_I = N_I$ and $m_e=n_e$ and use a stabilization method, such as the Dohrmann-Bochev method introduced in Sec.~\ref{s:stab}.
This latter approach is used with bi-quadratic Lagrange shape functions in all subsequent numerical examples.
It requires the interpolation of the auxiliary field $\check q$, written as
\eqb{l}
\check q^h = \ds\sum_{I=1}^{\check m_e}\check L_I(\zeta^\alpha)\,\check q_I(t)
= \mathbf{\check L}_e\,\mathbf{\check q}_e\,.
\label{e:qch}\eqe
A possible choice is to use $\check m_e = 3$ with 
\eqb{l}
\mathbf{\check L}_e = [1,\,\zeta^1,\,\zeta^2]\,.
\label{e:Lcheck}\eqe
It provides effective stabilization and allows to eliminate unknown $\mathbf{\check q}_e$ locally for each element, as is shown in Sec.~\ref{s:FEDB} below.
It is noted that for choice \eqref{e:Lcheck}, $\check q_I$ are not nodal values (that can be associated with nodes), but rather correspond to the coefficients of the first order polynomial induced by $\mathbf{\check L}_e$.

The variations $\delta\bx$, $\bw$, $\delta q$ and $\delta\check q$ are interpolated in the same way as $\bx$, $\bv_\mrm$,  $q$ and $\check q$, i.e.
\eqb{lll}
\delta\bx^h \is \mN_e\,\delta\mx_e\,, \\[1mm]
\bw^h \is \mN_e\, \mw_e\,, \\[1mm]
\delta q^h \is \mL_e\,\delta\mq_e\,, \\[1mm]
\delta\check q^h \is \mathbf{\check L}_e\,\delta\check\mq_e\,.
\label{e:dbxwq}\eqe
From this follows $\delta\ba_\alpha^h = \mN_{e,\alpha}\,\delta\mx_e$.

All the above equations show that the entire set of field variables is described exclusively by the parameters $\zeta^\alpha$ and $t$.
Parameters $\xi^\alpha$ and $\theta^\alpha$, introduced earlier, are not needed for the FE description.

To shorten notation, superscript $h$ is skipped in the subsequent presentation.

\subsection{Discretized weak form}\label{s:DWF}

Using the expressions from Sec.~\ref{s:FEI} the four weak form equations in Box~\ref{t:WF} are now discretized.

\subsubsection{Equation of motion}\label{s:f}

The first weak form in Box~\ref{t:WF} contains the three parts $G_\mathrm{in}$, $G_\mathrm{int}$ and $G_\mathrm{ext}$.
Discretizing them leads to the elemental contributions 
$G_\mathrm{in}^e = \delta\mx_e^\mrT\,\mf^e_\mathrm{in}$, 
$G_\mathrm{int}^e = \delta\mx_e^\mrT\,\mf^e_\mathrm{int}$ and 
$G_\mathrm{ext}^e = \delta\mx_e^\mrT\,\mf^e_\mathrm{ext}$,
where the vectors $\mf^e_\mathrm{in}$, $\mf^e_\mathrm{int}$ and $\mf^e_\mathrm{ext}$ contain the corresponding nodal forces of finite element $\Omega^e$.

Discretizing $G_\mathrm{in}$ leads to the elemental inertia vector $\mf^e_\mathrm{in} = \mf^e_\mathrm{trans} + \mf^e_\mathrm{conv}$ with
the transient part \eqb{l}
\mf^e_\mathrm{trans} := \mm_e\,\mv_e'\,,\quad
\mm_e := \ds\int_{\Omega^e}\rho\,\mN_e^\mrT\,\mN_e\,\dif a\,,
\label{e:mf_trans}\eqe
and convective part
\eqb{l}
\mf^e_\mathrm{conv} := \ds\int_{\Omega^e}\rho\,\mN_e^\mrT\,\bv_{\!,\alpha}\,u^\alpha\,\dif a\,.
\label{e:mf_conv}\eqe
Here $u^\alpha = \dot\zeta^\alpha$ and $\bv_{\!,\alpha}$ follow from \eqref{e:zetadoth} and \eqref{e:badha}.
The transient part is linear in $\mv'$.
The mass matrix $\mm_e$ is only constant as long as the mesh is fixed or described in a Lagrangian frame.
If the ALE frame is used, the mass can change, and hence $\mf^e_\mathrm{trans}$ needs to be linearized. 
Its linearization, along with that of $\mf^e_\mathrm{conv}$, is given in Appendix~\ref{s:Lfin}.

The discretization of $G_\mathrm{int}$ leads to the internal FE force vector \citep{membrane}
\eqb{l}
\mf^e_\mathrm{int} :=  \ds\int_{\Omega^e} \mN_{e,\alpha}^\mrT\,\sig^{\alpha\beta} \vaub\,\dif a\,.
\label{e:fint}
\eqe
In view of \eqref{e:sigab}, \eqref{e:2dabab} and \eqref{e:De}, this can be written as
\eqb{l}
\mf^e_\mathrm{int} = \mf^e_{\mathrm{int}q} + \mf^e_{\mathrm{int}v}
\label{e:felint}\eqe
with
\eqb{lll}
\mf^e_{\mathrm{int}q} \dis  \ds\int_{\Omega^e} \mD_e^\mrT\,q\,\dif a\,, \\[4mm]
\mf^e_{\mathrm{int}v} \dis \ds\int_{\Omega^e} \eta\,\mN_{e,\alpha}^\mrT\,\ba^{\alpha\beta}_\mrs \bv_{\!,\beta}\,\dif a\,.
\label{e:felvisc}
\eqe
Their linearization is given in Appendix \ref{s:Lfint}.

The discretization of $G_\mathrm{ext}$ leads to the two contributions
\eqb{lll}
\mf^e_f \dis \ds\int_{\Omega^e}\mN_e^\mrT\,\bff\,\dif a\,, \\[4mm]
\mf^e_T \dis \ds\int_{\partial_T\Omega^e}\mN_e^\mrT\,\bar\bT\,\dif s\,,
\label{e:fext}\eqe
due to external surface loads $\bff$ within the domain and boundary tractions $\bar\bT$ on Neumann boundaries.
The numerical examples in Secs.~\ref{s:Nexf}-\ref{s:Nex} either prescribe a given manufactured body force $\bff$, or prescribe pressure loading in the form $\bff = p\,\bn$.
The linearization of the latter case is given in Appendix \ref{s:Lfext}.
The examples in Sec.~\ref{s:Nexf} all have closed surfaces, such that there are no traction boundaries ($\partial_T\Omega^e = \emptyset$),
but the examples in Sec.~\ref{s:Nex} contain a Neumann boundary, where deformation dependent tractions are prescribed as described in Appendix \ref{s:Lft}. 

\begin{remark}\label{r:p_visc}
$\mf^e_{\mathrm{int}v}$ captures the in-plane resistance of the flow due to in-plane viscosity parameter $\eta$.
It can be advantageous to stabilize the surface motion by the out-of-plane resistance
\eqb{l}
p_\mathrm{visc} := -\eta_\mrn\,\bn\cdot\bv\,,
\label{e:pvisc}\eqe
where $\eta_\mrn\geq0$ physically corresponds to an out-of-plane viscosity parameter. 
Without \eqref{e:pvisc}, the out-of-plane surface motion is undamped and can become unbounded, which has been observed in the steady flow example of Sec.~\ref{s:sheardefo2} and the transient inflation example of Fig.~\ref{f:SoapBublTurb3}.
The pressure $p_\mathrm{visc}$ can be simply added to $\bff = p\,\bn$ within Eq.~\eqref{e:fext}.
Its linearization is given in \citet{dropslide}, see also Appendix \ref{s:Lfext}.
\end{remark}

\subsubsection{Area-incompressibility constraint}\label{s:FEDB}

The discretization of the second weak form in Box~\ref{t:WF} leads to the elemental pseudo force vector\footnote{Its unit is [m$^2$/s].}
\eqb{l}
\mbf^e = \mbf^e_\mathrm{div} - \mbf^e_\mathrm{DB}\,,
\label{e:mbf1}\eqe
with
\eqb{lll}
\mbf^e_\mathrm{div} \dis \ds\int_{\Omega^e}\mL_e^\mrT\,\divs\bv\,\dif a\,, \\[4mm]
\mbf^e_\mathrm{DB} \dis \ds\frac{\alpha_\mathrm{DB}}{\eta}\int_{\Omega^e}\mL_e^\mrT\,(q-\check q)\,\dif a\,,
\label{e:mbf2}\eqe
where $\divs\bv$, $q$ and $\check q$ are discretized by Eqs.~\eqref{e:sdivh}, \eqref{e:qha} and \eqref{e:qch}.
Further, the discretization of the third weak form in Box~\ref{t:WF} leads to
\eqb{l}
\mathbf{\check f}^e := \ds\frac{\alpha_\mathrm{DB}}{\eta}\int_{\Omega^e}\mathbf{\check L}_e(q-\check q)\,\dif a \,.
\eqe
Since $\mathbf{\check L}_e$ is defined locally, the third weak form leads to $\mathbf{\check f}^e = \mathbf{0}$, which can be used to eliminate $\mathbf{\check q}_e$ locally:
Inserting \eqref{e:qha} and \eqref{e:qch} into $\mathbf{\check f}^e = \mathbf{0}$ and solving for $\mathbf{\check q}_e$ then gives
\eqb{l}
\mathbf{\check q}_e = \mh_e^{-1}\,\mg_e\,\mq_e\,,
\eqe
with
\eqb{lll}
\mg_e \dis \ds\int_{\Omega^e}\mathbf{\check L}_e^\mrT\,\mL_e\,\dif a\,, \\[4mm]
\mh_e \dis \ds\int_{\Omega^e}\mathbf{\check L}_e^\mrT\,\mathbf{\check L}_e\,\dif a\,.
\eqe
Inserting this together with \eqref{e:qha} and \eqref{e:qch} into (\ref{e:mbf2}.2) then gives
\eqb{l}
\mbf^e_\mathrm{DB} = \mathbf{\bar d}_e\,\mq_e \,,
\label{e:mbf3}\eqe
with
\eqb{lll}
\mathbf{\bar d}_e \dis \ds\frac{\alpha_\mathrm{DB}}{\eta}\big[\mathbf{\bar m}_e -\mg^\mrT_e\,\mh^{-1}_e\mg_e \big]\,, \\[4mm]
\mathbf{\bar m}_e \dis \ds\int_{\Omega^e}\mL_e^\mrT\,\mL_\mre\,\dif a\,.
\label{e:mbf4}\eqe
The linearization of Eq.~\eqref{e:mbf1} based on this elimination is provided in Appendix \ref{s:Lfq}.
In the examples of Sec.~\ref{s:Nexf}-\ref{s:octodefo} the integration of $\mathbf{\bar m}_e$, $\mg_e$ and $\mh_e$ is chosen to be taken over $\dif A$, instead of $\dif a = J_\mrm\,\dif A$, in order to simplify the linearization.
This is justified, as $J_\mrm$ is very close to one in those examples.
In some of the examples of Sec.~\ref{s:bubble}, on the other hand, surface inflow leads to large $J_\mrm$, and integrating over $\dif a$ or $\dif A$ makes a difference as is shown.

\subsubsection{Mesh equations}\label{s:FEmesh}

The discretization of the fourth weak form in Box~\ref{t:WF} leads to the following elemental pseudo force vector.\footnote{Its unit is [m$^3$/s].}
For Eulerian mesh motion (option I), this vector becomes
\eqb{l}
\mtf^e := \mhm_e\,\mx'_e - \mhn_e\,\mv_e\,,
\label{e:mesh1}\eqe
with the constant matrix
\eqb{l}
\mtm_e := \alpha_\mrm\ds\int_{\Omega^e_0}\mN_e^\mrT\,\mN_e\,\dif A\,,
\eqe
and the surface-dependent matrix
\eqb{l}
\mtn_e := \alpha_\mrm\ds\int_{\Omega^e_0}\mN_e^\mrT(\bn\otimes\bn)\,\mN_e\,\dif A\,.
\label{e:mhatn}\eqe
This follows from inserting the elemental interpolations \eqref{e:bvmha}, \eqref{e:bvha} and (\ref{e:dbxwq}.2) into $\tilde G_0$ of Box~\ref{t:WF}.

For elastic mesh motion (option II), the elemental pseudo force vector becomes
\eqb{l}
\mtf^e := \mtf^e_\mri + \mtf^e_\mro\,,
\label{e:mesh2}\eqe
with
\eqb{lll}
\mtf^e_\mri \dis \ds\int_{\Omega_0^e}\big[\mN^\mrT_{e,\alpha}\,\ba_\beta + \mN_e^\mrT\bn\,b_{\alpha\beta}\big]\,\tau_\mrm^{\alpha\beta}\,\dif A\,, \\[4mm]
\mtf^e_\mro \dis \alpha_\mrm\,\mtn_e\,\big(\mx_e'-\mv_e\big) \,.
\label{e:fio}\eqe
This follows from inserting $w_{\alpha;\beta} = \mw_e^\mrT\big[\mN^\mrT_{e,\alpha}\,\ba_\beta + \mN_e^\mrT\bn\,b_{\alpha\beta} \big]$ \citep{membrane}, $w = \bw\cdot\bn$, \eqref{e:bvmha}, \eqref{e:bvha} and (\ref{e:dbxwq}.2) into $\tilde G_\mathrm{el}$ of Box~\ref{t:WF}. 
The in-plane part $\mtf^e_\mri$ was already used in \citet{droplet} to stabilize quasi-static droplet simulations.
Here, the stabilization stress $\tau_\mrm^{\alpha\beta}$ is taken from the simple elasticity model in \eqref{e:taum}.

The linearization of both options is provided in Appendix \ref{s:Lfh}.
It simplifies due to the fact that all the above contributions are integrated over the initial configuration. 
This integration simplification is justified as the mesh contributions all appear in a separate weak form equation.
This is different for the stabilization contributions of Sec.~\ref{s:FEDB}:
They are combined with the incompressibility constraint in Eq.~\eqref{e:mbf1}.

\subsubsection{Numerical quadrature}

All integrals are mapped to $\zeta^\alpha$-space and evaluated there by a numerical quadrature rule.
The mapping follows from the expressions in \eqref{e:da}.
Gaussian quadrature with $3 \times 3$ quadrature points is used for all quadratic elements.

\subsubsection{Coupled system of equations}

The system of equations to be solved -- after element assembly and elimination of Dirichlet boundary conditions -- is
\eqb{lllll}
\mf(\mv',\mv,\mx',\mx,\mq) \dis \mf_\mathrm{in}(\mv',\mv,\mx',\mx) + \mf_\mathrm{int}(\mv,\mx,\mq) - \mf_\mathrm{ext}(\mx) \is \mathbf{0}\,, \\[1mm]
\mtf(\mv,\mx',\mx) \dis \mtf_\mri(\mv,\mx',\mx) + \mtf_\mro(\mv,\mx',\mx) \is \mathbf{0}\,, \\[1mm]
\mbf(\mv,\mx,\mq) \dis \mbf_\mathrm{div}(\mv,\mx) - \mbf_\mathrm{DB}(\mq) \is \mathbf{0}\,.
\label{e:fsys}\eqe
These are the governing ordinary differential equations for the unknown nodal flow variables $\mv = [\bv_I]$, $\mq = [q_I]$ and surface position $\mx = [\mx_I]$.
In order to solve them, time integration needs to be used.
If the mesh motion $\mx(t)$ is known (e.g.~prescribed) the second equation in \eqref{e:fsys} becomes unneccessary.

\subsection{Time integration}\label{s:TI}

Using $\ma := \mv'$ and $\mv_\mrm = \mx'$ the system of coupled ODEs \eqref{e:fsys}
can be written as
\eqb{lllll}
\mf_\mrv \dis \mf(\ma,\,\mv,\,\mv_\mrm,\,\mx,\,\mq) \is \mathbf{0}\,, \\[1mm]
\mf_\mrx \dis \mtf(\mv,\,\mv_\mrm,\,\mx) \is \mathbf{0}\,, \\[1mm]
\mf_\mrq \dis \mbf(\mv,\,\mx,\,\mq) \is \mathbf{0}\,.
\label{e:fsys2}\eqe
Here, $\mf_\mrv=\mathbf{0}$ is a first order ODE, predominately for $\mv$, $\mf_\mrx=\mathbf{0}$ is a first order ODE, predominantly for $\mx$, and $\mf_\mrq=\mathbf{0}$ is an algebraic equation, predominantly for $\mq$.
Only in the Lagrangian case (where $\mv_\mrm=\mv$ eliminates the second equation) does $\mf_\mrv=\mathbf{0}$ become a second order ODE.
System \eqref{e:fsys2} is discretized in time by classical time stepping: 
Given $\ma_n$, $\mv_n$, $\mv^n_\mrm$, $\mx_n$ and $\mq_n$ at time step $t_n$ the new values $\ma_{n+1}$, $\mv_{n+1}$, $\mv^{n+1}_\mrm$, $\mx_{n+1}$ and $\mq_{n+1}$ at $t_{n+1}$ are determined.
Since these are five unknowns, two extra equations are needed.
These are provided through finite difference approximations of $\mv' = \ma$ and $\mx' = \mv_\mrm$, in the form
\eqb{lll}
\mv_{n+1} \is \mv_n + \Delta t_{n+1}\,\big((1-\gamma)\,\ma_n + \gamma\,\ma_{n+1}\big)\,, \\[2mm]
\mx_{n+1} \is \mx_n + \Delta t_{n+1}\,\big((1-\gamma)\,\mv_\mrm^n + \gamma\,\mv_\mrm^{n+1}\big)\,.
\label{e:vxnp1}\eqe
Taking $\gamma = 0$ and $\gamma = 1$ here, gives the classical explicit and implicit Euler scheme, respectively.
However those are only 1st order accurate in time.
The choice $\gamma = 1/2$ corresponds to the trapezoidal rule, which is used in all subsequent examples, since it is 2nd order accurate in time and implicit, making it suitable for strongly coupled systems.

The time integration can be easily extended to the generalized-$\alpha$ scheme \citep{chung93,jansen99}.
This is shown in Appendix~\ref{s:tang} even if the following examples do not use this.

\subsection{Solution procedure}\label{s:NR}

Using Eq.~\eqref{e:vxnp1} to eliminate $\ma_{n+1}$ and $\mv^{n+1}_\mrm$ gives
\eqb{lllll}
\mf_{n+1} \dis \mf(\mv_{n+1},\,\mx_{n+1},\,\mq_{n+1}) \is \mathbf{0}\,, \\[1mm]
\mtf_{n+1} \dis \mtf(\mv_{n+1},\,\mx_{n+1}) \is \mathbf{0}\,, \\[1mm]
\mbf_{n+1} \dis \mbf(\mv_{n+1},\,\mx_{n+1},\,\mq_{n+1}) \is \mathbf{0}\,,
\label{e:fsys3}\eqe
or more compactly, 
\eqb{l}
\mr_{n+1}(\muu_{n+1}) = \mathbf{0}\,, 
\label{e:fsys4}\eqe
with the vector of unknowns and the residual vector defined by
\eqb{l}
\muu := \begin{bmatrix}
\mv \\
\mx \\
\mq
\end{bmatrix},\quad 
\mr := \begin{bmatrix}
\mf \\
\mtf \\
\mbf
\end{bmatrix}.
\eqe
The nonlinear equation system \eqref{e:fsys4} is solved with the Newton-Raphson method.
It is based on the linearization of \eqref{e:fsys4}, which leads to the iterative solution of $\mK_n\,\Delta\muu_{n+1} = -\mr_n$, where
\eqb{l}
\mK_n := \begin{bmatrix}
\pa{\mf_n}{\mv_n} & \pa{\mf_n}{\mx_n} & \pa{\mf_n}{\mq_n} \\[1.5mm]
\pa{\mtf_n}{\mv_n} & \pa{\mtf_n}{\mx_n} & \pa{\mtf_n}{\mq_n} \\[1.5mm]
\pa{\mbf_n}{\mv_n} & \pa{\mbf_n}{\mx_n} & \pa{\mbf_n}{\mq_n}
\end{bmatrix} =: \begin{bmatrix}
\mK_{\mrv\mrv} & \mK_{\mrv\mrx} & \mK_{\mrv\mrq}\, \\[.5mm]
\mK_{\mrx\mrv} & \mK_{\mrx\mrx} & \mK_{\mrx\mrq}\, \\[.5mm]
\mK_{\mrq\mrv} & \mK_{\mrq\mrx} & \mK_{\mrq\mrq}\,
\end{bmatrix}_{\!n}
\label{e:K}\eqe
is the tangent matrix at $t_n$.
Obviously $\mK_{\mrx\mrq}^n=\partial\mtf_n/\partial\mq_n = \mathbf{0}$.
The remaining eight blocks of $\mK_n$ are given in Appendix \ref{s:tang}.

\subsection{Normalization}\label{s:Norm}

The above equations are implemented in normalized form.
A length scale $L_0$, time scale $T_0$ and force $F_0$ are used for this.
All lengths, times, forces and combined units are then expressed in terms of these quantities.
Pressures, for example are then multiples of $p_0 = F_0/L_0^2$.
The examples in Sec.~\ref{s:Nexf}-\ref{s:octodefo} use unit values for all quantities (i.e.~$L_0=T_0=F_0=1$), while the example in Sec.~\ref{s:SB} uses $L_0 = 1$mm, $T_0 = 1$ms and $p_0 = 1$Pa.

\section{Numerical examples for fixed surface flows}\label{s:Nexf}

The following two sections present several numerical examples, first for fixed surfaces (Sec.~\ref{s:Nexf}), then deforming surfaces (Sec.~\ref{s:Nex}).
All examples are full Navier-Stokes examples -- they contain the nonlinear convective part.
Their complexity is increased successively in order to test the computational formulation thoroughly. 
They use spheres as initial configuration, as these are stable configurations of free fluid films.
Other shapes, such as cylinders, are unstable beyond certain lengths \citep{droplet,ALE} and not considered here.

\subsection{Problem setup}\label{s:Nexsetup}
 
Sec.~\ref{s:Nexf} studies two simple verification examples for rotational flow on fixed spheres.
They have manufactured analytical solutions, see \citet{ALEtheo}.
The comparison between numerical and analytical solution allows to examine convergence rates and verify both FE implementation and manufactured solution. 

For fixed surfaces the out-of-plane velocity $v_\mrn$ is given, e.g.~as zero.
It makes a difference if this is explicitly enforced or not, which is illustrated below.
Correspondingly, two different out-of-plane boundary conditions are considered: (i) imposing zero out-of-plane velocity (Dirichlet BC), or (ii) imposing the surface pressure corresponding to zero $v_\mrn$ (Neumann BC).
Both cases are complemented with essential BCs to fix all six rigid body modes, as well as the datum value of the surface tension $q$.
The first BC is imposed by local projection (the dofs of each node are first transformed into the local spherical coordinate system and then the normal component is eliminated). 

Since no curvature-depending forces are present in the considered examples, classical $C^0$-continuous finite elements are sufficient.
In all subsequent examples quadrilateral bi-quadratic Lagrange FE are used.
The spheres are meshed with six square patches, containing $2m\times2m$ elements each.
Thus, $n_\mathrm{el} = 24 m^2$ is the total number of elements of the sphere and $n_\mathrm{eq}=8m$ is the number of elements around the equator. 
Further, the element size $h$ is proportional to $m^{-1}$ and $n_\mathrm{el}^{-2}$.
The sequence of meshes used here is shown in Table~\ref{t:shearflow}.
%-------------------------------------------------------------------------------------------------------------------------------
\begin{table}[h]
\centering
\begin{tabular}{|r|r|l|r|r|r|r|r|r|}
  \hline
   $m$ & $n_\mathrm{el}$ & $h_\mathrm{eq}/r$ & $n_\mathrm{no}$ & 3-dof & 4-dof & 7-dof & $n_{t1}/4$ & $n_{t2}/4$ \\[0mm] \hline 
   & & & & & & & & \\[-4mm]   
   1 &  24 & $\pi/4$ & 98 & 294 & 392 & 686 & 4 & 2 \\ [0mm] 
   2 & 96 & $\pi/8$ & 386 & 1,158 & 1,544 & 2,702 & 8 & 6 \\ [0mm] 
   4 & 384 & $\pi/16$ & 1,538 & 4,614 & 6,152 & 10,766 & 16 & 16 \\[0mm] 
   8 &  1,536 & $\pi/32$ & 6,146 & 18,438 & 24,584 & 43,022 & 32 & 45 \\[0mm] 
   16 & 6,144 & $\pi/64$ & 24,578 & 73,734 & 98,312 & 172,046 & 64 & 128 \\[0mm] 
   32 & 24,576 & $\pi/128$ & 98,306 & 294,918 & 393,224 & 688,142 & 128 & 362 \\[0mm]   
   \hline
\end{tabular}
\caption{Shear flow on spheres: Considered FE meshes based on bi-quadratic Lagrange elements and corresponding dofs for three ($v^\alpha$, $q$), four ($\bv$, $q$) and seven ($\bv$, $q$, $\bv_\mrm$) unknowns.
Three and four unknowns are used in the examples of Sec.~\ref{s:Nexf}, seven in the examples of Sec.~\ref{s:Nex}.
$h_\mathrm{eq} = 2\pi r/(8m)$ is the element length at the equator;
$n_{ti}/4$ are the number of time steps per quarter period used in ALE case d (see Table~\ref{t:shearflowcases}).\\[-2mm]
}
\label{t:shearflow}
\end{table}
%-------------------------------------------------------------------------------------------------------------------------------
They are used to examine the algebraic $L^2$ error measure
\eqb{l}
e(\muu) := \ds\frac{\norm{\muu_\mathrm{FE}-\muu_\mathrm{ana}}}{\norm{\muu_\mathrm{ana}}}
\label{e:L2}\eqe
applied to the nodal values of $\bv$, $\gamma=q$, $\omega$ and $p$.
The latter two are obtained by post-processing.

\begin{remark} 
All the following convergence figures plot the error against $n_\mathrm{el}$ on a logscale.
Plotted against $h$, simply makes all slopes twice as steep, since $h$ scales with $n_\mathrm{el}^{-2}$.
\end{remark}

\subsection{Simple shear flow}\label{s:shearflow}

The first example in Sec.~\ref{s:Nexf} considers the axisymmetric flow field
\eqb{l}
\bv = r\omega_0\sin\theta\cos\theta\,\be_\phi\,, 
\label{e:shearflowv}\eqe
on the sphere ($0<\phi<2\pi$ and $-\pi/2<\theta<\pi/2$). 
Here, $\phi$ and $\theta$ are the azimuthal and elevation angle, respectively, and $\be_\phi$ the unit basis vector along $\phi$; $\omega_0$ is constant so that the flow is steady. 
Mesh velocity $\bv_\mrm$ is considered given, such that no ALE ODE needs to be solved.
The four different ALE cases shown in Table~\ref{t:shearflowcases} are studied.
%-------------------------------------------------------------------------------------------------------------------------------
% TRANSLATION of ALE cases: code - paper
% 				  	      	   0	     a
% 					    	   a	     b
%					    	   b	     c
%					    	   c  	     d
\begin{table}[h]
\vspace{1mm}
\centering
\begin{tabular}{|l|ll|l|l|l|l|}
  \hline
    & & \hspace{-4mm} load case & 1 & 2 & 3 & 4 \\[0mm] \hline
   ALE & & & $f_2=0$ & $q=$ const. & $f_2=0$ & $q=$ const., \\[0mm]
   case & & & DBC $v_\mrn=0^{\,\ast}$\!\! & DBC $v_\mrn=0^{\,\ast}$\!\! & NBC $p=\bar p$  & NBC $p=\bar p$  \\[0mm] \hline
   & & & & & & \\[-4mm]   
   a &  $\bv_\mrm=\mathbf{0}$, & $\!\!\!\theta_0=0$ & Fig.~\ref{f:shearflow1} \& \ref{f:shearflow2} & Fig.~\ref{f:shearflowA1} & Fig.~\ref{f:shearflow6} & Fig.~\ref{f:shearflowA2} \\ [0mm] 
   b &  $\bv_\mrm=\bc_0$, & $\!\!\!\theta_0=0$ & Fig.~\ref{f:shearflow3} & Fig.~\ref{f:shearflowA1} & Fig.~\ref{f:shearflow6} & Fig.~\ref{f:shearflowA2} \\ [0mm] 
   c &  $\bv_\mrm=\mathbf{0}$, & $\!\!\!\theta_0=0.5$ & Fig.~\ref{f:shearflow4} & Fig.~\ref{f:shearflowA1} & Fig.~\ref{f:shearflow6} & Fig.~\ref{f:shearflowA2} \\ [0mm] 
   d &  $\bv_\mrm=\bv_\mrm(t)$, & $\!\!\!\theta_0=0.5$ & Fig.~\ref{f:shearflow5} & Fig.~\ref{f:shearflowA1} & Fig.~\ref{f:shearflow6} & Fig.~\ref{f:shearflowA2} \\ [0mm] 
   \hline
\end{tabular}
\caption{Simple shear flow on a rigid sphere: Considered ALE mesh motion cases and surface load cases.
The ALE cases consider either no mesh motion and distortion (case a), the constant mesh velocity of Eq.~\eqref{e:vm0} (case b), the fixed mesh distortion of Eq.~\eqref{e:xm1} (case c), or the transient mesh motion of \eqref{e:xm1} \& \eqref{e:tm1} (case d).
The load cases consider different body forces, and different out-of-plane boundary conditions (DBC = Dirichlet boundary condition, NBC = Neumann boundary condition).
Load cases 1 \& 2 have three unknowns ($v^\alpha,\,q$), while load cases 3 \& 4 have four ($\bv,\,q$).
$^\ast$In ALE case b, $v_\mrn = \bc_0\cdot\bn\neq0$.\\[-1mm]
}
\label{t:shearflowcases}
\end{table}
%-------------------------------------------------------------------------------------------------------------------------------
They are combined with four different load cases, and discussed in the following four sub-sub-sections.
Load cases 3 \& 4 have four unknowns $(\bv,q)$.
For load cases 1 \& 2, the normal velocity $v_n$ is eliminated by projection along the surface normal, leaving only three unknowns $(v^\alpha,q)$.
Flow \eqref{e:shearflowv} is caused by the body force field \citep{ALEtheo}
\eqb{l}
\bff = f_\alpha\,\ba^\alpha + p\,\bn\,, \quad\left\{\begin{array}{lll}
f_1 \is 4\eta\,\omega_0\sin\theta\cos^2\theta\,, \\[.5mm]
f_2 \is \rho r^2 \omega_0^2\sin^3\theta\cos\theta-q_{,2}\,,
\end{array}\right.
\label{e:shearflowf}\eqe
where surface tension $q$ is picked such that either $f_2=0$ or $q_{,2}=0$, and surface pressure $p$ either follows as the reaction to the Dirichlet boundary condition on $v_\mrn$ or is prescribed by the Neumann boundary condition $p=\bar p$, see Table~\ref{t:shearflowcases}.
Unit values are used for $r$, $\omega_0$, $\eta$ and $\rho$, while $\eta_\mrn = 0$ in all cases of Secs.~\ref{s:shearflow} and \ref{s:octflow}.

\subsubsection{Zero mesh velocity (ALE case a)}\label{s:shearALE0} 

The first ALE case considers no mesh motion and no mesh distortion ($\bv_\mrm=\mathbf{0}$), i.e.~the surface parameterization based on $\phi$ and $\theta$ is fixed in time.
%----------------------------------------------------------------------------------------------------------------------------------
\begin{figure}[h]
\begin{center} \unitlength1cm
\begin{picture}(0,5.6)
\put(-8.8,-.22){\includegraphics[height=60mm]{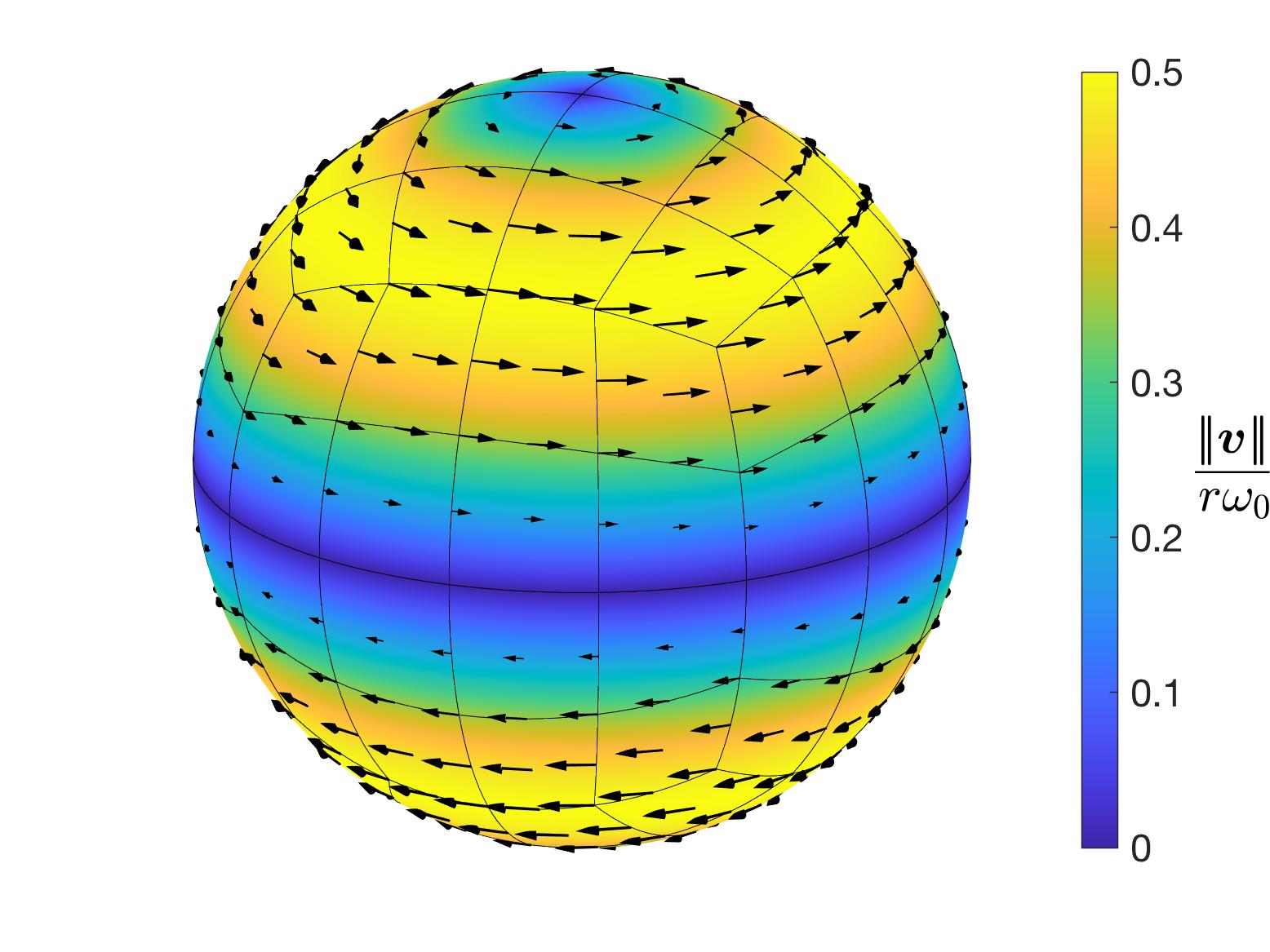}}
\put(0.2,-.3){\includegraphics[height=58mm]{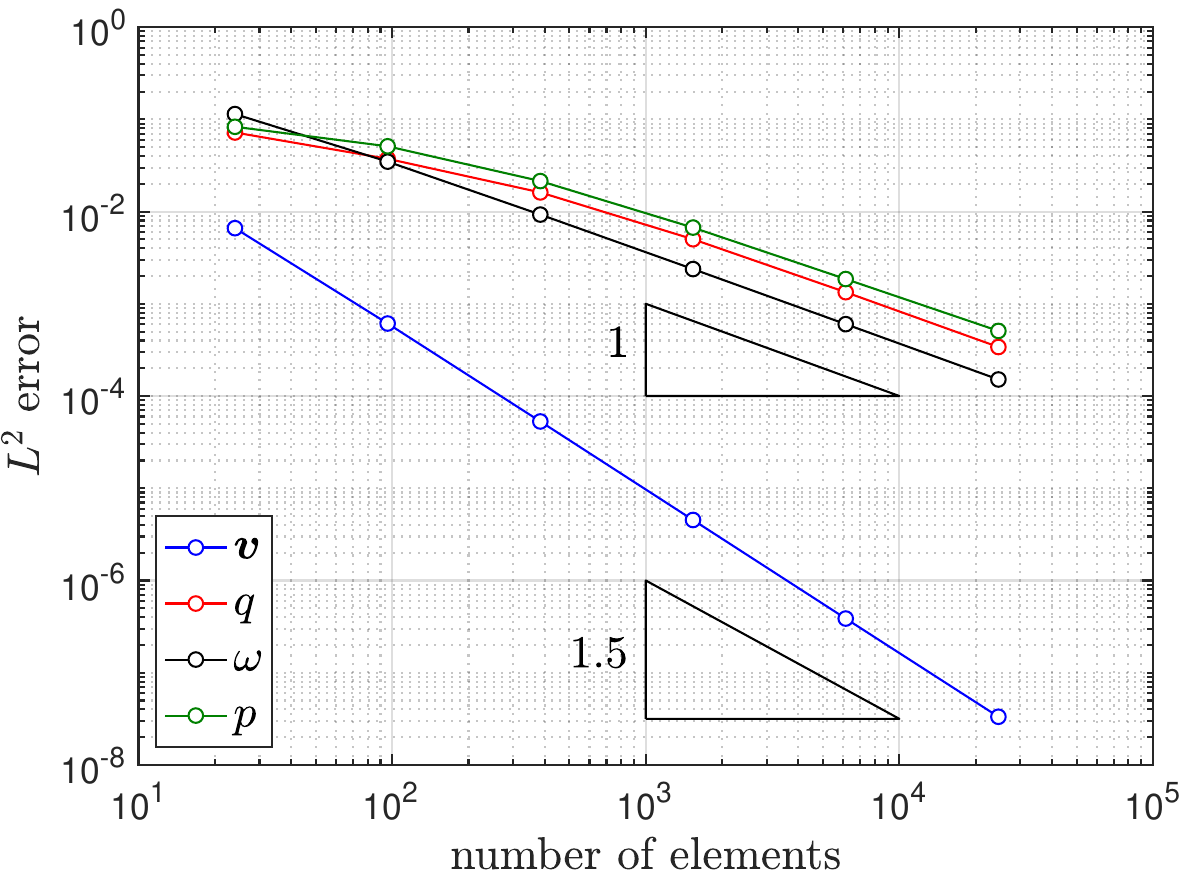}}
\put(-7.95,-.15){\footnotesize (a)}
\put(0.2,-.15){\footnotesize (b)}
\end{picture}
\caption{Simple shear flow on a rigid sphere: 
(a) Flow field $\bv$ and (b) convergence behavior for ALE case~a and load case 1.
Optimal convergence rates are obtained for all fields: At least $O(h^3)=O(n^{-1.5}_\mathrm{el})$ for $\bv$, and $O(h^2)=O(n^{-1}_\mathrm{el})$ for $q$, $p$ and $\omega$.}
\label{f:shearflow1}
\end{center}
\end{figure}
% run sShearFlow/ vShearFlowNS/ mat{1}.case = 1 & pShearFlow.m/ jALE = 0, jcase = 1 
%----------------------------------------------------------------------------------------------------------------------------------
Fig.~\ref{f:shearflow1}a shows the surface velocity for this case.
It is caused by a manufactured body force that admits two special cases: \\
\underline{Load case 1:} Picking $f_2=0$ (zero body force along elevation angle $\theta$) leads to varying surface tension $q$, as shown in Fig.~\ref{f:shearflow2}a.
%----------------------------------------------------------------------------------------------------------------------------------
\begin{figure}[h]
\begin{center} \unitlength1cm
\begin{picture}(0,3.7)
\put(2.1,-.6){\includegraphics[height=44mm]{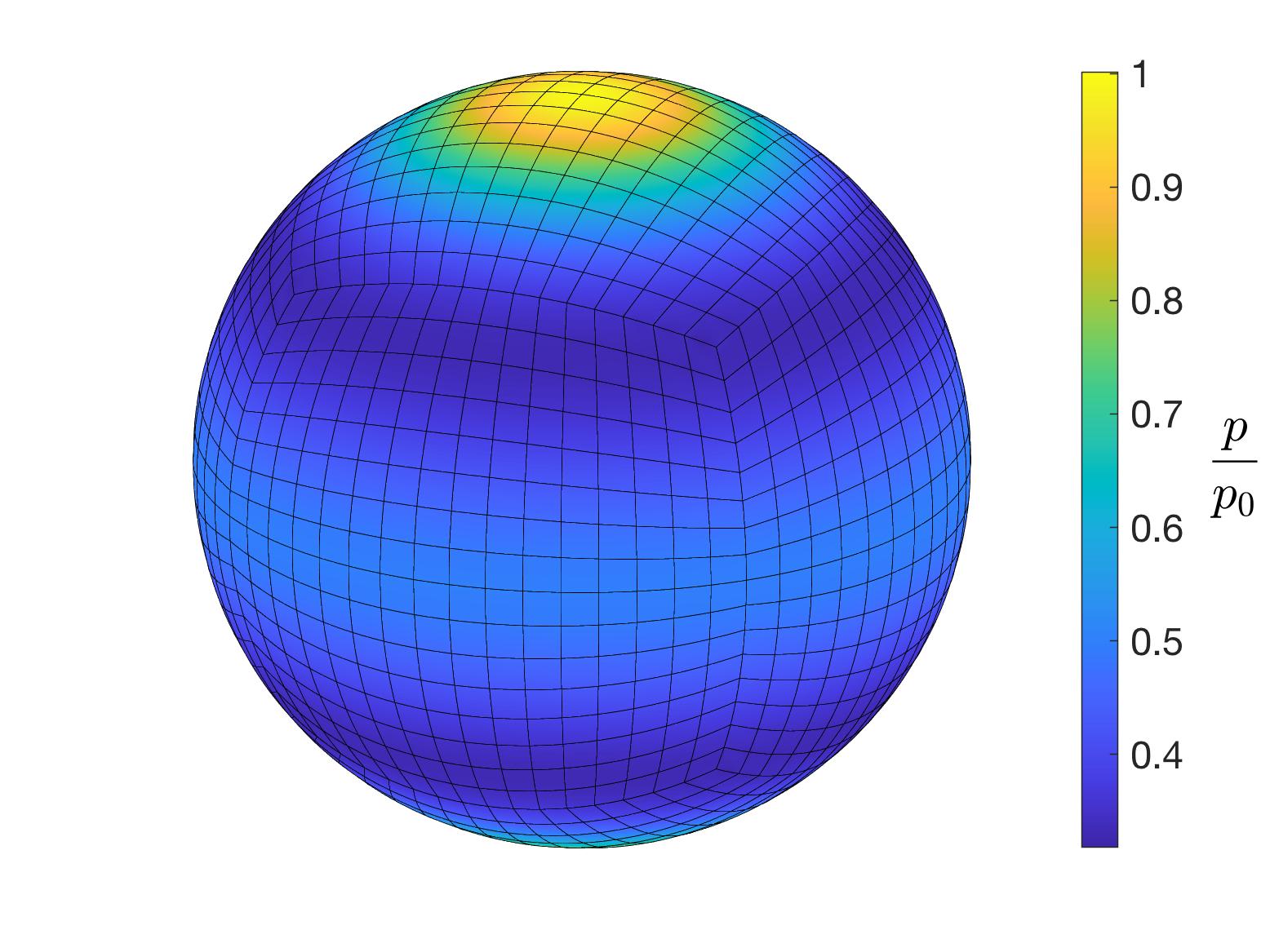}}
\put(-3.35,-.6){\includegraphics[height=44mm]{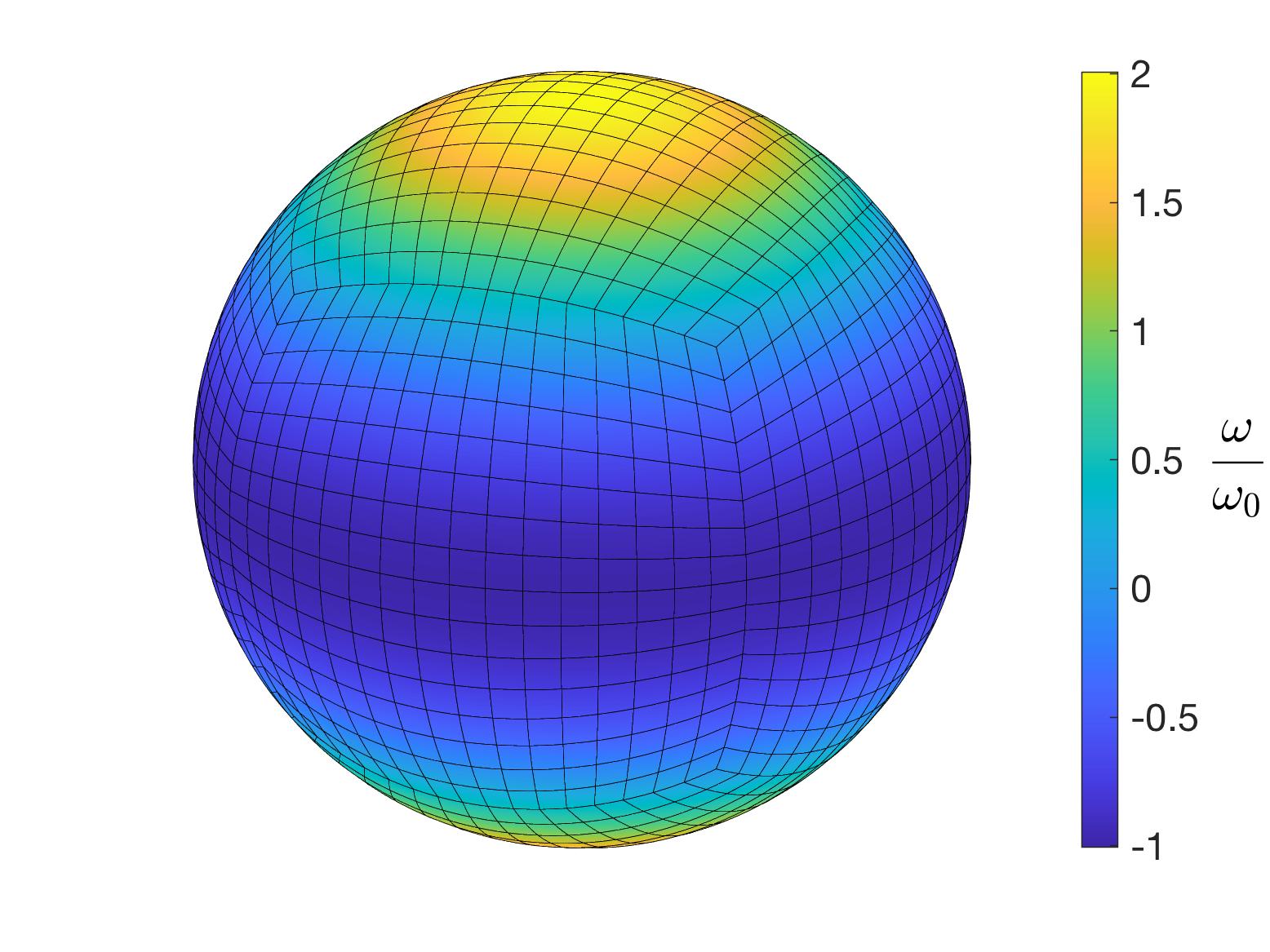}}
\put(-8.8,-.6){\includegraphics[height=44mm]{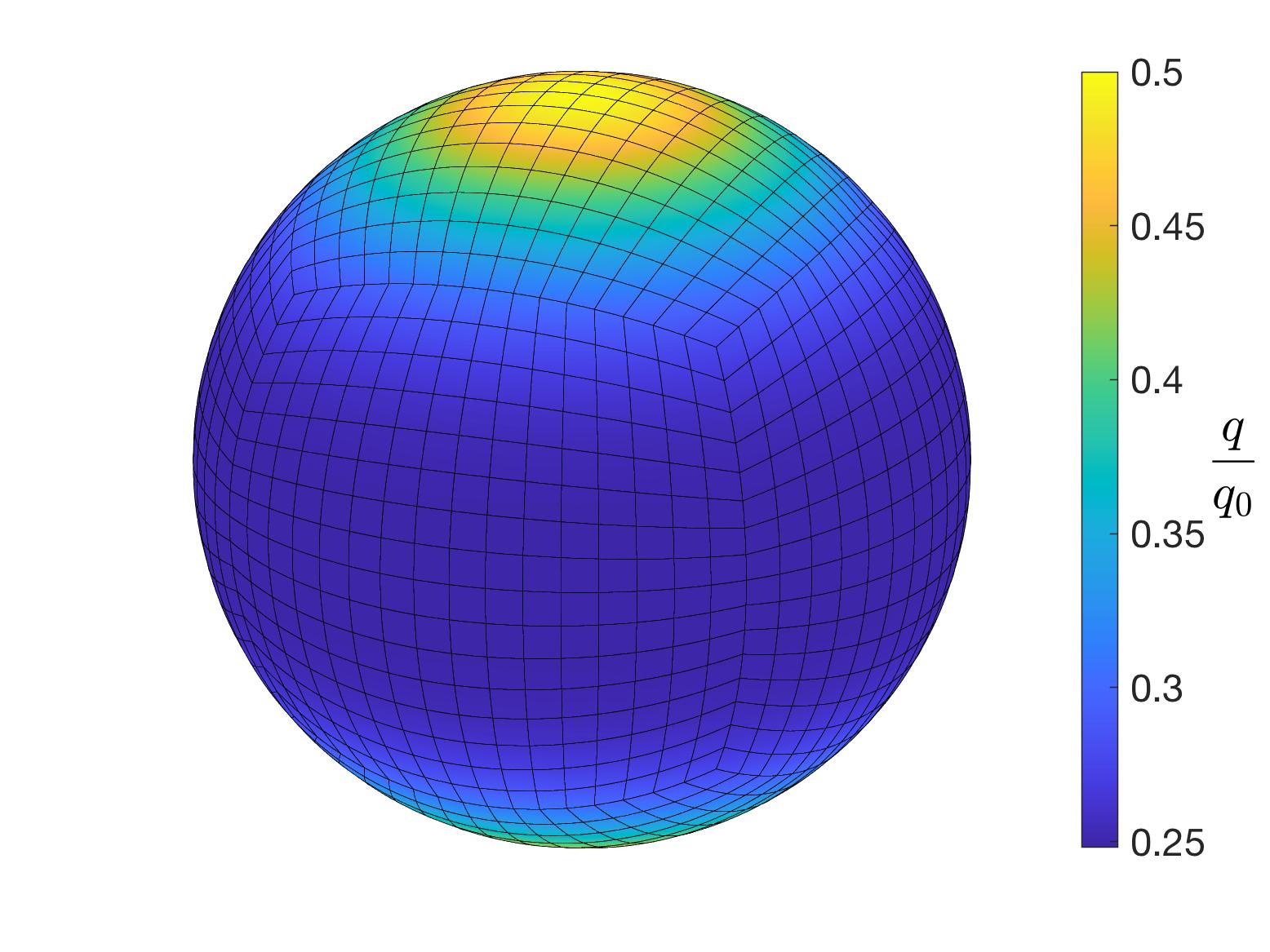}}
\put(-7.95,-.15){\footnotesize (a)}
\put(-2.5,-.15){\footnotesize (b)}
\put(2.95,-.15){\footnotesize (c)}
\end{picture}
\caption{Simple shear flow on a rigid sphere: 
(a) Surface tension $q$, (b) surface vorticity $\omega$ and (c) lateral surface pressure $p$ for ALE case a and load case 1.
The plots are normalized by $q_0 = \rho r^2\omega_0^2$, $w_0$ and $p_0 = \rho r\omega_0^2$.}
\label{f:shearflow2}
\end{center}
\end{figure}
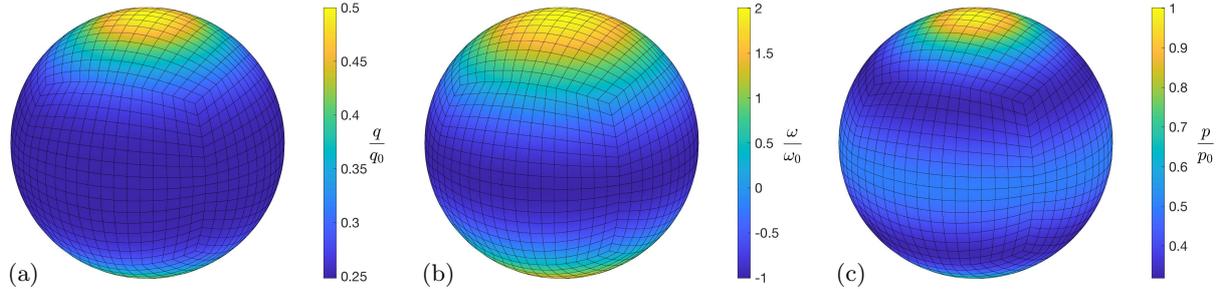
% run sShearFlow/ vShearFlowNS/ mat{1}.case = 1
%----------------------------------------------------------------------------------------------------------------------------------
Fig.~\ref{f:shearflow2} also shows the vorticity $\omega$ and the outward surface pressure $p$ required to hold the surface in place (i.e.~the lateral boundary reactions).
A much finer mesh is needed for $q$, $\omega$ and $p$ to yield comparable accuracy than for $\bv$.
The flow field is invariant w.r.t.~to a chosen datum pressure.
Here this is fixed by specifying the pole pressure as $p_\mrp=p_0=\rho r\omega_0^2$.
As shown in Fig.~\ref{f:shearflow1}b, the FE convergence rates for $\bv$, $q$, $\omega$ and $p$ are optimal (or better):
$O(h^3)=O(n^{-1.5}_\mathrm{el})$ for $\bv$, and $O(h^2)=O(n^{-1}_\mathrm{el})$ for $q$ according to \citet{dohrmann04}. \\
\underline{Load case 2:} Picking $f_2$ such that $q$ becomes constant, see \citet{ALEtheo}, leads to different $p$, but the same $\bv$ and $\omega$ as before.
The convergence rates for load case 2 are also optimal, as Fig.~\ref{f:shearflowA1}a in Appendix~\ref{s:add} shows.

So far the mesh motion was zero.
The next three ALE cases consider three different prescribed mesh motions/distortions.

\begin{remark} 
The solution above contains surface Stokes flow for the special choice $\rho=0$.
For Stokes flow, both $q$ and $p$ remain constant in both load cases, see \citet{ALEtheo}.
\end{remark}

\subsubsection{Constant mesh velocity (ALE case b)}\label{s:shearALEa}

\begin{figure}[h]
\begin{center} \unitlength1cm
\begin{picture}(0,5.6)
\put(-8.8,-.22){\includegraphics[height=60mm]{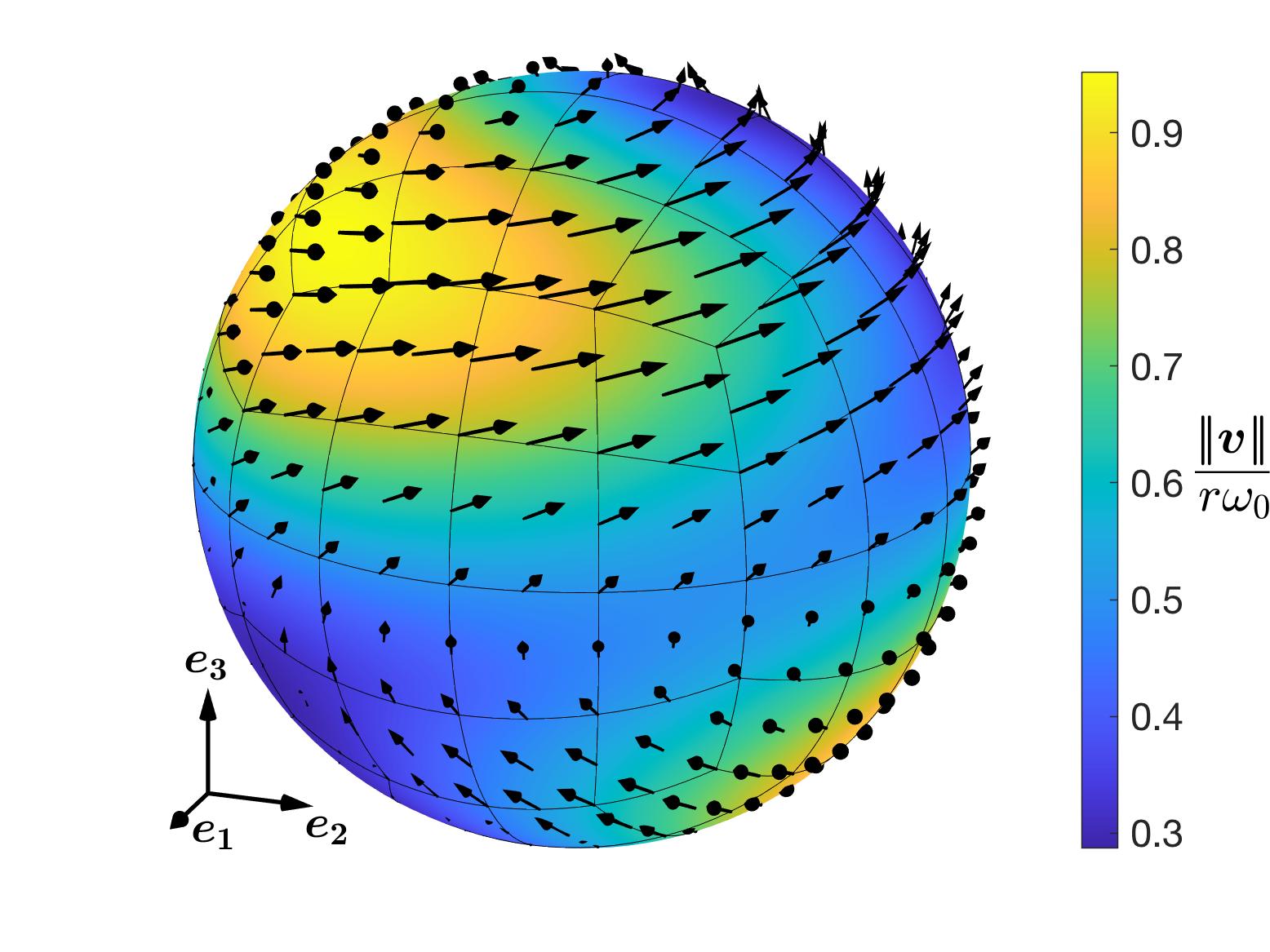}}
\put(0.2,-.3){\includegraphics[height=58mm]{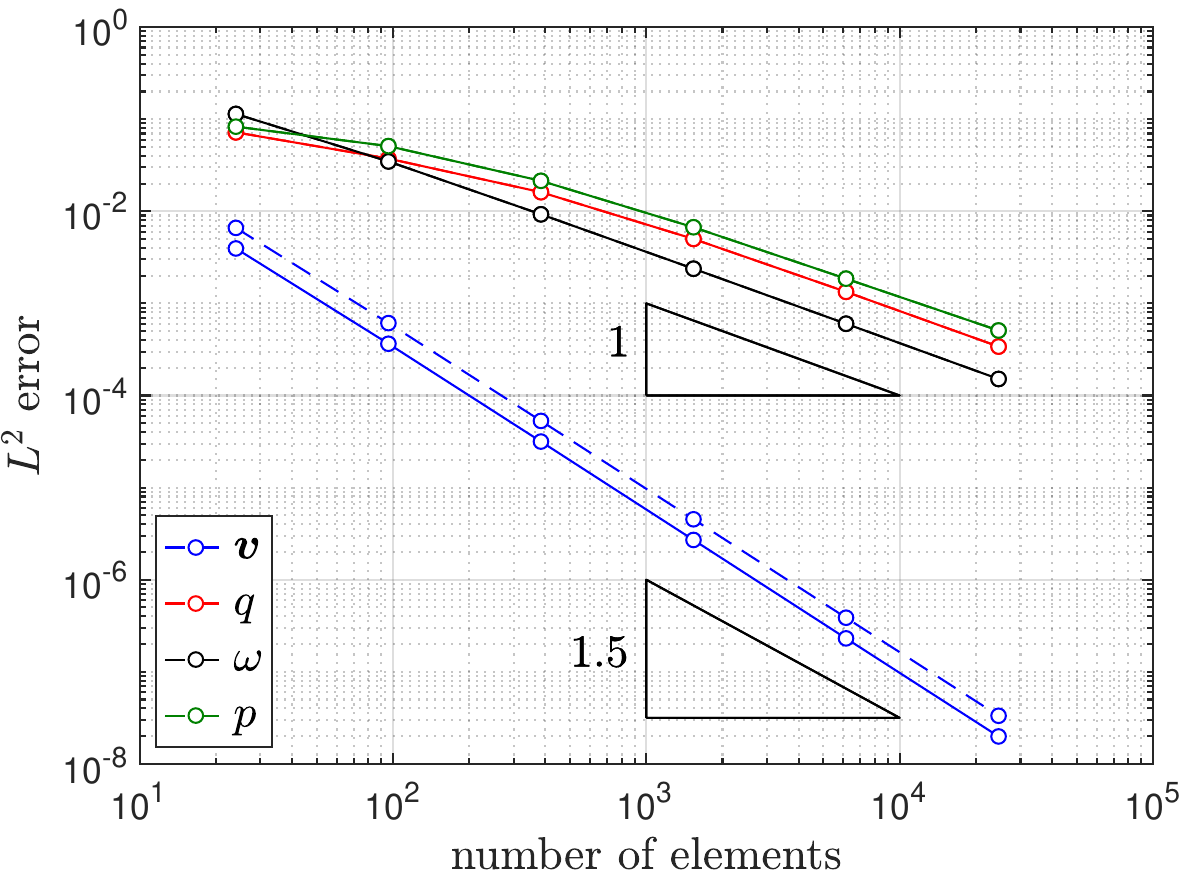}}
\put(-7.95,-.15){\footnotesize (a)}
\put(0.2,-.15){\footnotesize (b)}
\end{picture}
\caption{Simple shear flow on a rigid sphere: 
(a) Flow field $\bv$ and (b) convergence behavior for the prescribed constant mesh velocity $\bv_\mrm=\bc_0$ given in \eqref{e:vm0} (ALE case b, load case 1).
In this case, $\bv_\mrm$, and hence also $\bv$, have out-of-plane components. 
As in Fig.~\ref{f:shearflow1}, optimal convergence rates are obtained for all fields.
The dashed line in (b) is the line from Fig.~\ref{f:shearflow1}b.}
\label{f:shearflow3}
\end{center}
\end{figure}

Now the constant mesh velocity
\eqb{l}
\bv_\mrm = \bc_0
\label{e:vm0}\eqe
is considered, where the constant $\bc_0$ is chosen as $\bc_0 = v_{\mrm0}(\be_1 + \be_2 + \be_2)/\sqrt{3}$ with $v_{\mrm0} = r\omega_0/2$.
Thus $\bv$ has out-of-plane components, as Fig.~\ref{f:shearflow3}a shows.
%----------------------------------------------------------------------------------------------------------------------------------
% run sShearFlow/ vShearFlowALEa mat{1}.case = 1 &  pShearFlow.m/ jALE = 1, jcase = 1 
%----------------------------------------------------------------------------------------------------------------------------------
This $\bv_\mrm$ represents a constant translation that tests all three components in $\bbR^3$.
The case works well -- the optimal convergence rates are maintained as Fig.~\ref{f:shearflow3}b shows.
Load case 2 behaves similarly, as Fig.~\ref{f:shearflowA1}b in Appendix~\ref{s:add} shows: 
Again optimal convergence rates are obtained.

\subsubsection{Fixed mesh distortion (ALE case c)}\label{s:shearALEb}

\begin{figure}[h]
\begin{center} \unitlength1cm
\begin{picture}(0,5.6)
\put(-8.8,-.22){\includegraphics[height=60mm]{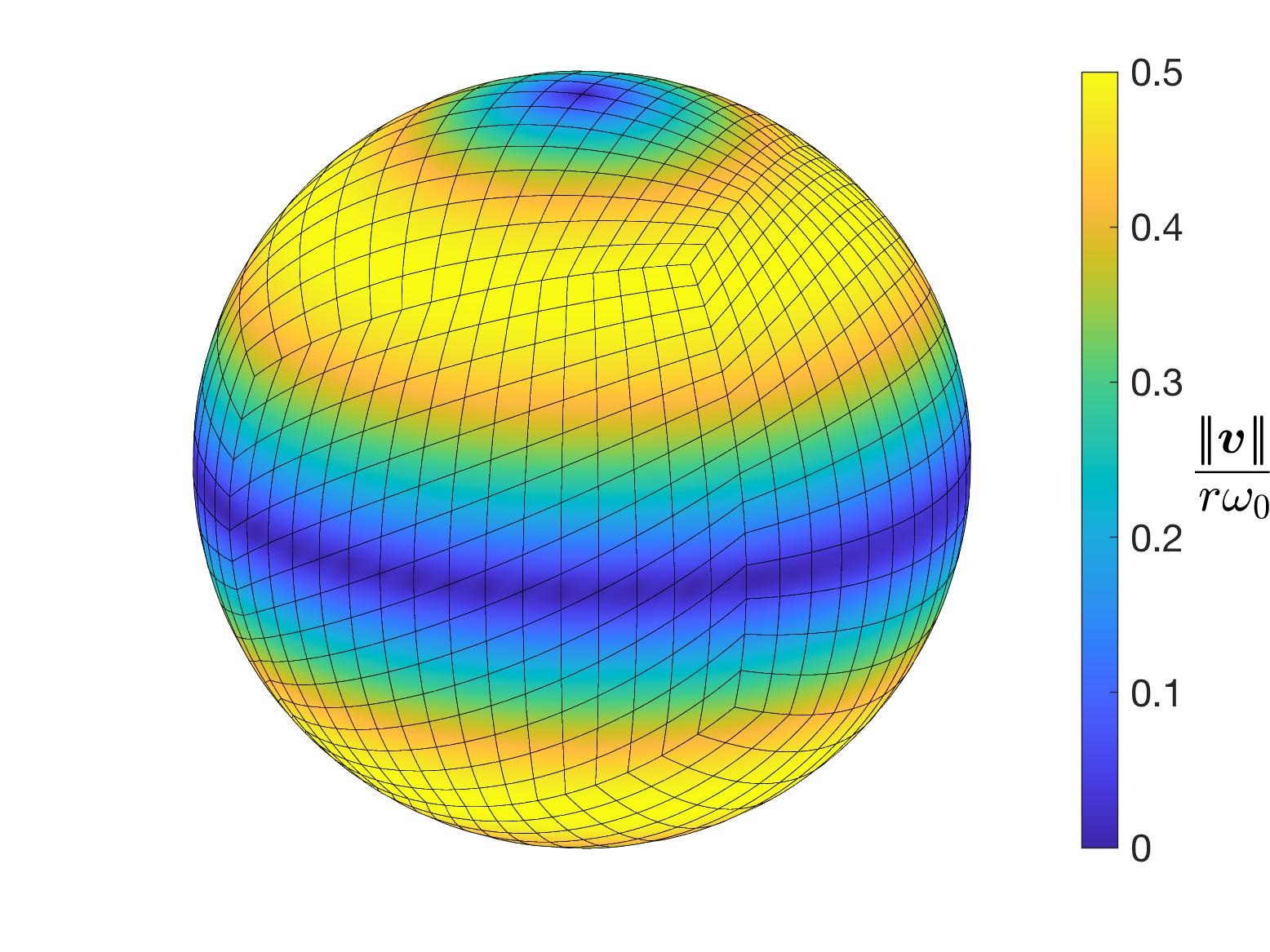}}
\put(0.2,-.3){\includegraphics[height=58mm]{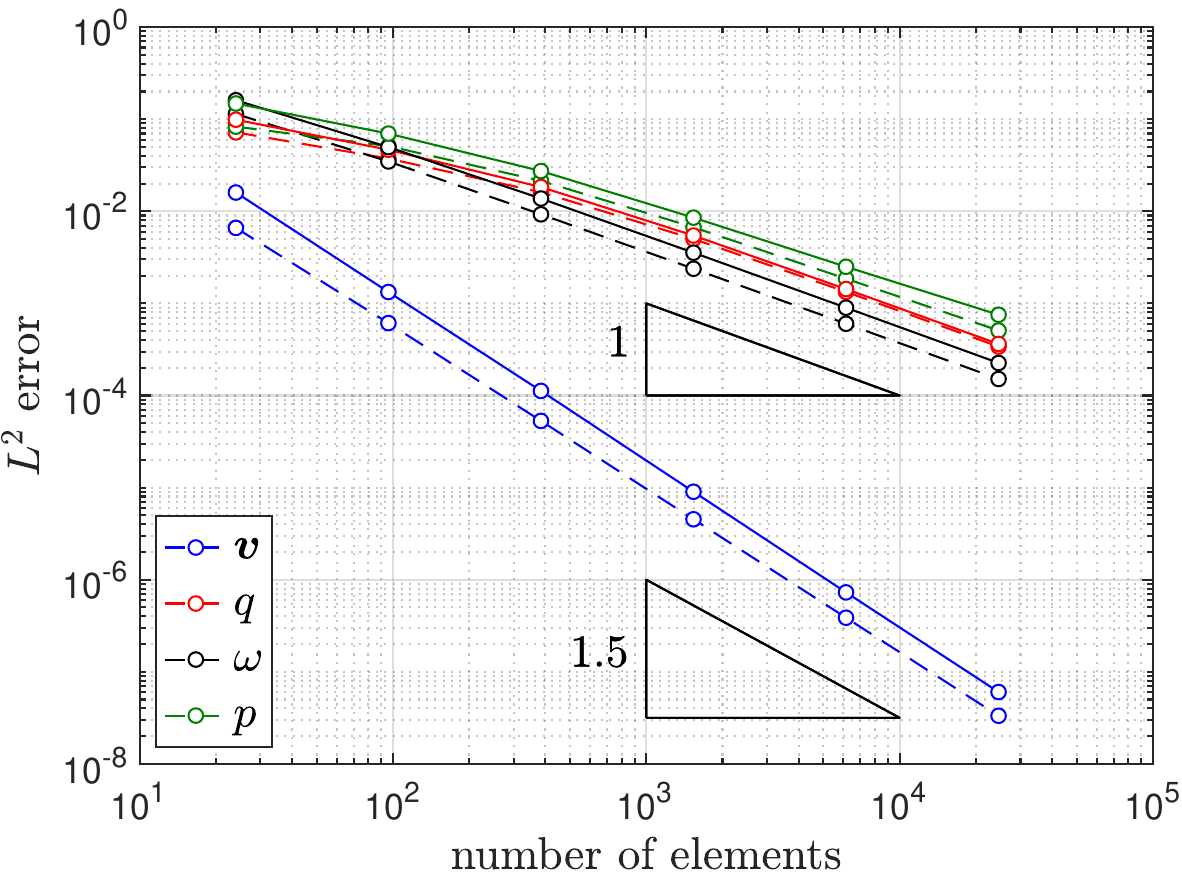}}
\put(-7.95,-.15){\footnotesize (a)}
\put(0.2,-.15){\footnotesize (b)}
\end{picture}
\caption{Simple shear flow on a rigid sphere: 
(a) Flow field $\bv$ and (b) convergence behavior due to the prescribed fixed mesh distortion $\bx_\mrm$ given in \eqref{e:xm1} \& \eqref{e:tm0} (ALE case c, load case 1).
Again, optimal convergence rates are obtained for all fields.
The dashed lines in (b) are the lines from Fig.~\ref{f:shearflow1}b.
}
\label{f:shearflow4}
\end{center}
\end{figure}

Next, the fixed mesh distortion
\eqb{l}
\bx = r\,\be_r(\theta)\,,
\label{e:xm1}\eqe
with 
\eqb{l}
\theta = \Theta + \theta_0 \sin\phi\cos^2\Theta\,,
\label{e:tm0}\eqe
is considered.
It leads to $\bv_\mrm=\mathbf{0}$ and is shown in Fig.~\ref{f:shearflow4}a for $\theta_0 = 1/2$ and load case 1.
%----------------------------------------------------------------------------------------------------------------------------------
% run sShearFlow/ vShearFlowALEb mat{1}.case = 1 &  pShearFlow.m/ jALE = 2, jcase = 1 
%----------------------------------------------------------------------------------------------------------------------------------
Also this case works well -- the optimal convergence rates are maintained as Fig.~\ref{f:shearflow4}b shows.
They are compared here to the ones in Fig.~\ref{f:shearflow1}, which was the case of no mesh distortion ($\theta_0=0$).
Load case 2 behaves similarly, as Fig.~\ref{f:shearflowA1}c in Appendix~\ref{s:add} shows: 
Again optimal convergence rates are obtained.

\subsubsection{Varying mesh velocity (ALE case d)}\label{s:shearALEc} 

\begin{figure}[h]
\begin{center} \unitlength1cm
\begin{picture}(0,5.6)
\put(-8.8,-.22){\includegraphics[height=60mm]{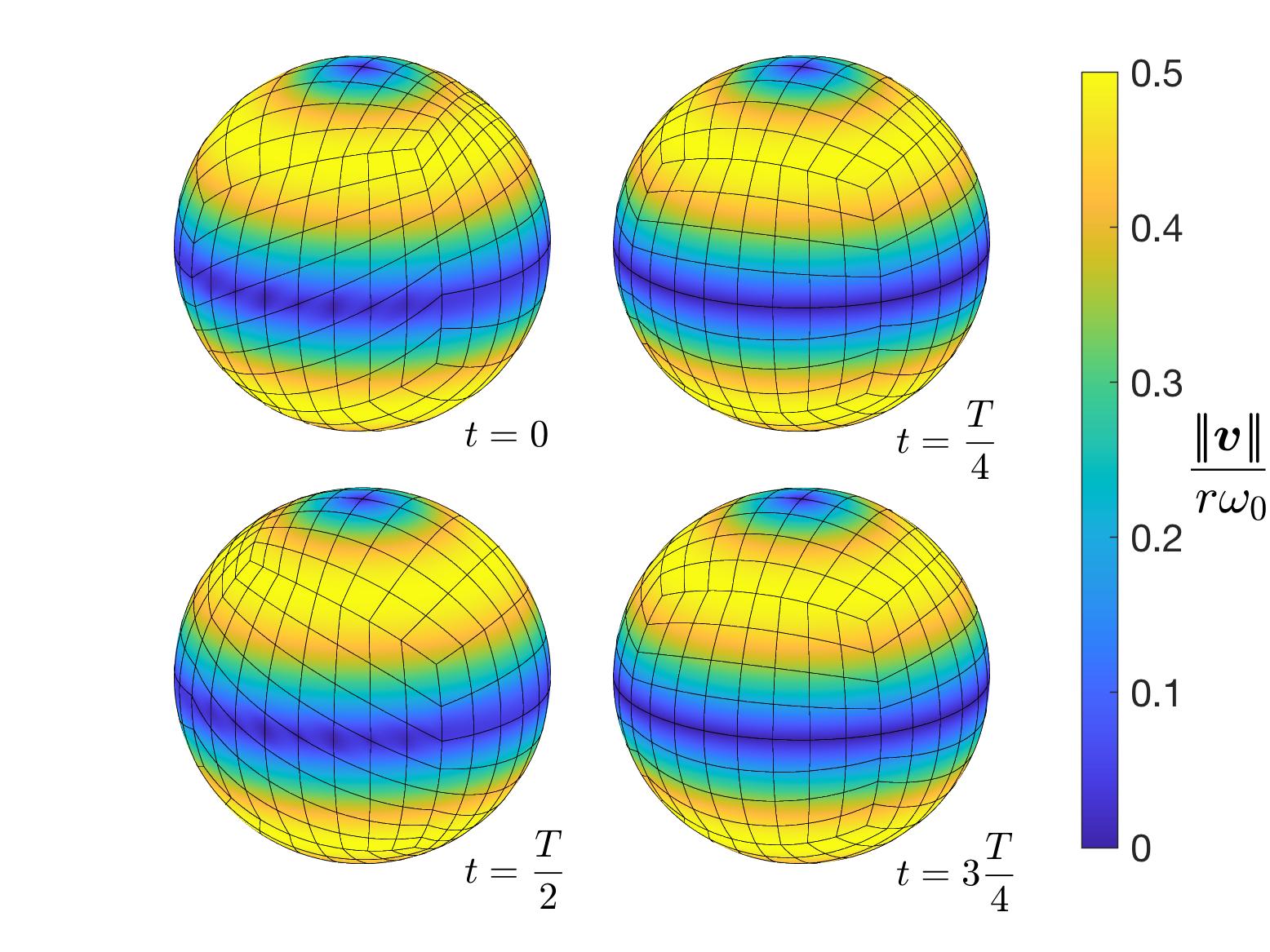}}
\put(0.2,-.3){\includegraphics[height=58mm]{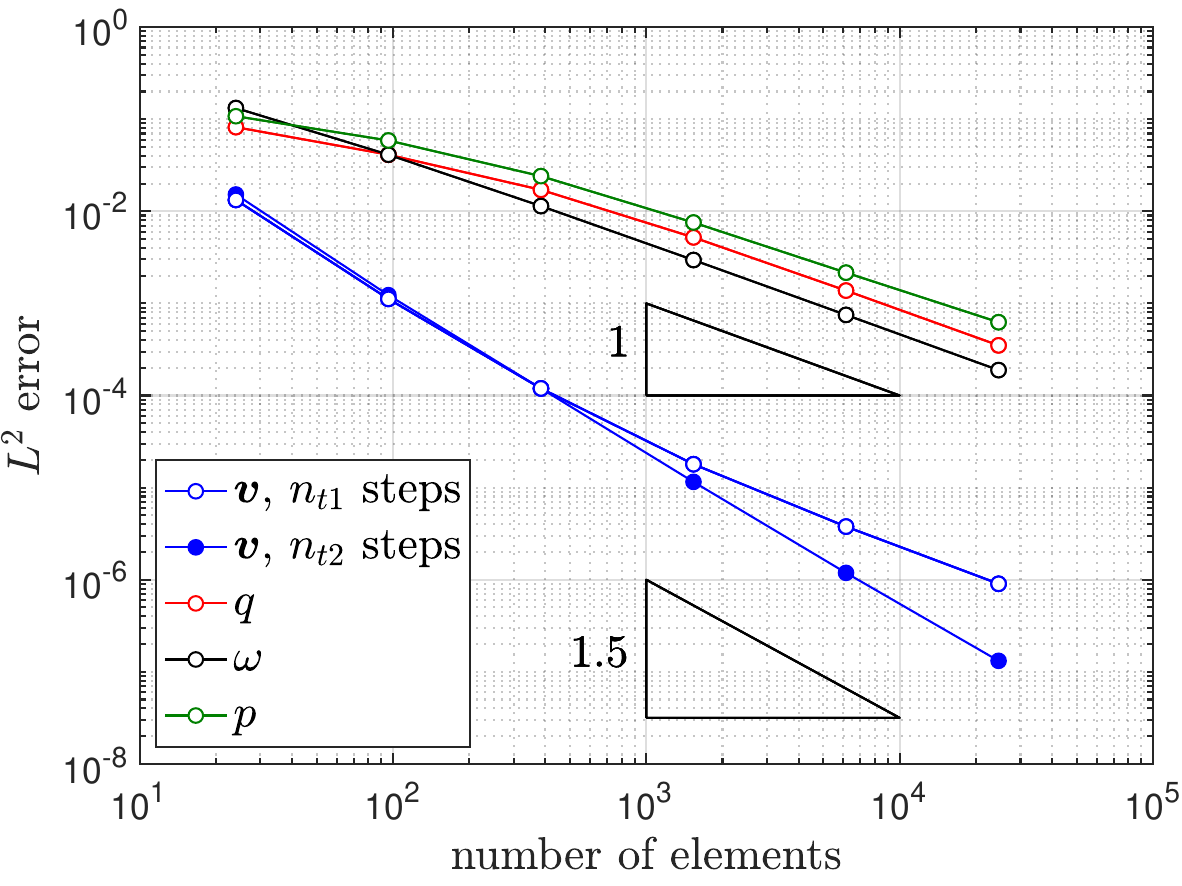}}
\put(-7.95,-.15){\footnotesize (a)}
\put(0.2,-.15){\footnotesize (b)}
\end{picture}
\caption{Simple shear flow on a rigid sphere: 
(a) Flow field $\bv$ and (b) convergence behavior due to the prescribed transient mesh motion $\bx_\mrm(t)$ given in \eqref{e:xm1} \& \eqref{e:tm1} (ALE case d, load case 1). 
Again, optimal convergence rates are obtained if sufficiently many time steps are used.
Here, the $L^2$ error from Eq.~\eqref{e:L2} is averaged over $n_t/4$ time steps.}
\label{f:shearflow5}
\end{center}
\end{figure}

Next, the transient mesh motion
\eqb{l}
\theta(t) = \Theta + \theta_0 \cos(\omega_\mrm t) \sin\phi\cos^2\Theta
\label{e:tm1}\eqe
is used inside Eq.~\eqref{e:xm1}.
It is shown at quarter intervals of the period $T = 2\pi/\omega_\mrm$ in Fig.~\ref{f:shearflow5}a for $\theta_0 = 1/2$ and $\omega_\mrm = \omega_0$ (the parameter from Eq.~\eqref{e:shearflowv}).
%----------------------------------------------------------------------------------------------------------------------------------
% run sShearFlow/ vShearFlowALEc mat{1}.case = 1 & pShearFlowALEc2.m & pShearFlowALEc.m
%----------------------------------------------------------------------------------------------------------------------------------
Since $\be_{r,\theta} = \be_\theta$ (see \citet{ALEtheo}, Appendix E.2), this mesh motion leads to the spatially and temporally varying mesh velocity
\eqb{l}
\bv_\mrm = \theta' r\, \be_\theta\,,
\label{e:vm1}\eqe
with
\eqb{l}
\theta'(t) = - \theta_0\,\omega_\mrm \sin(\omega_\mrm t) \sin\phi\cos^2\Theta\,,
\label{e:tpm1}\eqe
that has no out-of-plane component.
The maximum mesh velocity for $\theta_0 = 1/2$ is $v_{\mrm0} = \omega_\mrm r/2$.
The convergence behavior of load case 1 with this superposed mesh motion is shown in Fig.~\ref{f:shearflow5}b.
The trapezoidal rule is used for time integration, which is second order accurate in time.
If the time step size $\Delta t$ is refined proportionally to the element size, see $n_{t1}$ in Table~\ref{t:shearflow}, the convergence rate in $\bv$ is therefore only quadratic, i.e.~$O(h^2) = O(n^{-1}_\mathrm{el})$.
In order to maintain $O(h^3)$ convergence, the time step needs to be proportional to $h^{1.5}$, i.e.~proportional to $m^{-1.5}$.
This is confirmed in Fig.~\ref{f:shearflow5}b for the time step number $n_{t2}$ given in Table~\ref{t:shearflow}. 

Load case 2 behaves similarly, as Fig.~\ref{f:shearflowA1}d in Appendix~\ref{s:add} shows: 
Again optimal convergence rates are obtained.

Thus it is possible to maintain ideal convergence rates even for transient mesh motions.
It is believed that this only works because in this example $\bv'$ lies in the tangent plane: 
From \citet{ALEtheo}, Eq.~(153), follows
\eqb{l}
\bv' = \bv_{\!,\alpha} v_\mrm^\alpha = \theta'\bv_{\!,2} =  \theta' r \omega_0 (2\cos^2\theta-1)\,\be_\phi
\eqe 
for the velocity in Eq.~\eqref{e:shearflowv}.
This is different for the example of Sec.~\ref{s:octflow}, where $\bv'$ has out-of-plane components -- even for a fixed surface -- and where the convergence rate for $\bv$ is only $O(h^2)$, see Fig.~\ref{f:octoflow5}.

\subsubsection{Unknown normal velocity $v_\mrn$ (Load cases 3 \& 4)}\label{s:NBC1}

The preceding results of Secs.~\ref{s:shearALE0}-\ref{s:shearALEc} are for out-of-plane Dirichlet BC (load cases 1 \&~2), where the out-of-plane flow velocity is prescribed through the given mesh velocity ($v_\mrn = v_{\mrm\mrn}$). 

Next we consider out-of-plane Neumann conditions, where the out-of-plane flow velocity $v_\mrn$ is unknown (load case 3 \& 4, see Table~\ref{t:shearflowcases}).
In this case the surface velocity is unconstrained and thus a full 3D vector that can have out-of-plane components in the iterative solution process, even for cases with solution $v_\mrn=0$, as is considered here.
It is thus a test to check whether unknown $v_\mrn$ can be handled by the proposed numerical scheme. 

\begin{figure}[h]
\begin{center} \unitlength1cm
\begin{picture}(0,11.5)
\put(-8,5.7){\includegraphics[height=58mm]{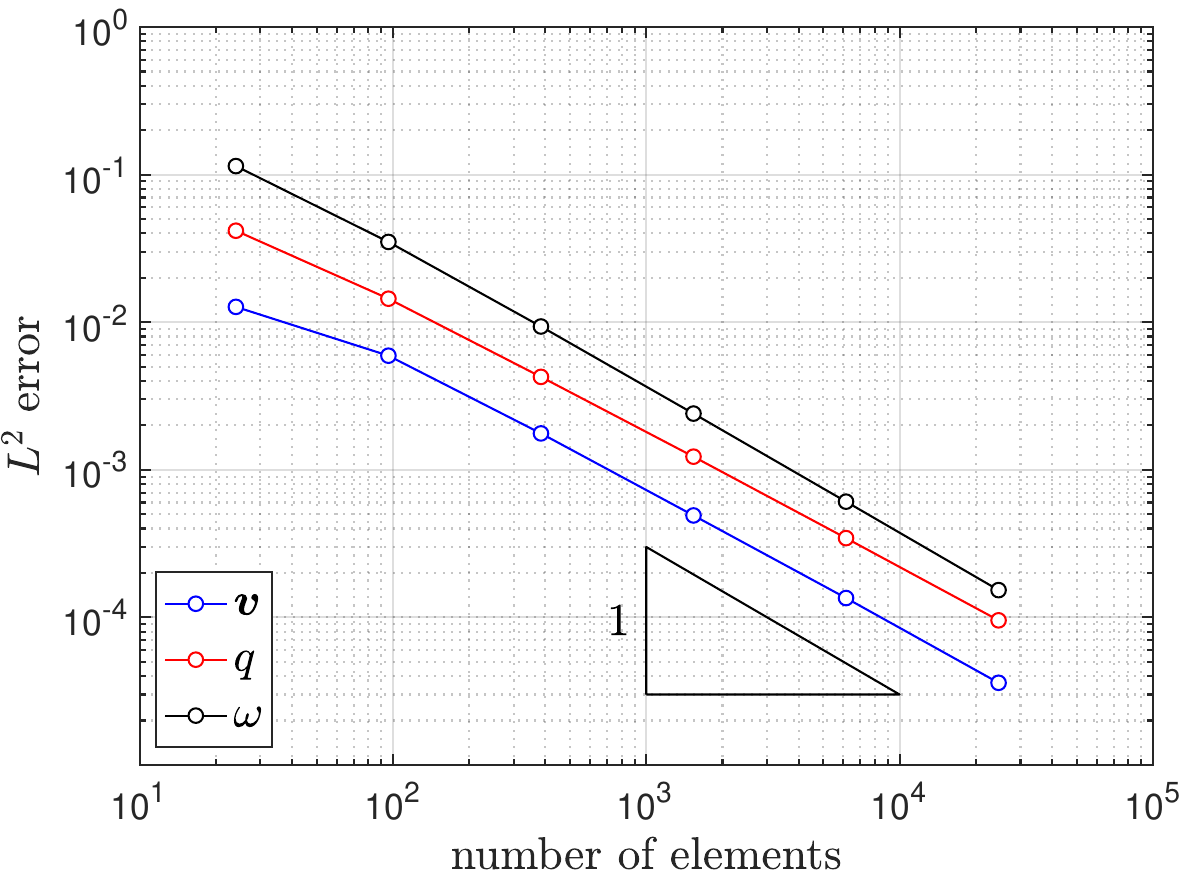}}
\put(0.2,5.7){\includegraphics[height=58mm]{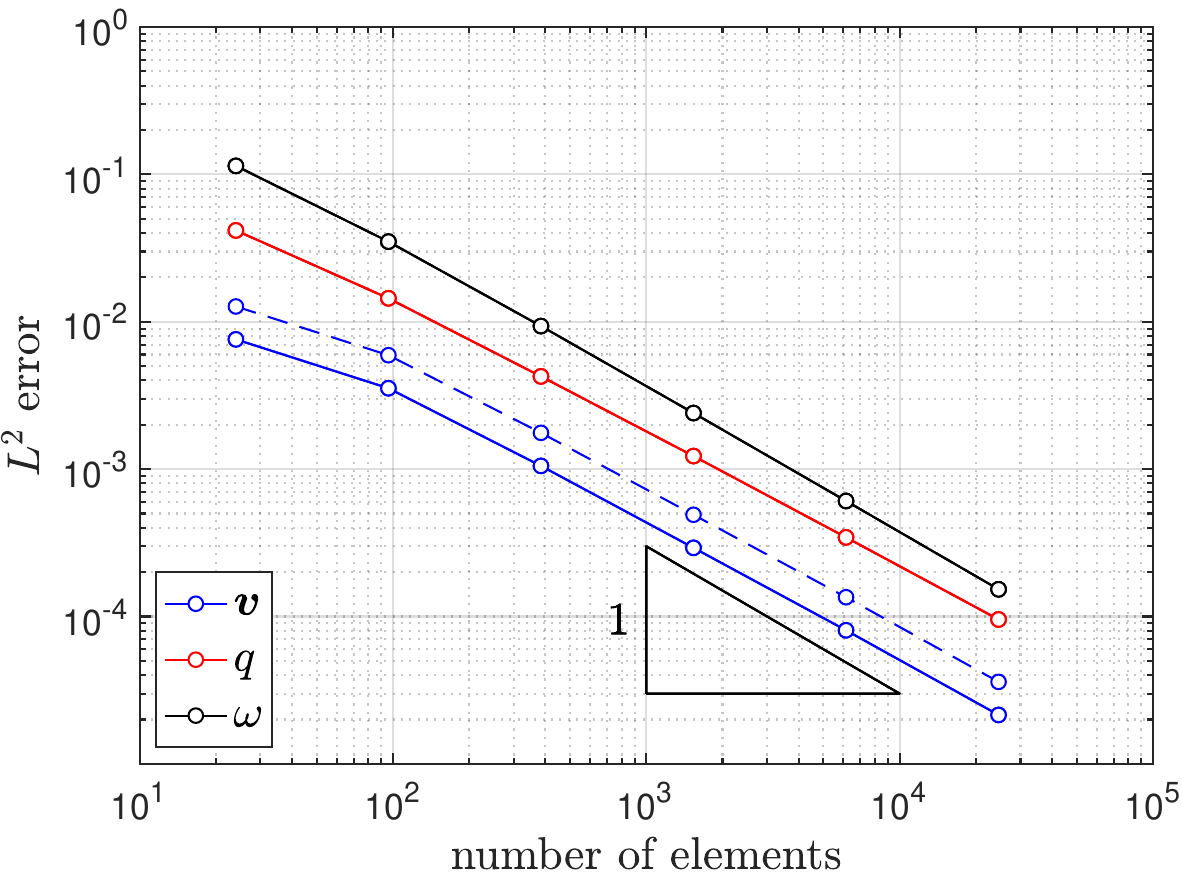}}
\put(-8,-.3){\includegraphics[height=58mm]{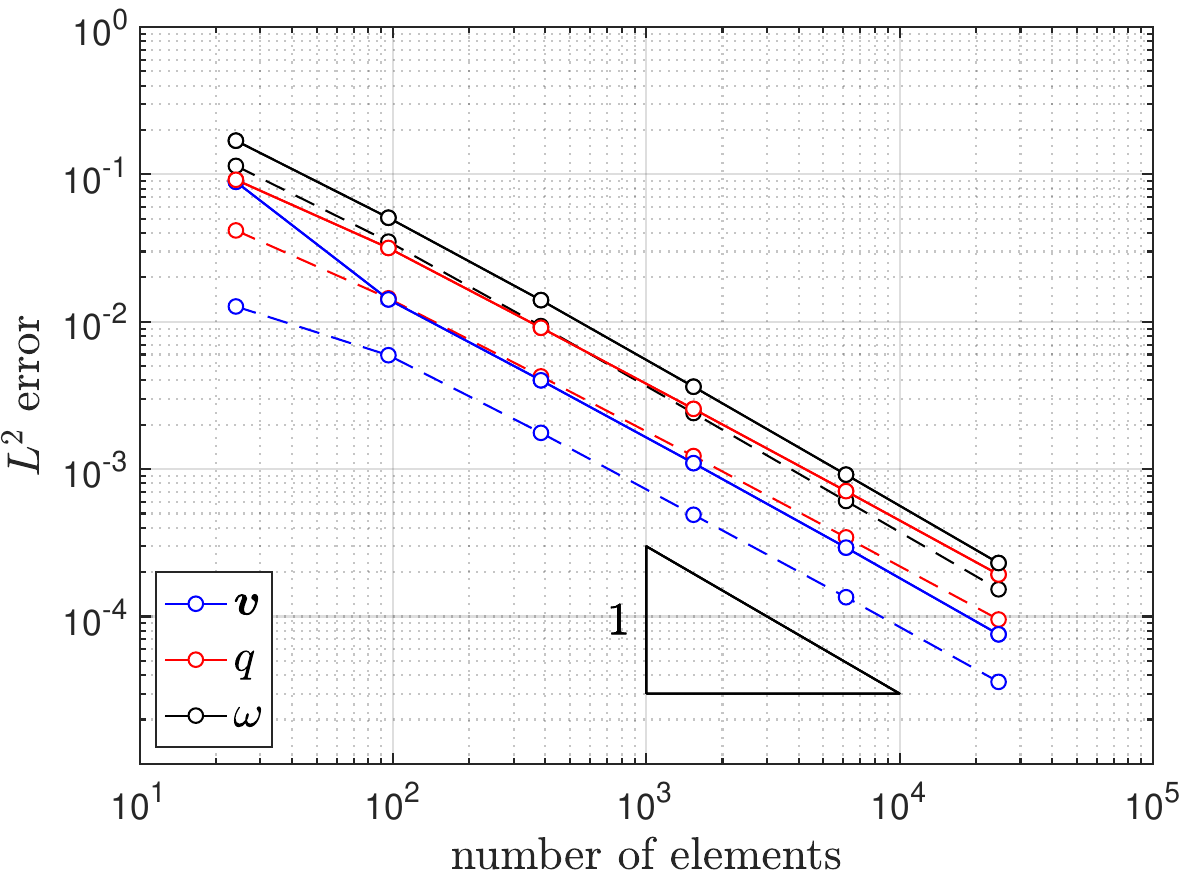}}
\put(0.2,-.3){\includegraphics[height=58mm]{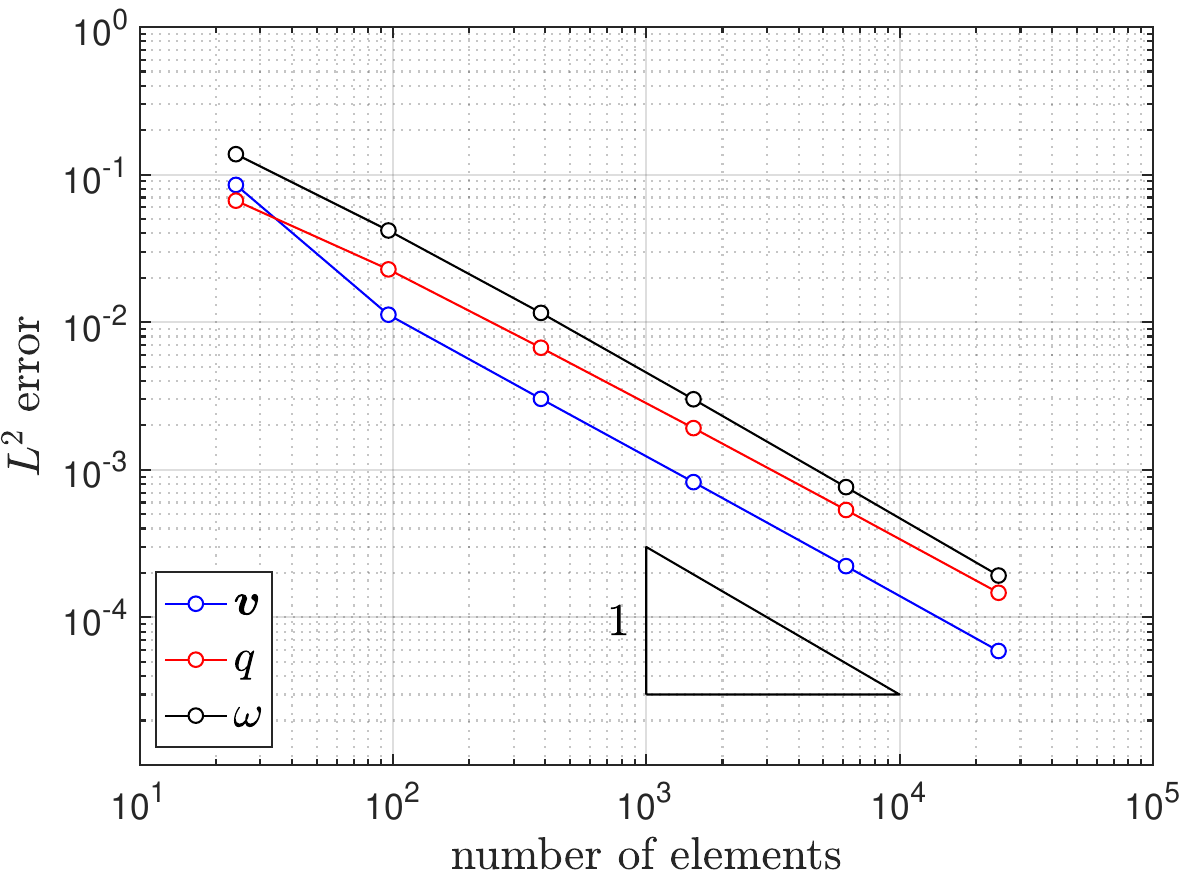}}
\put(-7.95,5.85){\footnotesize (a)}
\put(0.2,5.85){\footnotesize (b)}
\put(-7.95,-.15){\footnotesize (c)}
\put(0.2,-.15){\footnotesize (d)}
\end{picture}
\caption{Simple shear flow on a rigid sphere for load case 3: 
Convergence of velocity, surface tension and vorticity for the four ALE cases (a-d) in the $L^2$ error norm.
All convergence rates are $O(h^2)=O(n_\mathrm{el}^{-1})$.
The surface pressure error is at machine precision and hence not shown.
The dashed lines in (b) \& (c) are the results from (a).}
\label{f:shearflow6}
\end{center}
\end{figure}
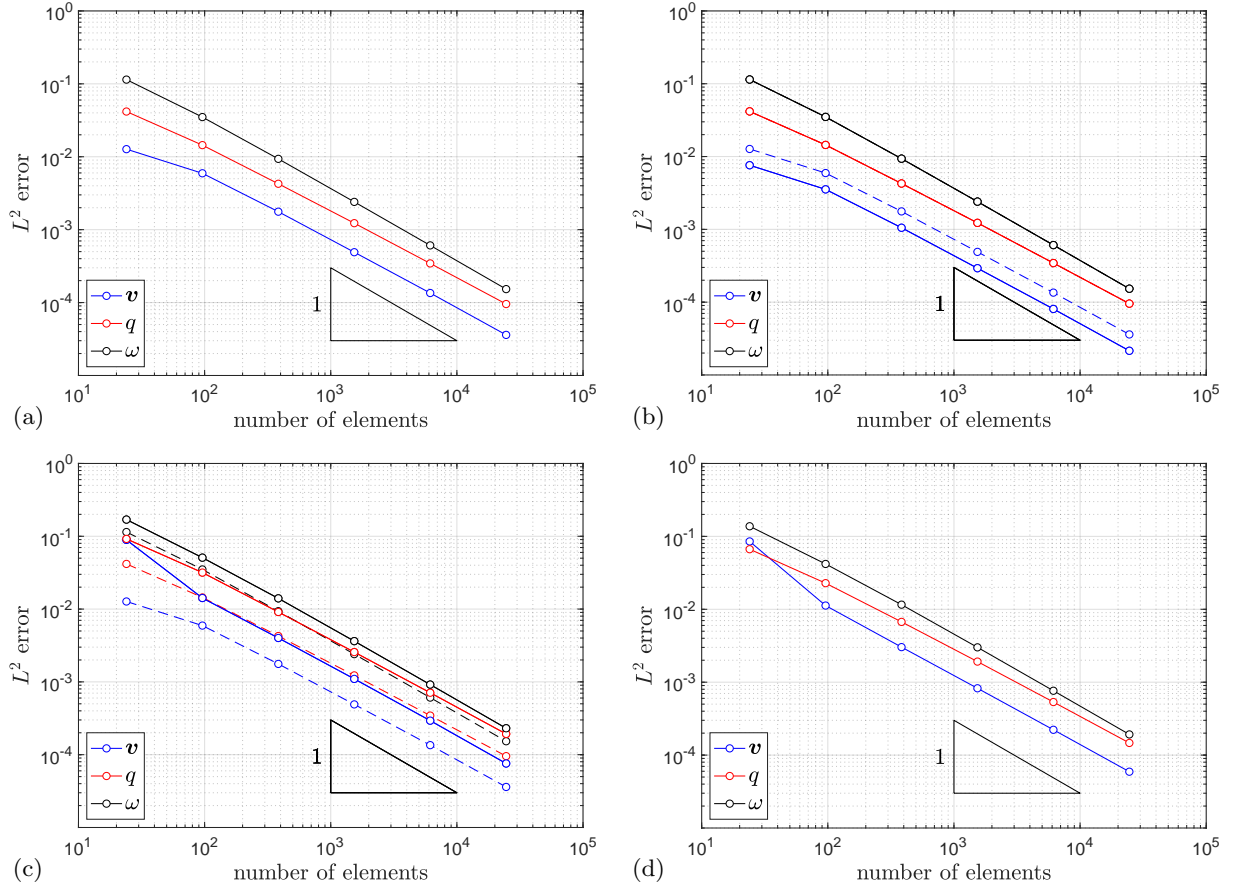

The first surface pressure that leads to zero\footnote{or $v_\mrn=v_\mathrm{mn}$ in ALE case b, see Table~\ref{t:shearflowcases}} $v_\mrn$ is 
\eqb{l}
\bar p = p_\mrp + \rho r \omega_0^2\bigg(\ds\frac{3}{2}\sin^4\theta -\sin^2\theta - \frac{1}{2}\bigg)\,,\quad p_\mrp = \ds\frac{2q_\mrp}{r}
\label{e:shearflowp3}\eqe
\citep{ALEtheo}, where $p_\mrp=\bar p(\theta=\pm\pi/2)$ denotes the pole pressure.
It is denoted load case~3.
Here, $q_\mrp = \rho\,r^2\,\omega_0^2/2$ is chosen such that the pole pressure becomes $p_\mrp = \rho r \omega_0^2$, while the equatorial pressure is $p_\mathrm{eq} = \bar p(\theta=0) = p_\mrp/2$.
The pressure distribution of Eq.~\eqref{e:shearflowp3} is then equal to the one shown in Fig.~\ref{f:shearflow2}c.
The convergence rates for this load case are shown in Fig.~\ref{f:shearflow6} for the four ALE cases.
%----------------------------------------------------------------------------------------------------------------------------------
% run pShearFlow.m & pShearFlowALEc.m
% data from sShearFlow/ vShearFlow*p0 mat{1}.case = 3
%----------------------------------------------------------------------------------------------------------------------------------
The velocity rates are only $O(h^2)$ now, while the other rates are as before.
It is believed this is due to the fact that the weak form for the unknown fluid velocity $\bv$ depends on the surface derivatives $\ba_\alpha=\bx_{,\alpha}$, e.g.~through $\mD_e$ and $\ba^{\alpha\beta}_\mrs$ in $\mf^e_\mathrm{int}$, see Eqs.~\eqref{e:felint}-\eqref{e:felvisc}. 
Since the accuracy of surface $\bx$ is $O(h^3)$, the surface derivatives $\ba_\alpha$ are only $O(h^2)$, as differentiation reduces an order.
(The $O(h^3)$ convergence rate of $\bx$ can be seen in the example of Fig.~\ref{f:evolveflow1}.)
For ALE case c, the time step $\Delta t$ is refined proportionally to $h$ so that all convergence rates are optimal, i.e.~$O(h^2)$ for the trapezoidal rule (and the surface order reduction noted above). 

The second surface pressure that leads to zero$^3$ $v_\mrn$ is
\eqb{l}
\bar p = p_\mrp - \rho r \omega_0^2\sin^2\theta\cos^2\theta\,,\quad p_\mrp = \ds\frac{2q_\mrp}{r}
\eqe
\citep{ALEtheo}.
It has constant surface tension across the surface and is denoted load case 4, see Table~\ref{t:shearflowcases}.
Picking $q_\mrp = \rho\,r^2\,\omega_0^2/2$ leads to the polar and equatorial pressures $p_\mrp = p_\mathrm{eq} = \rho r \omega_0^2$. 
The convergence rates for this load case are very similar to the previous load case, as Fig.~\ref{f:shearflowA2} shows.
Again, convergence rates are optimal and the surface pressure error is at machine precision.

\subsection{Octahedral vortex flow}\label{s:octflow}

The second example in Sec.~\ref{s:Nexf} considers surface flow defined by the stream function
\eqb{l}
\psi = v_0\,r\,\sin2\phi\,\sin\theta\,\cos^2\!\theta\,.
\label{e7:psi}\eqe
The resulting velocity field $\bv$ and required body force field are derived analytically in \citet{ALEtheo} assuming constant surface tension $q$.
Contrary to the preceding example, $\bv$ is now non-axisymmetric and has components along both $\phi$ and $\theta$, and those vary along both $\phi$ and $\theta$.
The same parametrization and meshes as before are used, see Table~\ref{t:shearflow}.
Combining different boundary conditions and ALE cases leads to the seven cases shown in Table~\ref{t:octoflowcases}.
%-------------------------------------------------------------------------------------------------------------------------------
% TRANSLATION of ALE cases: code - paper
% 				  	      	   0	     a
% 					    	   a	     b
%					    	   b	     c
%					    	   c  	     d
\begin{table}[h]
\centering
\begin{tabular}{|l|ll|l|l|}
  \hline
    & & \hspace{-4mm} load case & 2 & 4 \\[0mm] \hline
   ALE & & & $q=$ const. & $q=$ const., \\[0mm]
   case & &  & DBC $v_\mrn=0$ & NBC $p=\bar p$  \\[0mm] \hline
   & & & & \\[-4mm]   
   a &  $\bv_\mrm=\mathbf{0}$, & $\!\!\!\theta_0=0$ & Fig.~\ref{f:octoflow1} & Fig.~\ref{f:octoflow1}  \\ [0mm] 
   b &  $\bv_\mrm=\bc_0$, & $\!\!\!\theta_0=0$ & Fig.~\ref{f:octoflow3} & Fig.~\ref{f:octoflow3} \\ [0mm] 
   c &  $\bv_\mrm=\mathbf{0}$, & $\!\!\!\theta_0=0.5$ & Fig.~\ref{f:octoflow4} & Fig.~\ref{f:octoflow4} \\ [0mm] 
   d &  $\bv_\mrm=\bv_\mrm(t)$, & $\!\!\!\theta_0=0.5$ & -- & Fig.~\ref{f:octoflow5} \\ [0mm] 
   \hline
\end{tabular}
\caption{Octahedral vortex flow on a rigid sphere: Considered ALE mesh motion cases and surface load cases.
The flow field $(\bv,q)$ is the same for both load cases, but only in load case 4, $v_\mrn$ is treated unknown.
It is nonzero in ALE case b.
The ALE cases are the same as used in the previous example (see Sec.~\ref{s:shearflow}).
DBC = Dirichlet boundary condition, NBC = Neumann boundary condition. \\
}
\label{t:octoflowcases}
\end{table}
%-------------------------------------------------------------------------------------------------------------------------------
For out-of-plane NBC (load case 4), $v_\mrn$ is an unknown that needs to be solved for, even if the applied surface pressure $\bar p$ is the one that gives $v_\mrn=0$.
According to \citet{ALEtheo}, that pressure follows from the velocity field $\bv$ as 
\eqb{l}
\bar p = p_\mrp - \ds\frac{\rho}{r}\bv\cdot\bv\,,
\label{e:pocto}\eqe
where $p_\mrp$ is the pole pressure related to the constant surface tension $q$ by the Young-Laplace equation $p_\mrp = 2q/r$.
The cases of Table~\ref{t:octoflowcases} are examined in the following sub-sub-sections using a unit value for $q$.

\subsubsection{Zero mesh velocity (ALE case a)}\label{s:octoALE0} 

\begin{figure}[h]
\begin{center} \unitlength1cm
\begin{picture}(0,5.6)	
\put(-8.8,-.22){\includegraphics[height=60mm]{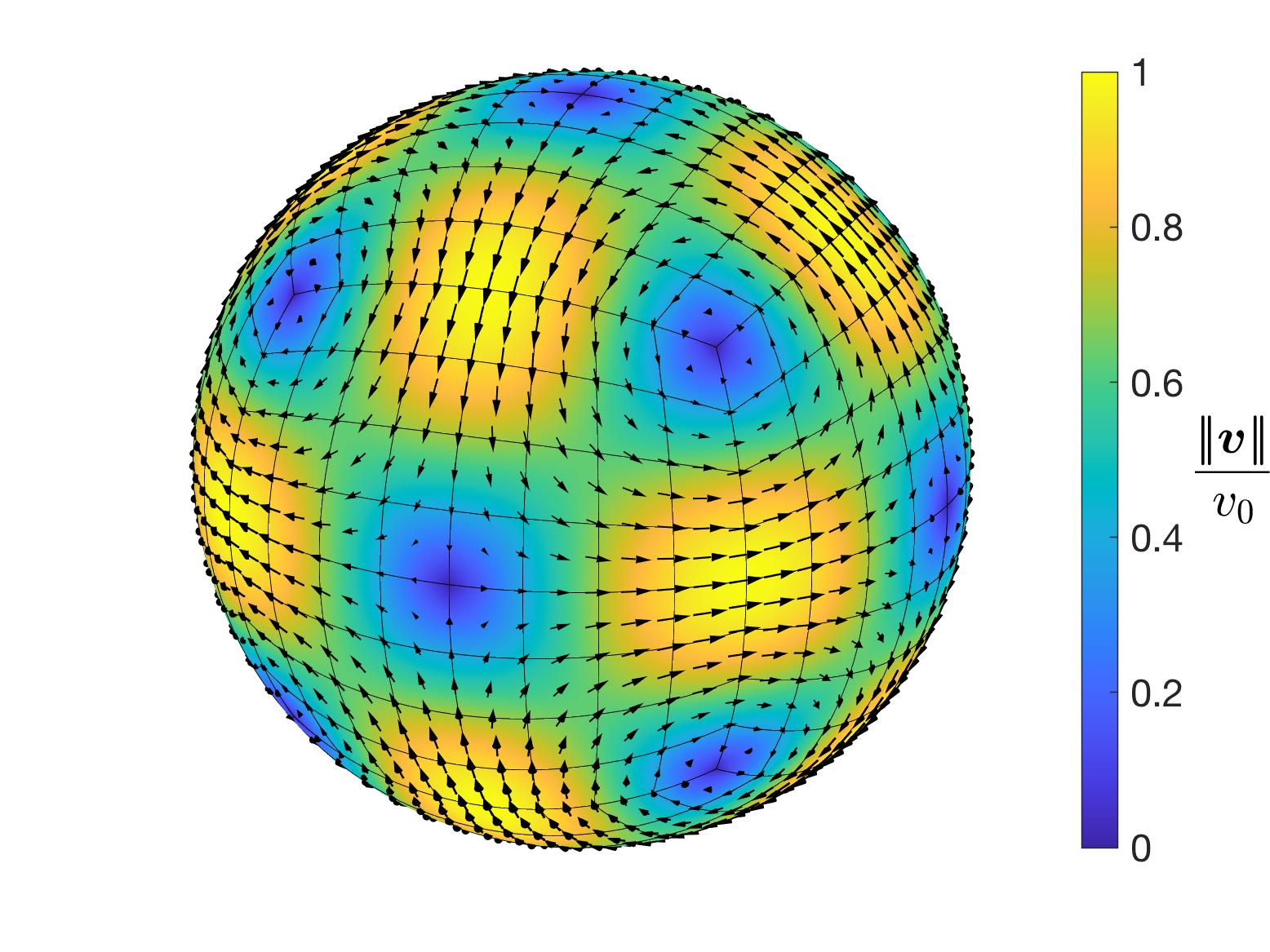}}
\put(0.2,-.3){\includegraphics[height=58mm]{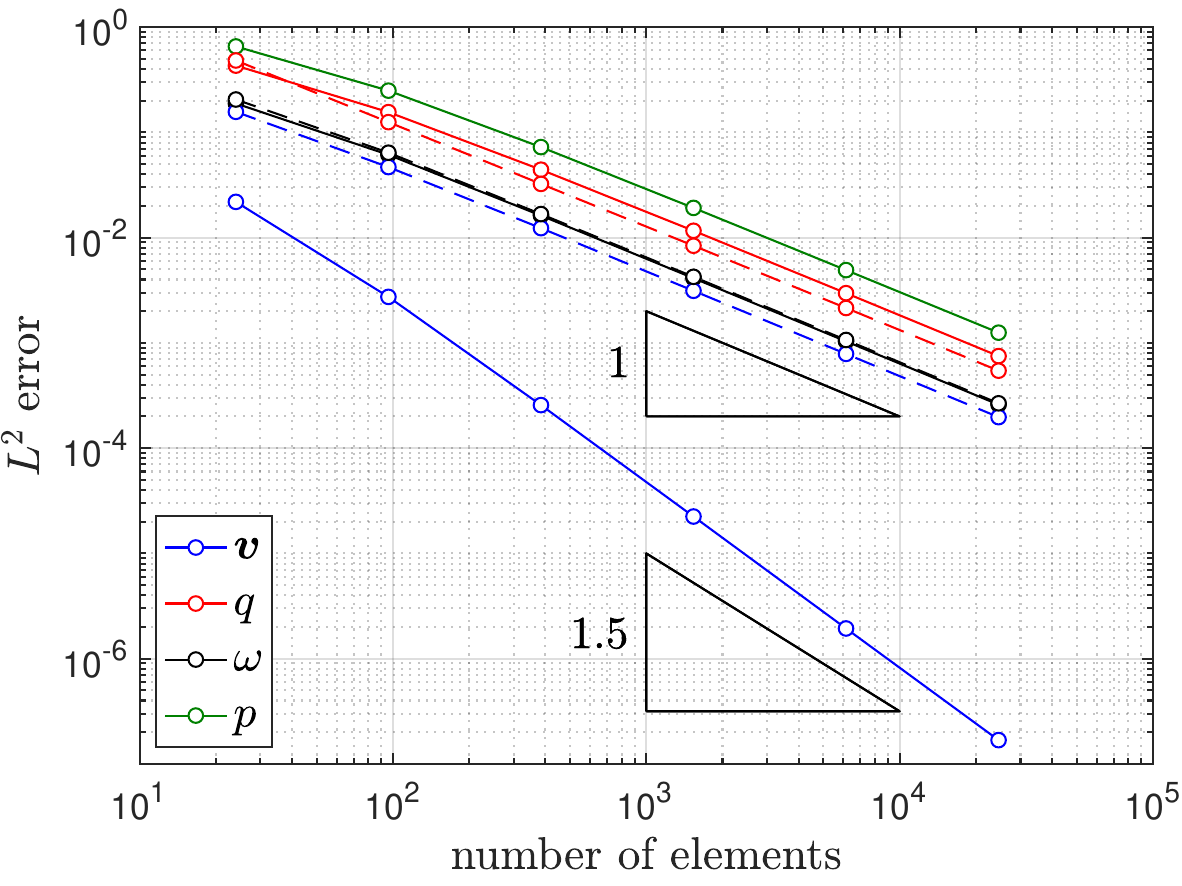}}
\put(-7.95,-.15){\footnotesize (a)}
\put(0.2,-.15){\footnotesize (b)}
\end{picture}
\caption{Octahedral vortex flow on a rigid sphere: 
(a) Flow field $\bv$ and (b) convergence behavior for ALE case~a.
Solid lines are for load case 2; dashed lines for load case 4. ($p$ is prescribed in the latter case and hence does not show up in the plot.)
All convergence rates are optimal: At least $O(h^3)=O(n^{-1.5}_\mathrm{el})$ for $\bv$ in load case 2, and $O(h^2)=O(n^{-1}_\mathrm{el})$ for all other fields.
The convergence rate in $\bv$ reduces to $O(h^2)$ for load case 4 due to its surface gradient dependency.}
\label{f:octoflow1}
\end{center}
\end{figure}

\begin{figure}[h]
\begin{center} \unitlength1cm
\begin{picture}(0,5.2)
\put(-8.8,-.7){\includegraphics[height=60mm]{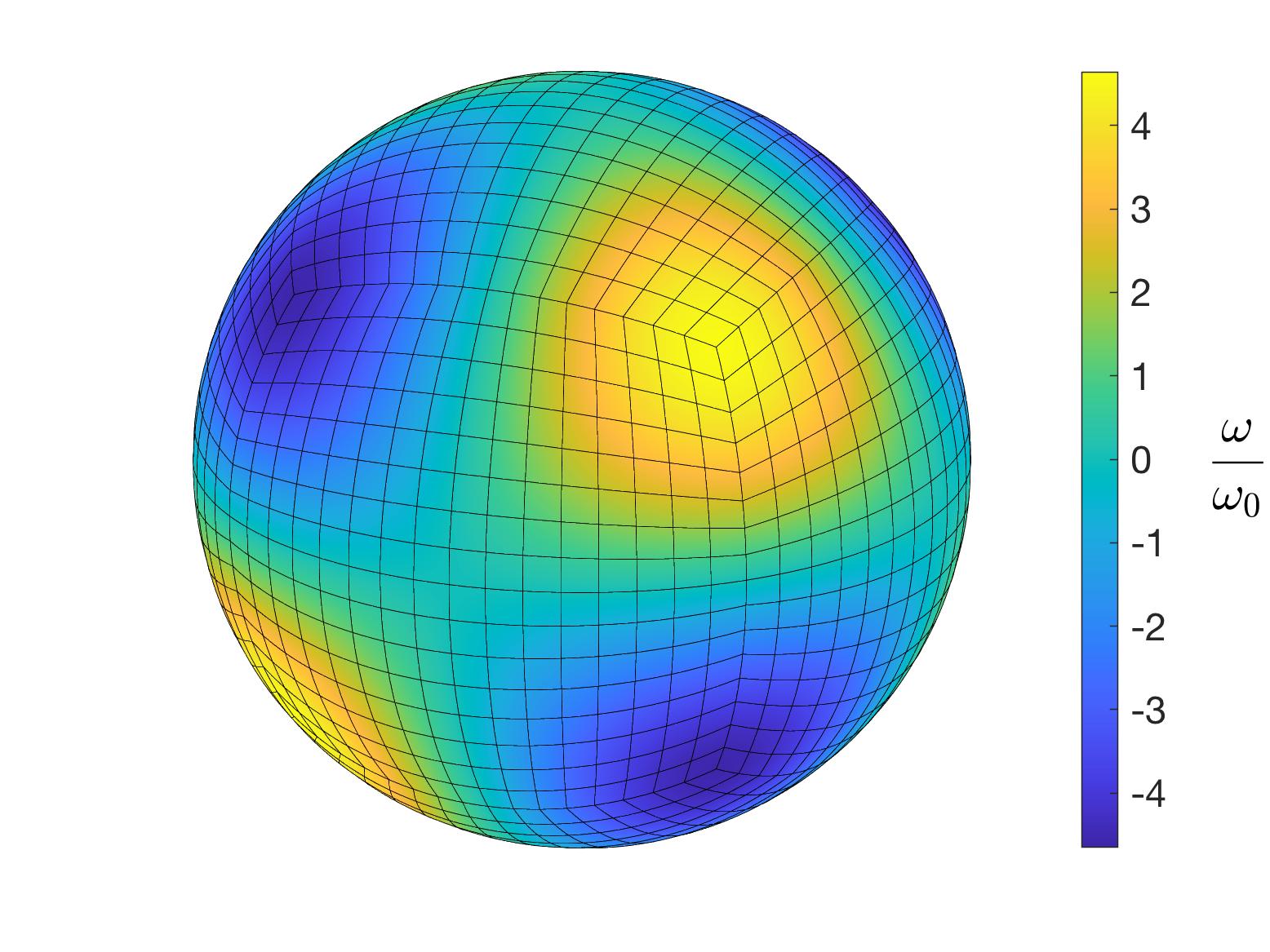}}
\put(-0.4,-.7){\includegraphics[height=60mm]{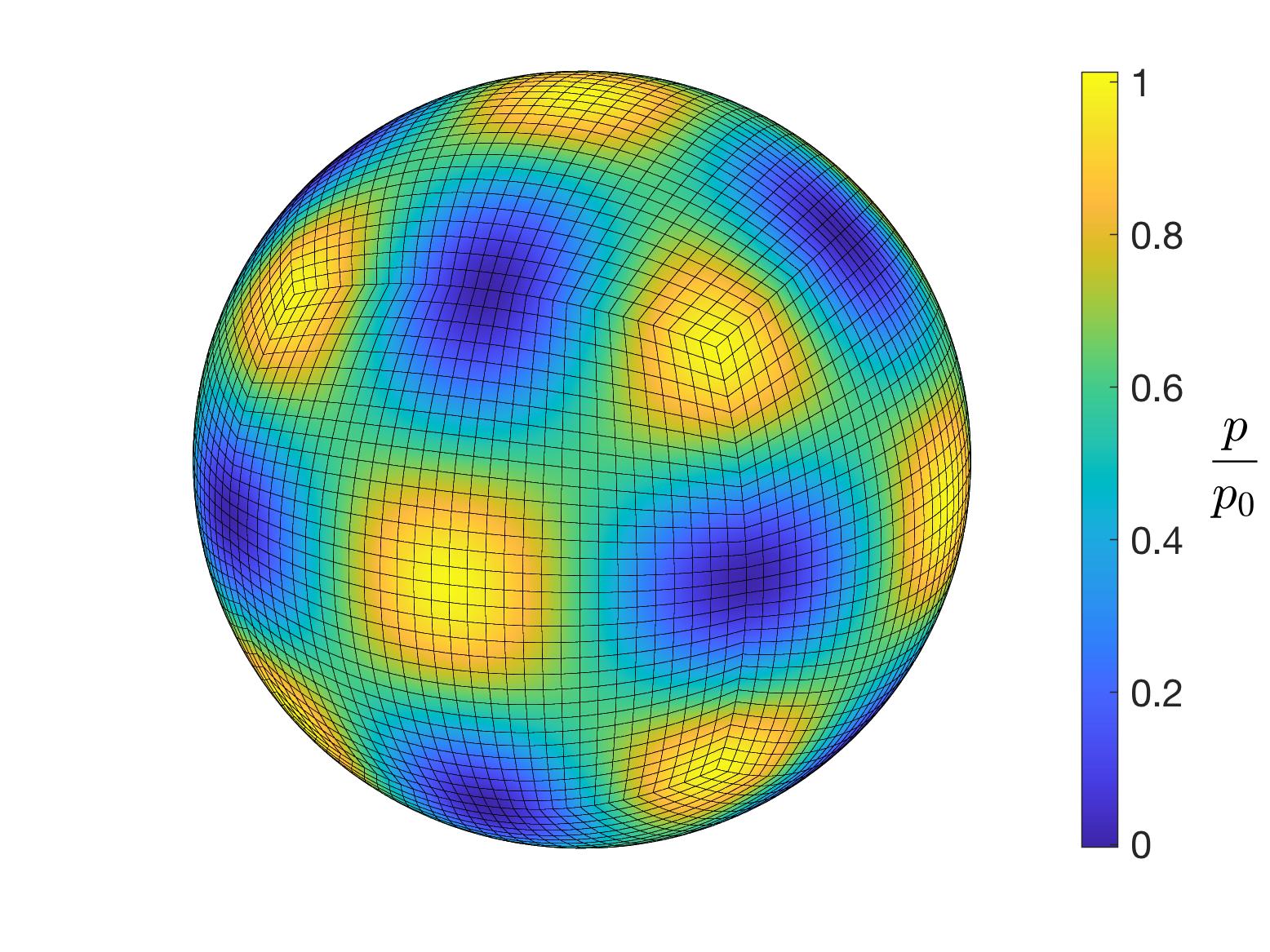}}
\put(-7.95,-.15){\footnotesize (a)}
\put(0.45,-.15){\footnotesize (b)}
\end{picture}
\caption{Octahedral vortex flow on a rigid sphere: 
(a) Vorticity $\omega$ and (b) lateral surface pressure~$p$. 
A much finer mesh is required for $\omega$ and $p$ to be of comparable accuracy as $\bv$.
Here $\omega_0 = v_0/r$ and $p_0 = p_\mrp$.}
\label{f:octoflow2}
\end{center}
\end{figure}

The first ALE case considers no mesh motion and no mesh distortion.
Fig.~\ref{f:octoflow1}a shows the velocity field for this case.
%-----------------------------------------------------------------
% run sOctoFlow
%-----------------------------------------------------------------
It is caused by an elaborate manufactured surface force \citep{ALEtheo}.
Fig.~\ref{f:octoflow2} shows the vorticity $\omega$ and the outward surface pressure $p$ required to hold the surface in place 
($p$ is the lateral boundary reaction for case 2, and the prescribed pressure for case 4; it is equal to $\bar p$ in \eqref{e:pocto}). 
%-----------------------------------------------------------------
% run sOctoFlow
%-----------------------------------------------------------------
A much finer mesh is needed for $\omega$ and $p$ to be of comparable accuracy as $\bv$, especially for load case 2.
This can be seen from the convergence plots of Fig.~\ref{f:octoflow1}b.
The convergence rates are as in the previous example (see Sec.~\ref{s:shearflow}): 
For the Dirichlet case (load case 2, marked in solid lines) $\bv$ converges at least cubically, while for the Neumann case (load case 4, marked dashed) $\bv$ converges only quadratically due to the surface gradient dependency noted before.
The direct comparison between the two load cases, shows that $q$ is slightly more accurate for the Neumann case.

\subsubsection{Constant mesh velocity (ALE case b)}\label{s:octoALEa}

\begin{figure}[h]
\begin{center} \unitlength1cm
\begin{picture}(0,5.6)
\put(-8.8,-.22){\includegraphics[height=60mm]{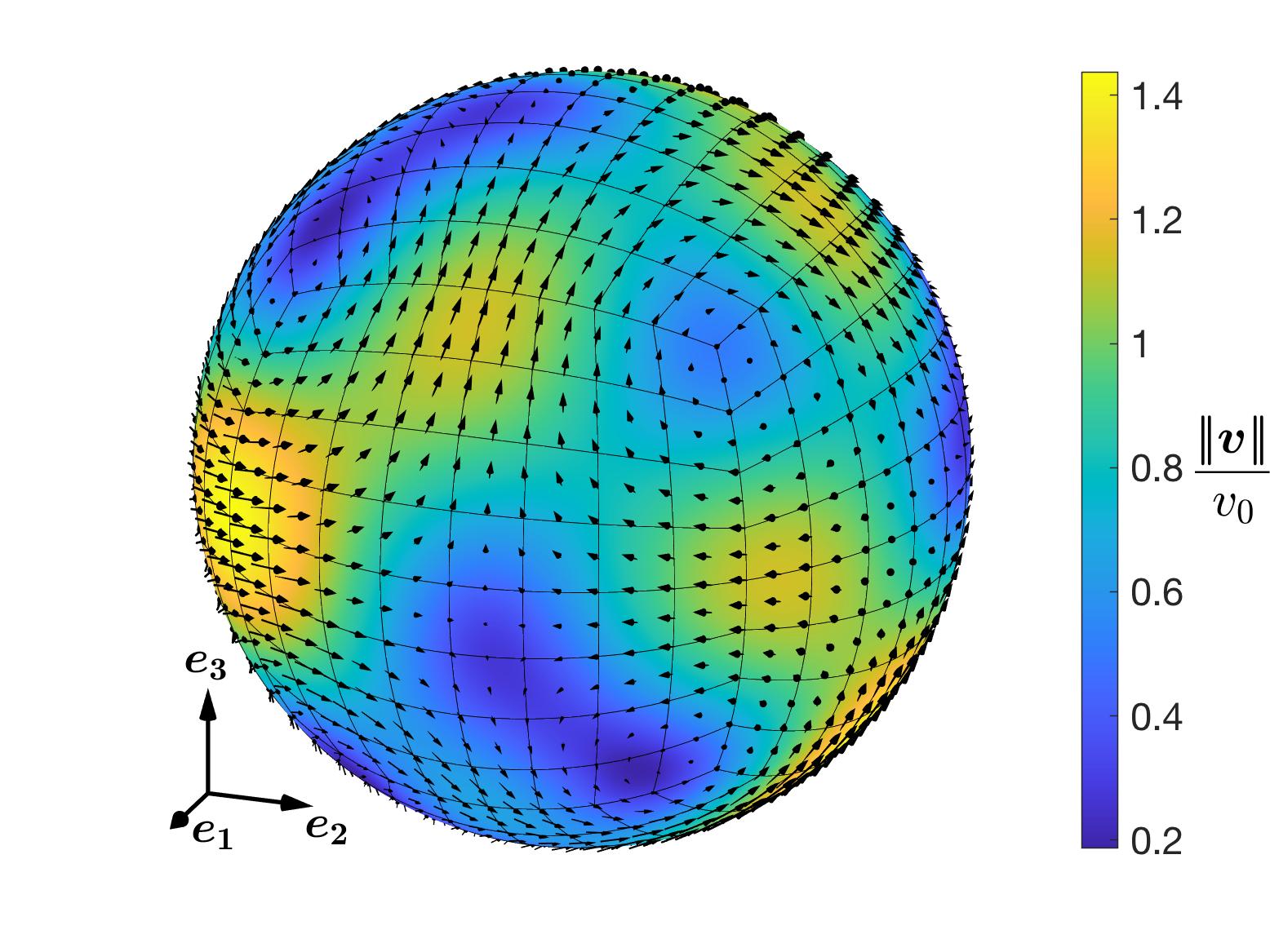}}
\put(0.2,-.3){\includegraphics[height=58mm]{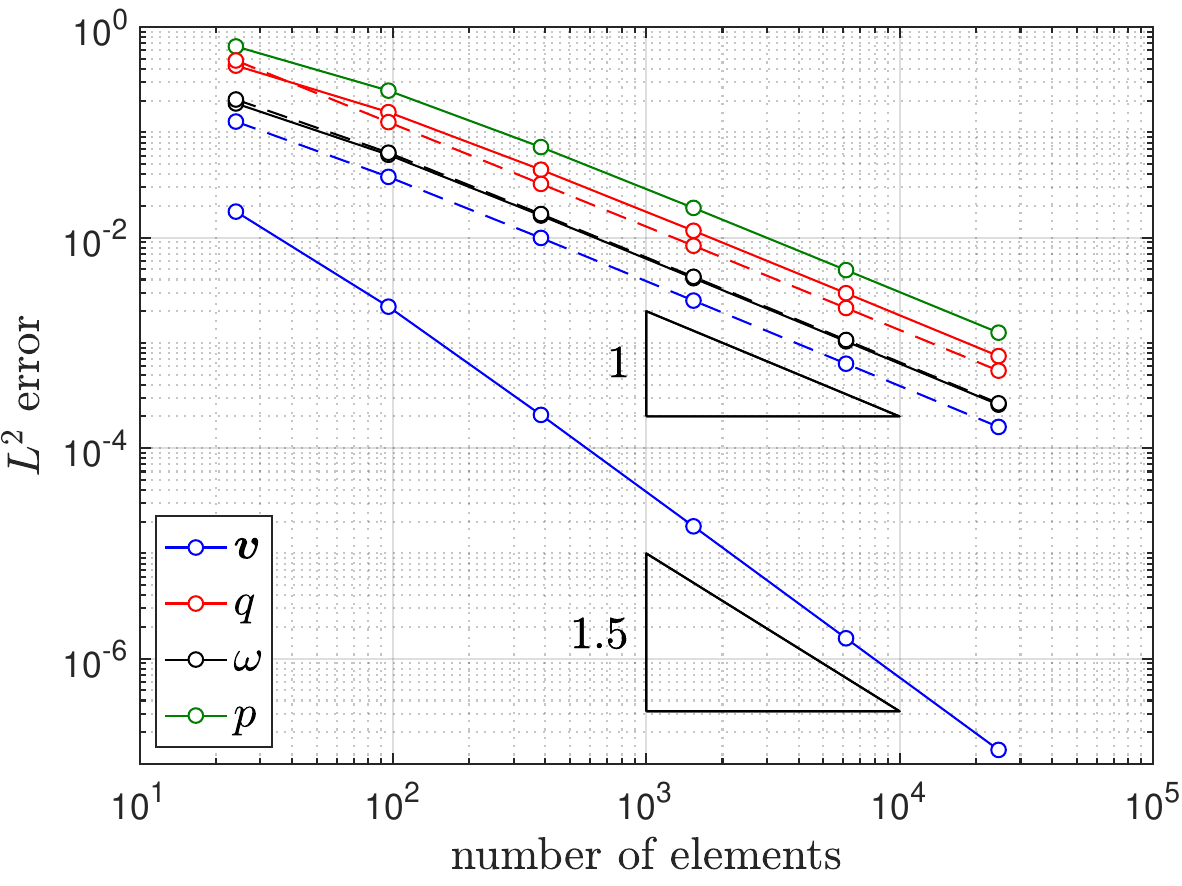}}
\put(-7.95,-.15){\footnotesize (a)}
\put(0.2,-.15){\footnotesize (b)}
\end{picture}
\caption{Octahedral vortex flow on a rigid sphere: 
(a) Flow field $\bv$ and (b) convergence behavior due to the prescribed constant mesh velocity $\bv_\mrm=\bc_0$ given in \eqref{e:vm0} (ALE case b).
In this case, $\bv_\mrm$, and hence also $\bv$, have out-of-plane components.
Solid lines are for load case 2; dashed lines for load case 4.
All convergence rates are optimal or better, as was also seen in Fig.~\ref{f:octoflow1}.}
\label{f:octoflow3}
\end{center}
\end{figure}

The second ALE case (defined by the constant mesh velocity of Eq.~\eqref{e:vm0}) has an out-of-plane velocity component.
Fig.~\ref{f:octoflow3} shows the resulting flow field and its convergence rates. 
%-----------------------------------------------------------------
% run sOctoFlow &  pOctoFlow.m
%-----------------------------------------------------------------
Again, only load case 2 attains at least cubic velocity convergence.
Compared to ALE case a, the error in fields $q$, $\omega$ and $p$ is equal, while $\bv$ has a slightly lower error now.
The added constant translation thus slightly improves the relative accuracy.
The same was observed in the example of Sec.~\ref{s:shearflow}.

\subsubsection{Fixed mesh distortion (ALE case c)}\label{s:octoALEb}

\begin{figure}[h!]
\begin{center} \unitlength1cm
\begin{picture}(0,5.6)
\put(-8.8,-.22){\includegraphics[height=60mm]{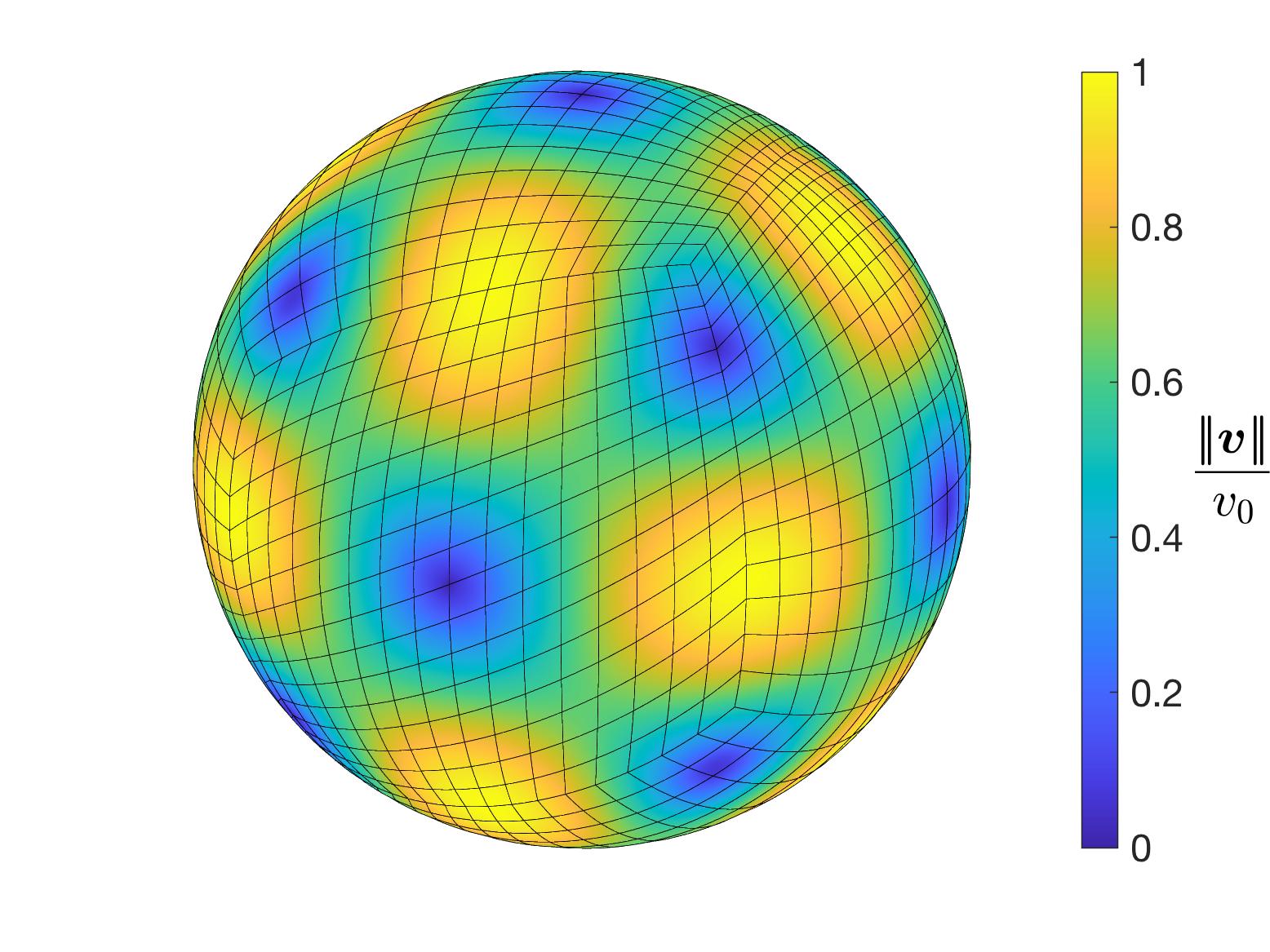}}
\put(0.2,-.3){\includegraphics[height=58mm]{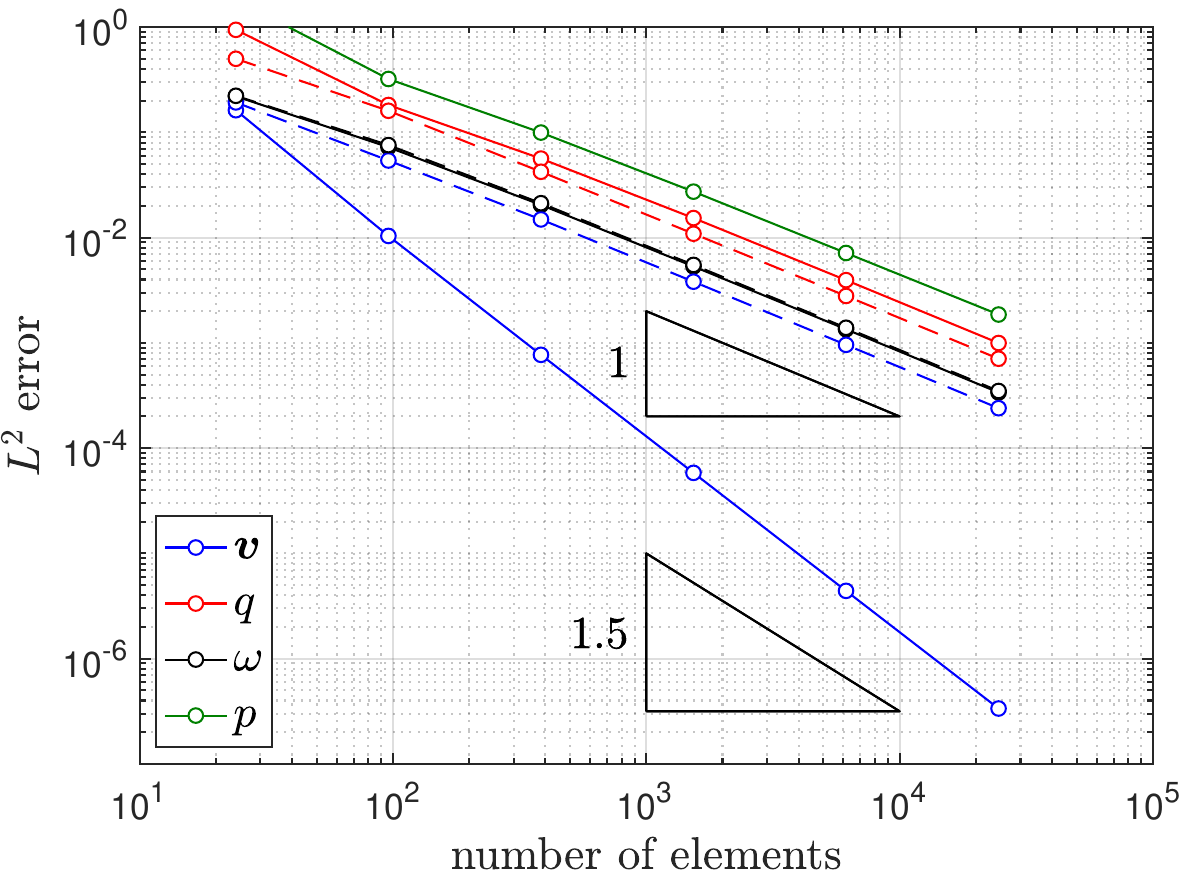}}
\put(-7.95,-.15){\footnotesize (a)}
\put(0.2,-.15){\footnotesize (b)}
\end{picture}
\caption{Octahedral vortex flow on a rigid sphere: 
(a) Flow field $\bv$ and (b) convergence behavior due to the prescribed fixed mesh distortion $\bx_\mrm$ given in \eqref{e:xm1} \& \eqref{e:tm0} (ALE case c).
Solid lines are for load case 2; dashed lines for load case 4.
All convergence rates are optimal or better, as seen in Fig.~\ref{f:octoflow1} before.}
\label{f:octoflow4}
\end{center}
\end{figure}

The third ALE case considers a fixed mesh distortion (defined by Eqs.~\eqref{e:xm1} \& \eqref{e:tm0}).
Fig.~\ref{f:octoflow4} shows the flow field and its convergence rates. 
%-----------------------------------------------------------------
% run sOctoFlow & pOctoFlow.m
%-----------------------------------------------------------------
Again, only load case 2 attains cubic velocity convergence.
Compared to ALE case a, the errors in all fields is slightly larger (i.e.~they are slightly offset).

\subsubsection{Varying mesh velocity (ALE case d)}\label{s:octoALEc}

\begin{figure}[h!]
\begin{center} \unitlength1cm
\begin{picture}(0,5.6)
\put(-8.8,-.22){\includegraphics[height=60mm]{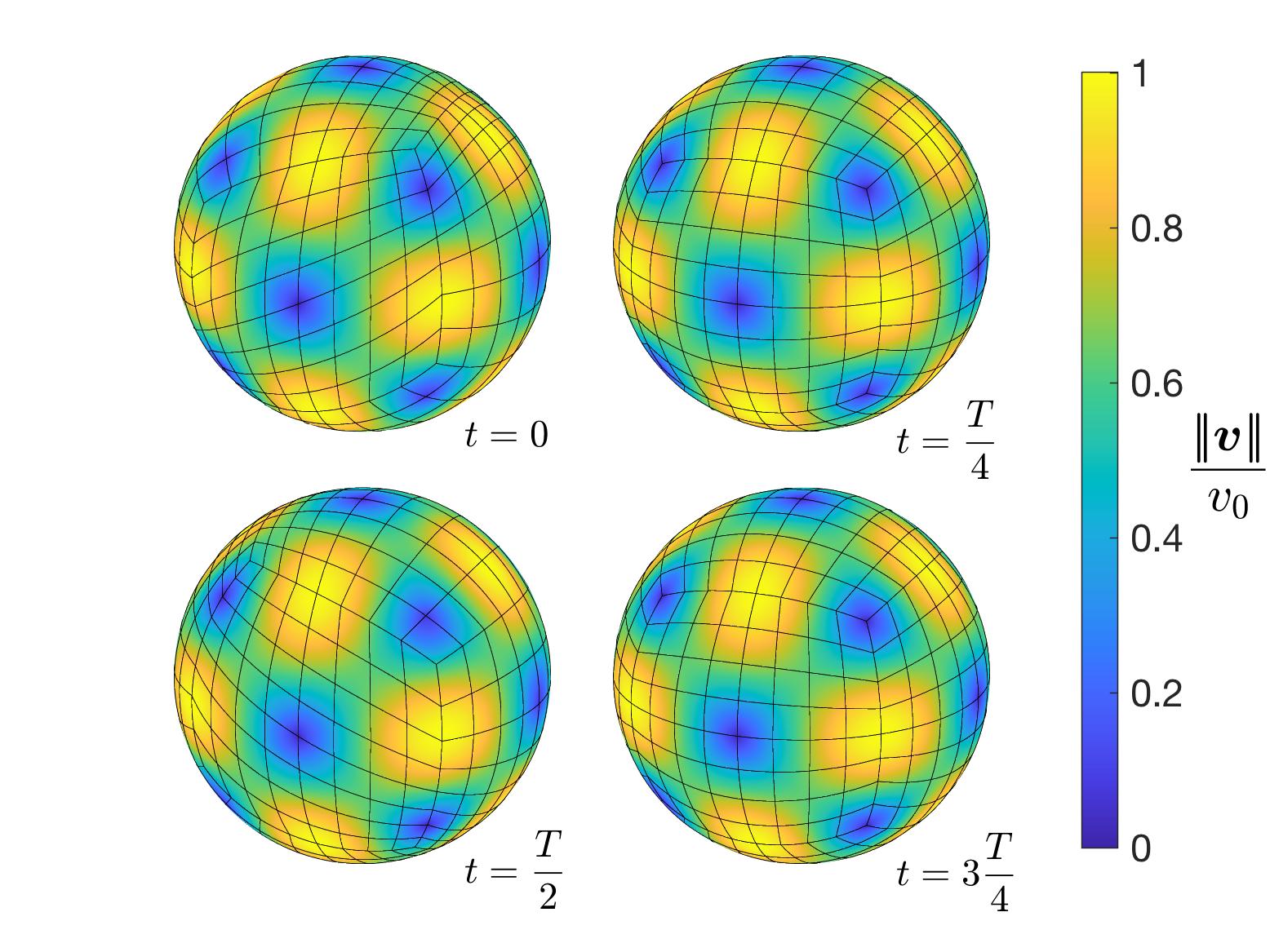}}
\put(0.2,-.3){\includegraphics[height=58mm]{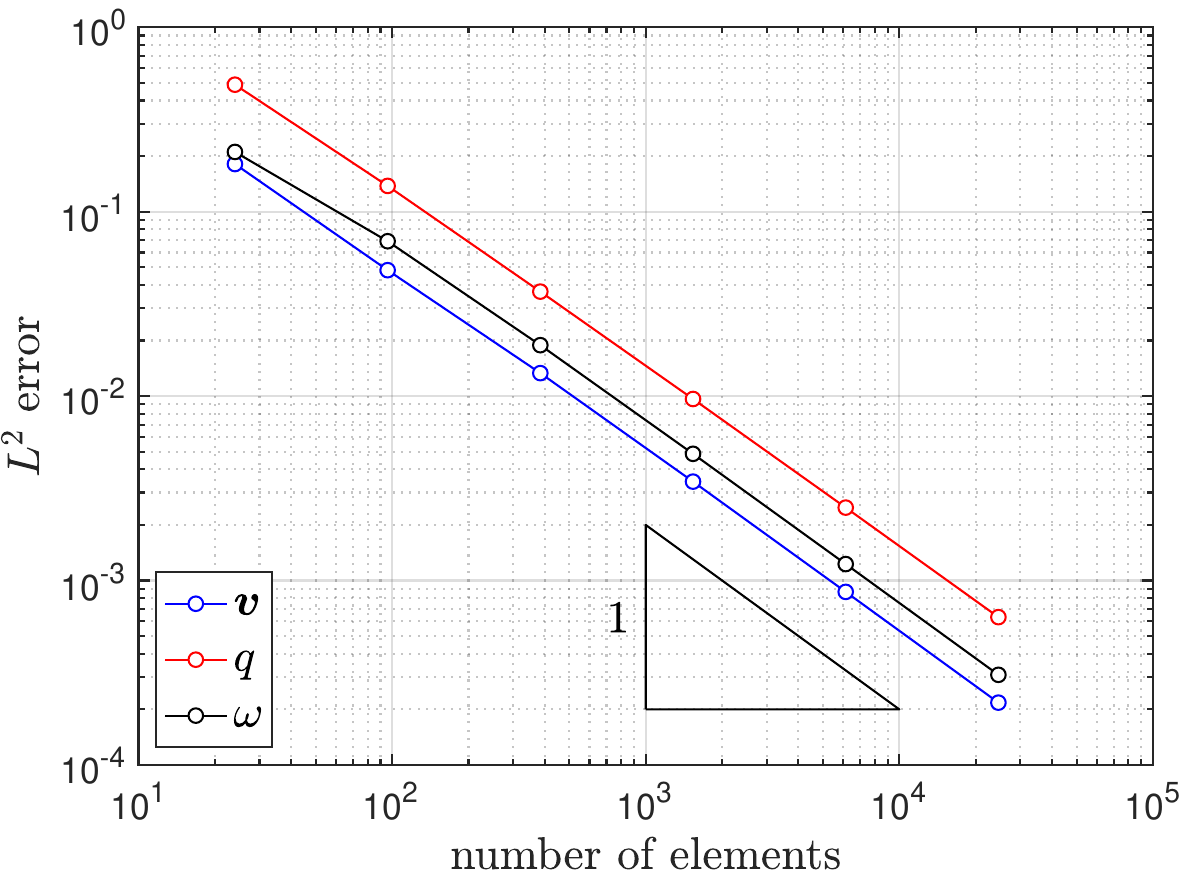}}
\put(-7.95,-.15){\footnotesize (a)}
\put(0.2,-.15){\footnotesize (b)}
\end{picture}
\caption{Octahedral vortex flow on a rigid sphere: 
(a) Flow field $\bv$ and (b) convergence behavior due to the prescribed mesh motion $\bx_\mrm(t)$ given in \eqref{e:xm1} \& \eqref{e:tm1} (ALE case d). 
This mesh motion causes out-of-plane components in $\bv'$ and hence can only be solved accurately with Neumann BC (load case~4).
All convergence rates are $O(h^2)=O(n_\mathrm{el}^{-1})$.
Here, the $L^2$ error from \eqref{e:L2} is averaged over $n_{t1}/4$ steps.}
\label{f:octoflow5}
\end{center}
\end{figure}

The last ALE case considers transient mesh motion (defined by $\bx_\mrm(t)$ given in \eqref{e:xm1} \& \eqref{e:tm1}).
Fig.~\ref{f:octoflow5} shows the flow field and its convergence rates.
%-----------------------------------------------------------------
% run pShearFlowALEc2.m & pShearFlowALEc.m
%-----------------------------------------------------------------
As in the previous example the transient part is $\bv' = \theta'\bv_{\!,2}$.
But now $\bv_{\!,2}$, and hence $\bv'$, has an out-of-plane component \citep{ALEtheo}. 
As a result, the problem cannot be solved accurately anymore by projection (load case 2).
Thus only load case 4 gives accurate results.
Also attributed to the out-of-plane component of $\bv'$ is that fluid velocity $\bv$ converges only quadratically in $h$, even for time step $n_{t2}$.
This was different in the previous example (see Fig.~\ref{f:shearflow5}), where $O(h^3)$-convergence was obtained for $n_{t2}$.
The convergence rates for $q$ and $\omega$ are not affected, and the same as before.

\section{Numerical examples for deforming surface flows}\label{s:Nex}

This section now considers the case when the surface deforms due to the flow.
Thus there are seven primary unknowns now: The components of $\bv$, $q$ and $\bv_\mrm$.
In Sec.~\ref{s:Nexf} there were never more than four ($\bv$, $q$).
First, the two examples from the previous section are revisited in Secs.~\ref{s:sheardefo} and \ref{s:octodefo}.
Due to area-incompressibility and a closed surface, the total surface area does not change during deformation.
This is different in the final example (Sec.~\ref{s:SB}) due to an inflow boundary. %\\
Since $\bv_\mrm$ is an unknown now, the mesh equations from Sec.~\ref{s:FEmesh} need to be included.
For this, the examples in Secs.~\ref{s:sheardefo} and \ref{s:octodefo} use Eulerian mesh motion, while the example of Sec.~\ref{s:SB} uses both Eulerian and elastic mesh motion.
The surface (mesh) $\bx$ then follows from integrating $\bv_\mrm$ using the update formula (\ref{e:vxnp1}.2), here with $\gamma = 1/2$.
Thus time stepping is still needed, even when considering steady flows (with negligible inertia $\rho\bv'$), which are the first two sub-cases considered in both Sec.~\ref{s:sheardefo} and \ref{s:octodefo}.

\subsection{Simple shear flow on a deformable sphere}\label{s:sheardefo}

First the axisymmetric shear flow example from Sec.~\ref{s:shearflow} is examined using the problem setup from Sec.~\ref{s:Nexsetup} (with the meshes from Table~\ref{t:shearflow} and unit values for $r$, $\omega_0$, $\eta$ and $\rho$), unless otherwise noted.
Three sub-cases are studied in the following.

\subsubsection{Fixed surface solution}\label{s:sheardefo1}

The first sub-case uses $\bff=\bff(\bX)$ from \eqref{e:shearflowf} and \eqref{e:shearflowp3}, which is the fixed sphere solution for which the surface is not supposed to change.
This is obtained correctly as Fig.~\ref{f:evolveflow1} shows:
%-----------------------------------------------------------------
\begin{figure}[h!]
\begin{center} \unitlength1cm
\begin{picture}(0,5.8)
\put(-8.8,-.22){\includegraphics[height=60mm]{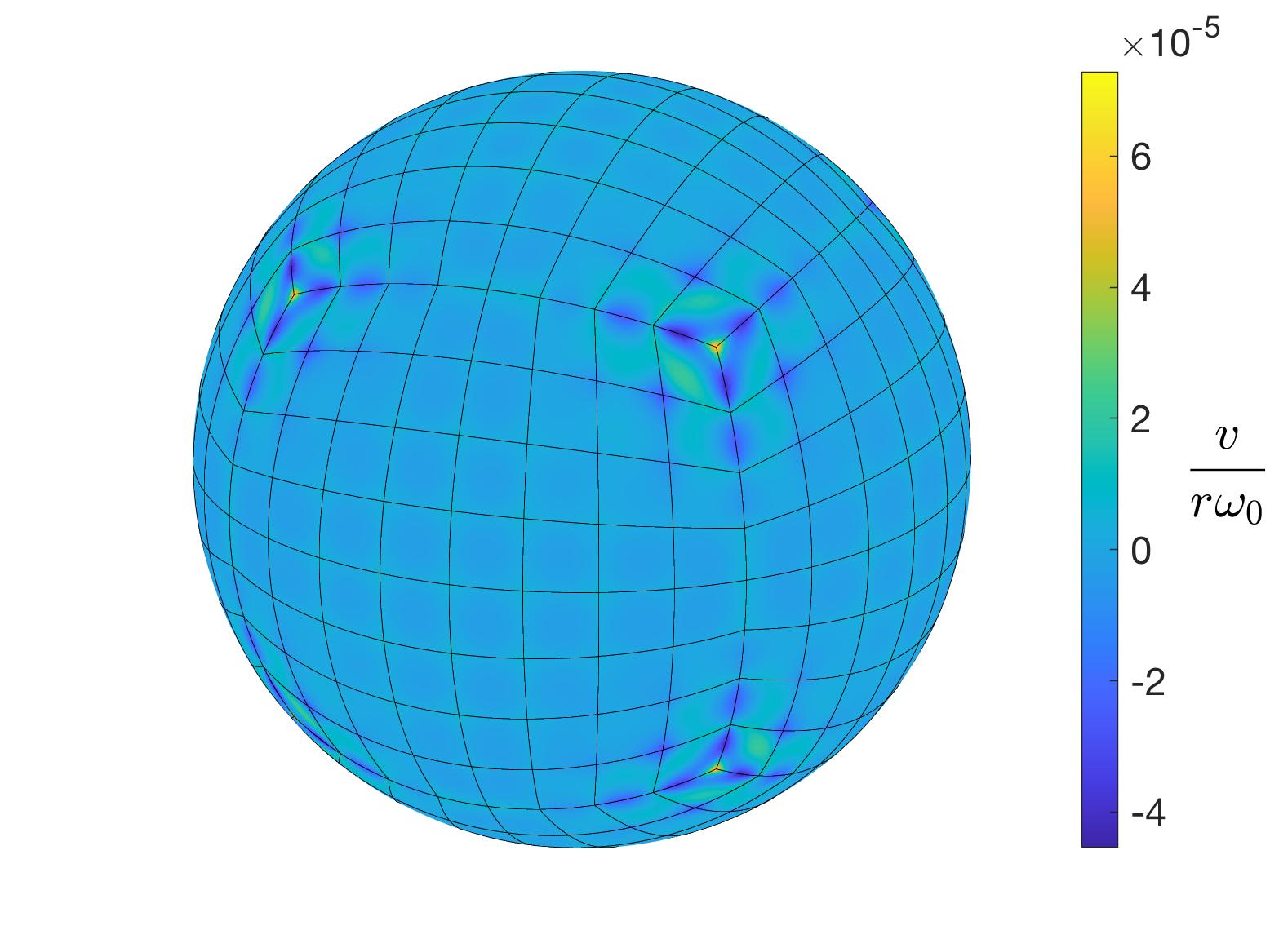}}
\put(0.2,-.3){\includegraphics[height=58mm]{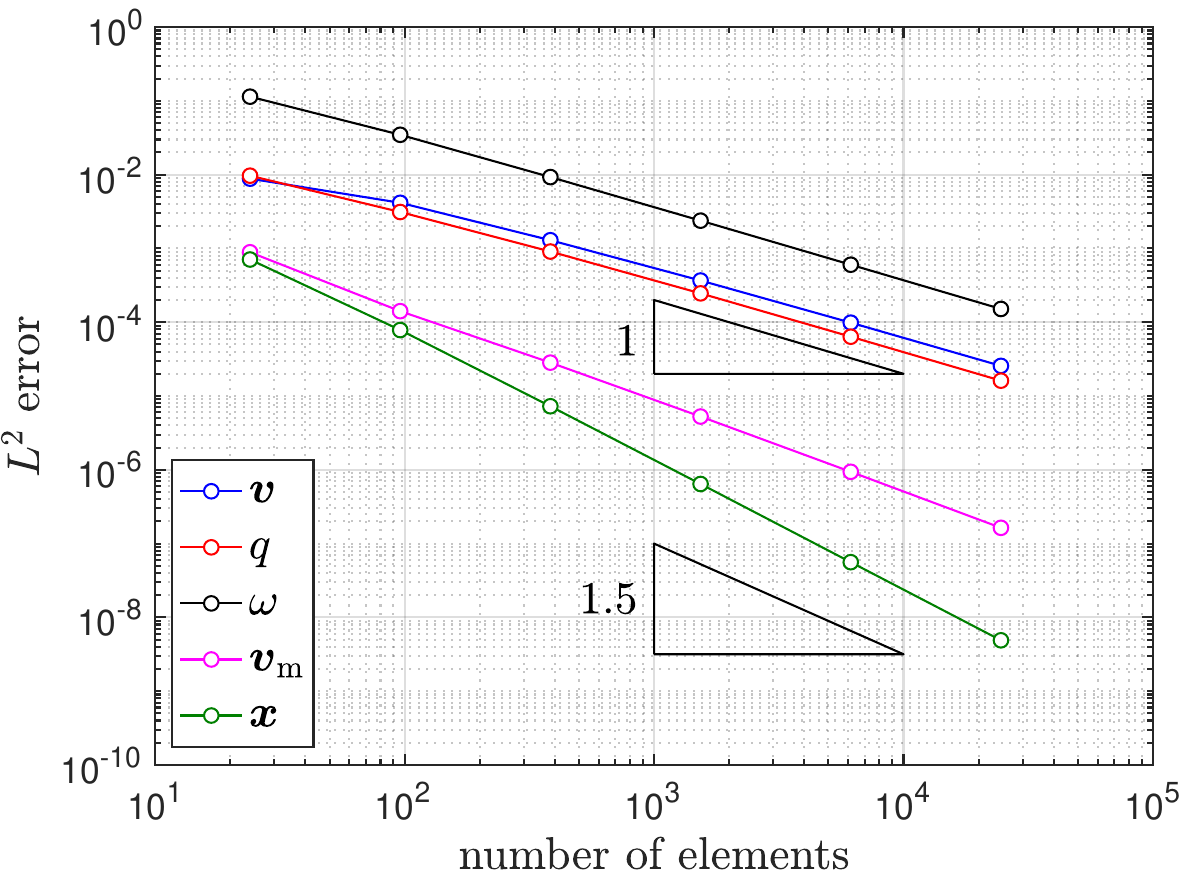}}
\put(-7.95,-.15){\footnotesize (a)}
\put(0.2,-.15){\footnotesize (b)}
\end{picture}
\caption{Simple shear flow on a deformable sphere with balancing pressure applied: 
(a) normal surface velocity $v=\bv\cdot\bn$ and (b) convergence behavior.
All convergence rates are optimal: 
The primary unknowns $\bv$, $q$, $\bv_\mrm$ converge with $O(h^2)=O(n_\mathrm{el}^{-1})$, while the integrated variable $\bx$ converges at least with $O(h^3)=O(n_\mathrm{el}^{-1.5})$.}
\label{f:evolveflow1}
\end{center}
\end{figure}
% run sShearEvolve/ vShearEvolveNS0 & pShearEvolve
%-----------------------------------------------------------------
The problem converges to the correct solution at ideal rates.
In particular, the normal velocity decays to zero with mesh refinement.
As Fig.~\ref{f:evolveflow1}a shows, for $m=4$ it is already four magnitudes below the tangential velocity (seen in Fig.~\ref{f:shearflow1}a).
In the simulations here, the time step $\Delta t = 1/m/\omega_0$ is used, while the polar pressure is taken as $p_\mrp = 4 \rho r \omega_0^2$.
The pressure is chosen larger than before since it has a stabilizing effect on the surface shape.
Conversely, too low surface pressures (esp.~negative ones) do not admit stable solution, which is seen in the following sub-case.

\subsubsection{Deformable surface solution for $\rho\bv'=\mathbf{0}$}\label{s:sheardefo2}

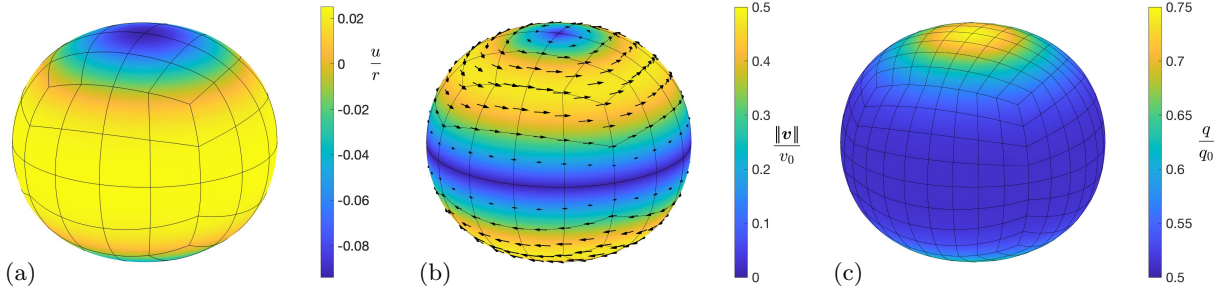
\begin{figure}[h]
\begin{center} \unitlength1cm
\begin{picture}(0,3.7)
\put(2.1,-.6){\includegraphics[height=44mm]{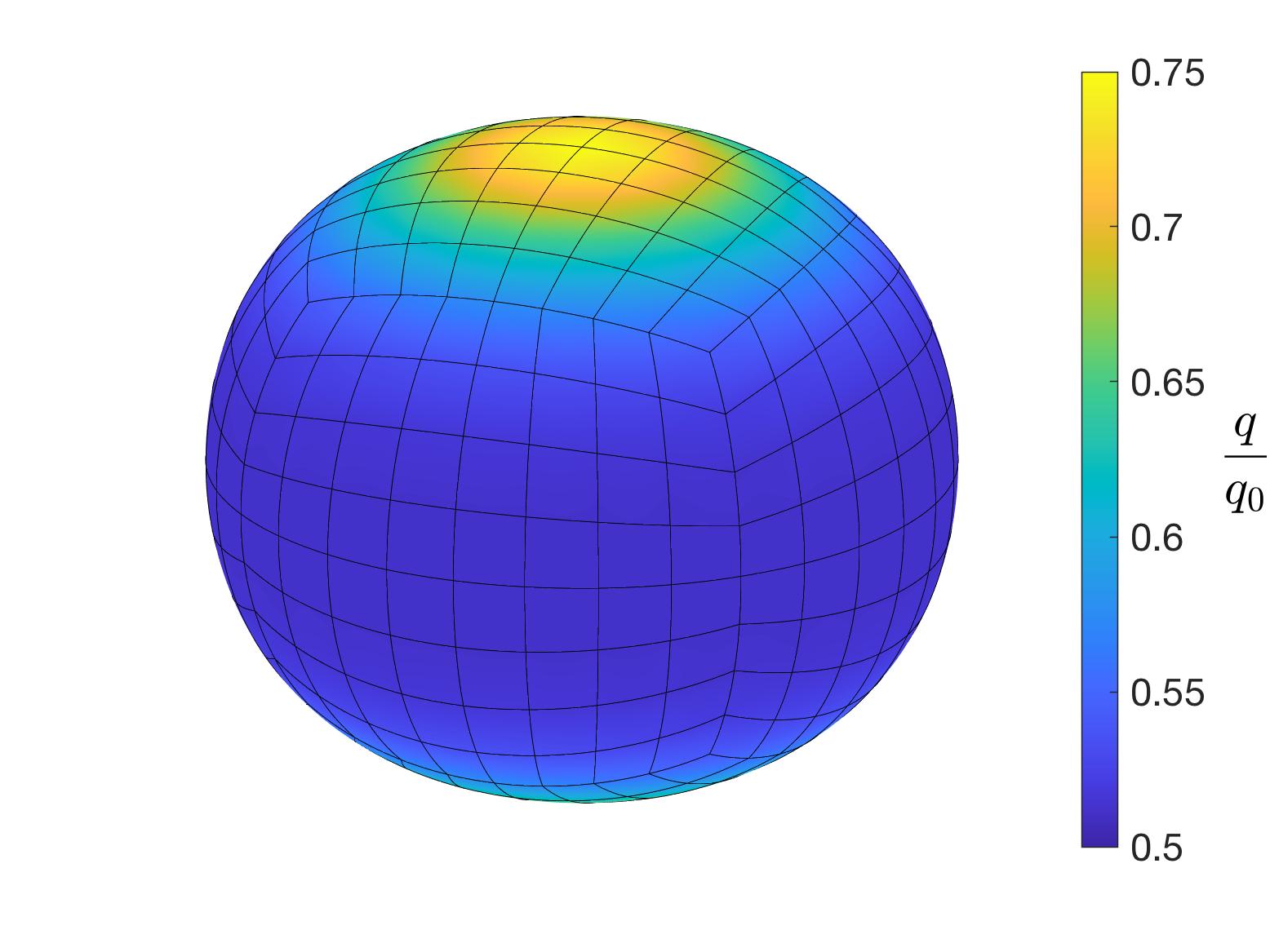}}
\put(-3.35,-.6){\includegraphics[height=44mm]{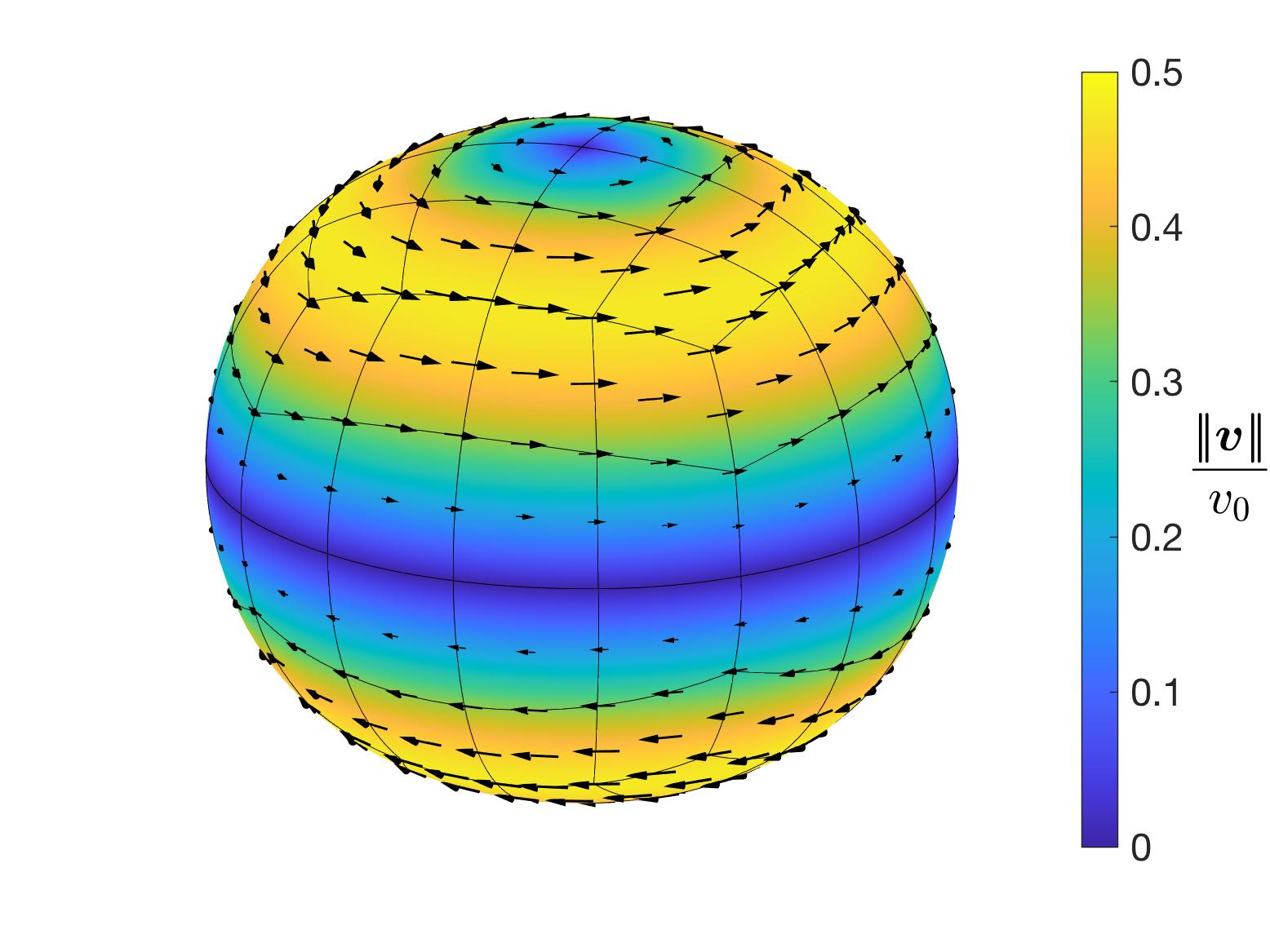}}
\put(-8.8,-.6){\includegraphics[height=44mm]{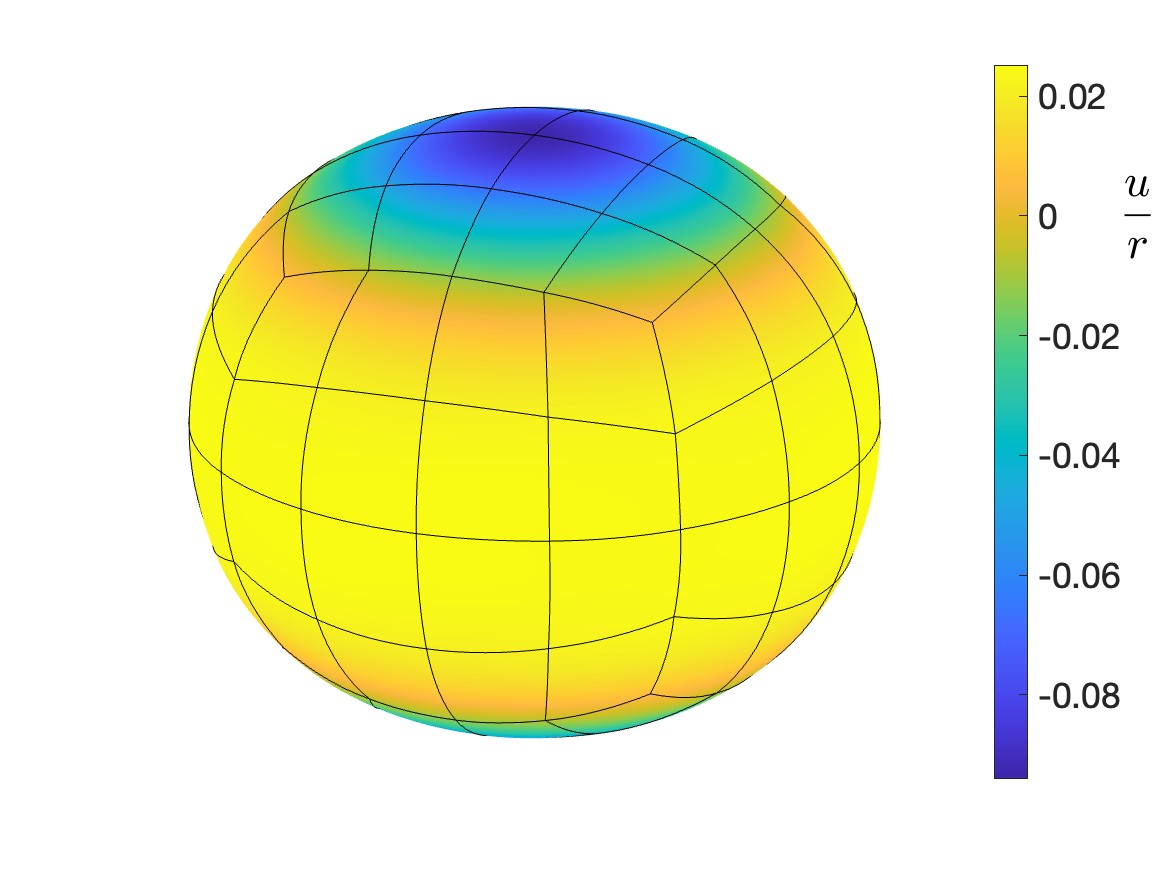}}   
\put(-7.95,-.15){\footnotesize (a)}
\put(-2.5,-.15){\footnotesize (b)}
\put(2.95,-.15){\footnotesize (c)}
\end{picture}
\caption{Simple shear flow on a deforming sphere with constant pressure $\bar p=\rho r \omega_0^2$ applied: 
(a)~radial surface displacement $u=\bu\cdot\be_r$, (b) velocity field $\bv$ and (c) surface tension $q$ plotted on the deformed sphere at time $t=40/\omega_0$.
The equatorial and polar distances change by 2.44\% and 9.41\%, respectively.
The velocity and surface tension range between $\norm{\bv}\in[0,\,0.476]\,v_0$ and $q\in[0.512,\,0.747]\,q_0$, 
where $v_0:=r\omega_0$ and $q_0 := r \bar p$.}
\label{f:shearevolve2}
\end{center}
\end{figure}

\begin{figure}[h]
\begin{center} \unitlength1cm
\begin{picture}(0,5.8)
\put(-8,-.12){\includegraphics[height=58mm]{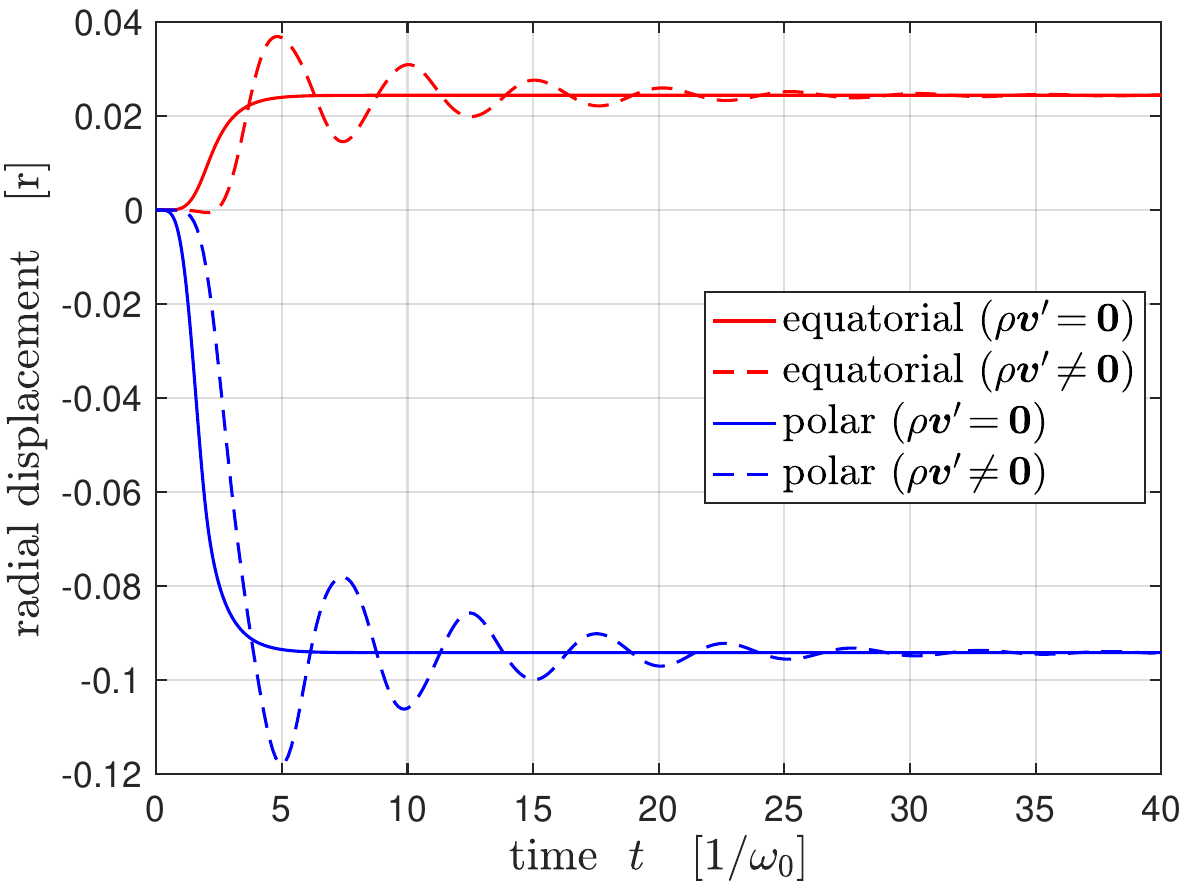}}
\put(0.2,-.2){\includegraphics[height=58mm]{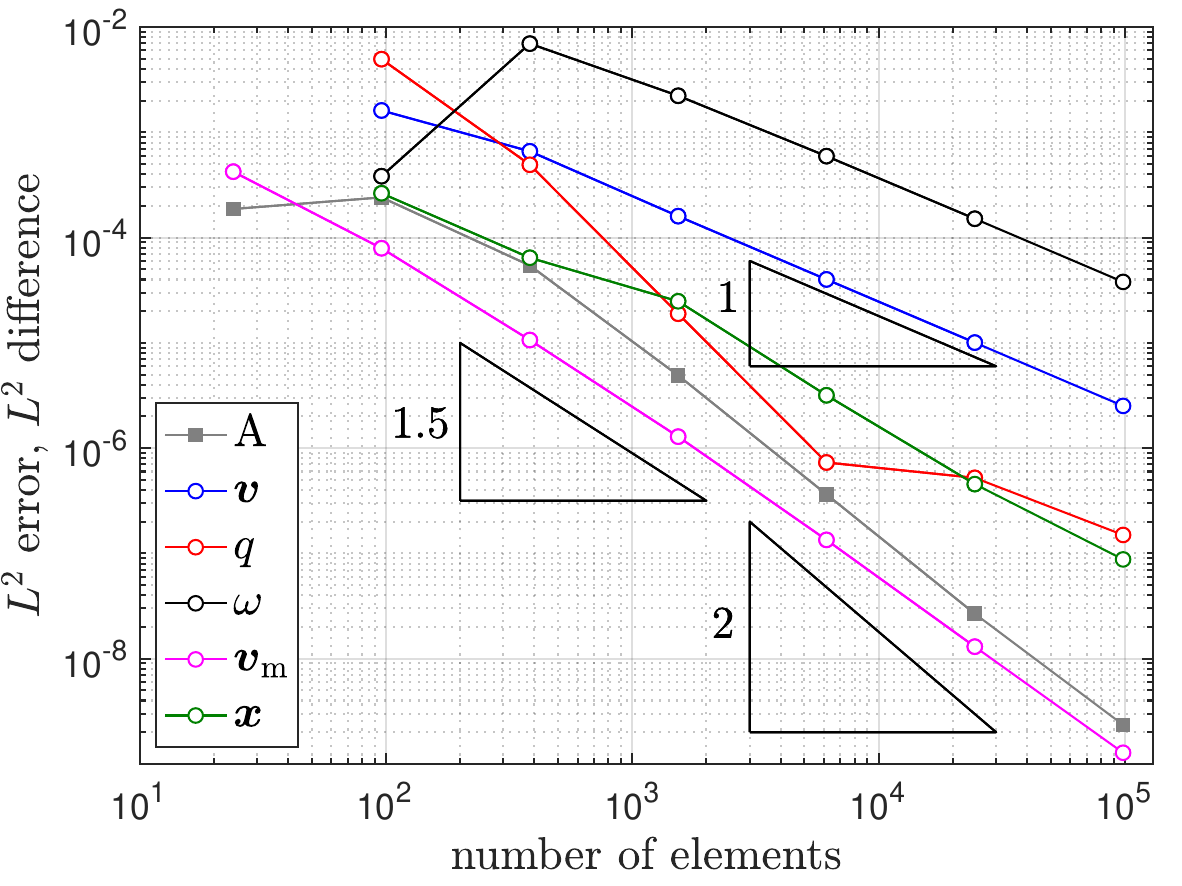}}
\put(-7.95,-.15){\footnotesize (a)}
\put(0.2,-.15){\footnotesize (b)}
\end{picture}
\caption{Simple shear flow on a deformable sphere with constant pressure $\bar p=\rho r \omega_0^2$: 
(a) shape change over time and (b) convergence behavior at $t=40/\omega_0$ for $\rho\bv'=\mathbf{0}$. 
All convergence rates are at least optimal: 
The primary unknowns $\bv$ and $q$ converge with $O(h^2)=O(n_\mathrm{el}^{-1})$, 
and $\bv_\mrm$ even with $O(h^3)=O(n_\mathrm{el}^{-1.5})$.
The integrated variables $\bx$ and $A$ (the surface area) converge with $O(h^3)=O(n_\mathrm{el}^{-1.5})$ and $O(h^4)=O(n_\mathrm{el}^{-2})$, respectively.
Here, the plots for $A$ and $\bv_m$ are $L^2$ errors (compared to the exact values), while the remaining plots are $L^2$ differences (between subsequent meshes).}
\label{f:shearevolve3}
\end{center}
\end{figure}

The second sub-case uses the prescribed uniform surface tractions $f_2 = 0$ and $p = \bar p = \rho r \omega_0^2$ within
Eq.~\eqref{e:shearflowf}, where $r$ still denotes the initial, undeformed sphere radius.
This internal pressure $p$ is larger at the equator than before, and so a flattening of the sphere can be expected.
This is confirmed by Fig.~\ref{f:shearevolve2}.
%-----------------------------------------------------------------
% run sShearEvolve/ vShearEvolveNS2 & pShearEvolve2
%-----------------------------------------------------------------
Here the equatorial radius has increased by about 2.4 \%, while the pole distance has decreased by about 9.4 \%.
The evolution of these deformations is shown in Fig.~\ref{f:shearevolve3}a.
%
%-----------------------------------------------------------------
% run pShearEvolve1
%-----------------------------------------------------------------
Here the prescribed shear traction $f_1$ has been ramped-up using
\eqb{l}
\bar\omega_0(t) = \ds\frac{\omega_0}{2}\,\left\{\begin{array}{ll} 
1 - \cos(\pi t/t_1) & $for$~~0\leq t < t_1 \\[.5mm]
2 & $for$~~t\geq t_1 
\end{array}\right.\,,
\label{e:rampup}\eqe
for $\omega_0$ in \eqref{e:shearflowf}.
The ramp-up time is chosen as $t_1=2/\omega_0$, while the model parameters $r$, $\omega_0$, $\rho$ and $\eta_\mrn$ are again taken as unity apart from $\eta = 1/2$.
The time step $\Delta t = 2/m/\omega_0$ is used.
It is noted that the sphere volume changes but the surface area is conserved due to the considered area-incompressibility.
The computations run stable as long as sufficient out-of-plane pressure $\bar p$ and out-of-plane viscosity $\eta_\mrn$ are applied, as they stabilize the surface motion.
The latter is applied according to \eqref{e:pvisc} with $\eta_\mrn$ taken as unity in this example.
Without such stabilization, the flow may cause surface instabilities such as out-of-plane surface buckling.

Fig.~\ref{f:shearevolve3}b shows the $L^2$ errors of the total surface area $A = 4\pi r^2$ and the mesh velocity $\bv_\mrm$ (which vanishes at steady state), as well as the $L^2$ differences of $\bv$, $q$, $\omega$ and $\bx$ (for which no exact solution exists).
As seen, the convergence rate (w.r.t.~element size $h$) is quartic for $A$, cubic for $\bv_\mrm$ \& $\bx$, and at least linear for the other fields.
The exceptional behavior of $q$ is due to a zero-crossing between the third and fourth point.
It is noted, that under Stokes flow conditions (achieved by setting $\rho=0$ to eliminate the convective term), no surface deformations occur in this example.

\subsubsection{Deformable surface solution for $\rho\bv'\neq\mathbf{0}$}\label{s:sheardefo3}

The results discussed so far are the steady solution (where $\rho\bv'$ is neglected).
Now the transient case is examined, where $\rho\bv'\neq \mathbf{0}$.
It is also shown in Fig.~\ref{f:shearevolve3}a.
As can be expected, the surface inertia $\rho\bv'$ leads to oscillations in the deformation.
They decay due to viscosity, such that eventually the steady state from before is reached.
Here now the viscosity parameters are reduced to $\eta = 1/4$ and $\eta_\mrn = 0$ compared to before.
This works well, since inertia additionally stabilizes the surface motion.

\subsection{Octahedral vortex flow on a deformable sphere}\label{s:octodefo}

Second, the non-axisymmetric octahedral surface flow from Sec.~\ref{s:octflow} is examined using again the problem setup from Sec.~\ref{s:Nexsetup}.
The same three sub-cases as in Sec~\ref{s:sheardefo} are investigated here.

\subsubsection{Fixed surface solution}\label{s:octodefo1}

\begin{figure}[h!]
\begin{center} \unitlength1cm
\begin{picture}(0,5.6)
\put(-8.8,-.22){\includegraphics[height=60mm]{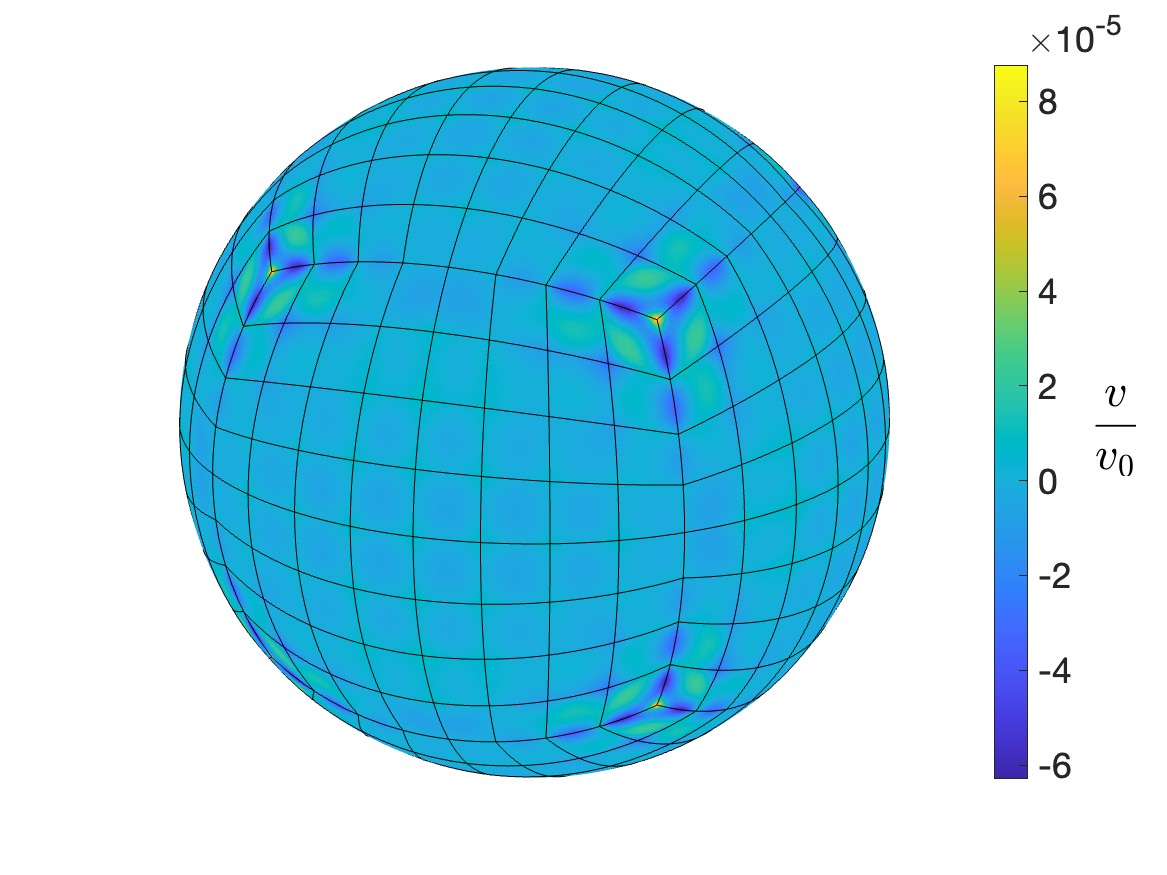}}
\put(0.2,-.3){\includegraphics[height=58mm]{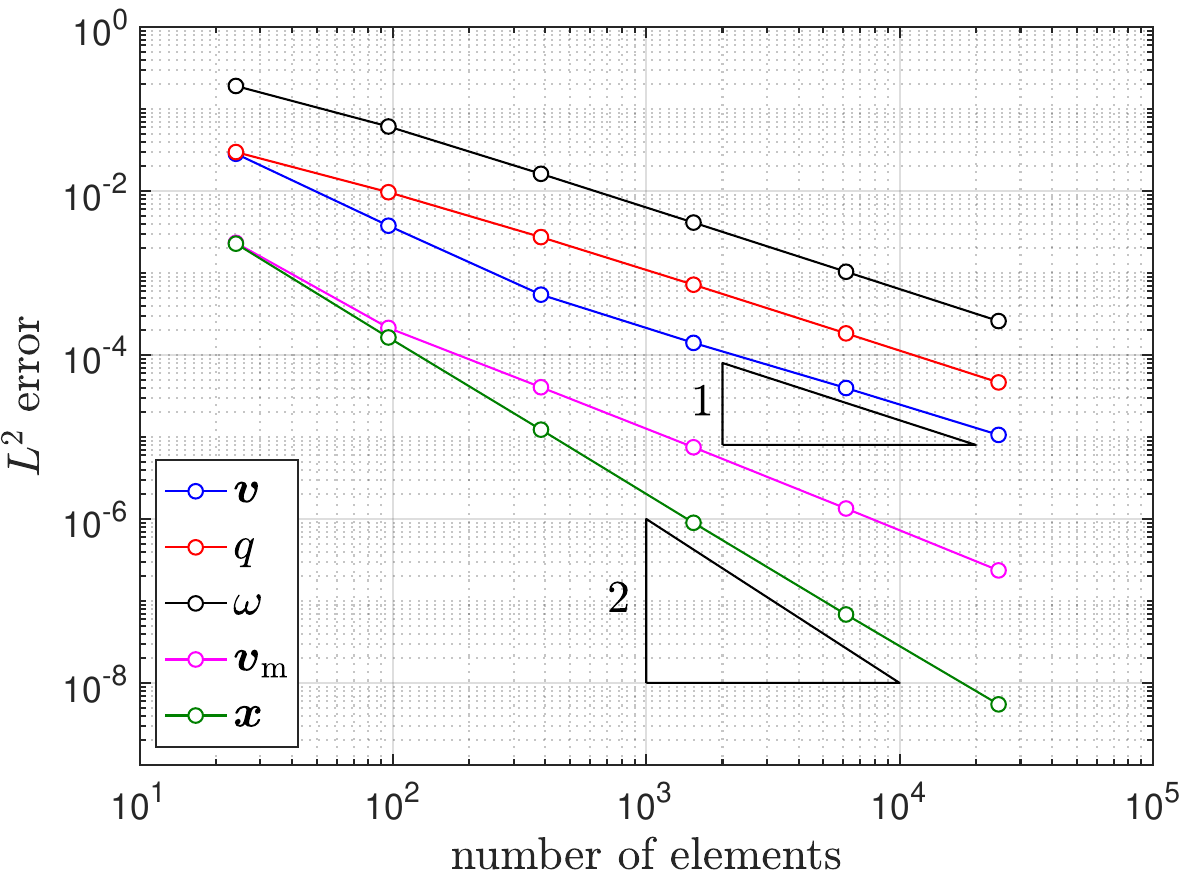}}
\put(-7.95,-.15){\footnotesize (a)}
\put(0.2,-.15){\footnotesize (b)}
\end{picture}
\caption{Octahedral vortex flow on a deformable sphere with balancing surface pressure applied: (a)~normal surface velocity $v=\bv\cdot\bn$ and (b) convergence behavior.
All convergence rates are optimal: 
At least $O(n_\mathrm{el}^{-1})$ for primary unknowns $\bv$, $q$, $\bv_\mrm$, and at least $O(n_\mathrm{el}^{-1.5})$ for the integrated variable $\bx$.} 
\label{f:octoevolve1}
\end{center}
\end{figure}

The first sub-case considers the fixed sphere solution that appears if the varying out-of-plane pressure $\bar p$ from Eq.~\eqref{e:pocto} is prescribed with $p_\mrp = 4p_0$, $p_0 := \rho v_0^2/r$.
As in Sec.~\ref{s:sheardefo1} before, the problem correctly converges to the analytical solution at ideal rates, as Fig.~\ref{f:octoevolve1} shows. 
The convergence rate for the surface shape $\bx$ is even close to $O(n_\mathrm{el}^{-2})$.
%-----------------------------------------------------------------
% run sOctoEvolveNS0 & pOctoEvolve
%-----------------------------------------------------------------

\subsubsection{Deformable surface solution}\label{s:octodefo2}

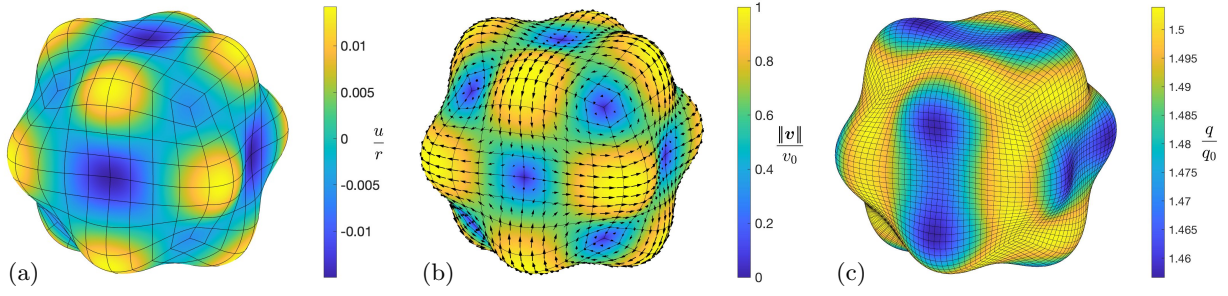
\begin{figure}[h]
\begin{center} \unitlength1cm
\begin{picture}(0,3.7)
\put(2.1,-.6){\includegraphics[height=44mm]{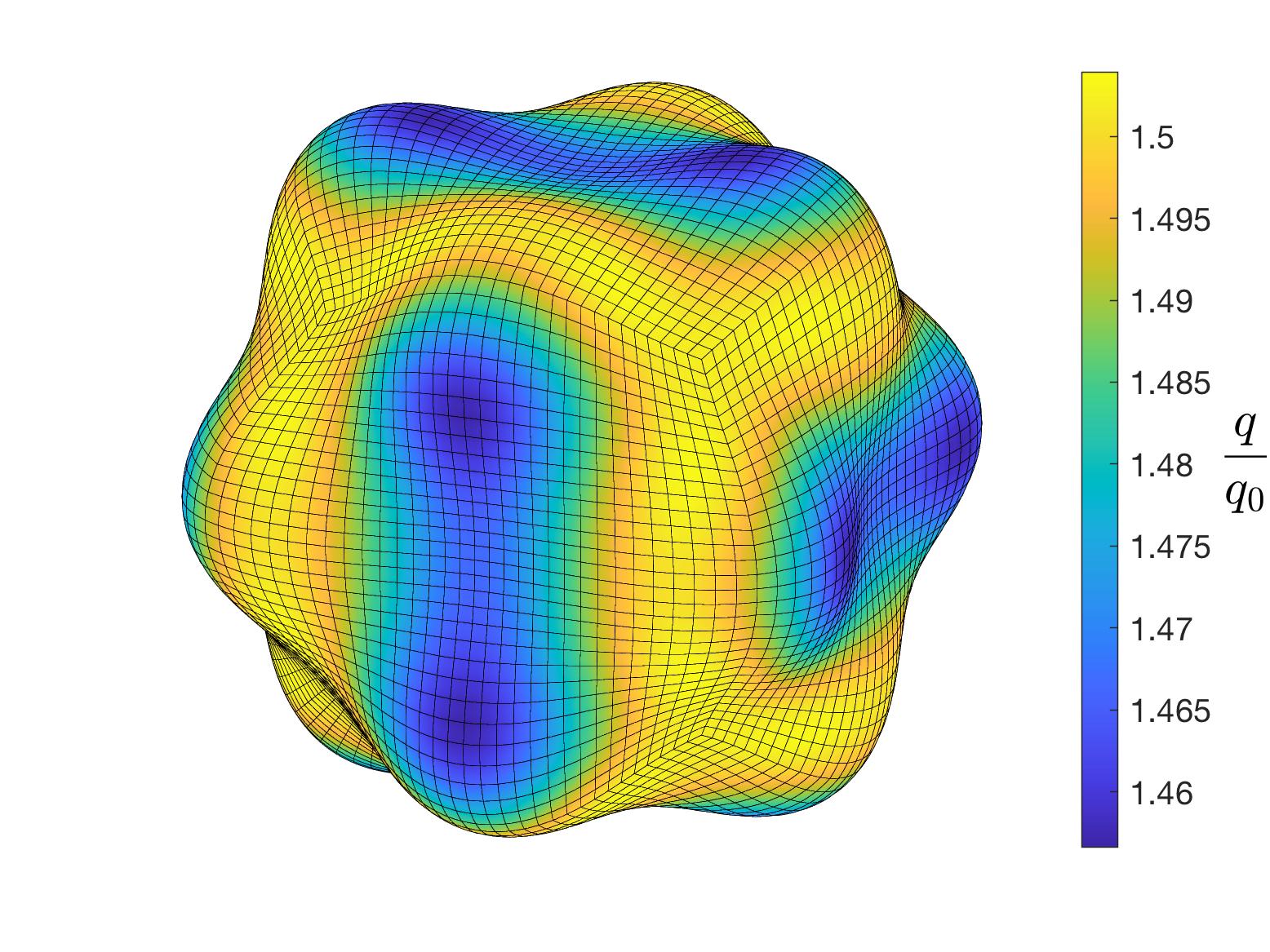}}
\put(-3.35,-.6){\includegraphics[height=44mm]{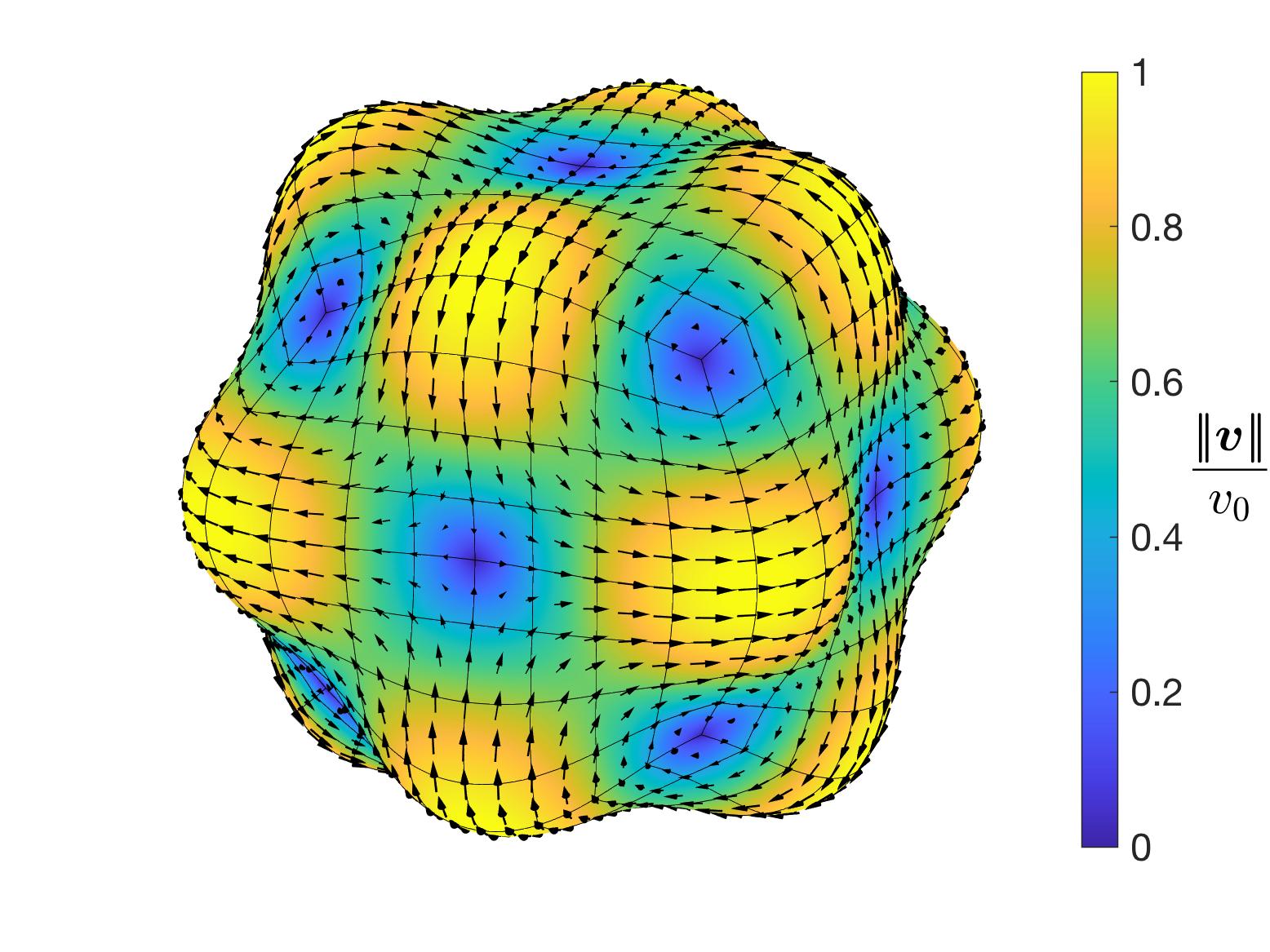}}
\put(-8.8,-.6){\includegraphics[height=44mm]{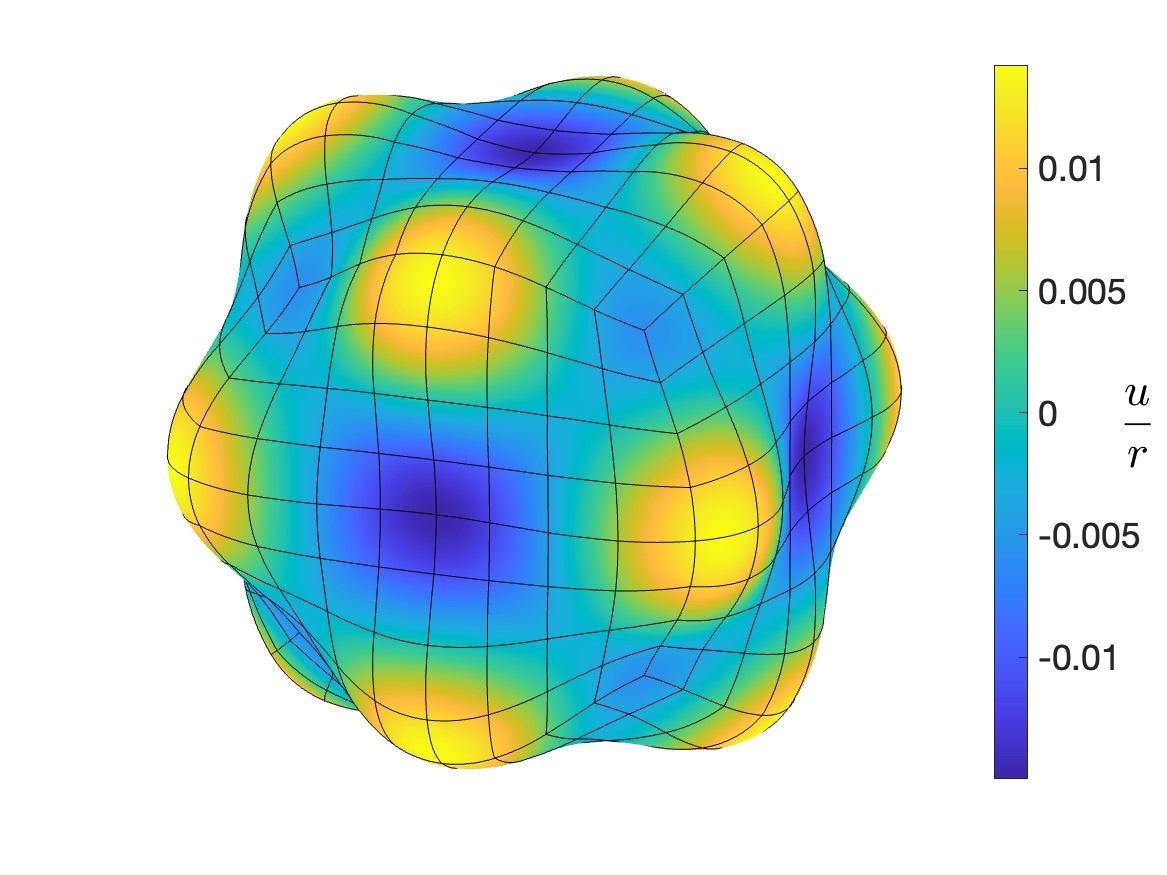}} % u = \bu \cdot \bN
\put(-7.95,-.15){\footnotesize (a)}
\put(-2.5,-.15){\footnotesize (b)}
\put(2.95,-.15){\footnotesize (c)}
\end{picture}
\caption{Octahedral vortex flow on a deformable sphere with constant pressure $\bar p=2.5p_0$ applied: 
(a) radial surface displacement $u=\bu\cdot\be_r$, (b) velocity field $\bv$ and (c) surface tension $q$ plotted on the deformed sphere.
The relative displacement ranges between -1.48\% and 1.42\%.
It is scaled-up by a factor of 10 in the three plots to increase its visibility.
The velocity and surface tension range between $\norm{\bv}\in[0,\,1.014]\,v_0$ and $q\in[1.457,\,1.504]\,q_0$, where $q_0 := rp_0$.}
\label{f:octoevolve2}
\end{center}
\end{figure}

\begin{figure}[h]
\begin{center} \unitlength1cm
\begin{picture}(0,5.8)
\put(-8,-.12){\includegraphics[height=58mm]{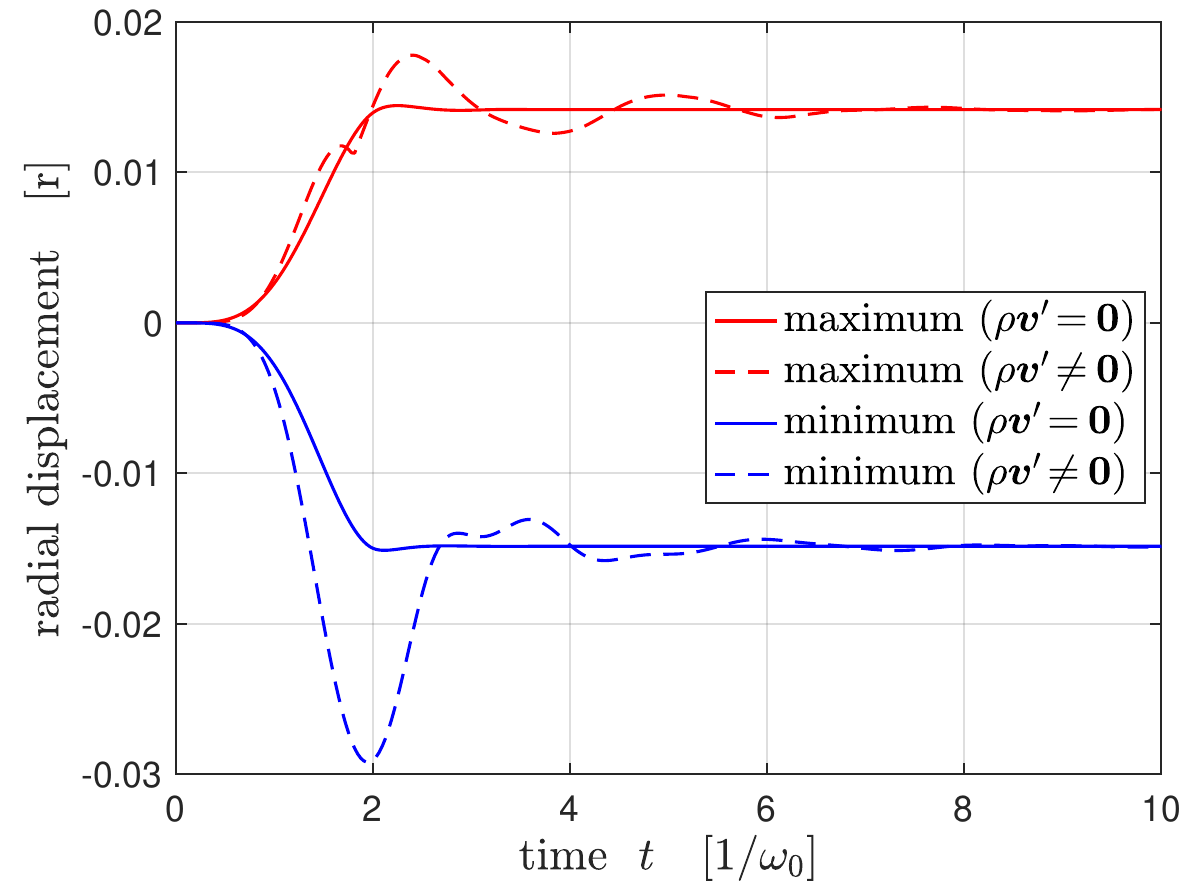}}
\put(0.2,-.2){\includegraphics[height=58mm]{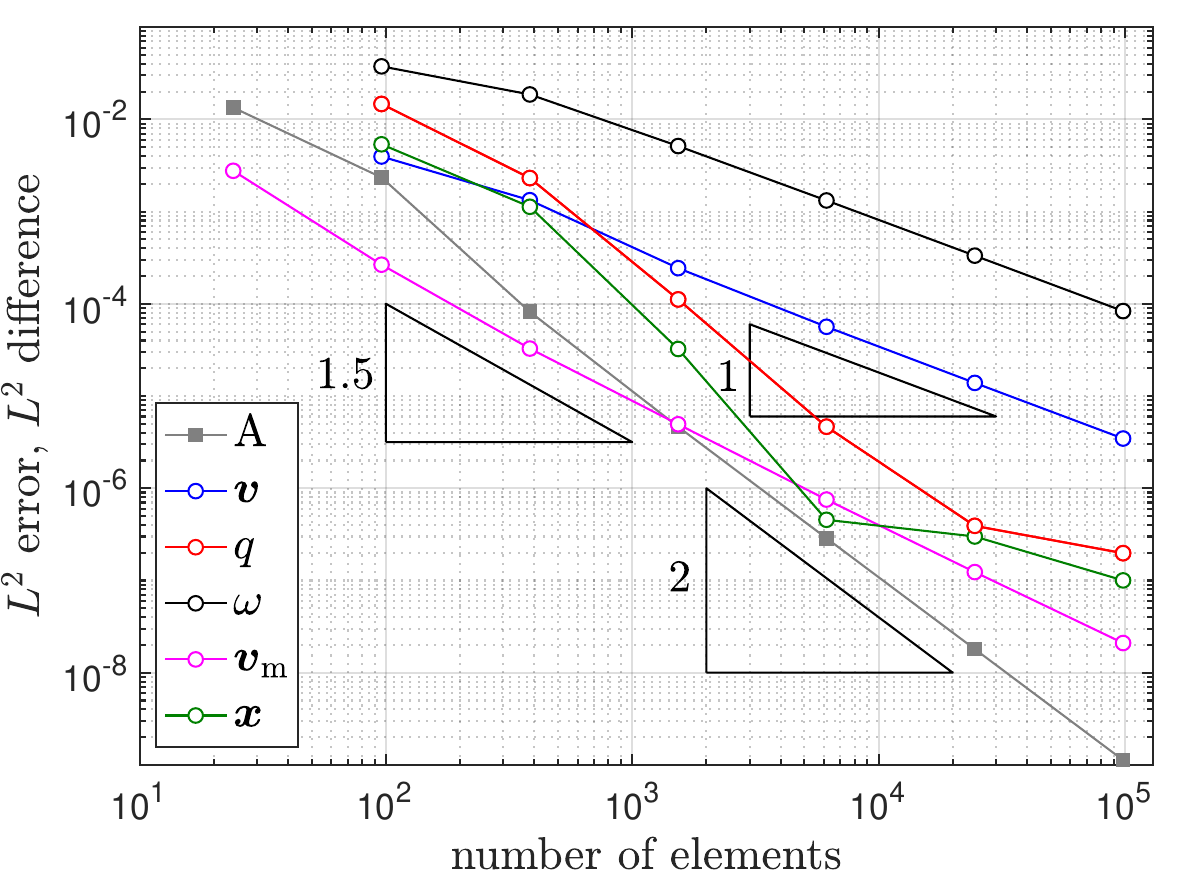}}
\put(-7.95,-.15){\footnotesize (a)}
\put(0.2,-.15){\footnotesize (b)}
\end{picture}
\caption{Octahedral vortex flow on a deformable sphere with constant pressure $\bar p=2.5p_0$: 
(a) shape change over time and (b) convergence behavior at $t=10/\omega_0$ for $\rho\bv'=\mathbf{0}$. 
Area $A$ and mesh velocity $\bv_m$ are $L^2$ errors converging at optimal rates.
The remaining fields are $L^2$ differences decreasing at least linearly.
The results for $q$ and $\bx$ have zero-crossings between the 4th and 5th point.
}
\label{f:octoevolve3}
\end{center}
\end{figure}

If a different surface pressure is used for $\bar p$, an out-of-plane motion occurs.
This is shown in Fig.~\ref{f:octoevolve2} for $\bar p = 2.5p_0$.
%-----------------------------------------------------------------
% run pOctoEvolve2
%-----------------------------------------------------------------
As seen the surface deformation aligns with the flow field.
The evolution of these deformations is shown in Fig.~\ref{f:octoevolve3}a both for the steady state ($\rho\bv'=\mathbf{0}$) and transient case ($\rho\bv'\neq\mathbf{0}$).
Here the flow has been ramped-up using Eq.~\eqref{e:rampup} with $t_1 = 2/\omega_0$ and $\Delta t =1/m/\omega_0$.
The convergence behavior seen in Fig.~\ref{f:octoevolve3}b is similar to the one of Fig.~\ref{f:shearevolve3}.
%-----------------------------------------------------------------
% run pOctoEvolve1
%-----------------------------------------------------------------

\subsection{Inflation of a soap bubble}\label{s:bubble}\label{s:SB}

\begin{figure}[h]
\begin{center} \unitlength1cm
\begin{picture}(0,4.3)
\put(-8,-.05){\includegraphics[height=44mm]{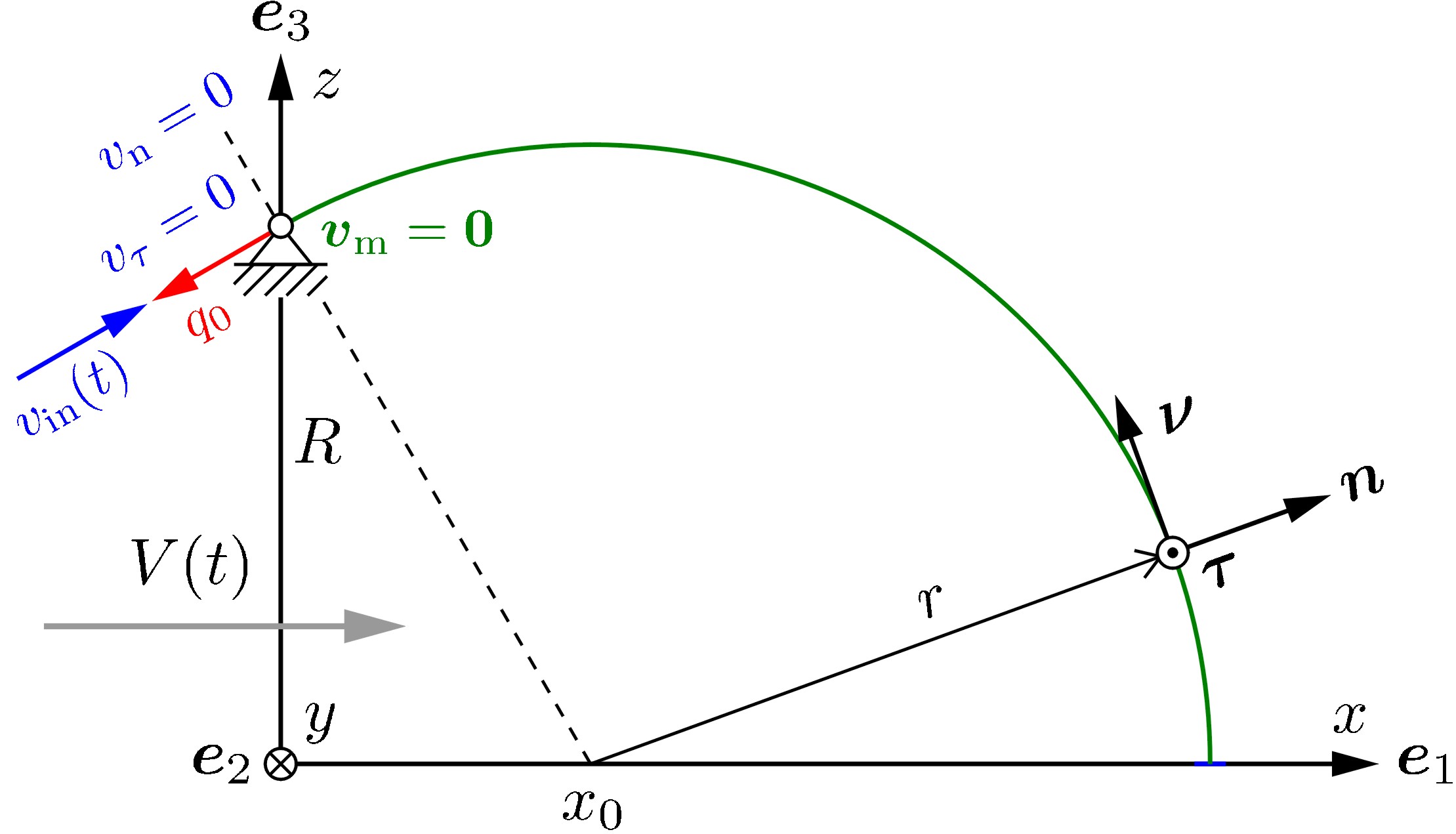}}
\put(0.35,-.05){\includegraphics[height=44mm]{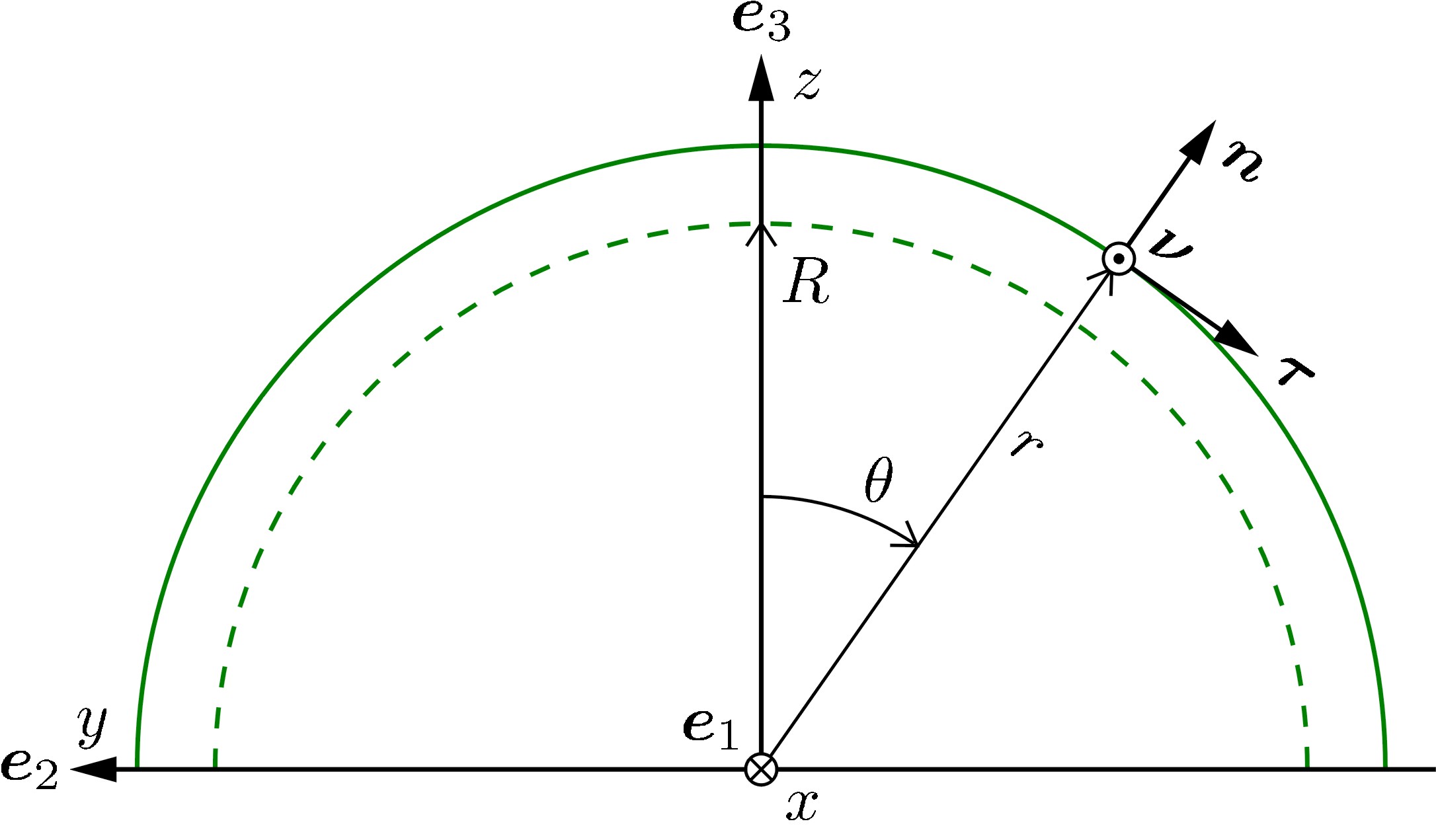}}
\put(-7.95,-.15){\footnotesize (a)}
\put(0.3,-.15){\footnotesize (b)}
\end{picture}
\caption{Soap bubble inflation: Problem setup, geometry and boundary conditions. 
(a) side view; (b) back view. 
At the inflow, the Dirichlet boundary conditions $v_\mrn=0$, $v_\tau = 0$, $q = q_0$, $\bv_\mrm = \mathbf{0}$ and the 
Neumann boundary condition $T_\nu = q$ are prescribed.}
\label{f:SoapBublSetup}
\end{center}
\end{figure}

The final example considers the inflation of a soap bubble.
It is an advanced flow example involving transient flow on deforming surfaces and introducing two new challenges: Large shape changes and deformation-dependent boundary conditions.
It is also used to examine $C^1$-continuous surface discretization and the influence of surface stretches on Eqs.~\eqref{e:mbf2}-\eqref{e:mbf4}.
The considered parameters are given in Table~\ref{t:soapbubble}.
The following computational setup is used, see Fig.~\ref{f:SoapBublSetup}:
%-------------------------------------------------------------------------------------------------------------------------------
\begin{table}[h!]
\vspace{1mm}
\centering
\begin{tabular}{|r|l|r|l|}
   \hline
   symbol & material parameter & value & unit \\[0mm] \hline 
   & & & \\[-3.5mm]   
   $R$ & initial bubble radius & 10 & mm \\ [.5mm] 
   $p_0$ & initial surface pressure & 5 & $\mu\mathrm{N}/\mathrm{mm}^2 = $ Pa \\ [1mm]    
   $q_0$ & initial surface tension & 25 & $\mu\mathrm{N}/\mathrm{mm}$ \\ [1mm]  
   $\eta$ & surface viscosity & 0.04 & $\mu\mathrm{N\,ms}/\mathrm{mm}$ \\ [1mm]   
   $\rho$ & surface density & 0.2 & $\mu\mathrm{N\,ms}^2/\mathrm{mm}^3$ \\ [1mm] 
   $\tmu$ & ALE mesh stiffness & $10^{-5}...\,10^{-9}$ & $\mu\mathrm{N}/\mathrm{mm}$ \\[.5mm]
   \hline
\end{tabular}
\caption{Soap bubble example: Considered geometry and material parameters.
Here, $q_0$ is estimated from \citet{sane17} for a water solution with 4\% soap, $\eta$ and $\rho$ are based on the bulk viscosity $\tilde\eta = 0.2$\,Pa$\cdot$s and density $\tilde\rho = 1$\,g/cm$^3$ multiplied by the soap film thickness $h = 0.2\,\mu$m.
From this follows $p_0 = 2q_0/R$.
Numerical parameter $\tmu$ is determined from computational tests.
$\eta_\mrn=0$ unless noted otherwise.}
\label{t:soapbubble}
\end{table}
%-------------------------------------------------------------------------------------------------------------------------------
%-----------------------------------------------------------------
%-----------------------------------------------------------------
An initially hemispherical bubble, pinned at the left boundary, is inflated by increasing the enclosed bubble volume according to
\eqb{l}
V(t) = V_0 + \Delta V \big( 1 - \cos(\pi t/t_1) \big)\,, \quad V_0 := \ds\frac{2\pi}{3}R^3\,.
\label{e:soapV}\eqe
This causes an inflow of new material at the left boundary.
The corresponding boundary conditions on the flow and mesh motion are illustrated in Fig.~\ref{f:SoapBublSetup}a. 
Explicitly these are $v_\mrn = 0$, $v_\tau = 0$ and $T_\nu = q$ for the fluid velocity, $q = q_0$ for the surface tension, and $\bv_\mrm = \mathbf{0}$ for the mesh motion.
For $\bv$, this is both a mixed BC -- Neumann along in-flow direction $\bnu$ and Dirichlet along normal and circumferential directions $\bn$ and $\btau$ -- and a follower BC, as directions $\bnu$ and $\bn$ change during inflation.
Details on its FE application and linearization are given in Appendix \ref{s:Lft}.
Volume constraint \eqref{e:soapV} is added to the system of equations by the Lagrange multiplier method, see \citet{membrane} for details.
The corresponding Lagrange multiplier -- the internal bubble pressure $p$ -- becomes a fourth unknown next to $\bv$, $q$ and $\bv_\mrm$.
The resulting transient system is solved monolithically with the implicit trapezoidal time integration scheme described in Sec.~\ref{s:TI}.
Table~\ref{t:soapbubblemesh} lists the FE meshes used for this.
%-------------------------------------------------------------------------------------------------------------------------------
\begin{table}[h]
\vspace{1.5mm}
\centering
\begin{tabular}{|r|r|r|r|r|r|r|r|}
  \hline
   $m$ & $n_\mathrm{sel}$ & $n_\mathrm{bel}$ & $n_\mathrm{no}$ (Q2) & $n_\mathrm{dof}$ (Q2) & $n_\mathrm{no}$ (N2) & $n_\mathrm{dof}$ (N2) & $n_{t}$ \\[0mm] \hline 
   & & & & & & & \\[-4mm]   
   4 & 96 & 16 &  417 &  2,920 & 160 & 1,121 & 50  \\[0mm] 
   8 & 384 & 32 & 1,601 &  11,208 & 504 & 3,529 & 100 \\[0mm] 
   16 & 1,536 & 64 & 6,273 & 43,912 & 1,768 & 12,377 & 200 \\[0mm] 
   32 & 6,144 & 128 & 24,833 & 173,832 & 6,600 & 46,201 & 400 \\[0mm]       
   64 & 24,576 & 256 & 98,817 & 691,720 & 25,480 & 178,361 &  800 \\[0mm]    
  128 & 98,304 & 512 & 394,247 & 2,759,730 & 100,104 & 700,729 & 1,600 \\[0mm]  
   \hline
\end{tabular}
\caption{Soap bubble inflation: Considered FE meshes based on bi-quadratic Lagrange (Q2) and bi-quadratic NURBS (N2) elements and corresponding dofs for unknown $\bv$, $q$, $\bv_\mrm$ (seven unknown components).
Here, $n_\mathrm{sel} = 6m^2$ is the number of surface FE, $n_\mathrm{bel} = 4m$ the number of boundary FE, $n_\mathrm{no}$ the number of nodes, and $n_\mathrm{dof} = 7n_\mathrm{no} + 1$ the number of dofs;
$n_t$ is the number of time steps used in the first example of Sec.~\ref{s:SBu}.}
\label{t:soapbubblemesh}
\end{table}
%-------------------------------------------------------------------------------------------------------------------------------

Two cases are considered in the following sections: uniform and non-uniform inflow.
They illustrate the inaccuracies resulting from an in-plane Eulerian surface description.
It is shown that these inaccuracies reduce substantially with the proposed in-plane mesh elasticity formulation of Eqs.~\eqref{e:fio} \& \eqref{e:taum}

\subsubsection{Uniform inflow}\label{s:SBu}

The first case considers a (spatially) uniform inflow velocity, such that the flow remains fully laminar.
The 2D analytical study of \citet{ALEtheo} showed that a purely Eulerian surface description leads to large mesh deformations at the tip of the bubble, while an ALE surface description allows for a uniform mesh deformation across the surface.
Therefore it is expected that the proposed ALE description based on mesh elasticity will yield better numerical results.
The inflation parameters in \eqref{e:soapV} are taken as $\Delta V = 9V_0$ and $t_1=2$s with time step size $\Delta t = 80\mathrm{ms}/m$.

\begin{figure}[h]
\begin{center} \unitlength1cm
\begin{picture}(0,6.6)
\put(-7.95,2.95){\includegraphics[height=39mm]{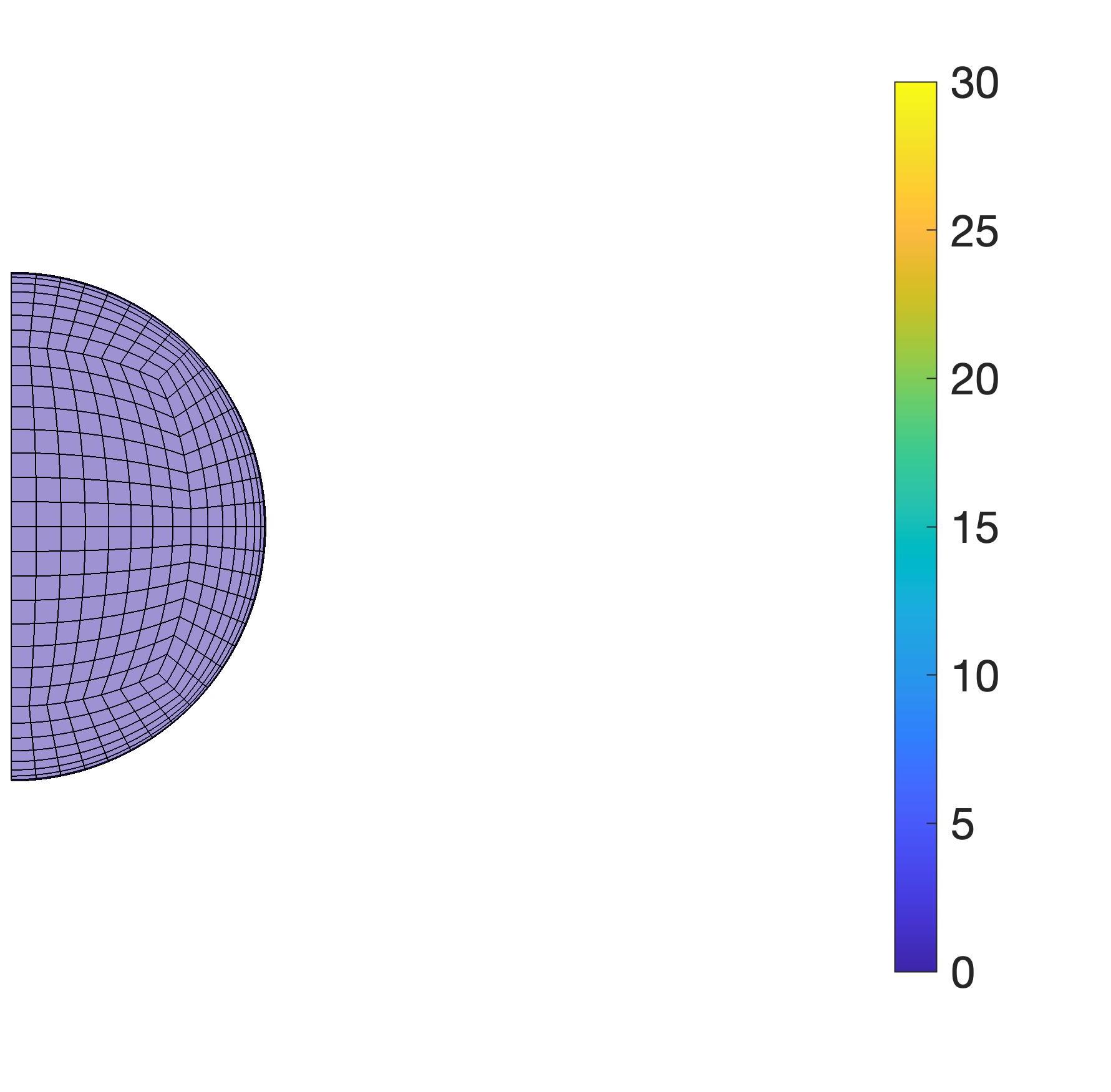}}
\put(-6.8,2.95){\includegraphics[height=39mm]{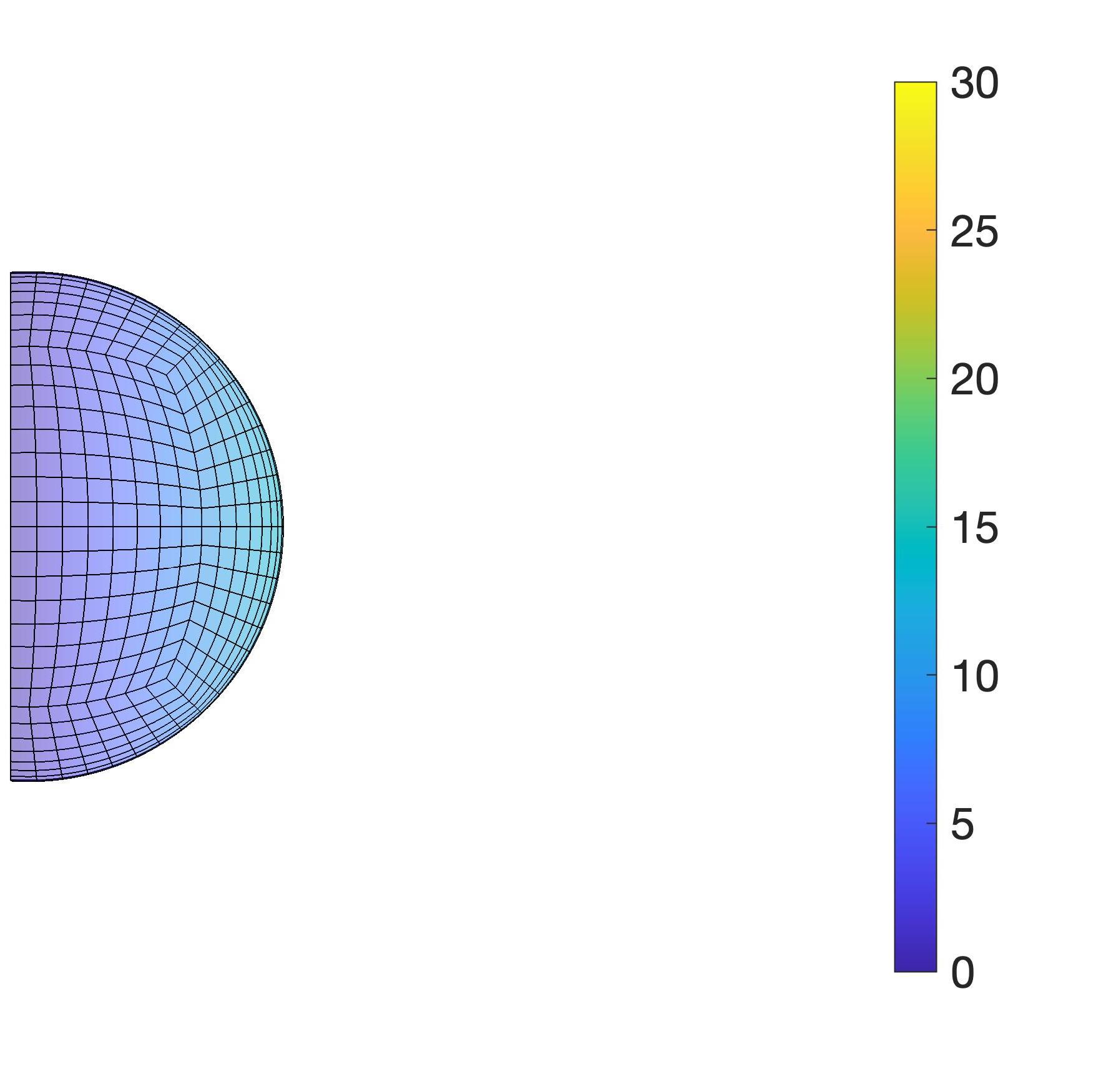}}
\put(-5.6,2.95){\includegraphics[height=39mm]{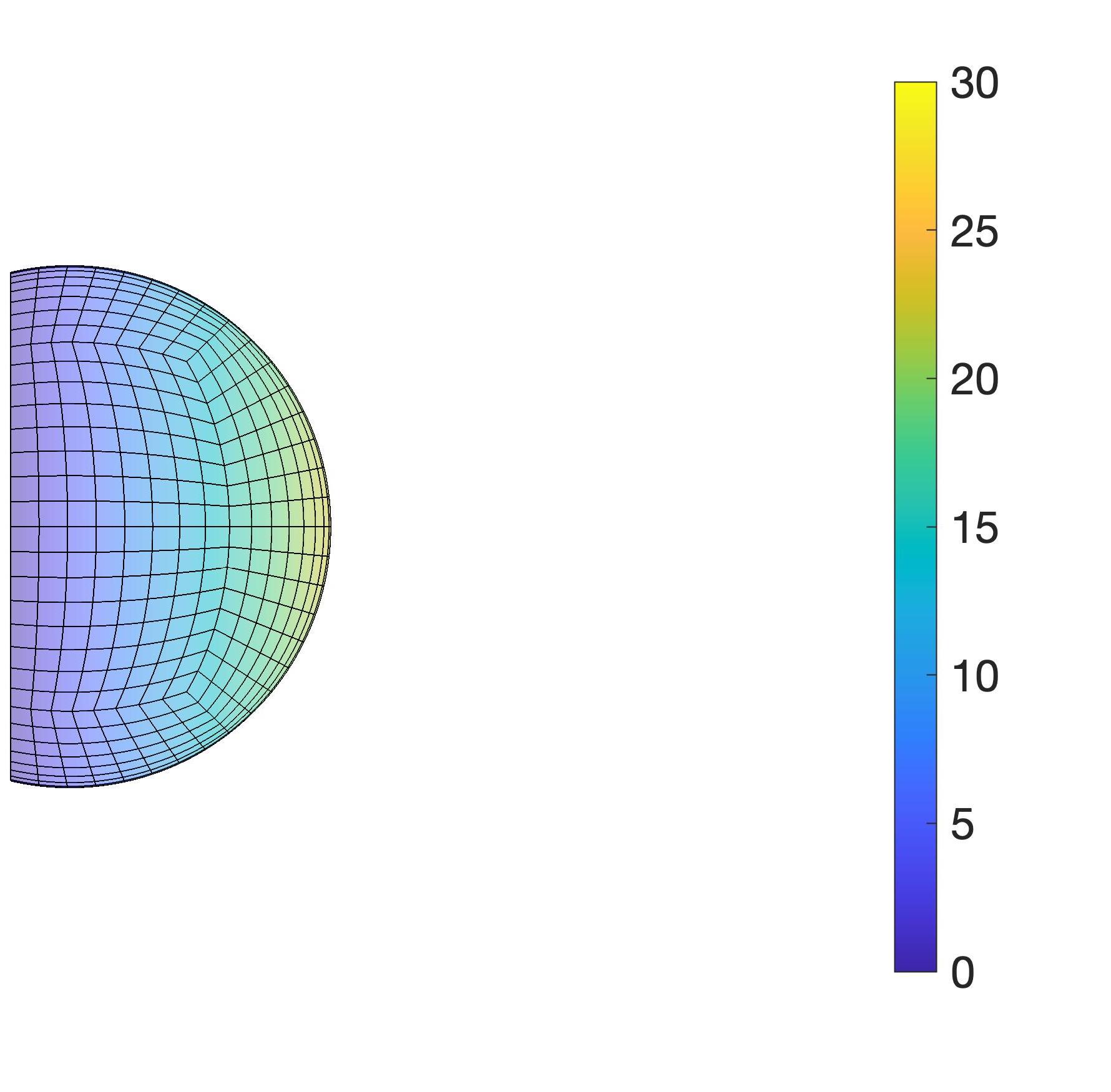}}
\put(-4.2,2.95){\includegraphics[height=39mm]{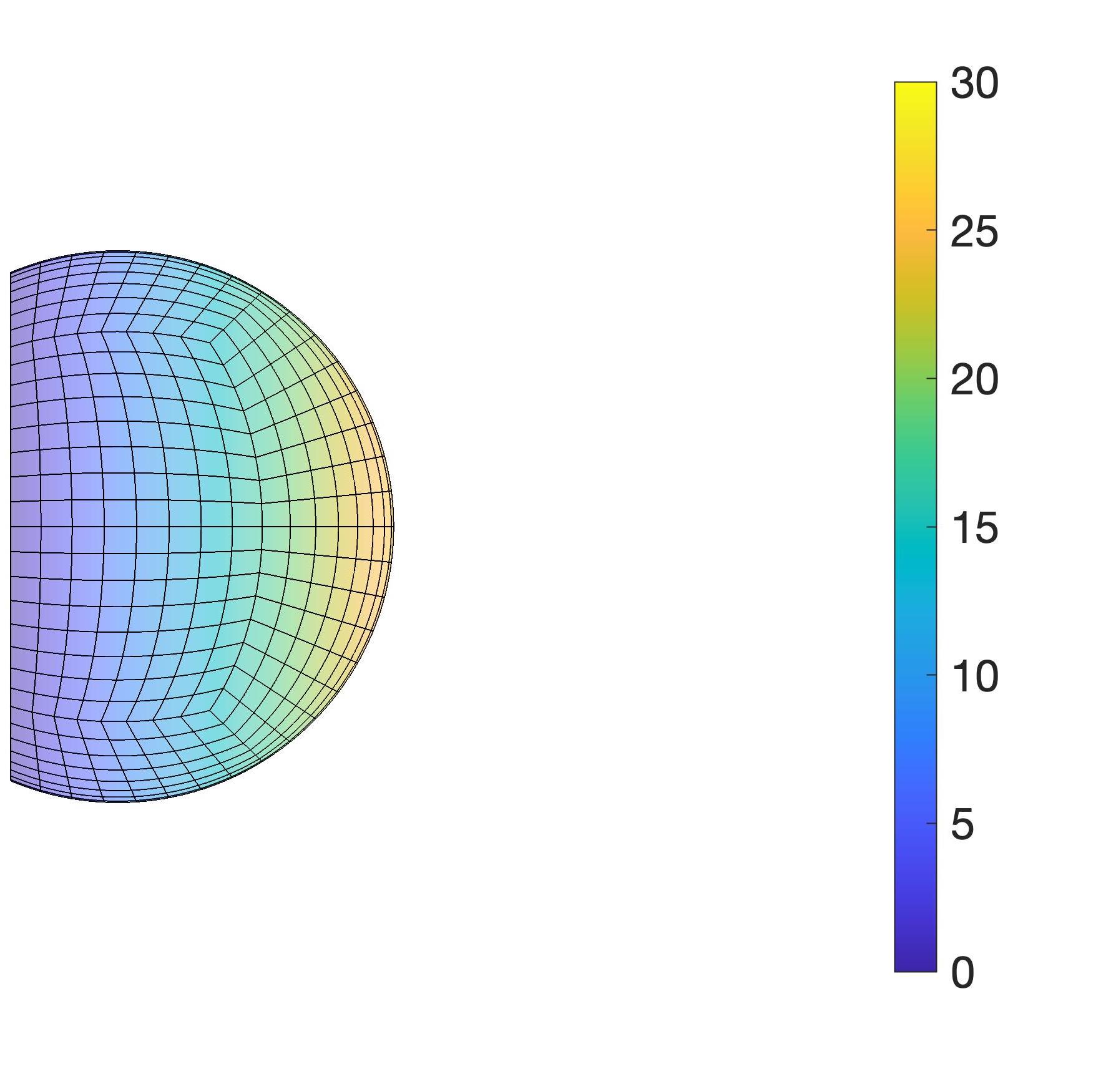}}
\put(-2.6,2.95){\includegraphics[height=39mm]{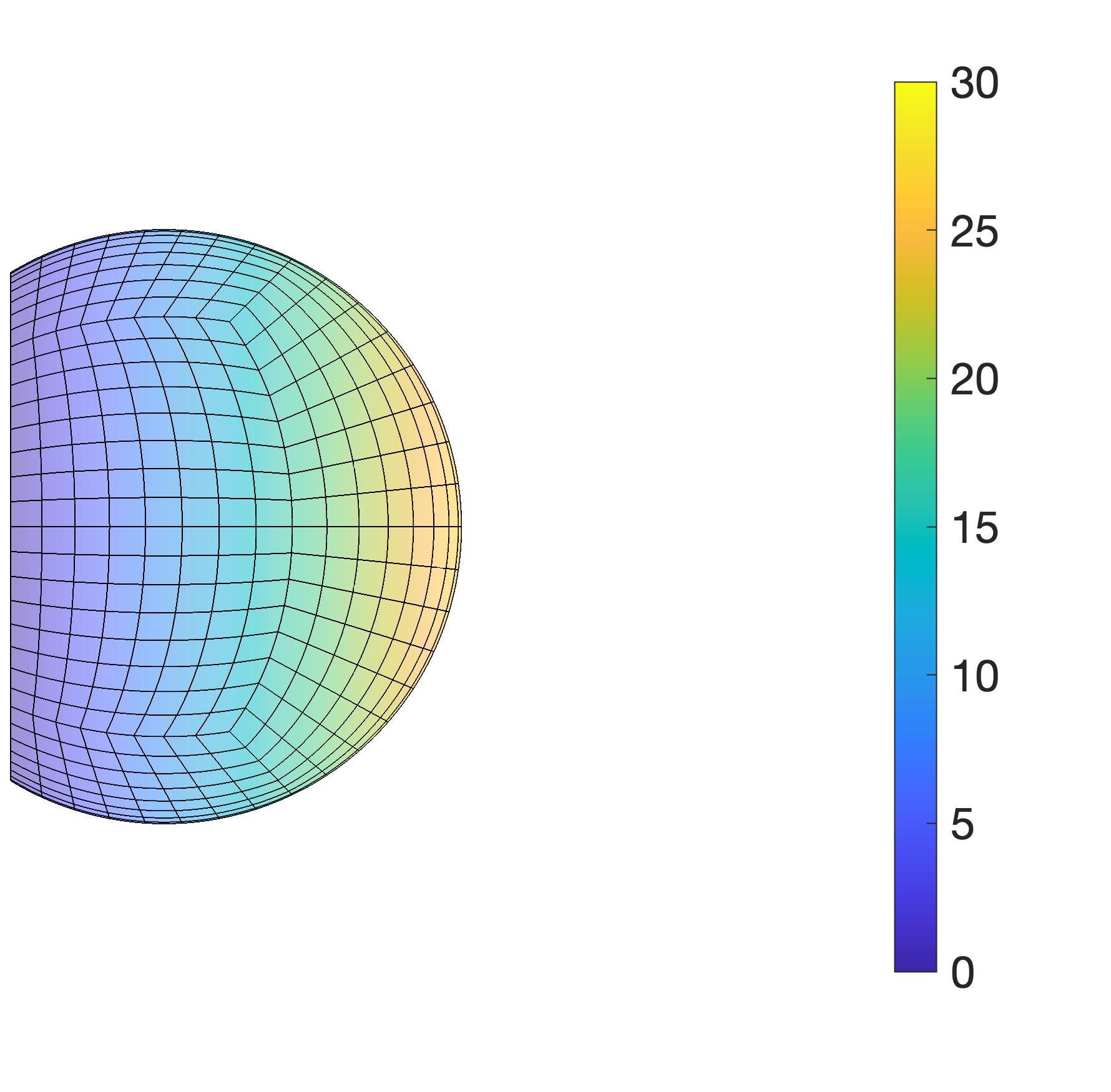}}
\put(-0.75,2.95){\includegraphics[height=39mm]{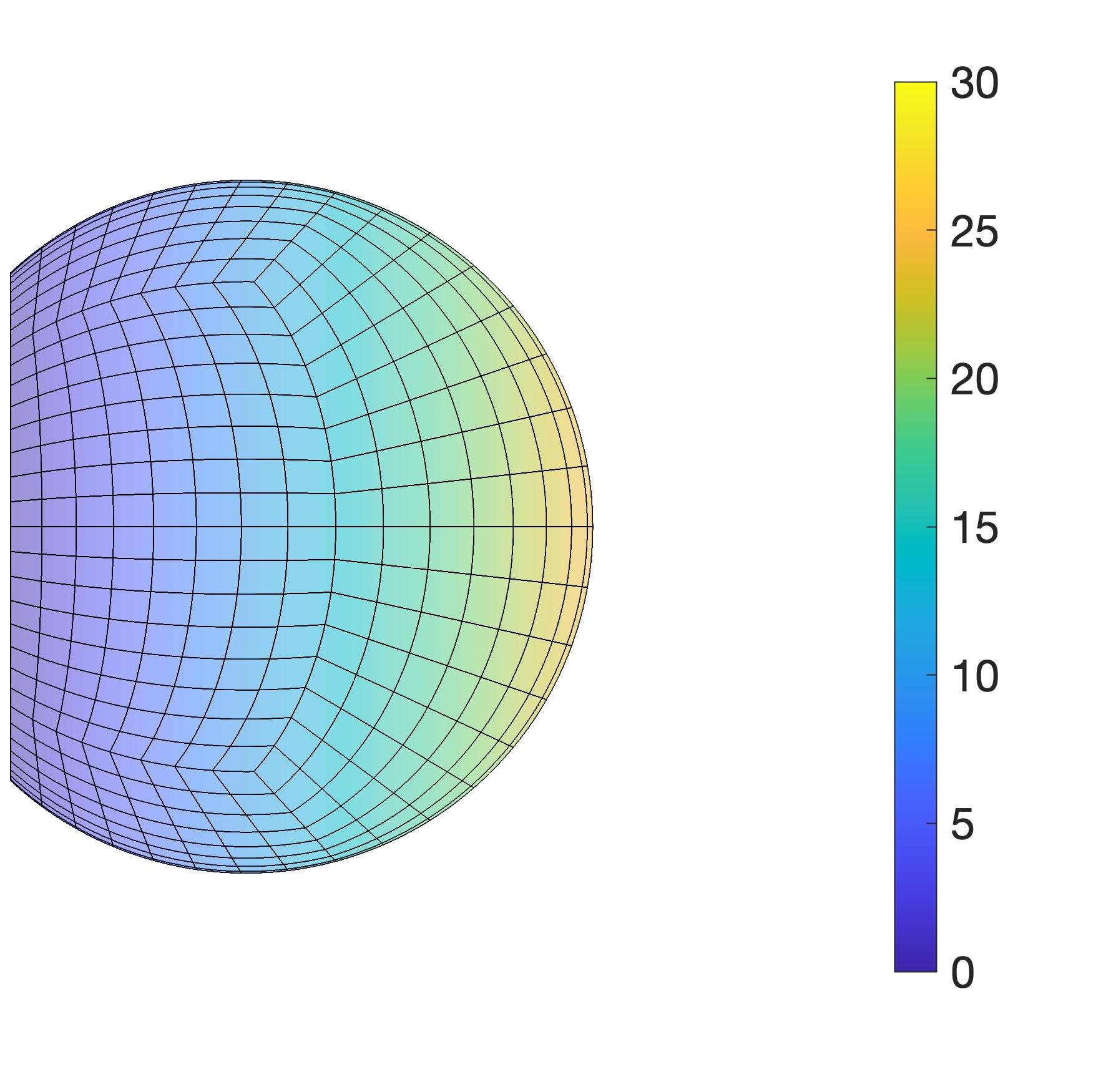}}
\put(1.55,2.95){\includegraphics[height=39mm]{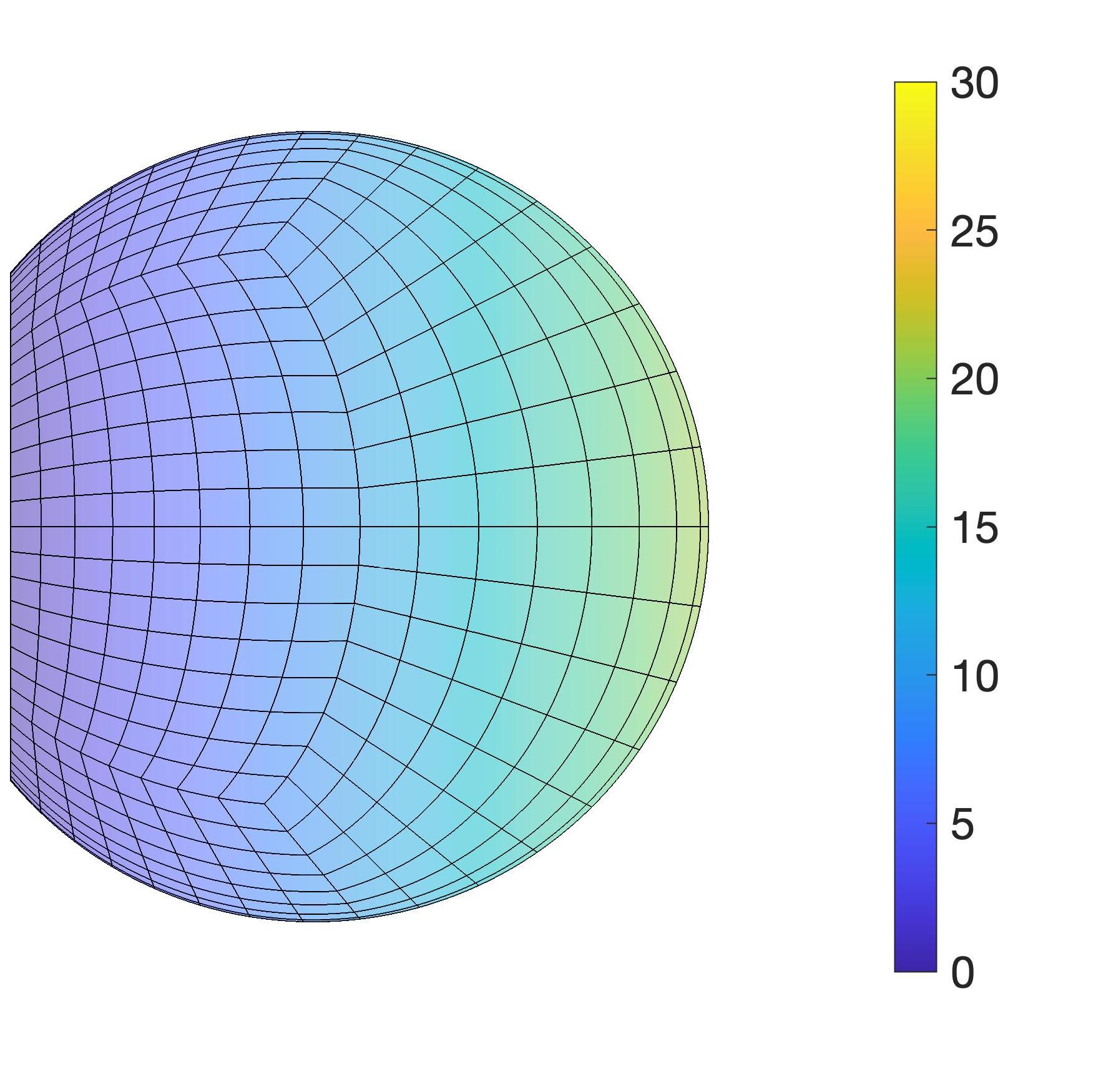}}
\put(4.3,2.95){\includegraphics[height=39mm]{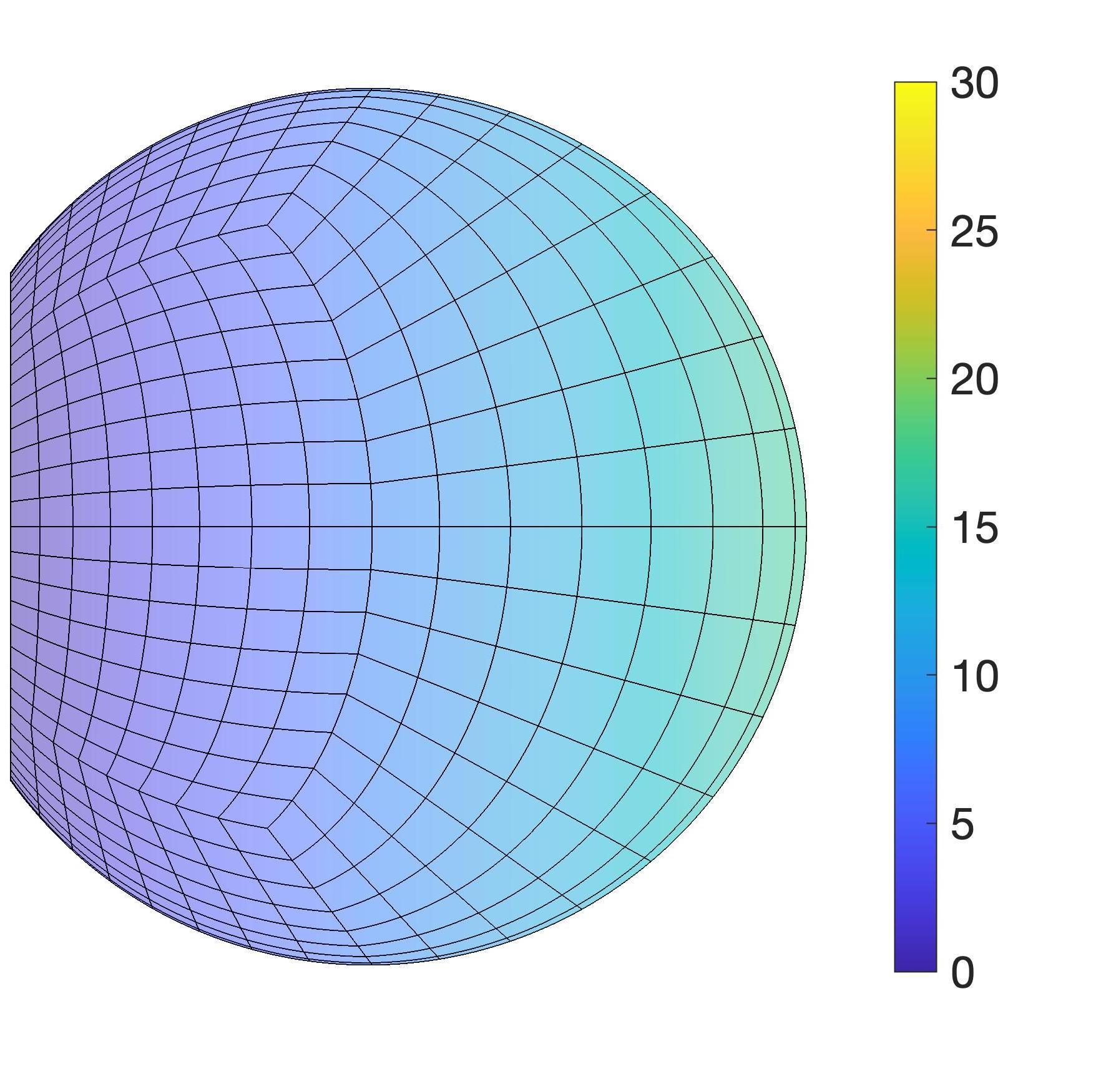}}
\put(-7.95,-.6){\includegraphics[height=39mm]{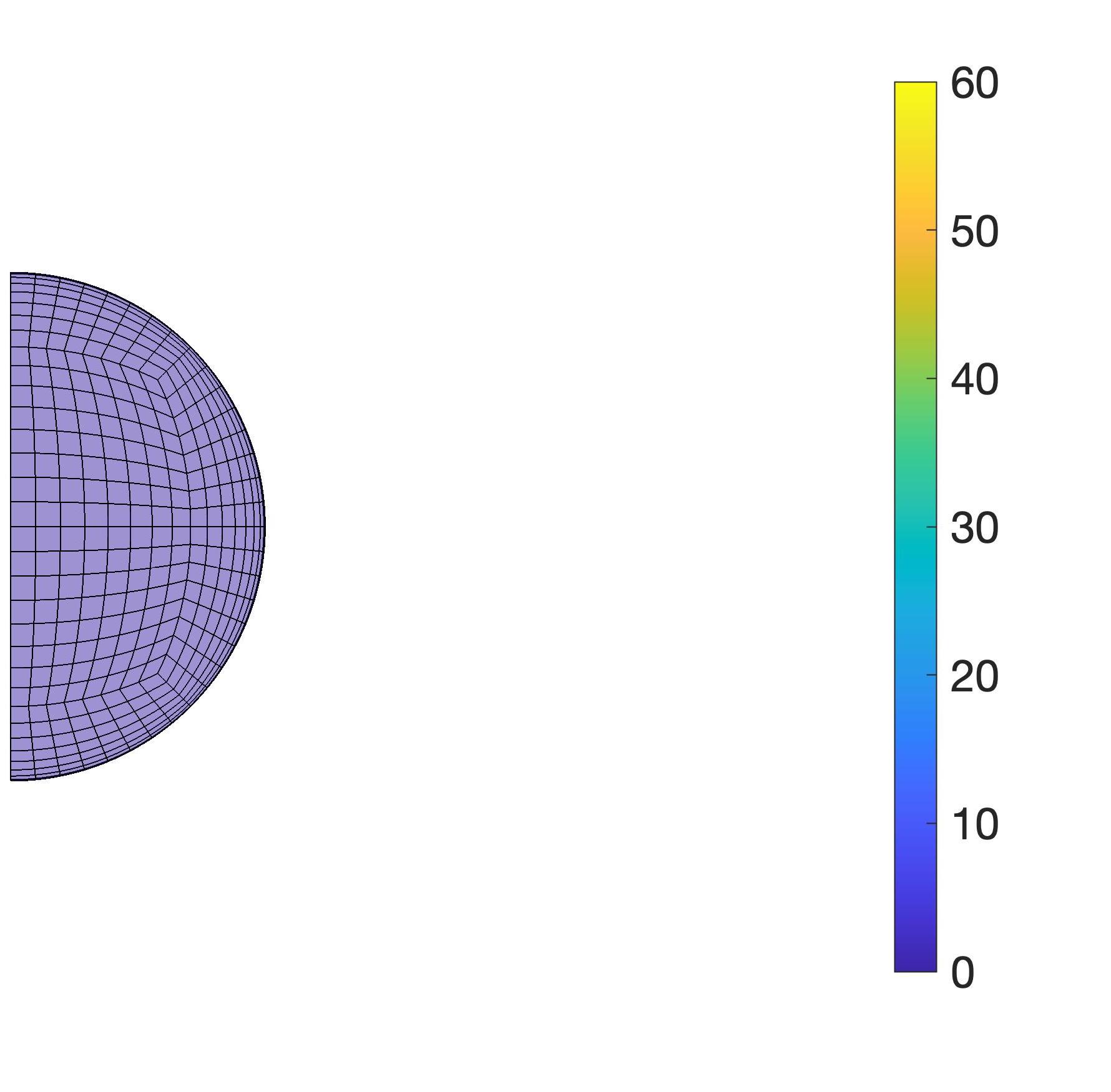}}
\put(-6.8,-.6){\includegraphics[height=39mm]{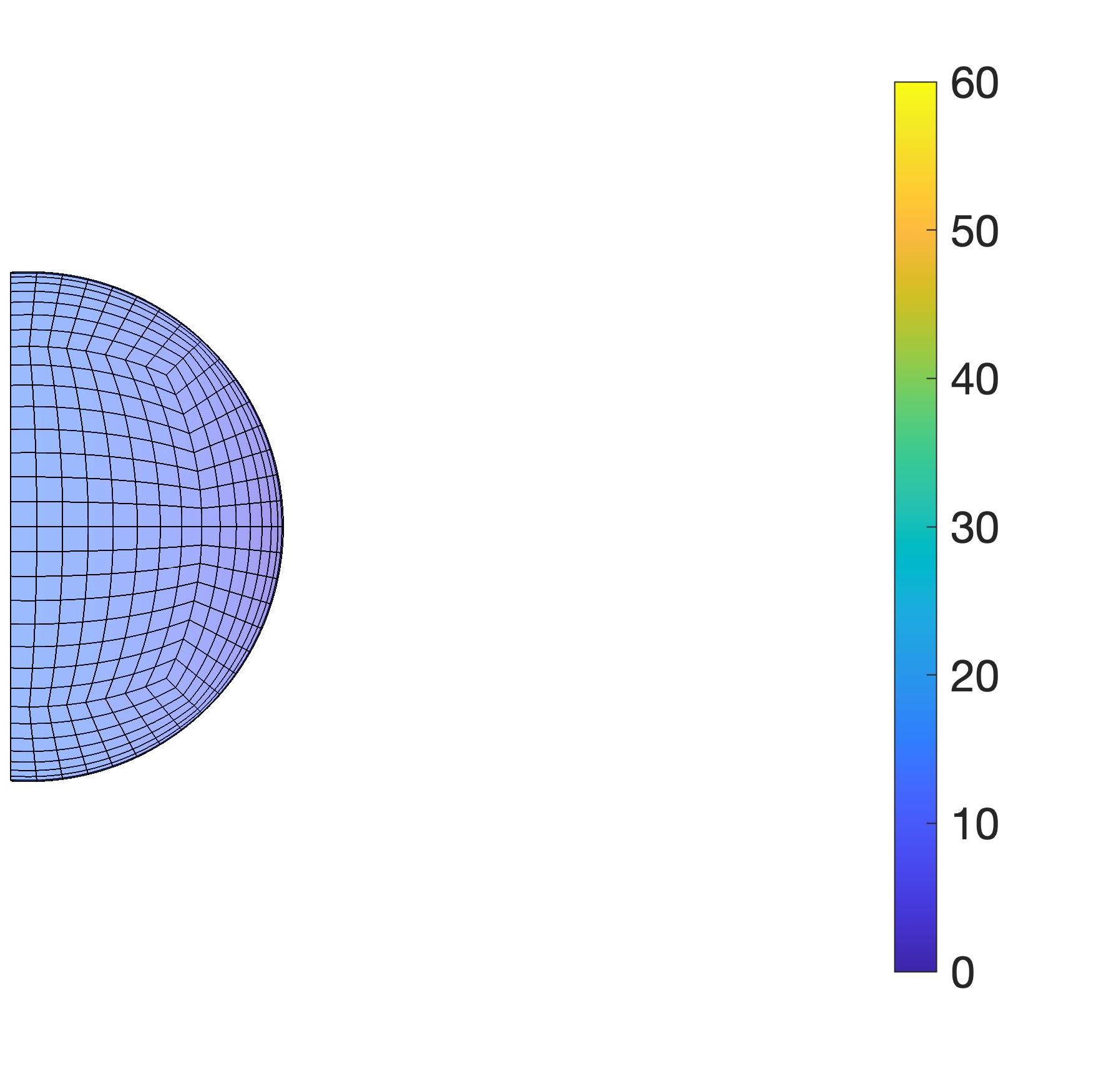}}
\put(-5.6,-.6){\includegraphics[height=39mm]{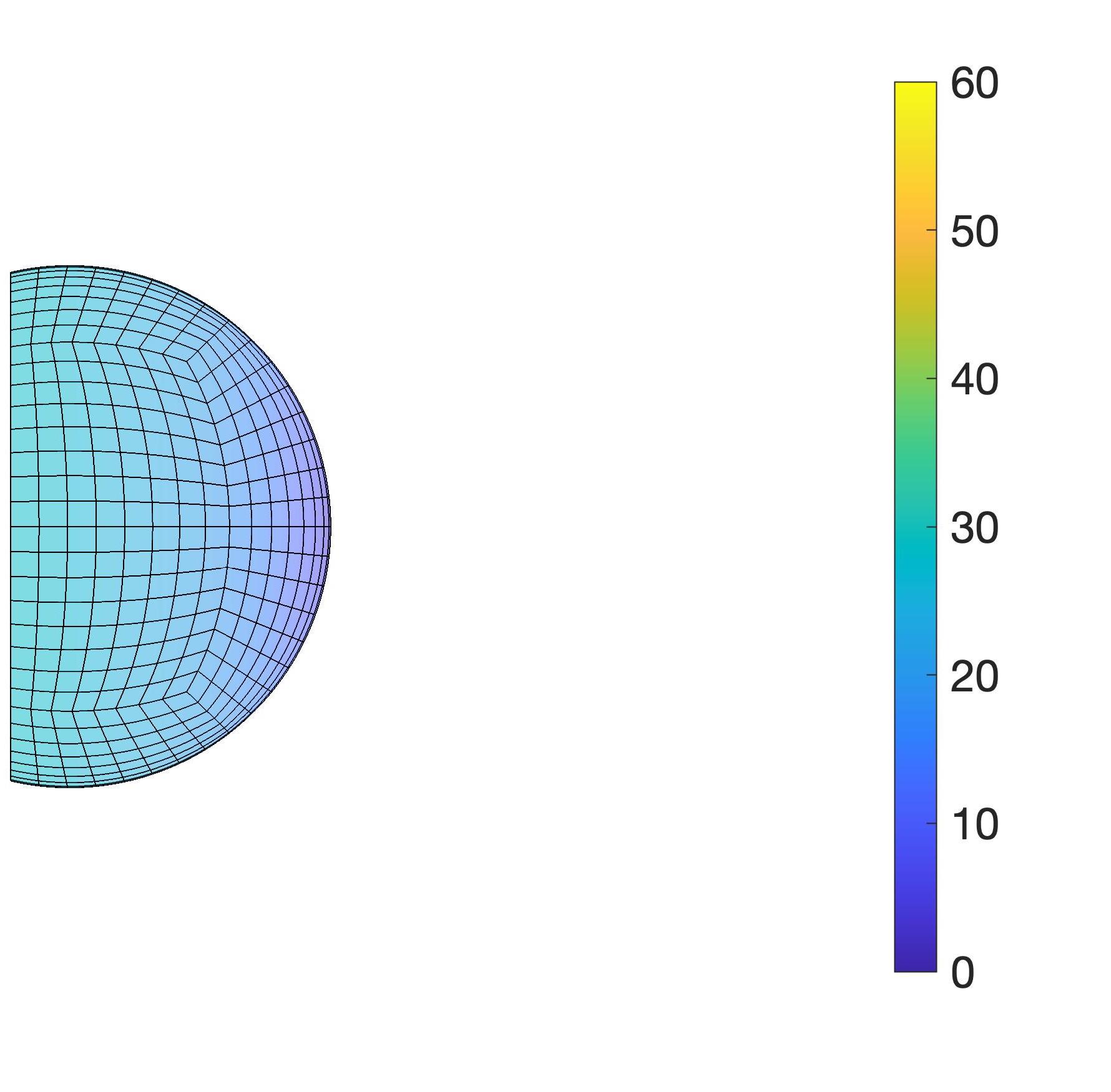}}
\put(-4.2,-.6){\includegraphics[height=39mm]{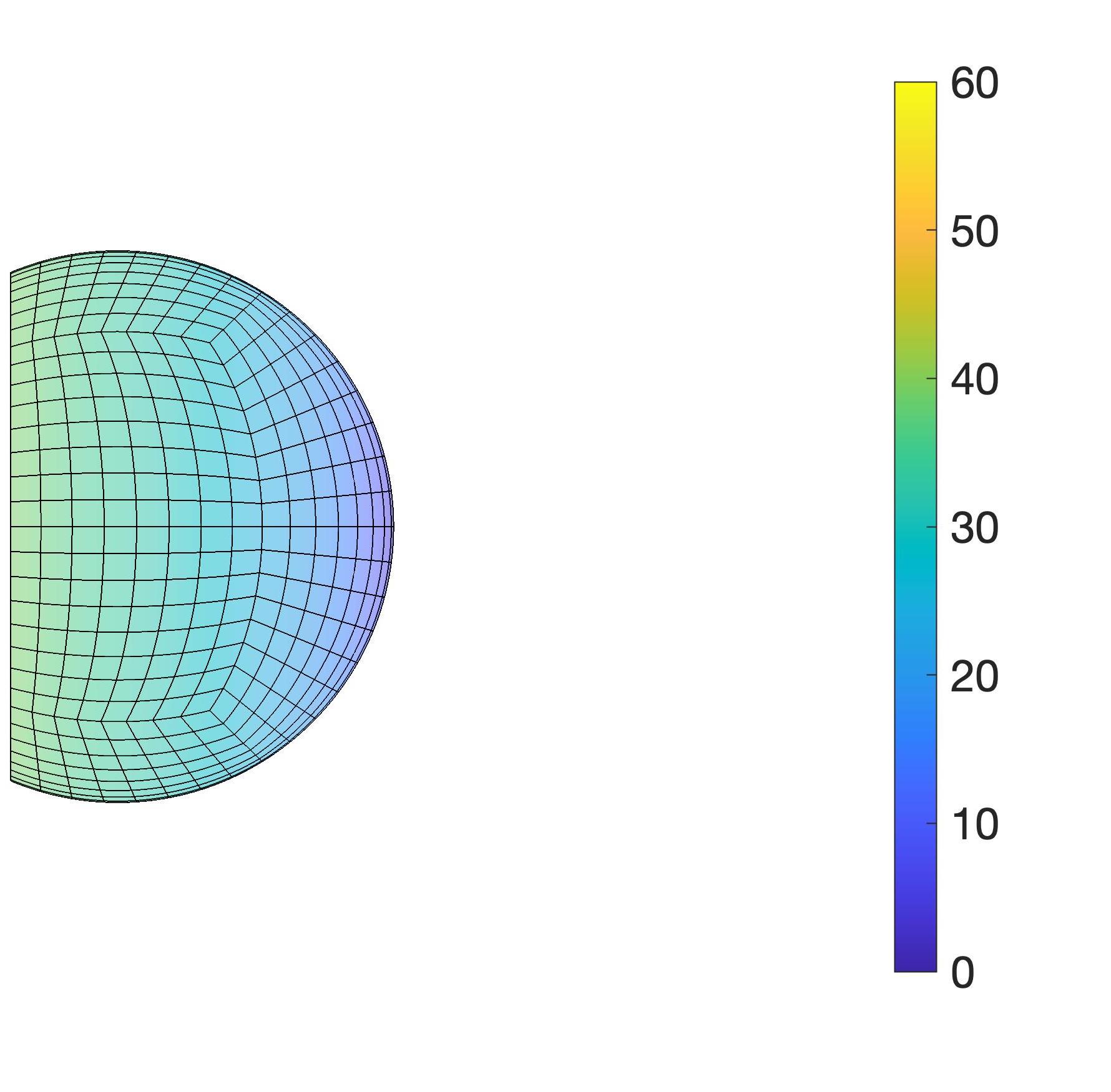}}
\put(-2.6,-.6){\includegraphics[height=39mm]{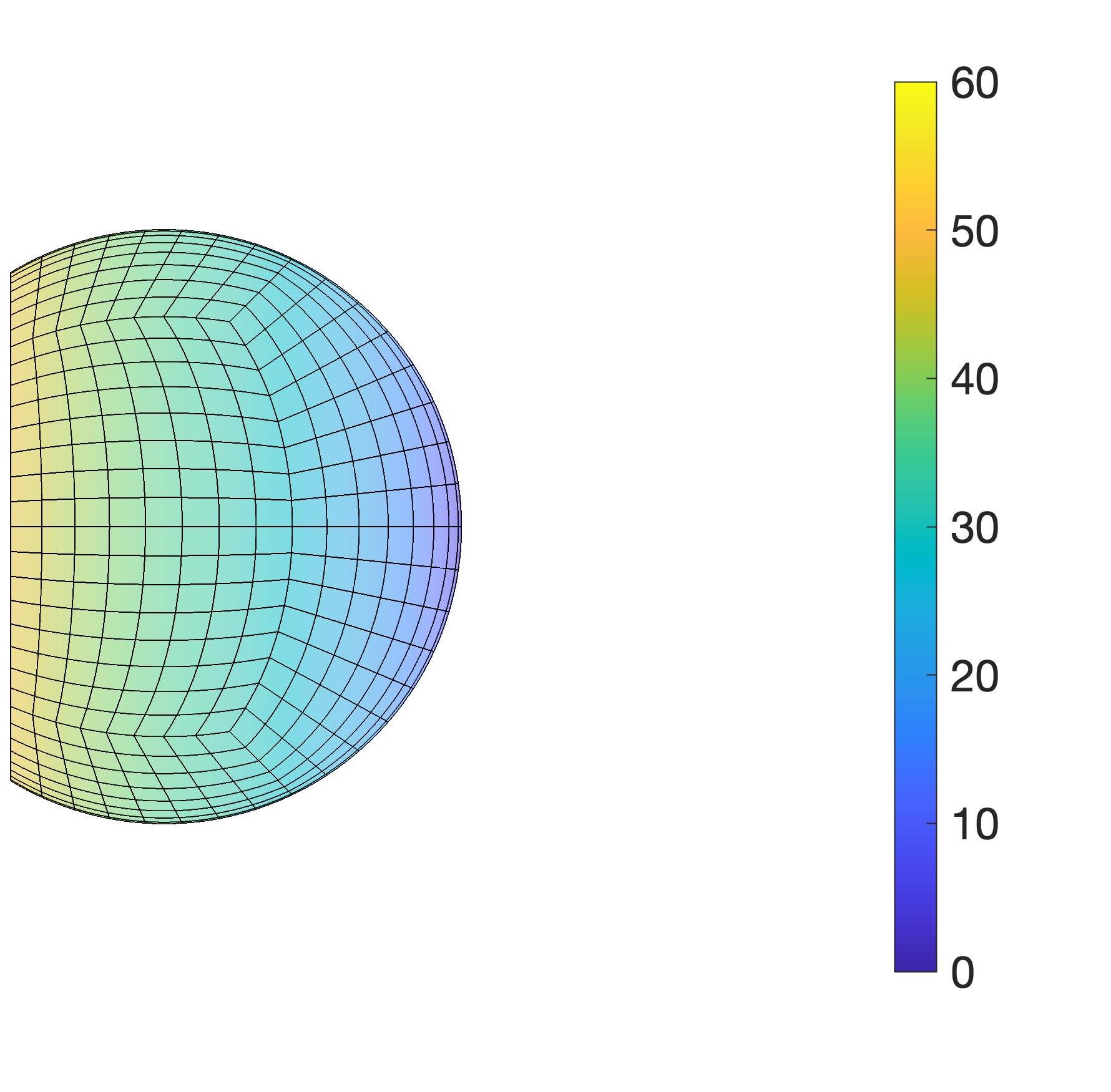}}
\put(-.75,-.6){\includegraphics[height=39mm]{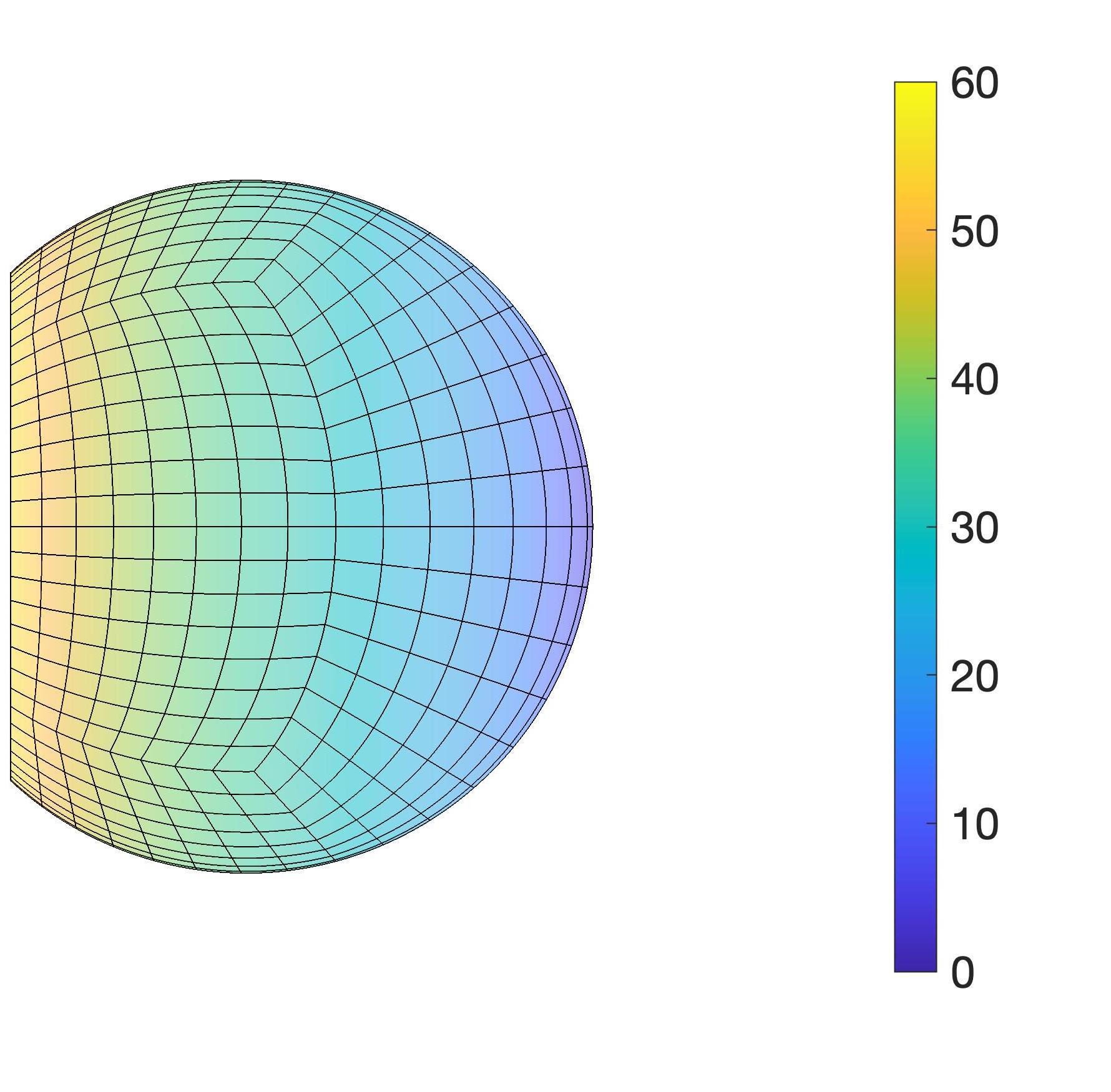}}
\put(1.55,-.6){\includegraphics[height=39mm]{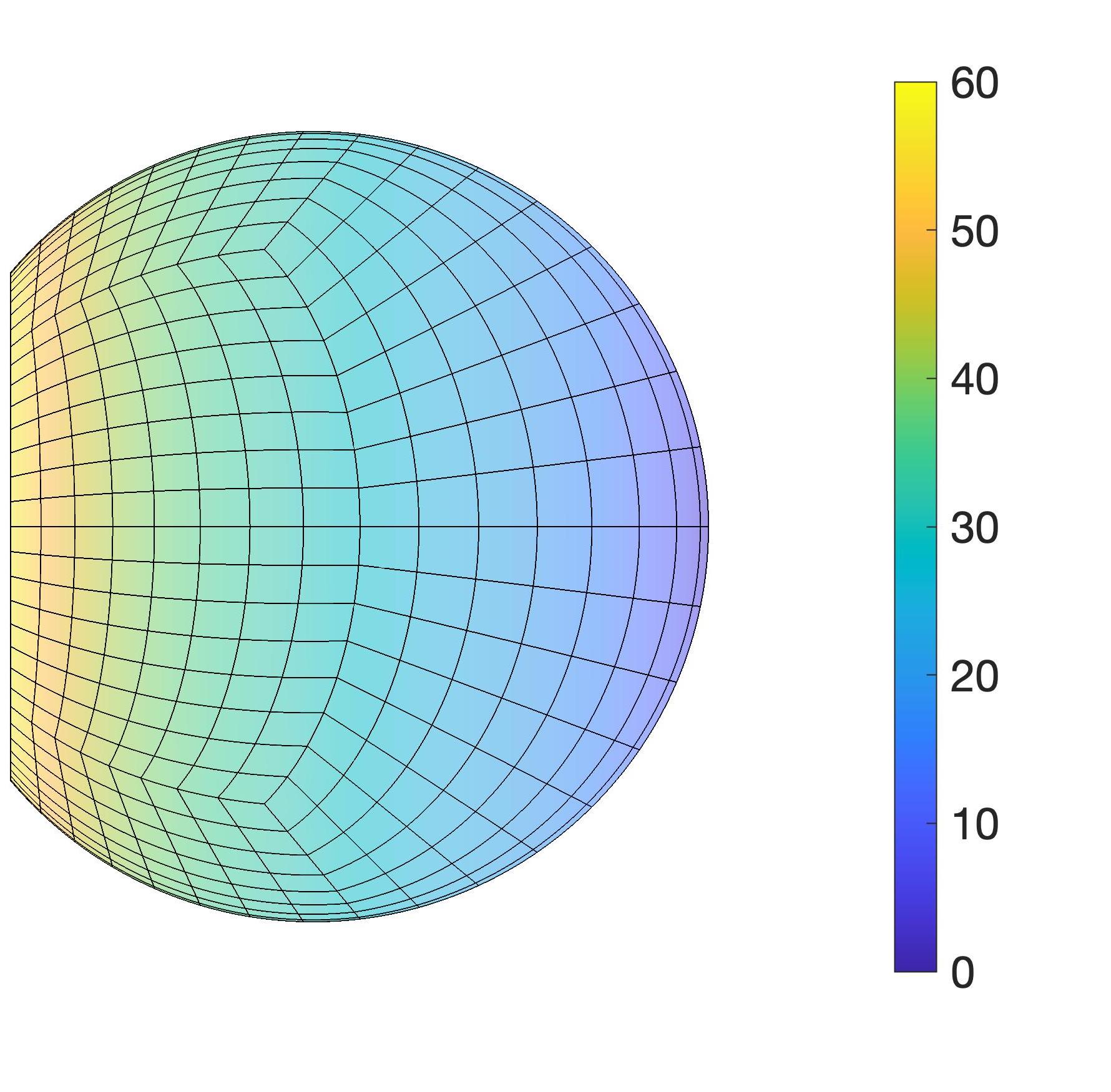}}
\put(4.3,-.6){\includegraphics[height=39mm]{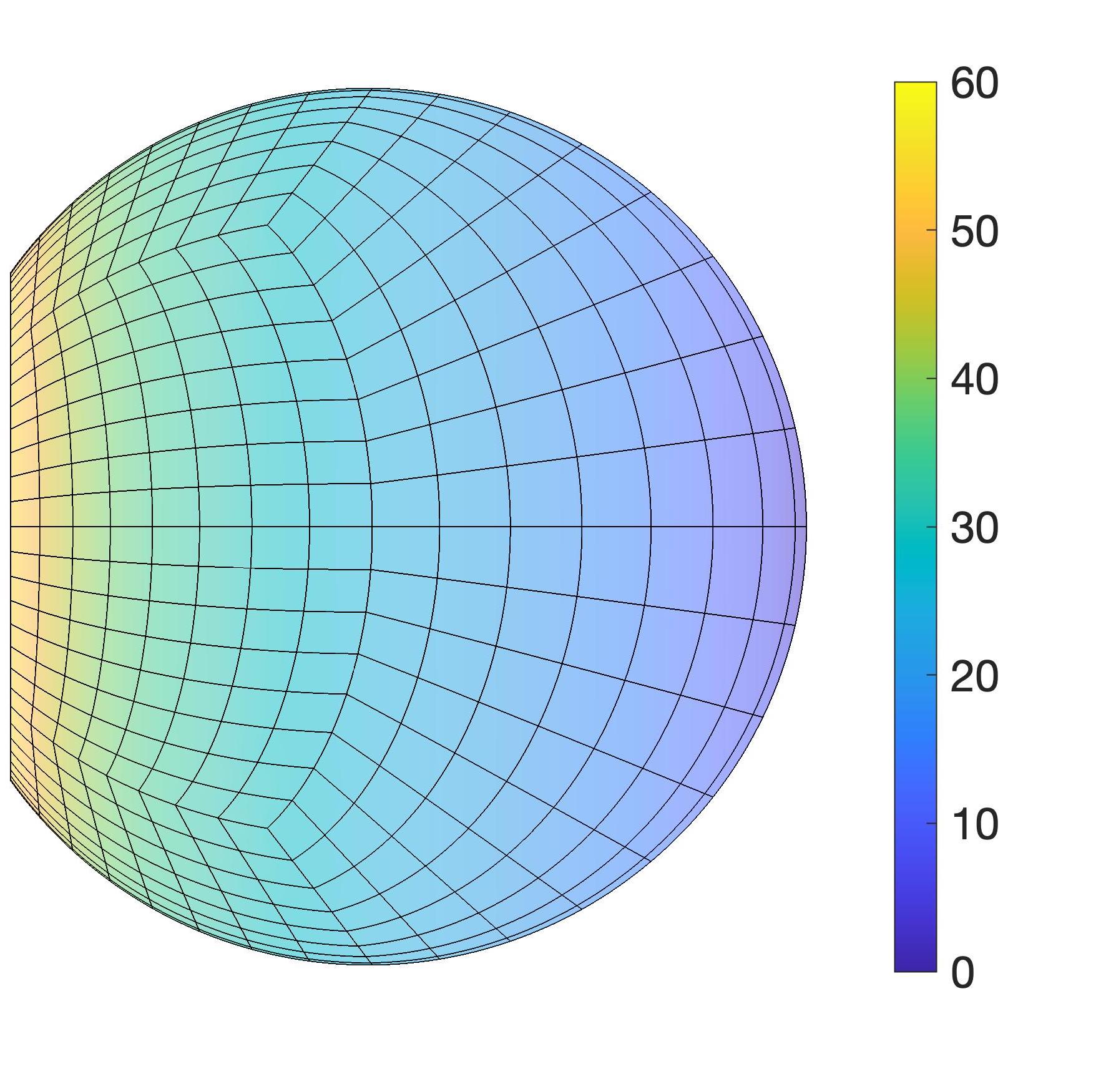}}
\end{picture}
\caption{Laminar inflation of a soap bubble: 
Bubble shape at $t = \{0,\,0.1,\,0.2,\,0.3,\,0.4,\,0.6,\,0.8,\,1\}$s, i.e.~$V \approx \{1,\,1.1,\,1.4,\,2.0,\,2.7,\,4.7,\,7.2,\,10\}V_0$ (left to right).
Top row: normal fluid velocity $v_\mrn$; bottom row: tangential fluid velocity $\norm{\bv_\mrt}$.
The results are from elastic mesh motion with $\tmu = 10^{-7}\mu$N/mm.} 
\label{f:SoapBublLam1}
\end{center}
\end{figure}

\begin{figure}[H]
\begin{center} \unitlength1cm
\begin{picture}(0,6.8)
\put(-7.95,3.1){\includegraphics[height=39mm]{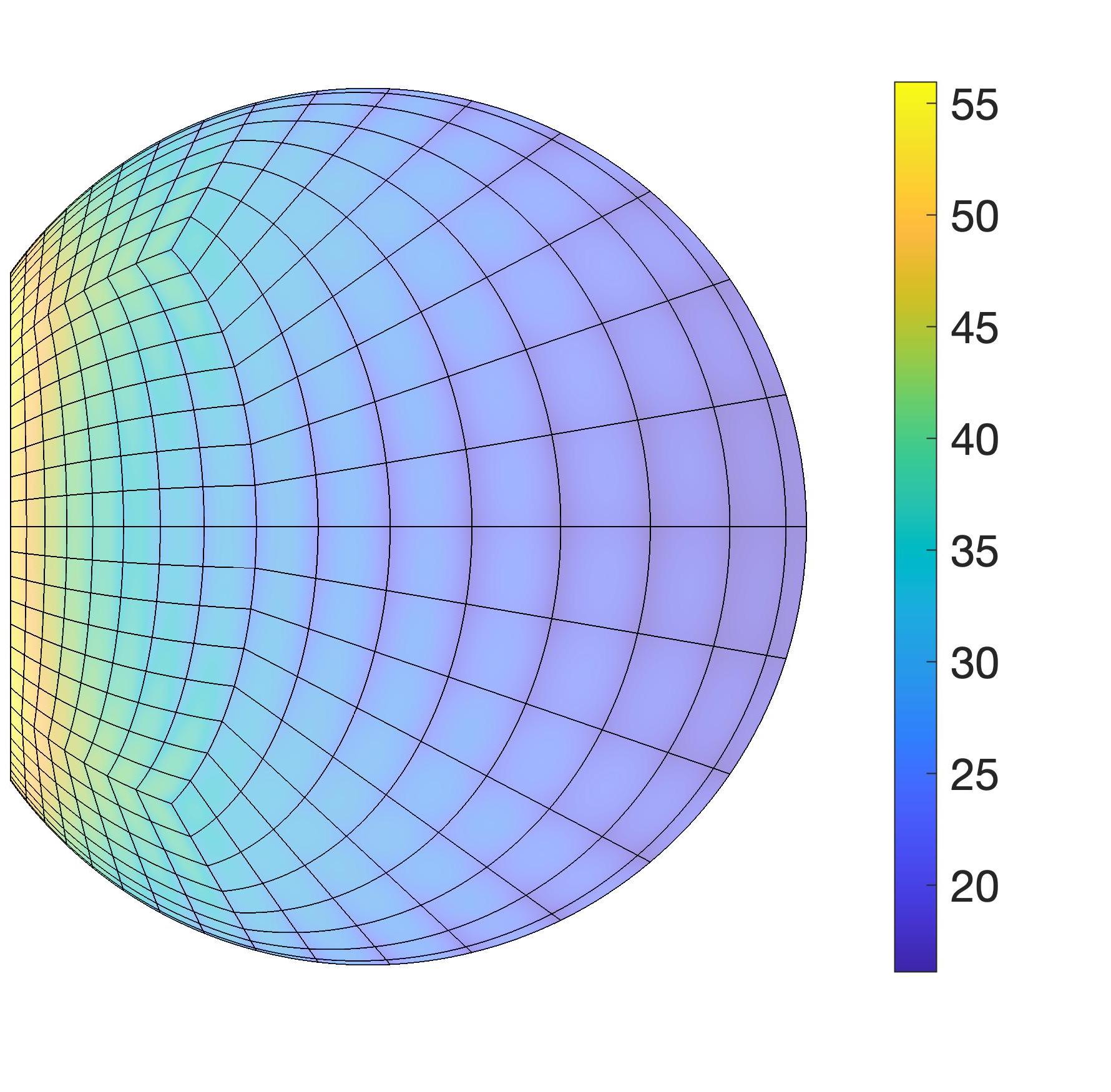}}
\put(-3.95,3.1){\includegraphics[height=39mm]{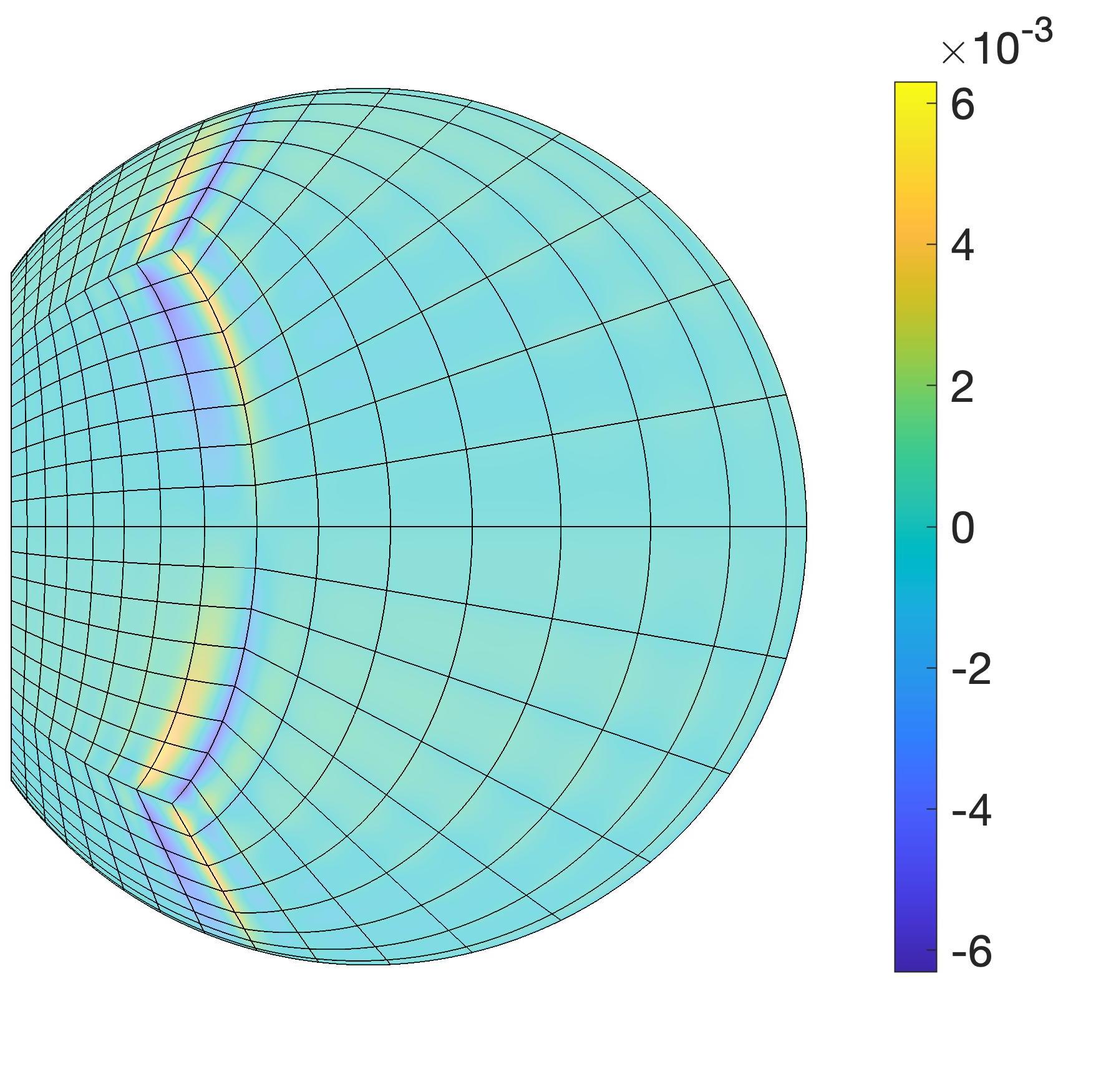}}
\put(.0,3.1){\includegraphics[height=39mm]{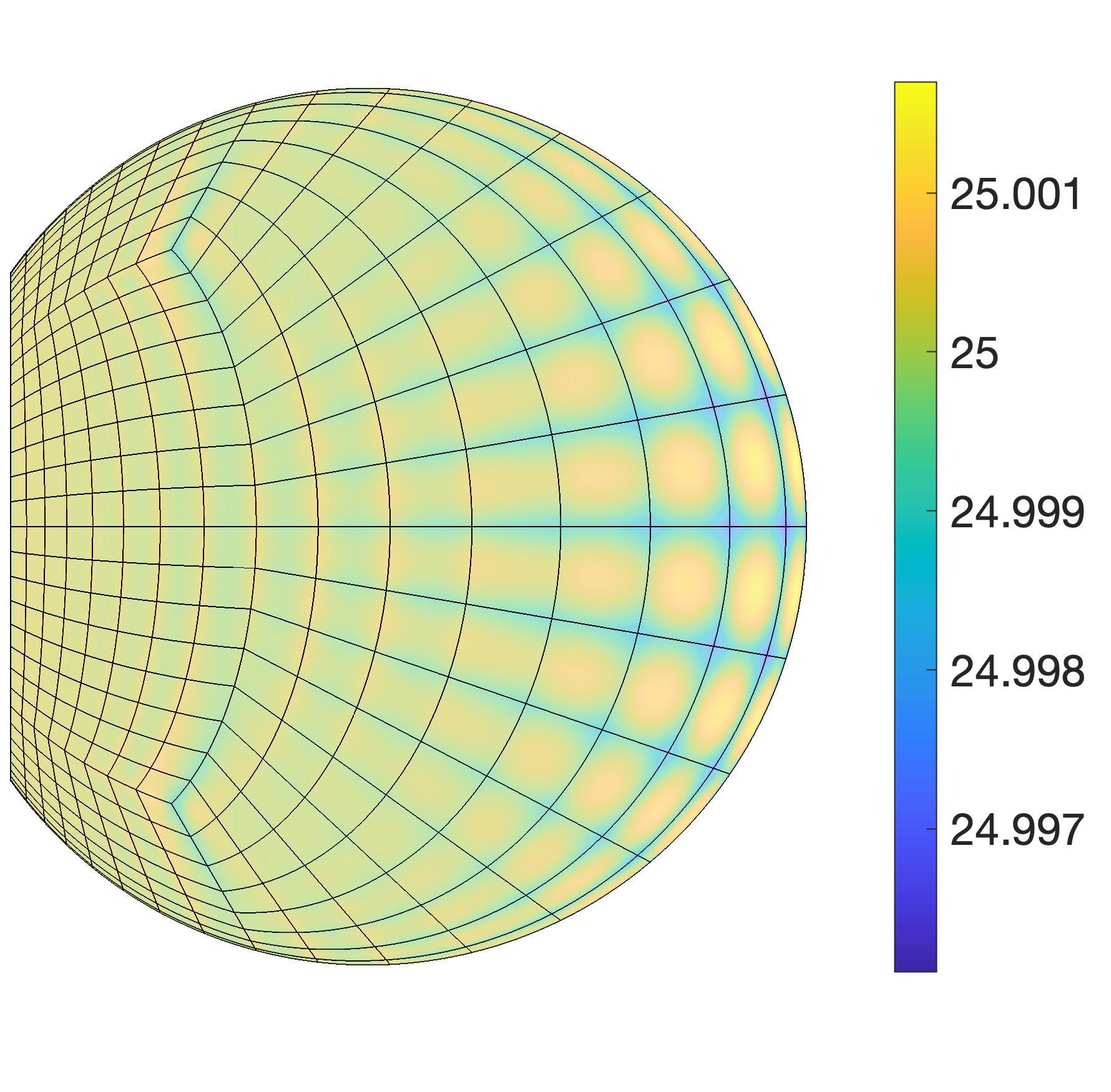}}
\put(4.3,3.1){\includegraphics[height=39mm]{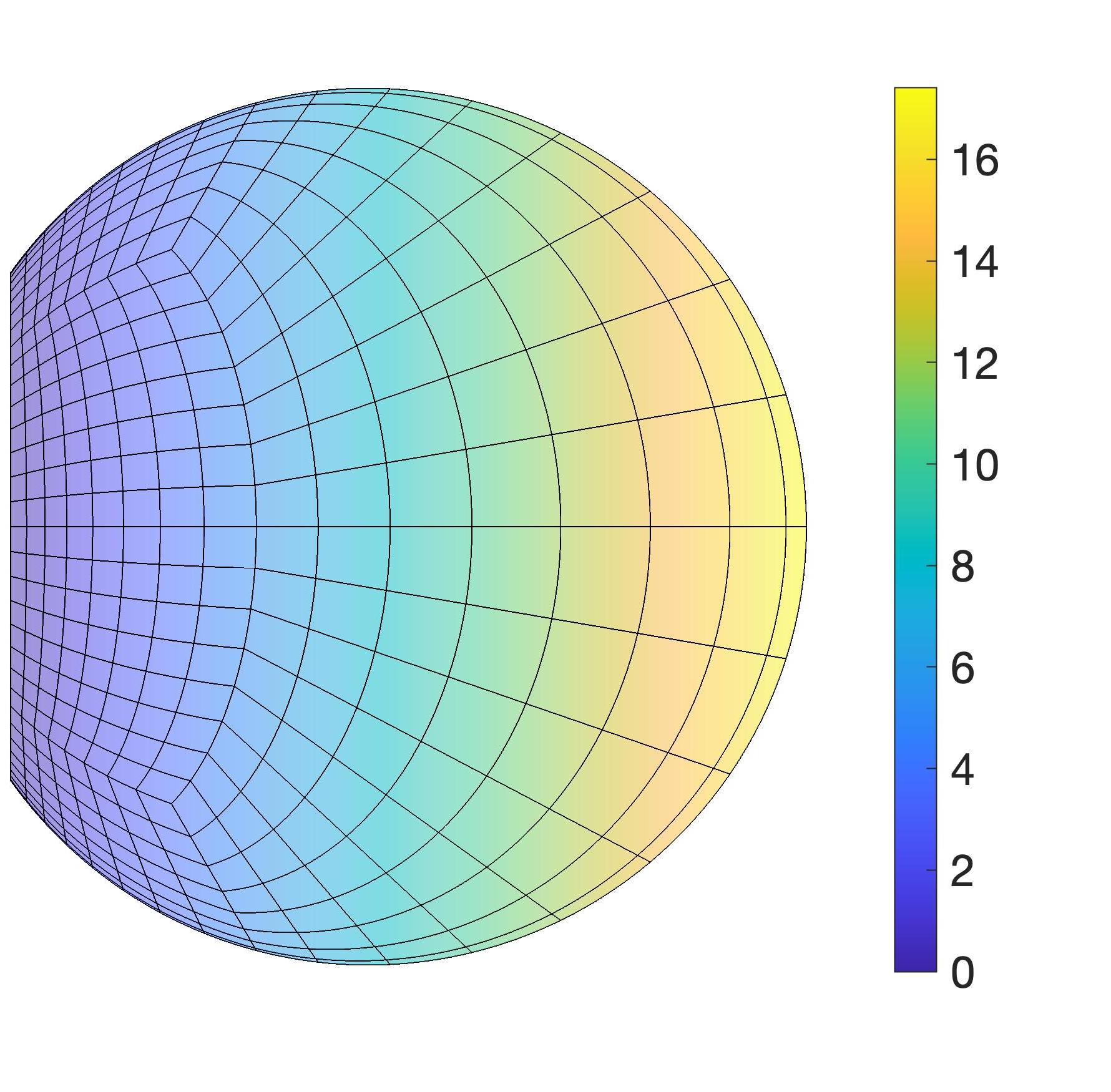}}
\put(-7.95,-.55){\includegraphics[height=39mm]{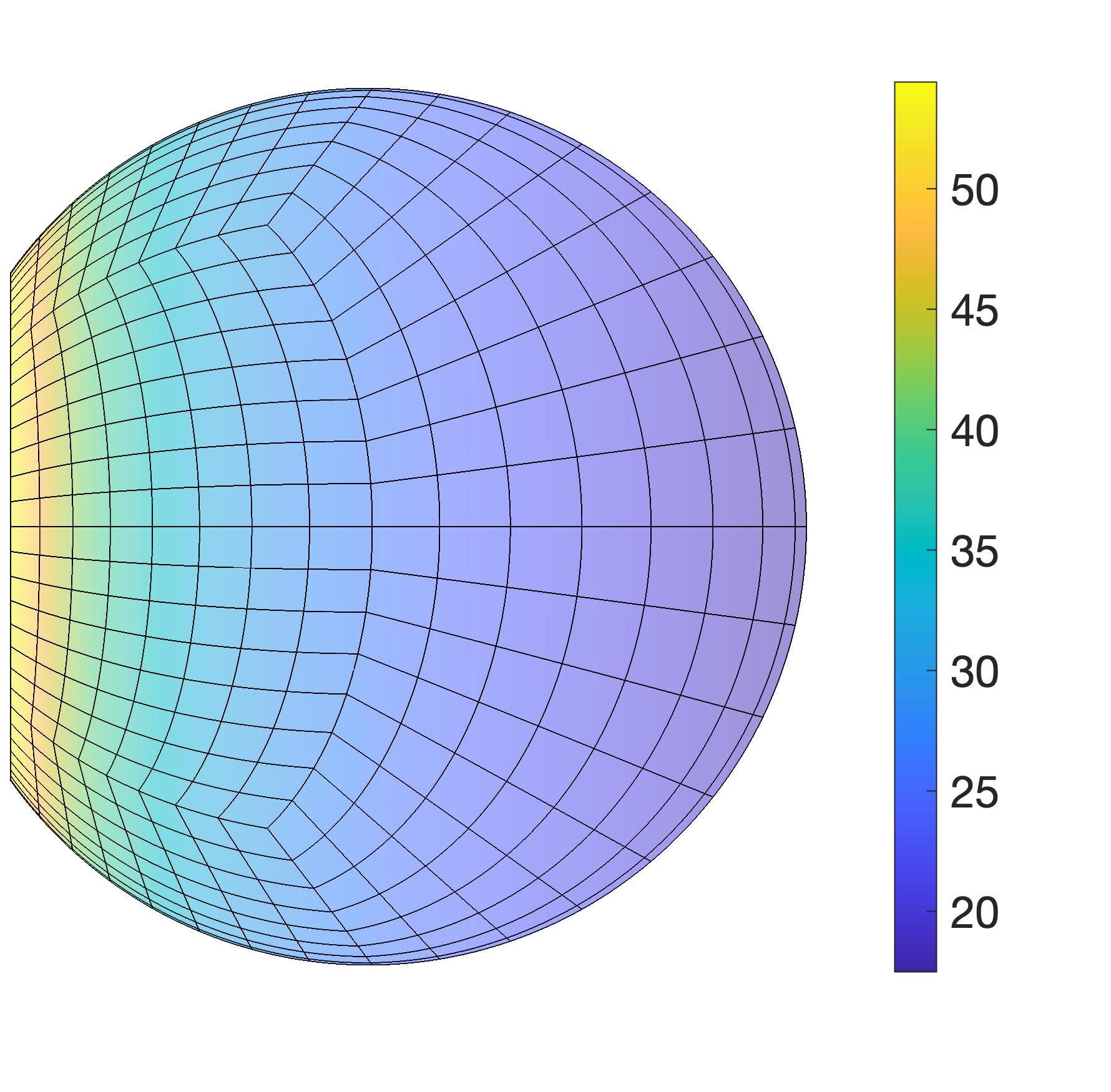}}
\put(-3.95,-.55){\includegraphics[height=39mm]{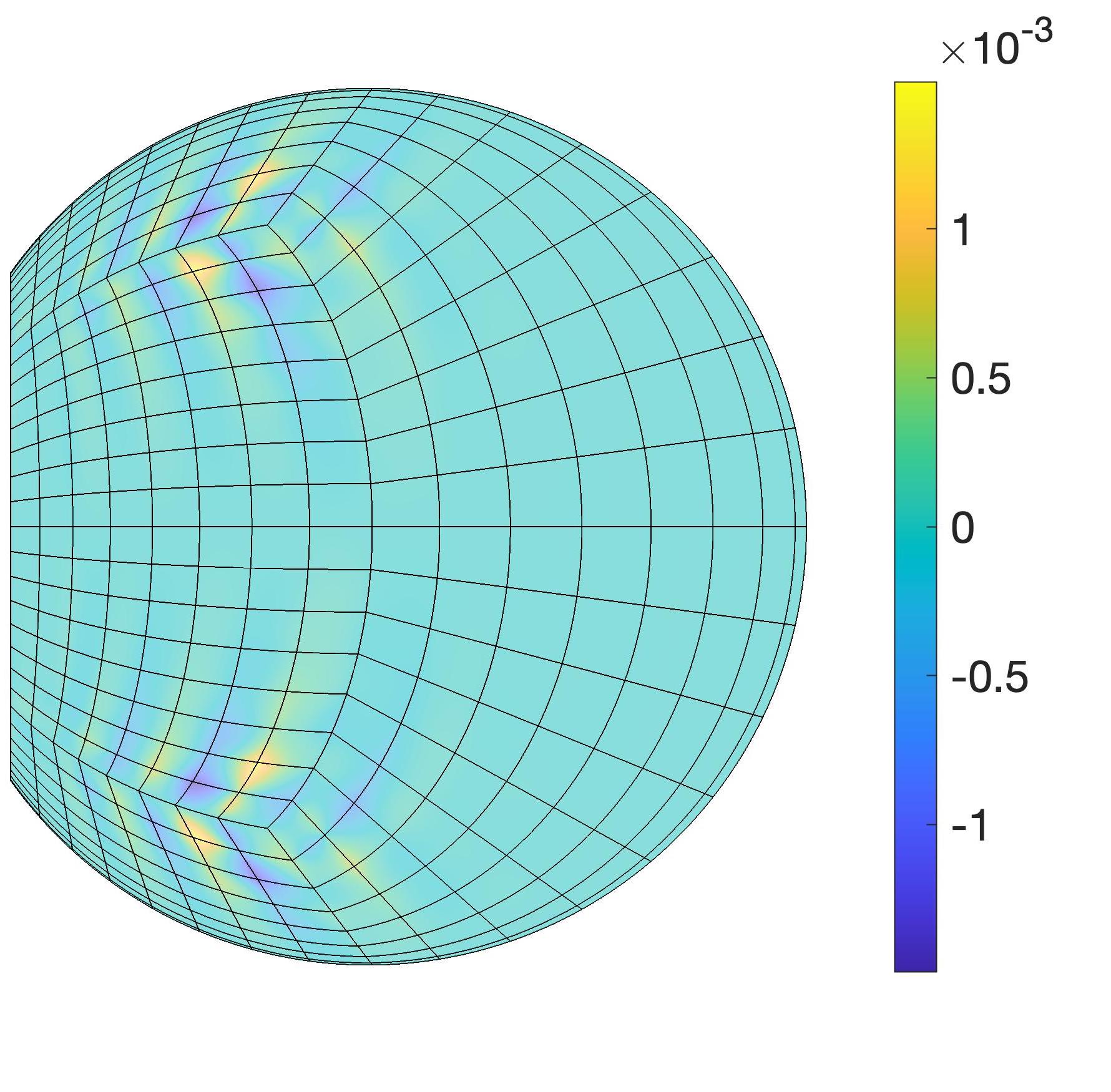}}
\put(.0,-.55){\includegraphics[height=39mm]{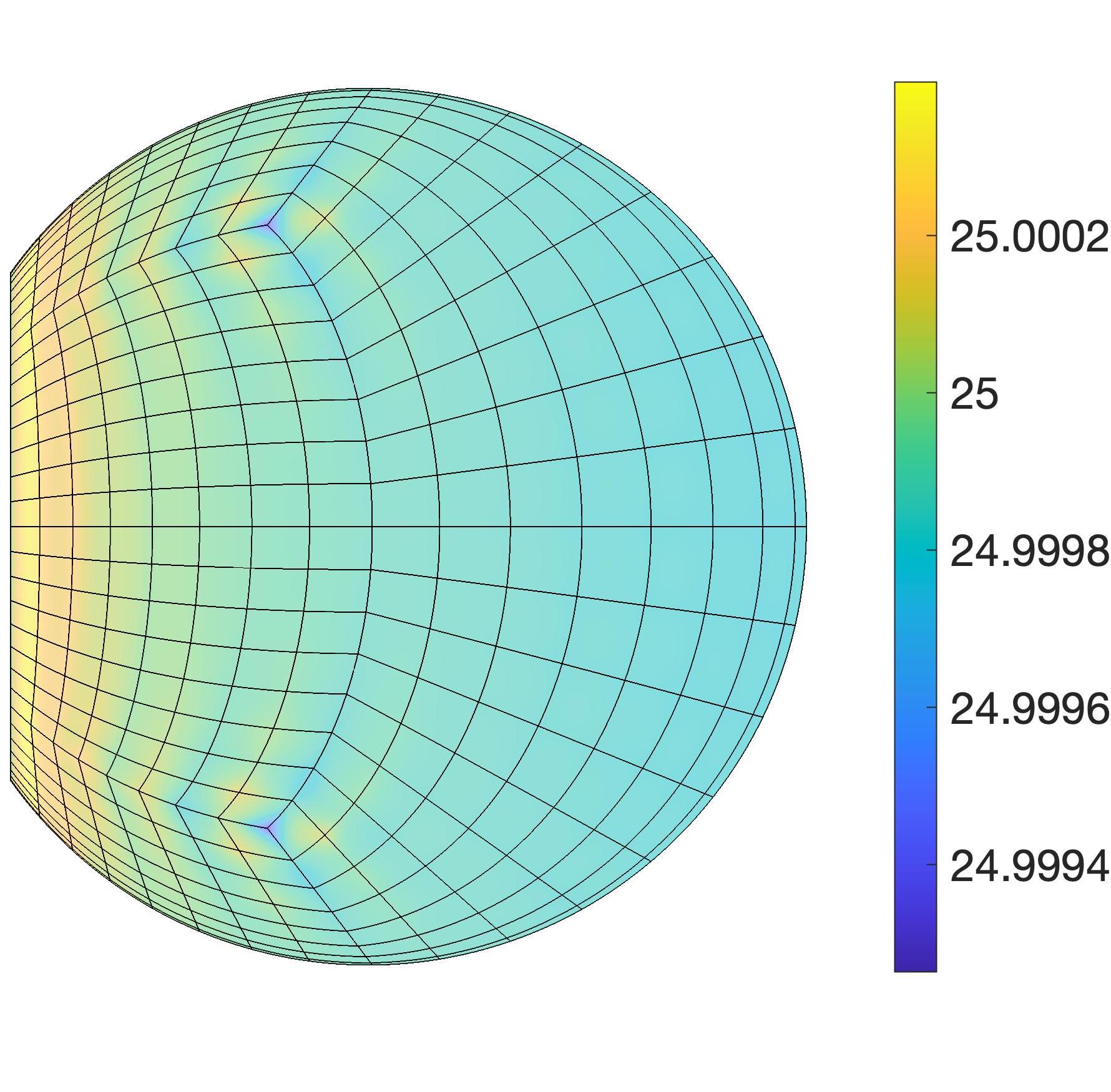}}
\put(4.3,-.55){\includegraphics[height=39mm]{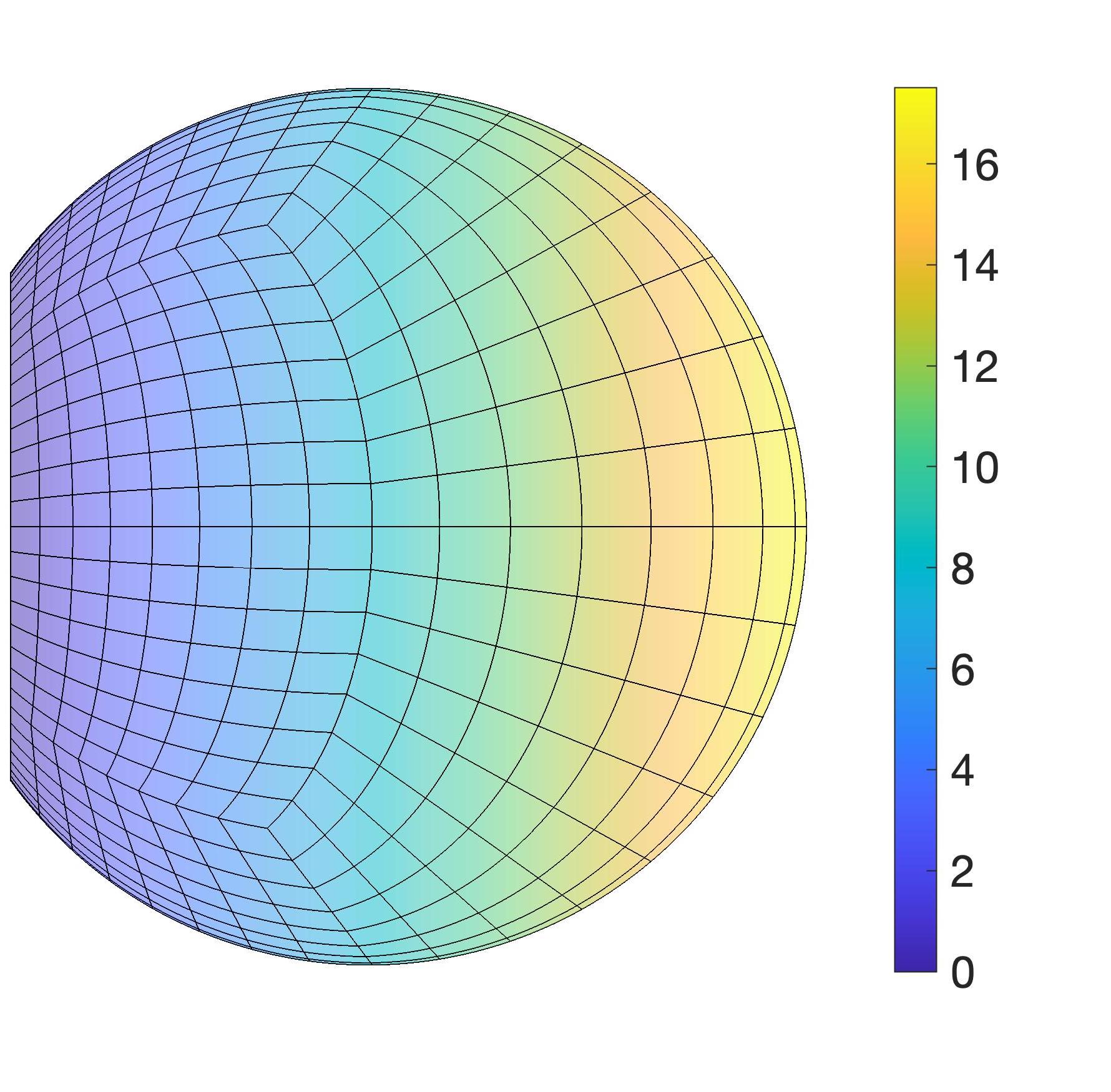}}
\end{picture}
\caption{Laminar inflation of a soap bubble:
Left to right: fluid velocity $\bv$ [mm/s], vorticity $\omega$ [1/s], surface tension $q$ [$\mu$N/mm] and mesh velocity $\bv_\mrm$ [mm/s] for $m=8$, $V = 10V_0$ and $\tmu = 10^{-7}\mu$N/mm.
Top row: Eulerian mesh motion; Bottom row: elastic mesh motion.
The latter is more accurate, especially for $\bv$, $\omega$ and $q$. 
It also shows much smaller mesh deformations towards the bubble tip.}
\label{f:SoapBublLam2}
\end{center}
\end{figure}

Fig.~\ref{f:SoapBublLam1} shows the resulting shape change of the bubble during inflation.
%-----------------------------------------------------------------
% run pSymBubble; dt = 80/m
%-----------------------------------------------------------------
As seen, for these parameters, the bubble remains spherical during inflation.
This is due to the small bubble radius and high surface tension.
The comparison between the in-plane Eulerian and in-plane elastic mesh motions is shown in Fig.~\ref{f:SoapBublLam2}.
%-----------------------------------------------------------------
% run pSymBubble
%-----------------------------------------------------------------

\begin{figure}[h]
\begin{center} \unitlength1cm
\begin{picture}(0,5.8)
\put(-8,-.12){\includegraphics[height=58mm]{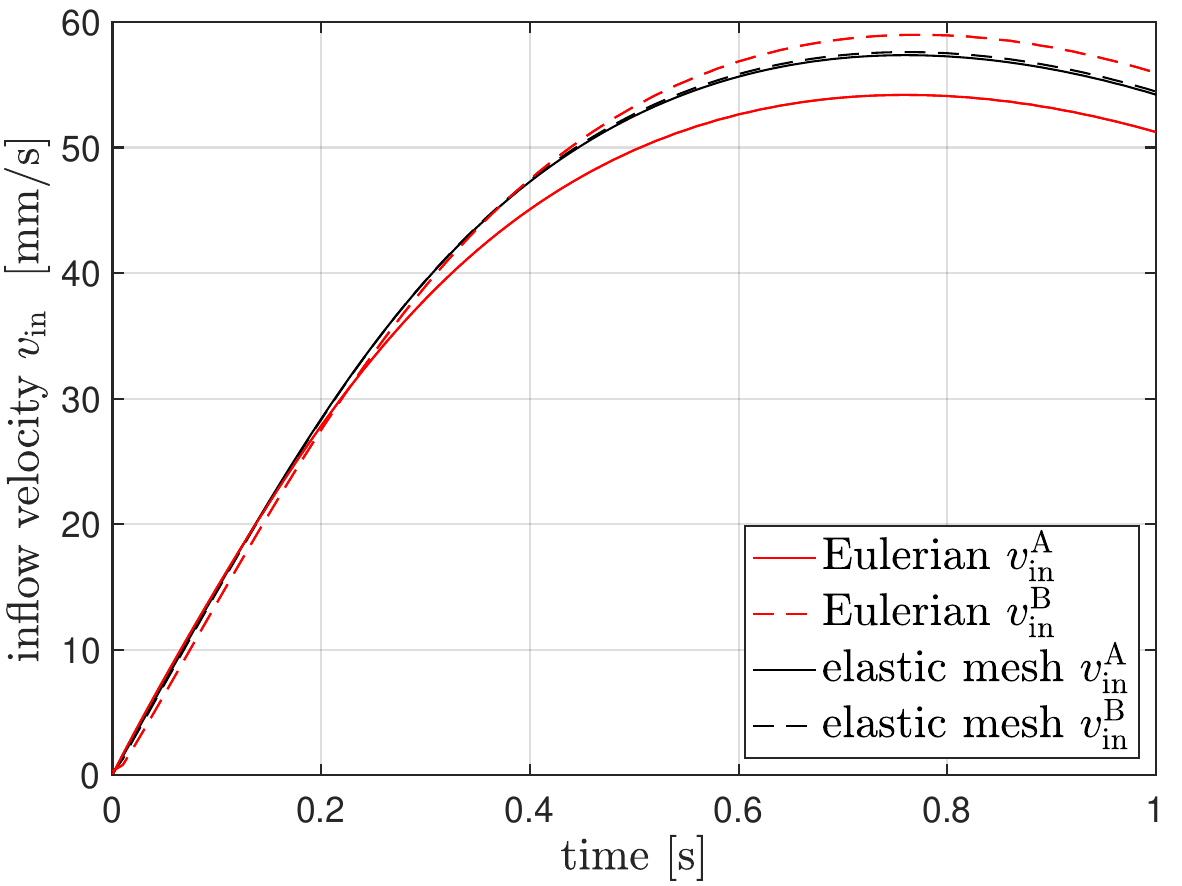}}
\put(0.2,-.2){\includegraphics[height=58mm]{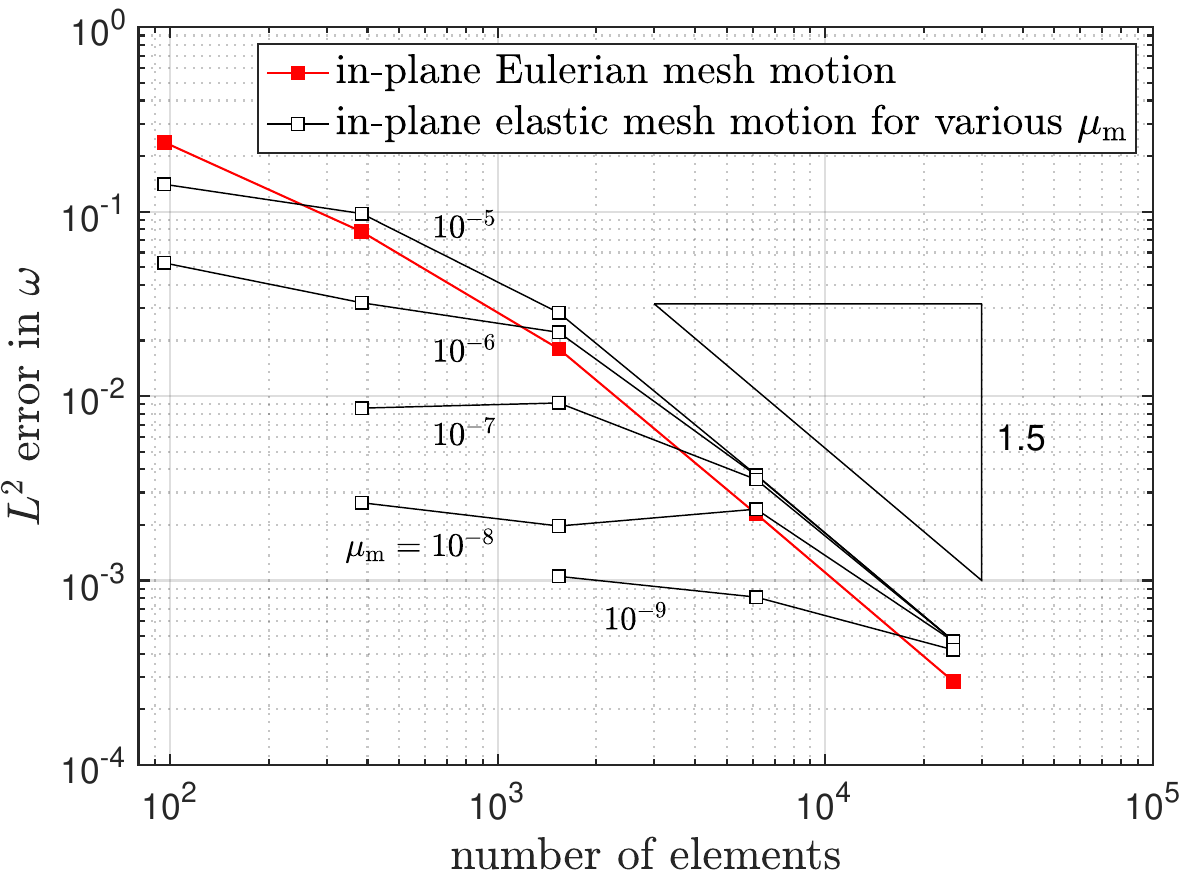}}
\put(-7.95,-.05){\footnotesize (a)}
\put(0.25,-.05){\footnotesize (b)}
\end{picture}
\caption{Laminar inflation of a soap bubble:
Accuracy comparison between the Eulerian and elastic mesh description. 
(a) Inflow velocity $v_\mathrm{in}(t)$ at inflow points A and B (at $\theta = 0$ and $\theta=\pi/4$) for $m=8$ and $\tmu = 10^{-7}\mu$N/mm.
(b) $L^2$ error of the surface vorticity (from Eq.~\eqref{e:L2w}) vs.~mesh refinement.
The elastic mesh description is much more accurate than the Eulerian mesh description, especially for coarse meshes and small $\tmu$.}
\label{f:SoapBublLam3}
\end{center}
\end{figure}

In the Eulerian case large errors are observed at the tip of the bubble.
In particular, $q$ and $\omega$ are quite off.
On the other hand, much more accurate results are obtained with elastic mesh ALE, as long as $\tmu$ is sufficiently small.
This is also seen in Fig.~\ref{f:SoapBublLam3}.
%-----------------------------------------------------------------
% run sSymBubbleNBC/*NBCnu10/*elastic, pSymBubble & pBubbleVorticity
%-----------------------------------------------------------------
Here, the left side shows the fluid velocities at two different inflow points, which should be identical.
As seen, the elastic mesh result is much better.
The right side shows the normalized $L^2$ error of the surface vorticity,
\eqb{l}
e_\omega := T\,\norm{\omega}_{L^2}\,,\quad
\norm{\omega}_{L^2} := \sqrt{\!\ds\int_\sS \omega^2\,\dif a}\,,\quad
\label{e:L2w}\eqe
normalized by the final time $T=1$s.
In theory, $e_\omega$ should be exactly zero, as there is no surface vorticity in this example.
Again, the elastic mesh results tend to be much better than the Eulerian ones, especially for small $\tmu$.\footnote{The smaller $\mu_\mrm$, the more accurate the results, especially for coarse meshes. But too small $\mu_\mrm$ cause ill-conditioning. The value $\mu_\mrm=10^{-6}\mu$N/mm seems to be a good compromise for the considered Lagrange meshes.}
As in the examples of Sec.~\ref{s:Nexf}, ideal convergence rates are obtained.
Interestingly, for each fixed $\tmu$, the error of the elastic mesh results starts at a very low level, then stagnates, and finally decreases together with the Eulerian case -- but falling slightly behind it.
The reason for this behavior is suspected to lie in the surface continuity:
The mesh elasticity formulation, contrary to the Eulerian formulation, relies on an accurate decomposition into in-plane and out-of-plane parts, as Eq.~\eqref{e:fio} shows.
Such a decomposition, however, is not represented very accurately with quadratic Lagrange elements, since they are only $C^0$-continuous at element boundaries.
$C^1$-continuous finite element discretizations can be expected to remedy this issue and yield further accuracy gains.
This is confirmed in the following example.
The superiority of $C^1$-continuous finite element discretizations was already seen for the quasi-static droplet simulations in \citet{droplet}, which also used
formulation \eqref{e:fio} to stabilize the mesh.

\begin{figure}[h]
\begin{center} \unitlength1cm
\begin{picture}(0,5.7)
\put(-8.5,-.15){\includegraphics[height=61mm]{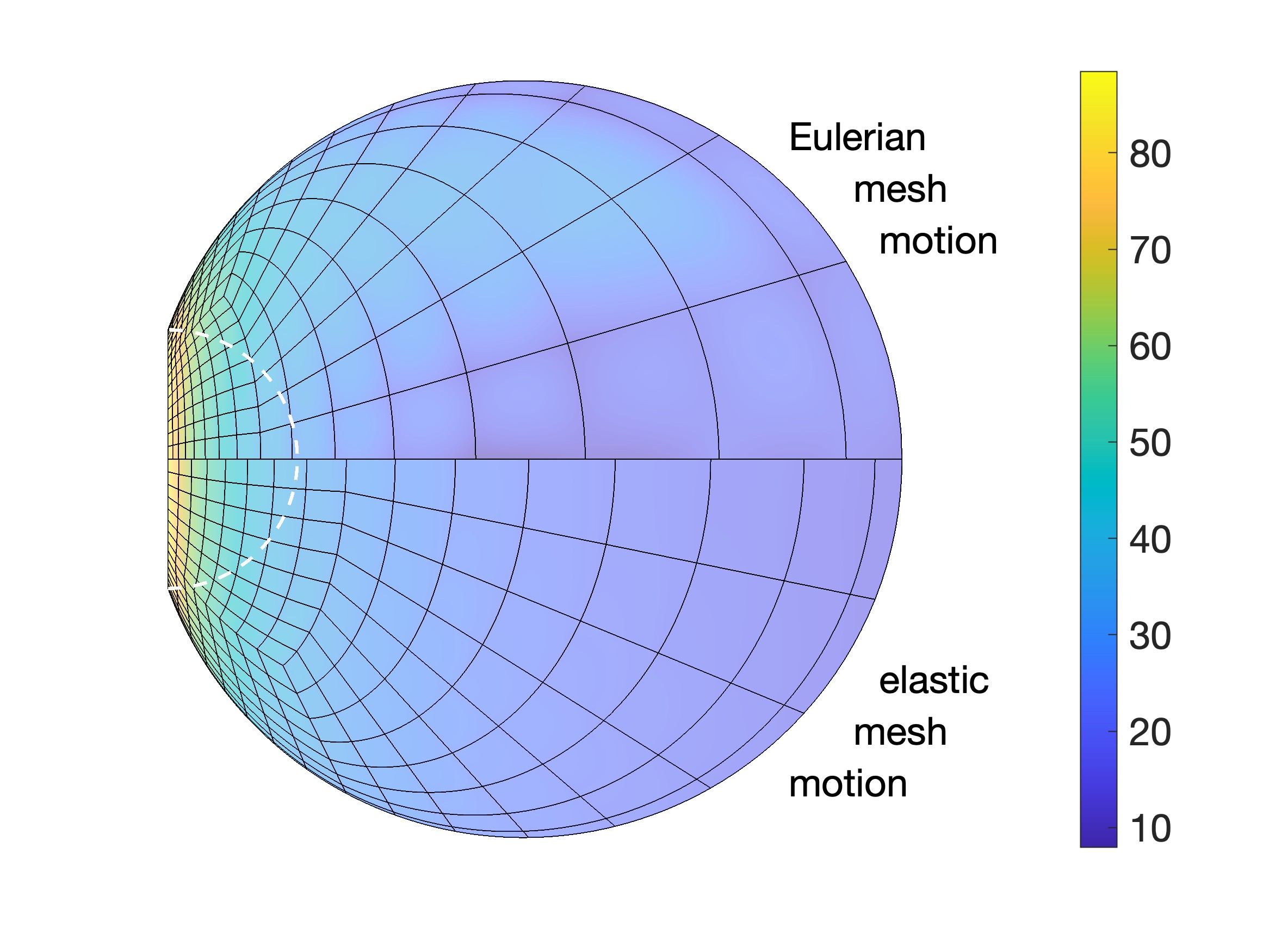}}
\put(0.2,-.2){\includegraphics[height=58mm]{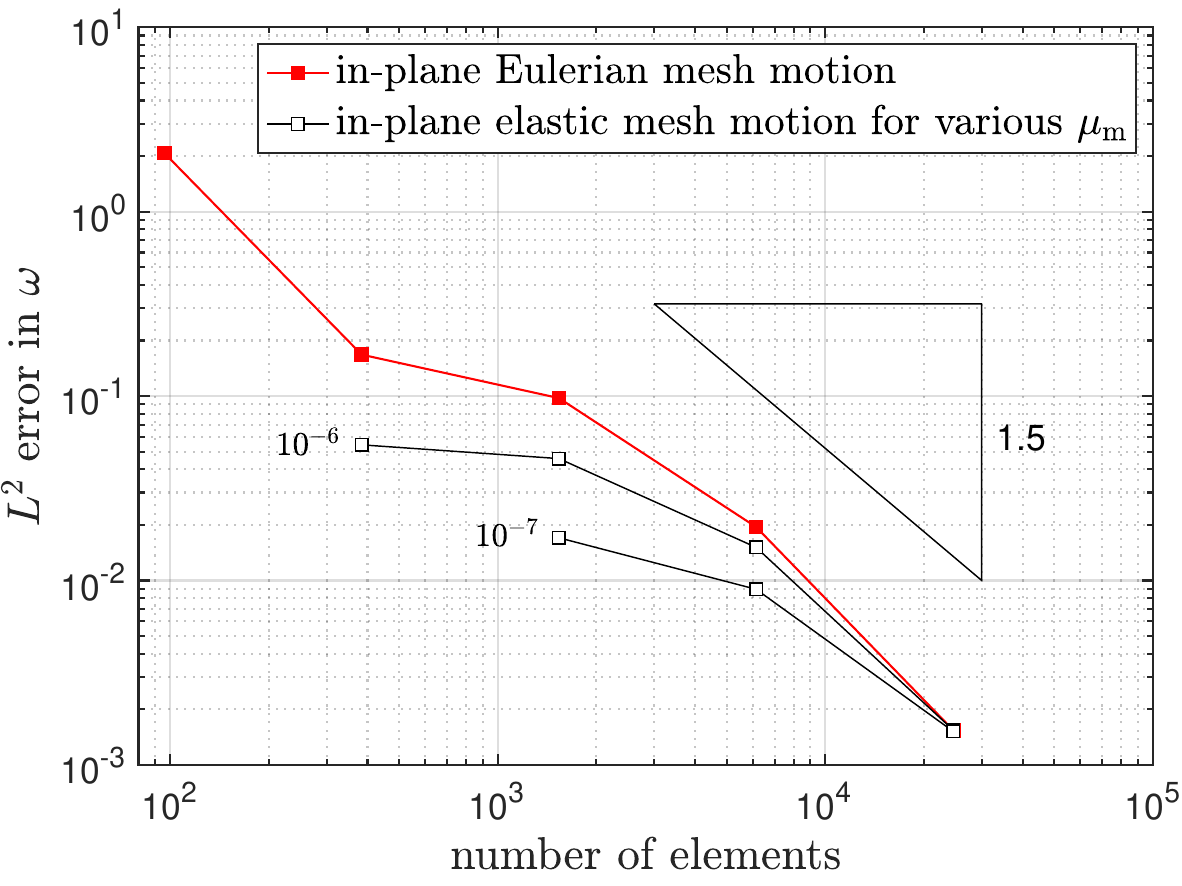}}
\put(-7.95,-.05){\footnotesize (a)}
\put(0.25,-.05){\footnotesize (b)}
\end{picture}
\caption{Laminar inflation of a soap bubble ($V = 50V_0$):
Accuracy comparison between Eulerian and elastic mesh motion for stabilization term integration over the \underline{initial} surface.
(a) Comparison of velocity field $\bv$ [mm/s] for $m = 8$ and $\tmu = 10^{-6}\mu$N/mm. 
(b) $L^2$ error of surface vorticity vs.~mesh refinement.
The accuracy gains of the elastic mesh formulation are bigger than in Fig.~\ref{f:SoapBublLam3} as they increase with bubble size.
The dashed white line shows the initial bubble shape.}
\label{f:SoapBublLam4}
\end{center}
\end{figure}

As the analytical example in \citet{ALEtheo} shows, the Eulerian mesh quality degrades with inflation size.
Therefore, if the bubble is inflated even further, the elastic mesh results should become even better than those in Fig.~\ref{f:SoapBublLam3}.
This is confirmed by the results in Fig.~\ref{f:SoapBublLam4}, which consider inflation up to $V = 50V_0$ (using $\Delta V = 49V_0$ and $t_1 = 4$s within Eq.~\eqref{e:soapV} and taking $\Delta t = 64\mathrm{ms}/m$). 
Now the elastic mesh results are consistently more accurate than the Eulerian ones. 
%-----------------------------------------------------------------
% run sSymBubbleNBC/*50, pSymBubble & pBubbleVorticity
%-----------------------------------------------------------------

The picture changes slightly when the stabilization terms of Sec.~\ref{s:FEDB} are integrated over the current instead of the initial mesh.
This is shown in Fig.~\ref{f:SoapBublLam5}:
%-----------------------------------------------------------------
\begin{figure}[h!]
\begin{center} \unitlength1cm
\begin{picture}(0,5.8)
\put(-8,-.15){\includegraphics[height=58mm]{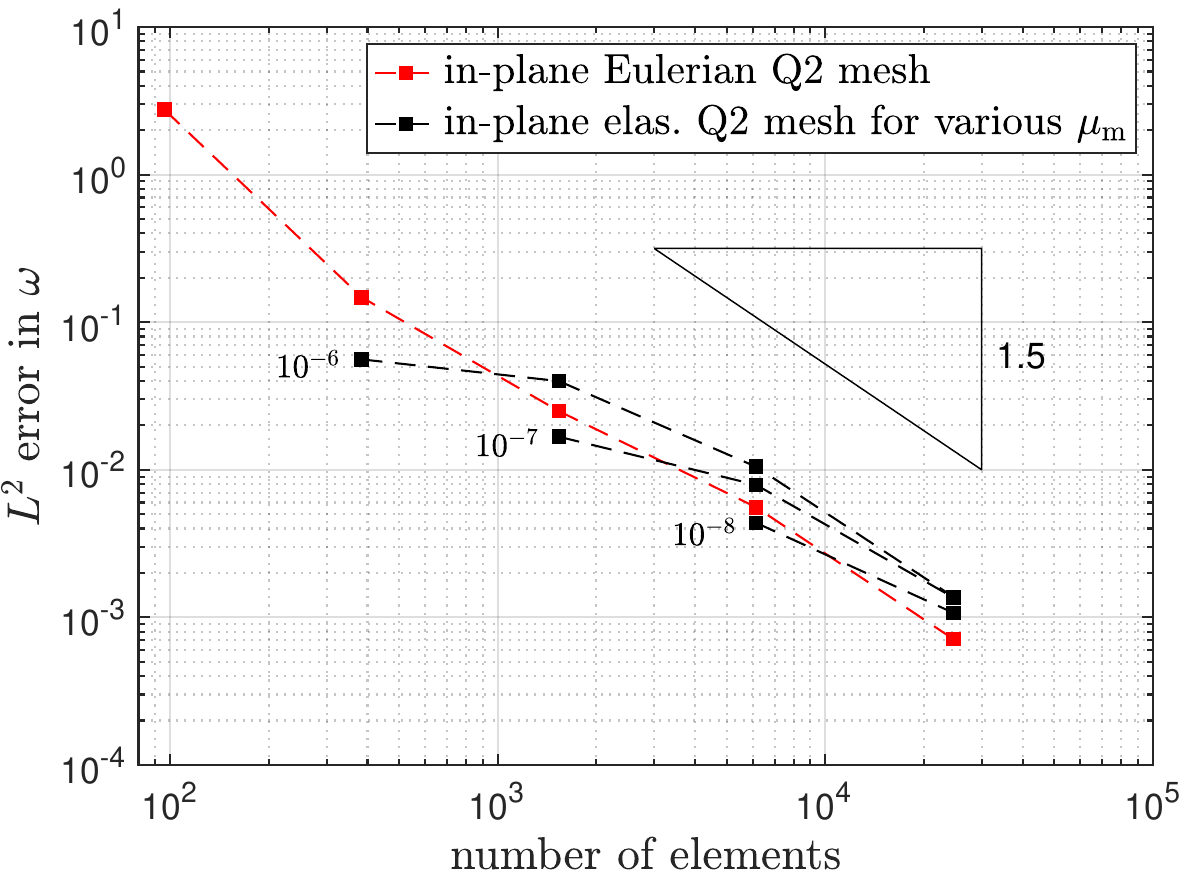}}
\put(0.2,-.2){\includegraphics[height=58mm]{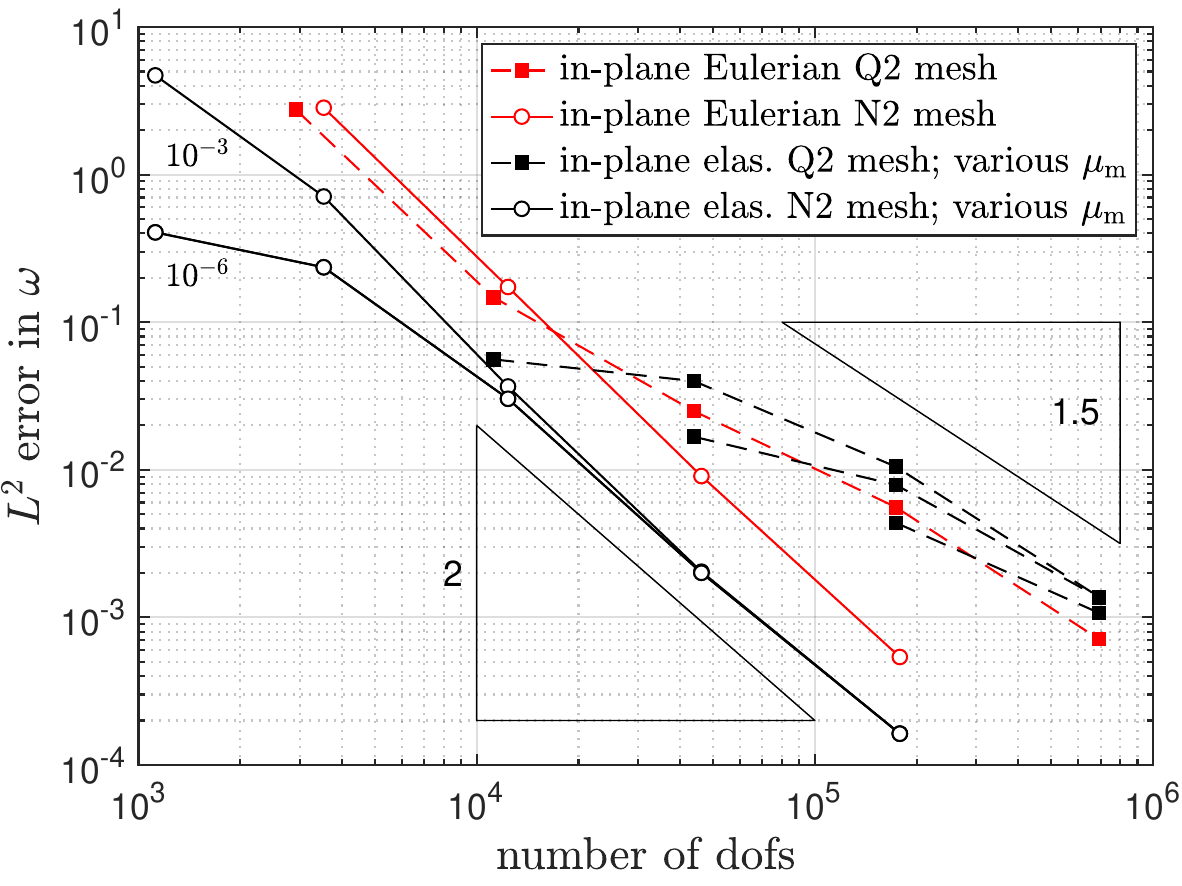}}
\put(-7.95,-.05){\footnotesize (a)}
\put(0.25,-.05){\footnotesize (b)}
\end{picture}
\caption{Laminar inflation of a soap bubble ($V = 50V_0$):
Accuracy comparison between Eulerian and elastic mesh motion for stabilization term integration over the \underline{current} surface.
$L^2$ error of surface vorticity vs.~mesh refinement for (a) quadratic Lagrange (Q2) discretization and (b) quadratic NURBS (N2) discretization.
The Q2 results of (a) are included in (b), but now plotted vs.~the number of dofs.
The current surface integration improves the accuracy, especially for the more distorted Eulerian mesh.
Thus, the latter becomes as accurate as the elastic mesh for Q2 elements.
But for N2 elements, the elastic mesh is still much more accurate.
Further, N2 achieves a better convergence rate than Q2.}
\label{f:SoapBublLam5}
\end{center}
\end{figure}
% run sSymBubbleNBC/*50, pSymBubble & pBubbleVorticity
%-----------------------------------------------------------------
Comparing Fig.~\ref{f:SoapBublLam5}a with \ref{f:SoapBublLam4}b, shows that the current surface integration improves accuracy, especially for the more distorted Eulerian mesh. 
Now Eulerian and elastic mesh descriptions are equally accurate for quadratic Lagrange elements.
For quadratic NURBS (non-uniform rational B-spline) elements, however, the elastic mesh description is still much more accurate than the Eulerian one, as Fig.~\ref{f:SoapBublLam5}b shows.
The figure also shows that the convergence rate of NURBS elements is higher than that of Lagrange elements.
The reason for the superior behavior of NURBS elements lies in their more accurate representation of surface curvatures, which appear explicitly in the elastic mesh integral of Eq.~(\ref{e:fio}.2).
For NURBS, mesh parameter $\mu_\mrm$ can also be chosen much larger, as Fig.~\ref{f:SoapBublLam5}b shows, and hence NURBS meshes are much less likely to become ill-conditioned.

The results of Fig.~\ref{f:SoapBublLam5} illustrate that the current surface integration of the stabilization terms can yield major accuracy gains.
This is because in that example the volume change and hence surface stretches are very large -- up to $J_\mrm = 250$ for Eulerian mesh motion, which is why accuracy gains are particularly high there.
In the previous (and remaining) examples, surface stretches are much smaller, and hence accuracy gains are much smaller.
Thus, integration over $\dif A$ becomes more convenient, as no extra linearization terms, that can contribute to ill-conditioning, are needed.

It is finally noted that in the examples of Sec.~\ref{s:SBu}, the bubble shape remains nearly spherical, which is why the surface tension (seen in Fig.~\ref{f:SoapBublLam2}) is nearly constant.
This can change for non-uniform inflow.

\subsubsection{Non-uniform inflow} \label{s:SBt}

\begin{figure}[h!]
\begin{center} \unitlength1cm
\begin{picture}(0,6.8)
\put(-7.95,3.25){\includegraphics[height=35mm]{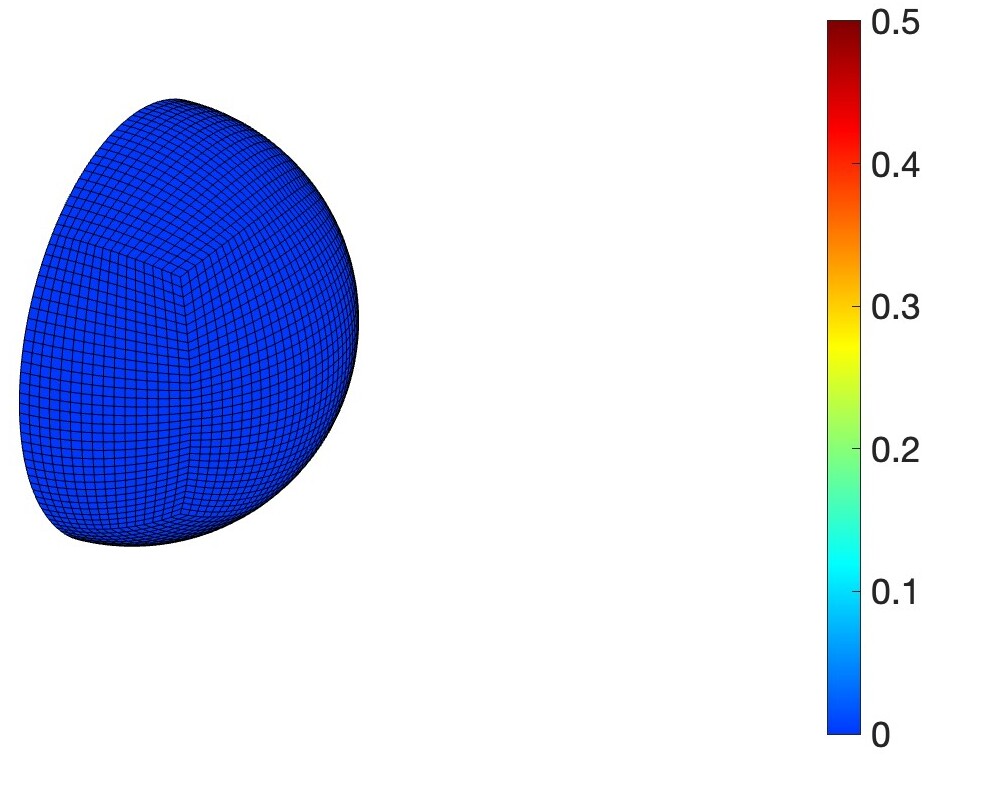}}
\put(-5.85,3.25){\includegraphics[height=35mm]{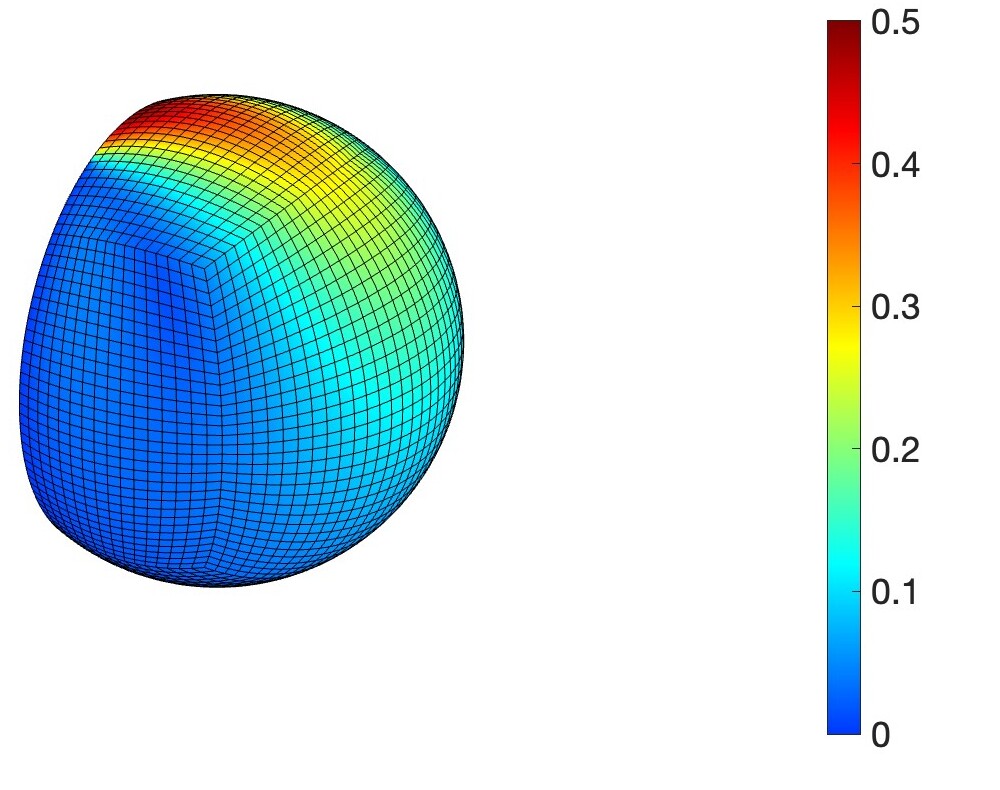}}
\put(-3.2,3.25){\includegraphics[height=35mm]{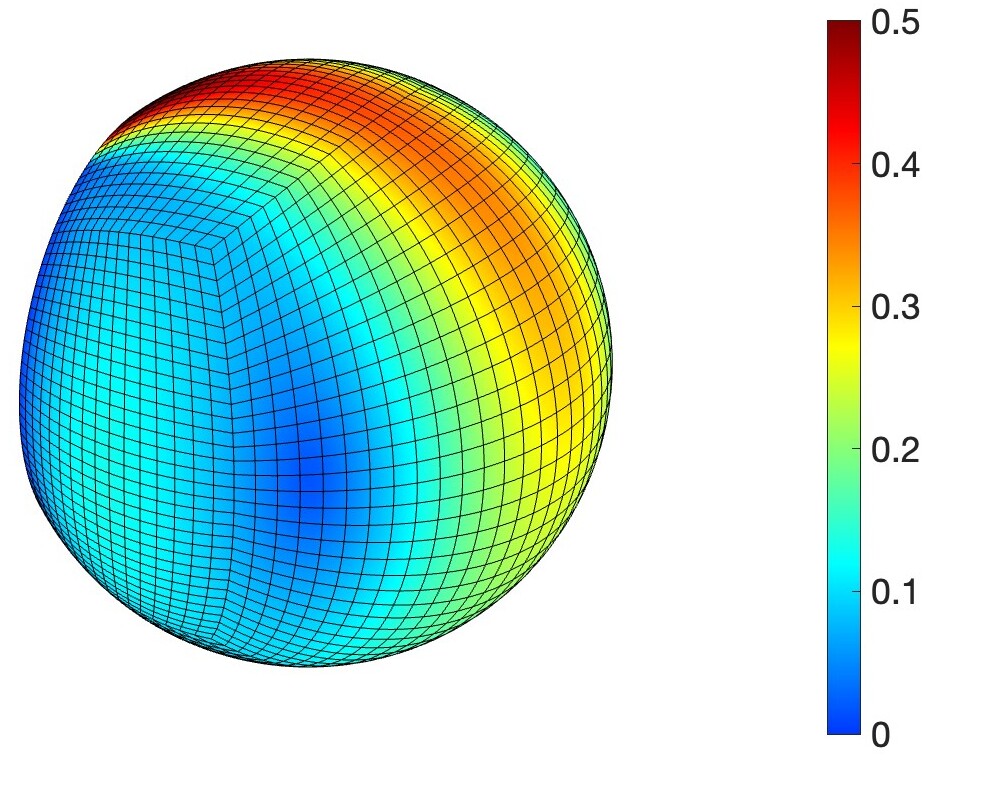}}
\put(0.1,3.25){\includegraphics[height=35mm]{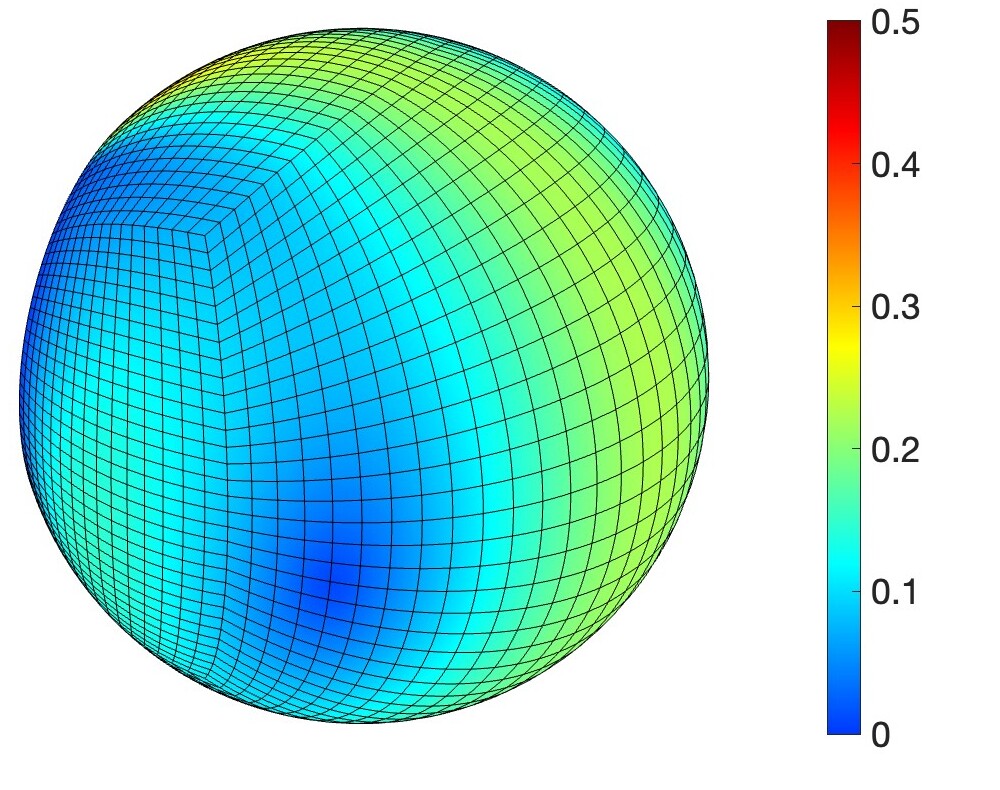}}
\put(3.7,3.25){\includegraphics[height=35mm]{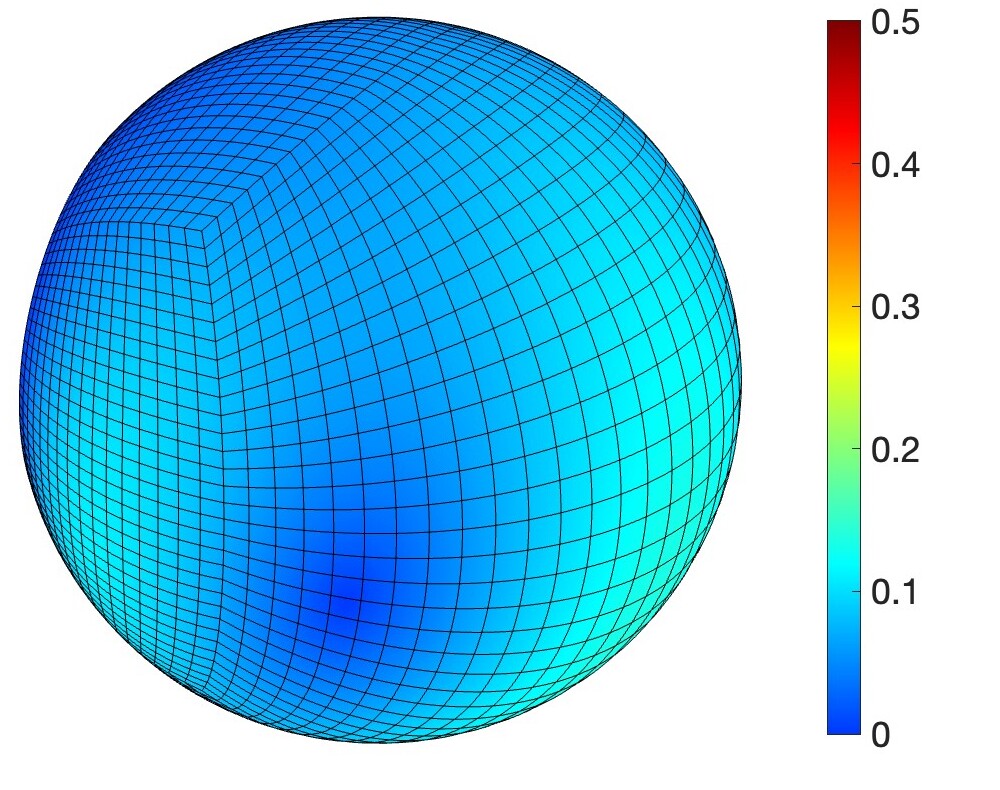}}
\put(-7.95,-.25){\includegraphics[height=35mm]{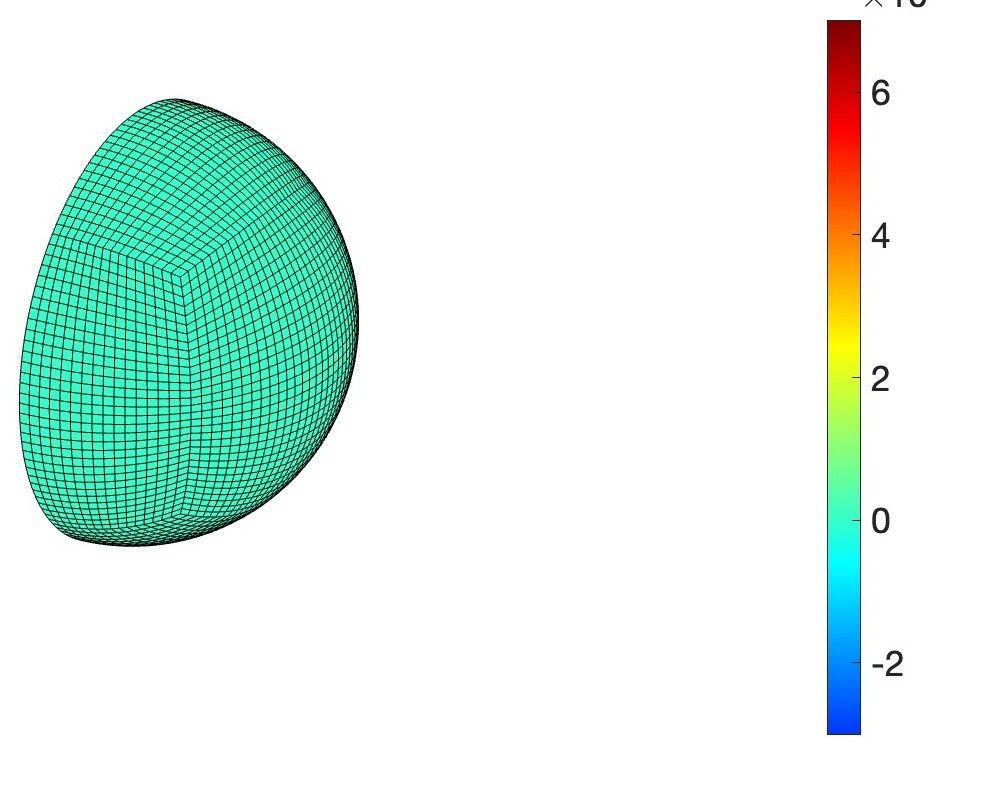}}
\put(-5.85,-.25){\includegraphics[height=35mm]{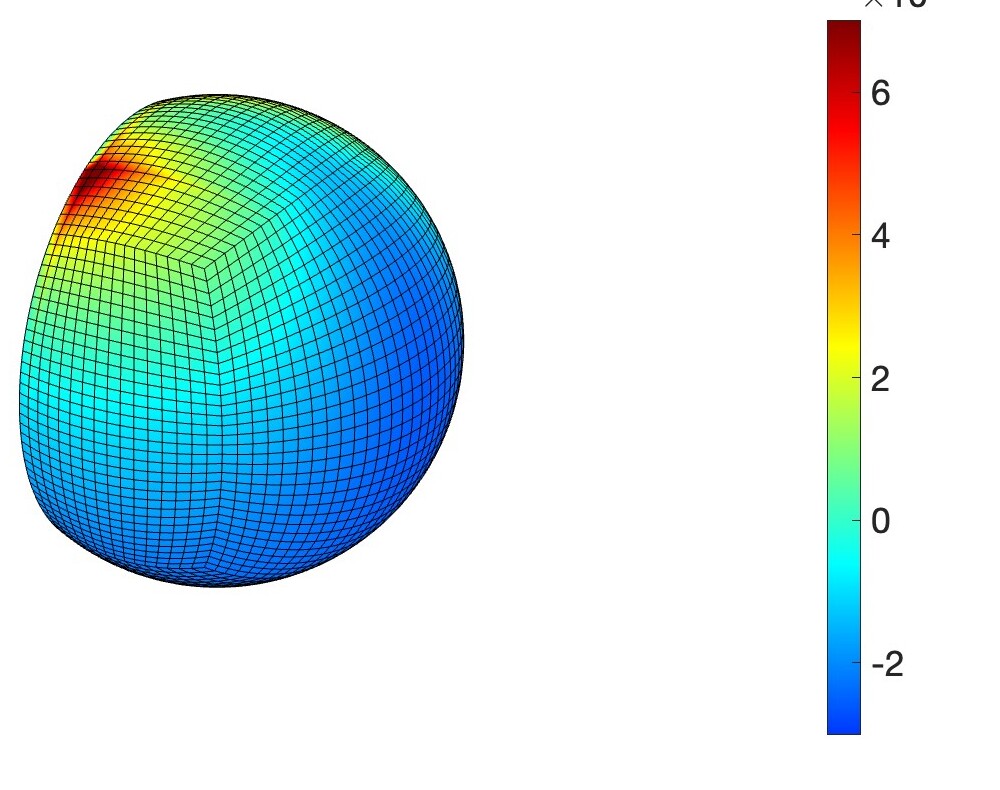}}
\put(-3.25,-.25){\includegraphics[height=35mm]{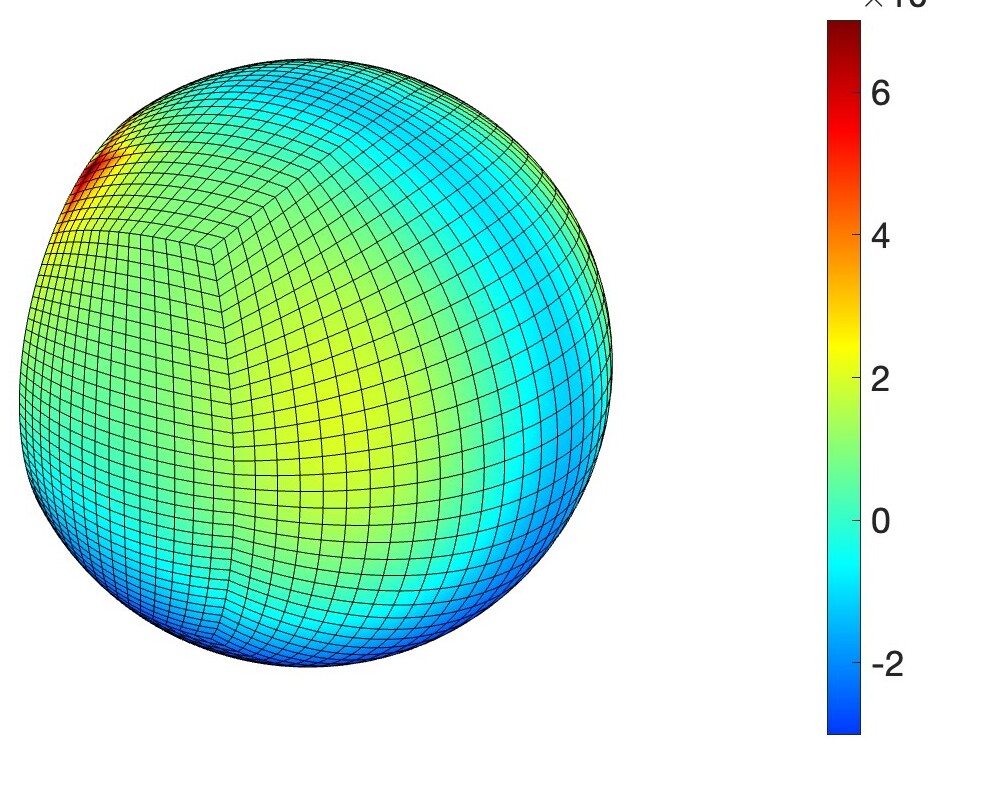}}
\put(0.05,-.25){\includegraphics[height=35mm]{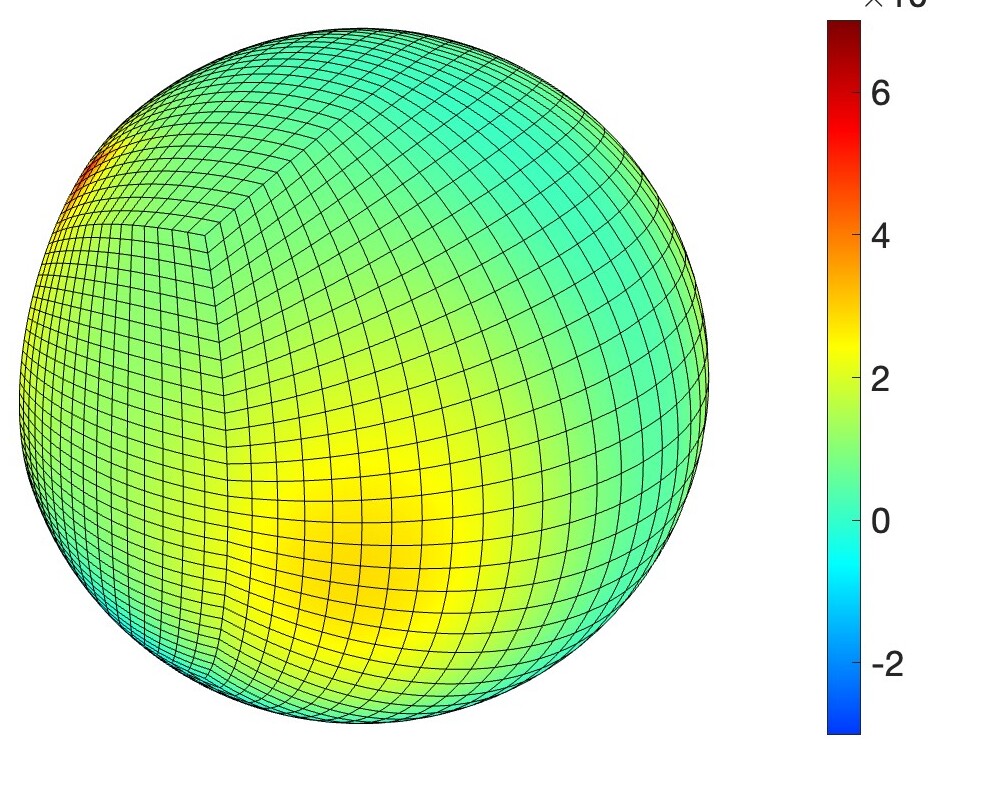}}
\put(3.7,-.25){\includegraphics[height=35mm]{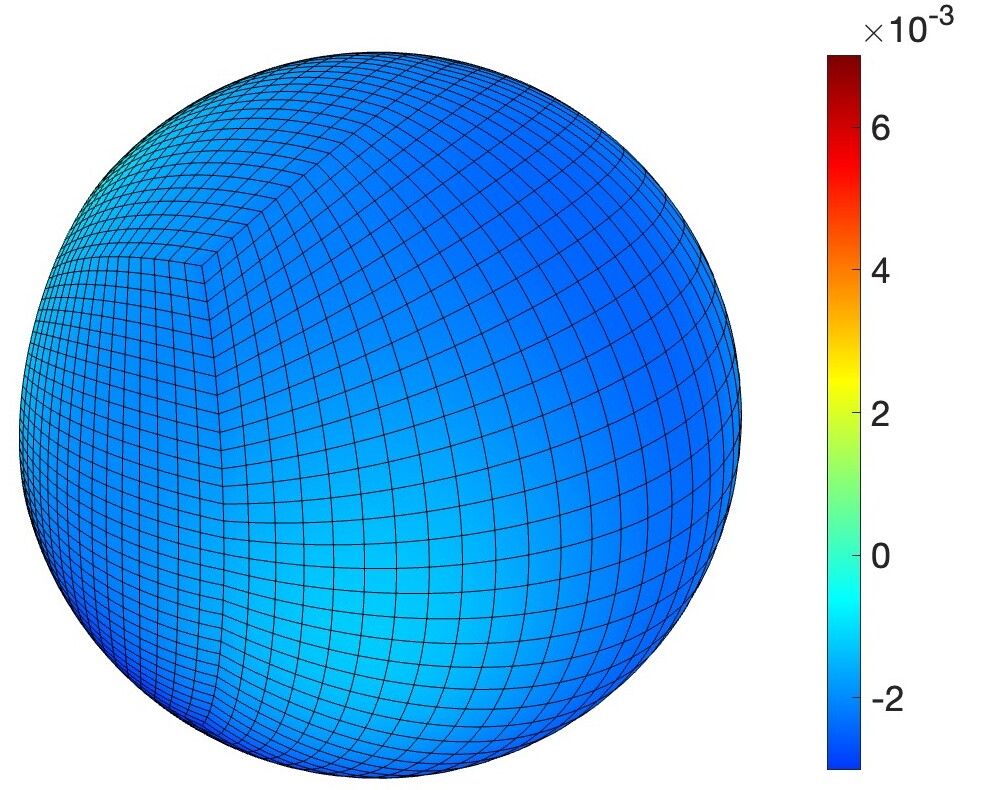}}
\end{picture}
\caption{Non-uniform inflation of a soap bubble with high surface tension: % (Case 1 = case 1s)
Left to right: Bubble shape at $t = \{0,\,0.25,\,0.5,\,0.75,\,1\}$s.
Top row: Velocity field $\bv$ [m/s]. 
Bottom row: Surface tension change $\Delta q$ [$\mu$N/mm].
The results are from elastic mesh motion with $m=16$, $\Delta t = 2.5$ms, $\tmu = 10^{-4}\mu$N/mm, $\alpha_\mathrm{DB}=1000$ and $R$ to $\rho$ from Table~\ref{t:soapbubble}. 
For these parameters, the bubble shape remains near spherical.
}
\label{f:SoapBublTurb1}
\end{center}
\end{figure}

\begin{figure}[h!]
\begin{center} \unitlength1cm
\begin{picture}(0,5.8)
\put(-8,-.15){\includegraphics[height=58mm]{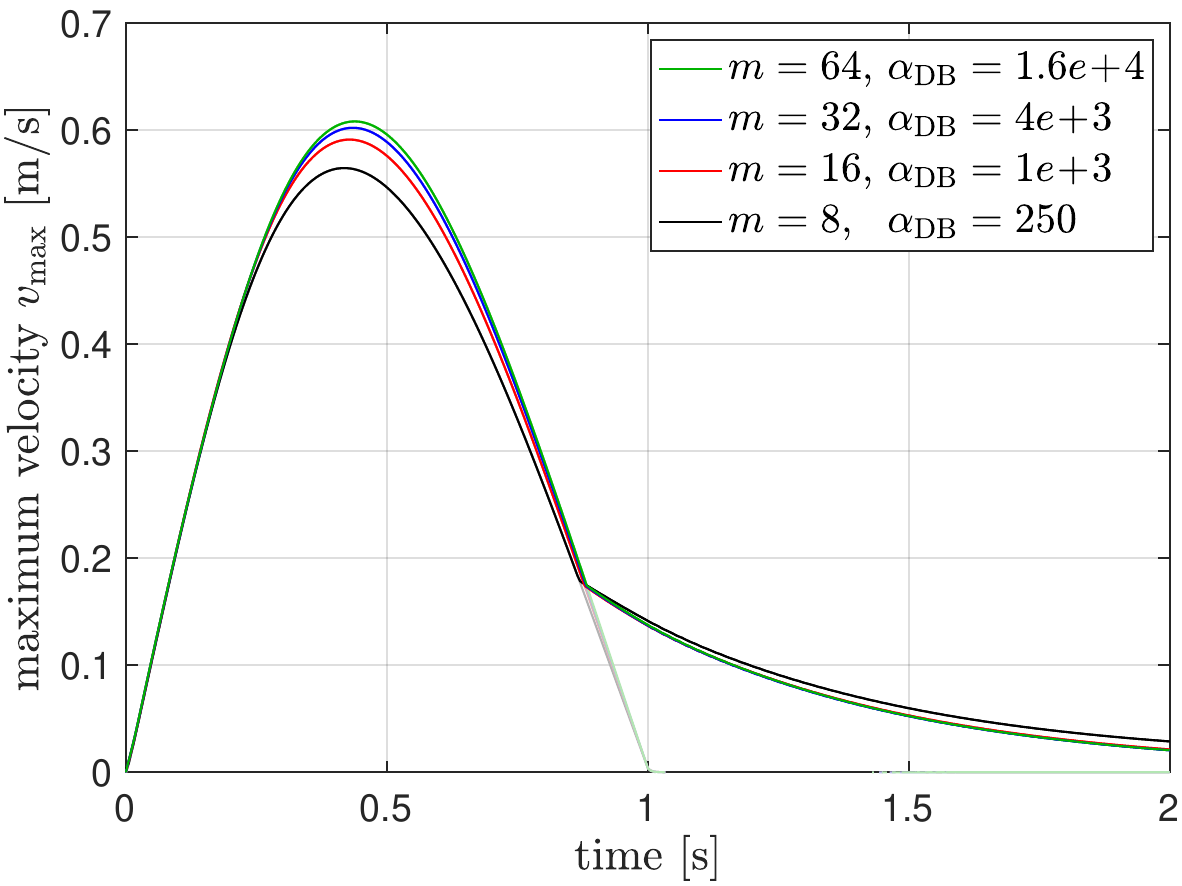}}
\put(0.2,-.2){\includegraphics[height=58mm]{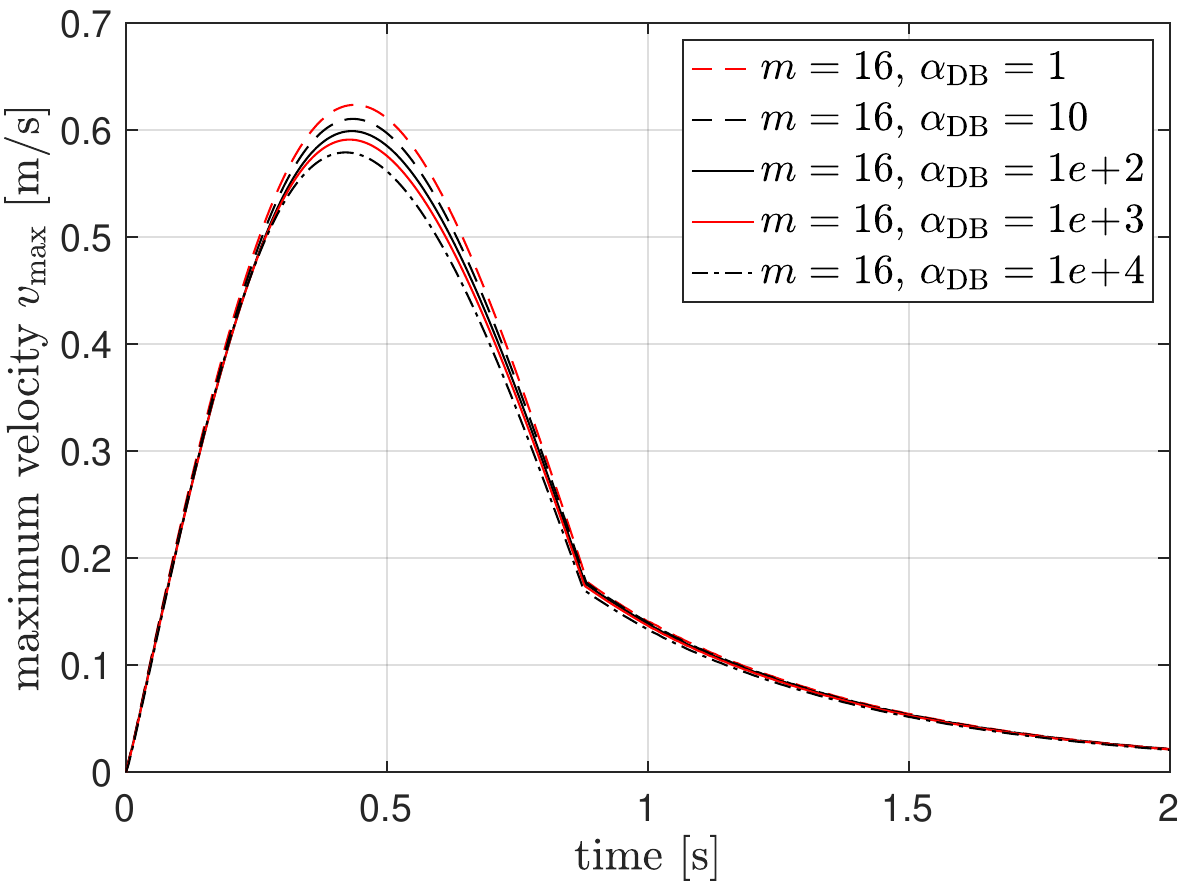}}
\put(-7.95,-.05){\footnotesize (a)}
\put(0.25,-.05){\footnotesize (b)}
\end{picture}
\caption{Non-uniform inflation of a soap bubble with high surface tension: % (Case 1)
(a) Mesh convergence and (b) $\alpha_\mathrm{DB}$-convergence of the maximum surface velocity.
The light lines in (a) show the inflow velocity at $\theta=0$, which is the maximum velocity until $t \approx 0.883$s.}
\label{f:SoapBublTurb2}
\end{center}
\end{figure}

The last case considers the situation when inflow is only allowed partially -- on the boundary between $|\theta|<\pi/8$ (see Fig.~\ref{f:SoapBublSetup}b).
On this boundary the boundary conditions are the same as before, while on the remaining boundary all velocities components are set to zero.
This partial inflow case causes two vortices to appear on the evolving surface, as Fig.~\ref{f:SoapBublTurb1} shows.
%-----------------------------------------------------------------
% run sSymBubbleTurb3/*1 pSymBubbleTurb
%-----------------------------------------------------------------
The corresponding surface velocity evolution is shown in Fig.~\ref{f:SoapBublTurb2}  for various meshes and $\alpha_\mathrm{DB}$ values. 
%-----------------------------------------------------------------
% run sSymBubbleTurb3/*1 pSymBubbleTurbPus
%-----------------------------------------------------------------
Considered are the inflow parameters $V = 8V_0$ and $t_1 = 1$s in Eq.~\eqref{e:soapV}.
The resulting peak value of $v_\mathrm{max}$ is about 0.61m/s leading to a Reynolds number of 23.95 (taking the inlet diameter $\pi R/4$ as characteristic length).
The example shows that the proposed ALE formulation with elastic mesh motion can capture non-trivial nonlinear flow fields on evolving surfaces.
It converges with mesh refinement as Fig.~\ref{f:SoapBublTurb2}a shows.
In this example, $\alpha_\mathrm{DB}$ has a significant influence on the solution, as Fig.~\ref{f:SoapBublTurb2}b shows. 
Larger $\alpha_\mathrm{DB}$ are observed to alleviate inaccuracies in $q$, but at the price of less accurate enforcement of area-incompressibility ($\divs\bv=0$).
The values used in Fig.~\ref{f:SoapBublTurb2}a are chosen based on examining fields $q$ and  $\divs\bv$ for different $\alpha_\mathrm{DB}$, and they
constitute a good compromise, but more rigorous guidelines should be investigated in future work.

\begin{figure}[h!]
\begin{center} \unitlength1cm
\begin{picture}(0,10.25)
\put(-7.95,6.75){\includegraphics[height=35mm]{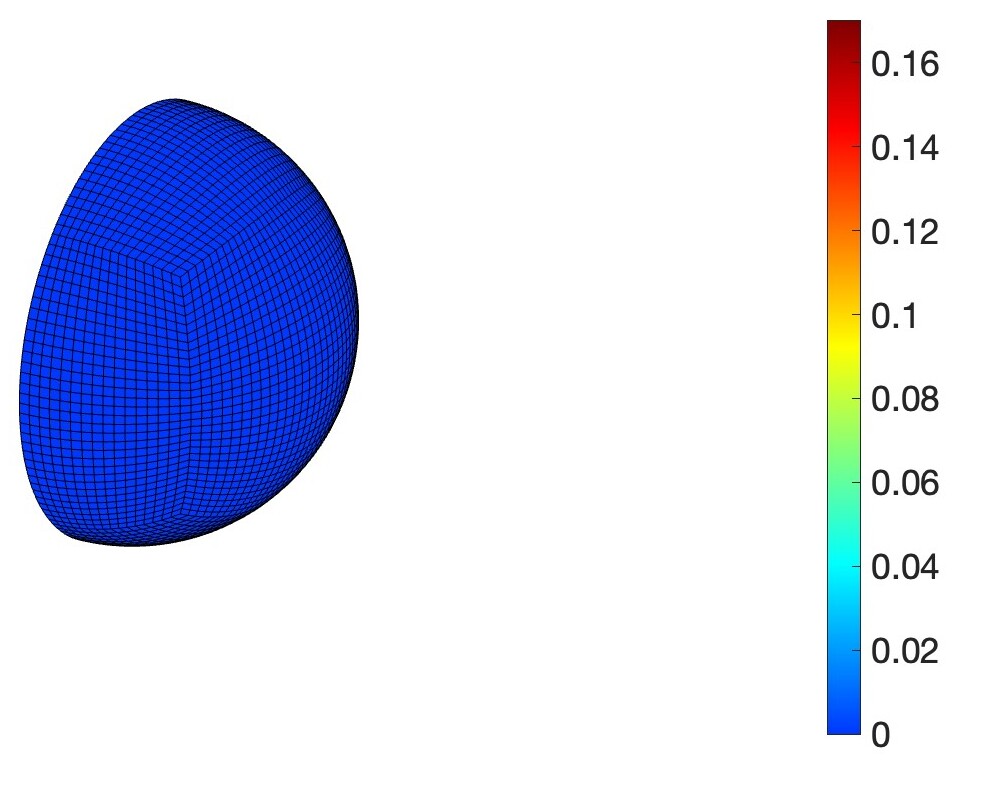}}
\put(-5.85,6.75){\includegraphics[height=35mm]{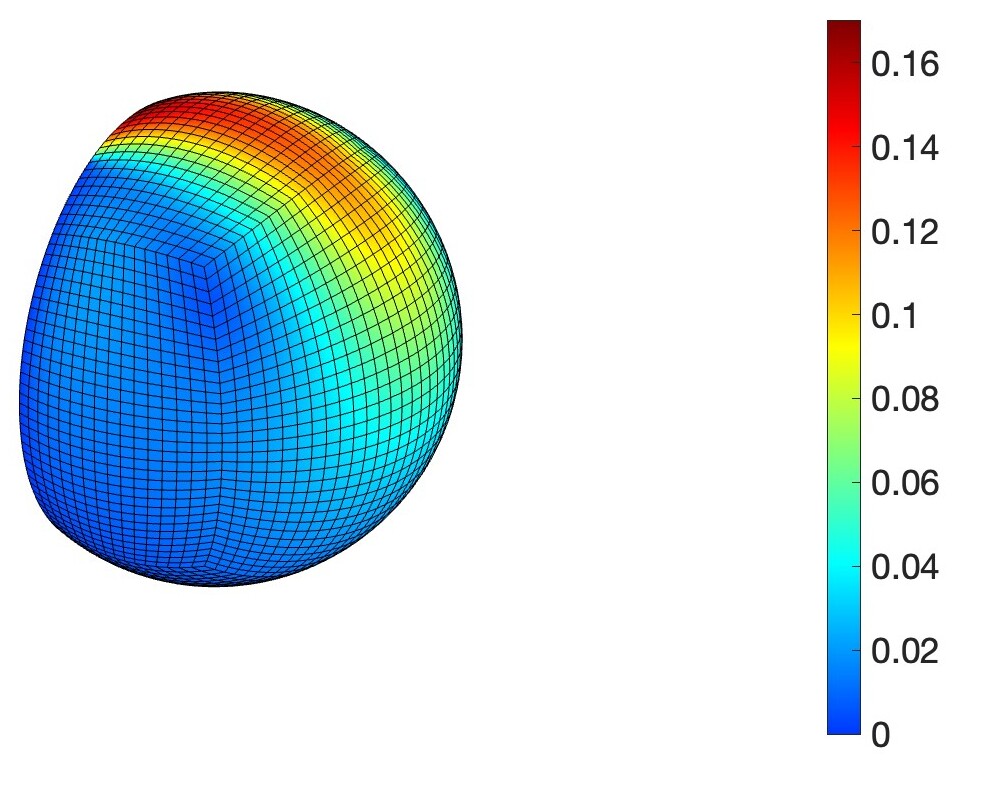}}
\put(-3.25,6.75){\includegraphics[height=35mm]{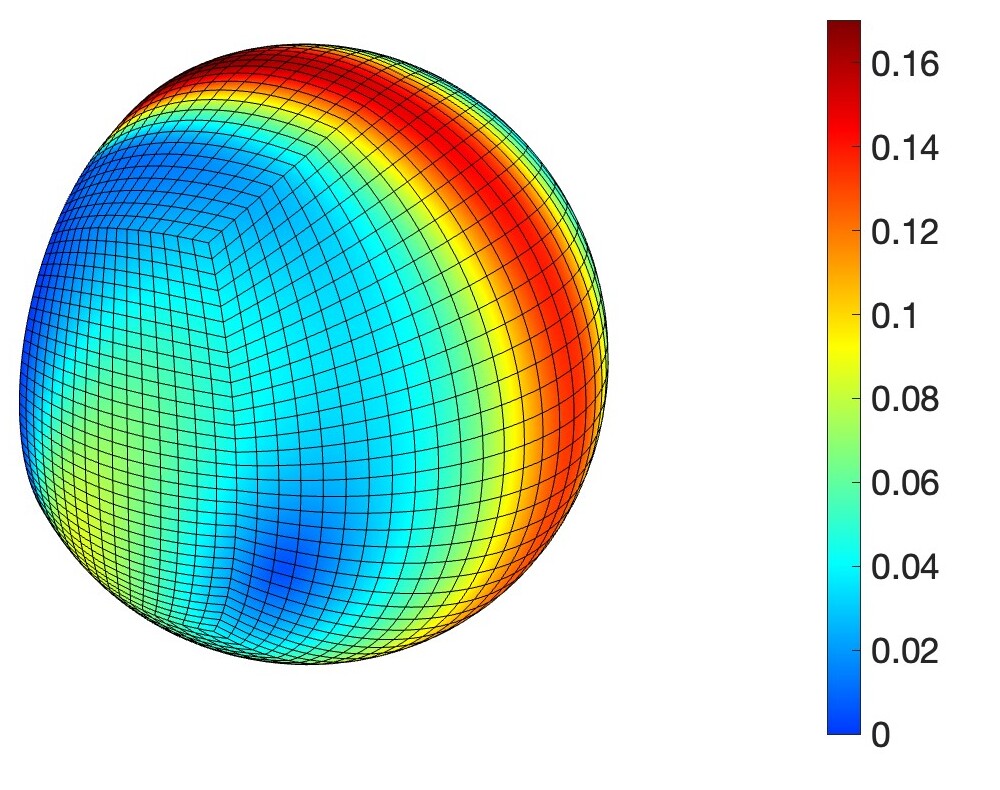}}
\put(0.05,6.75){\includegraphics[height=35mm]{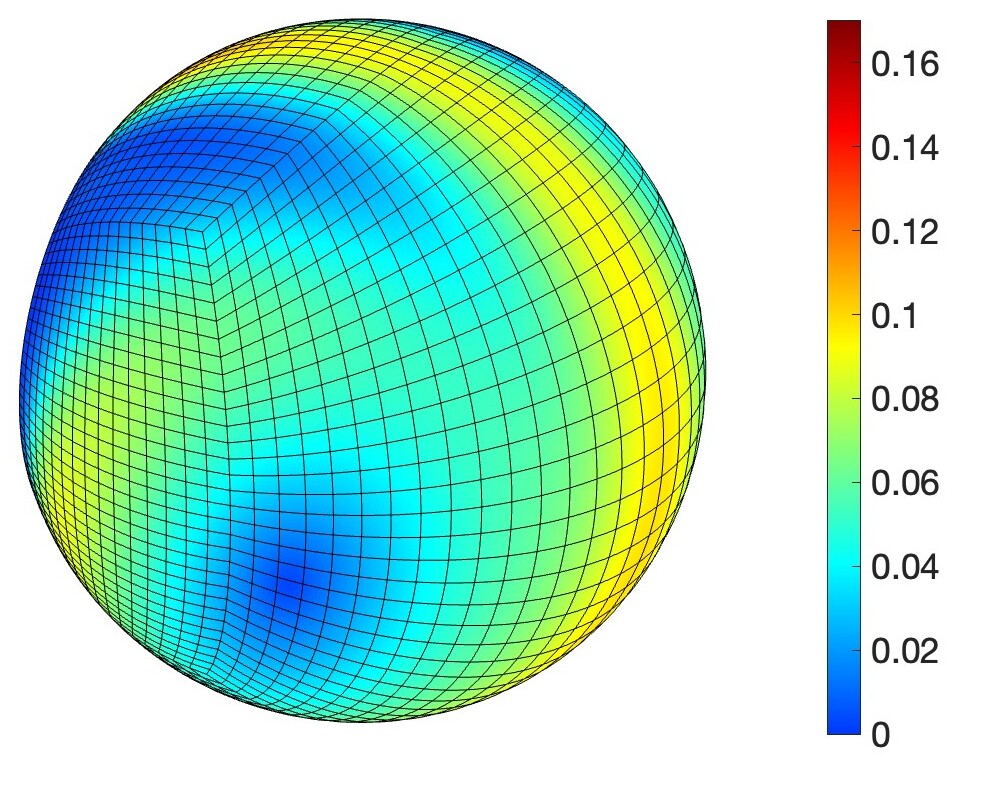}}
\put(3.7,6.75){\includegraphics[height=35mm]{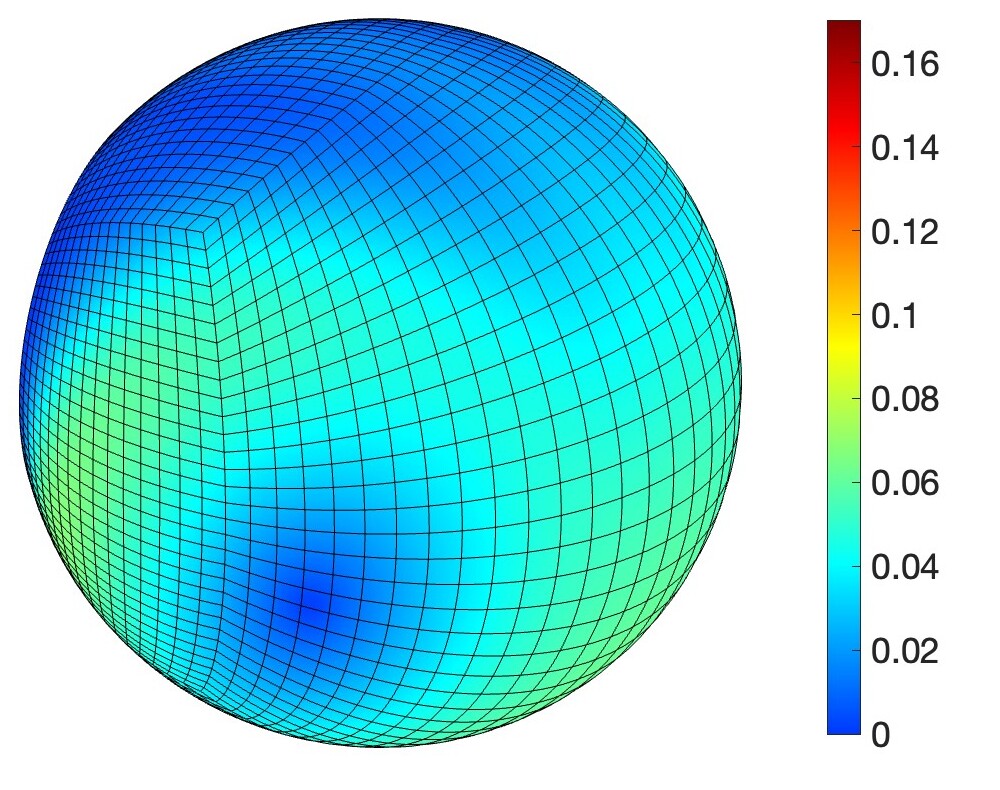}}
\put(-7.95,3.25){\includegraphics[height=35mm]{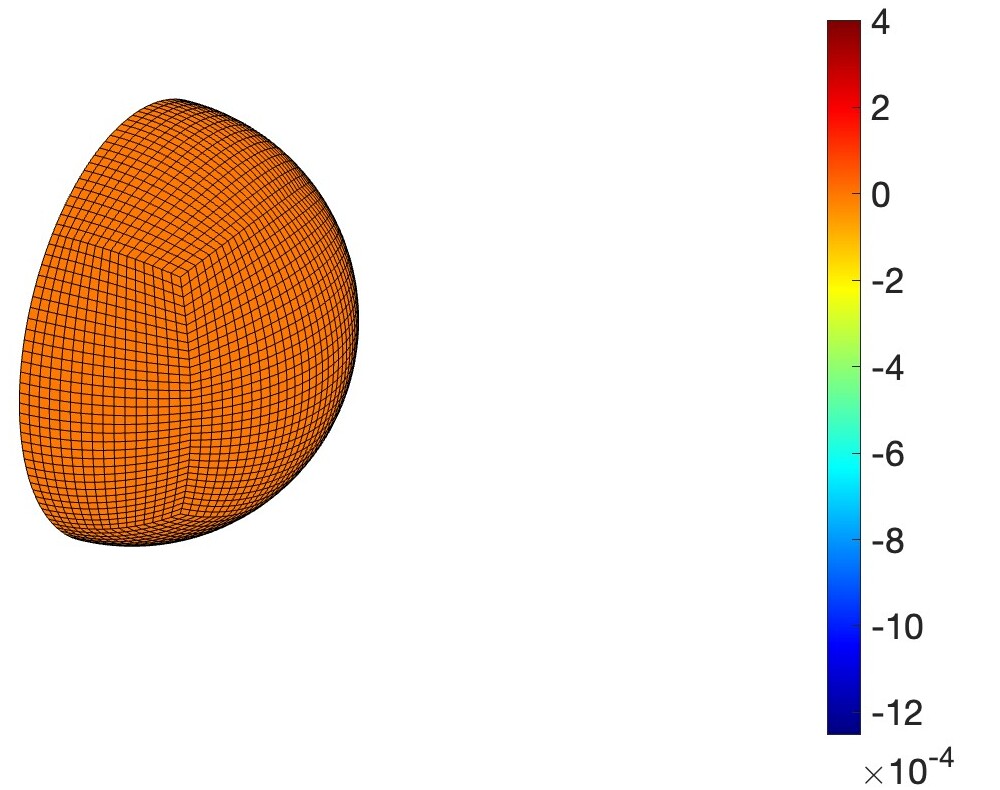}}
\put(-5.85,3.25){\includegraphics[height=35mm]{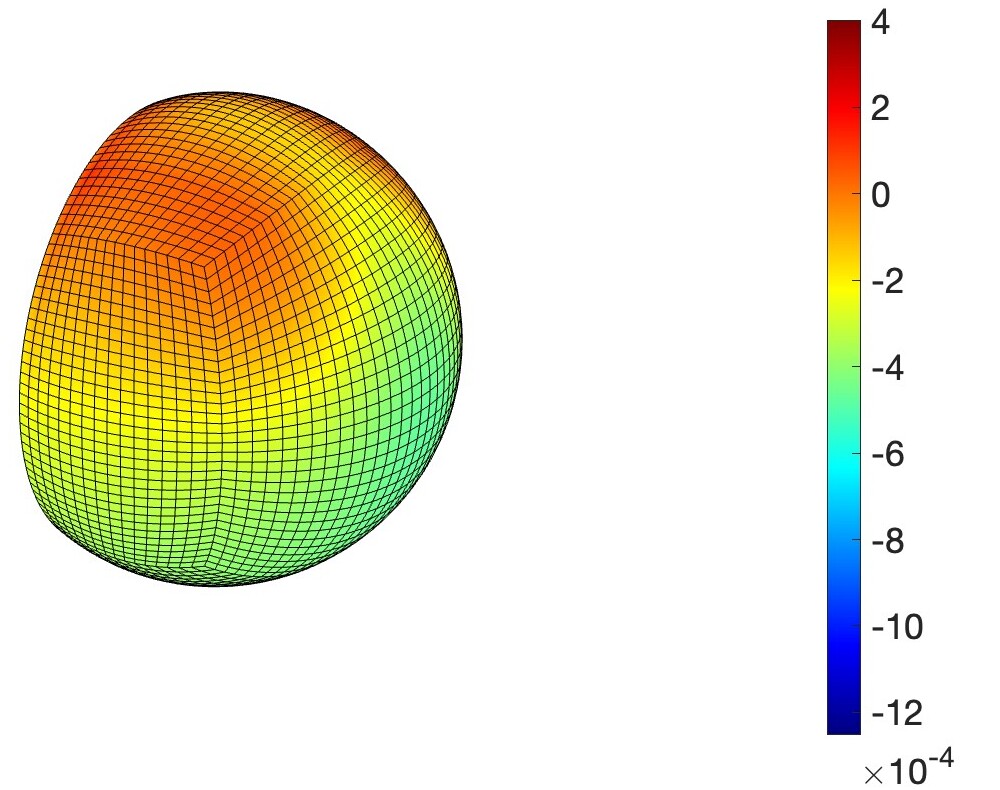}}
\put(-3.25,3.25){\includegraphics[height=35mm]{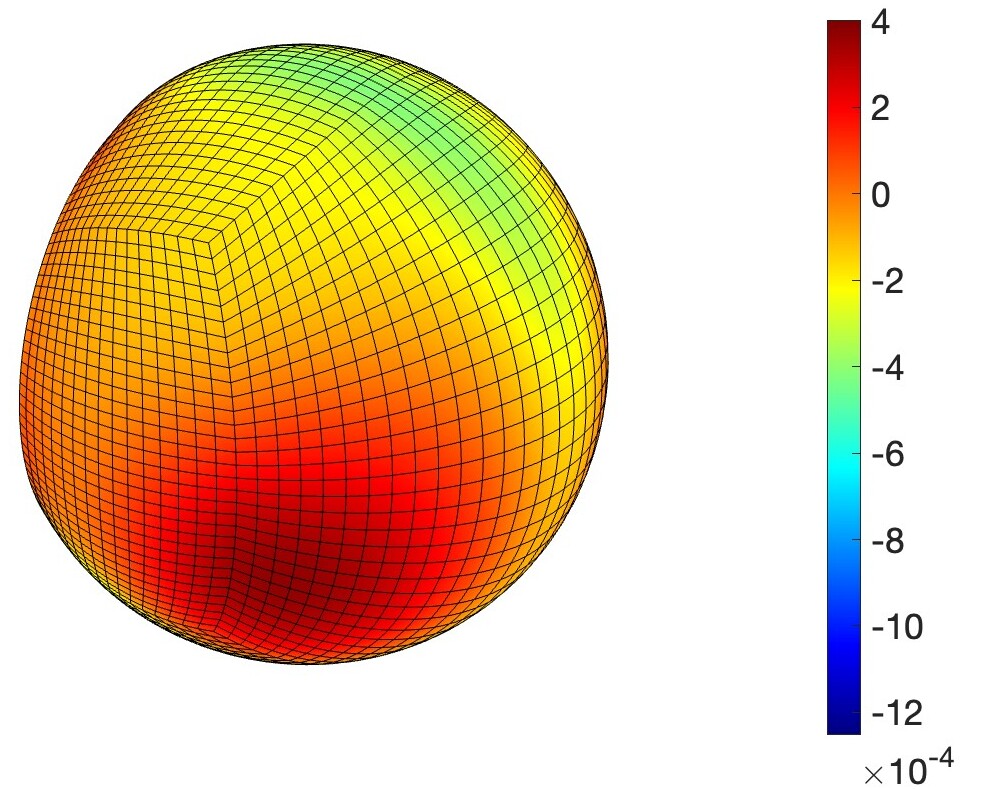}}
\put(0.05,3.25){\includegraphics[height=35mm]{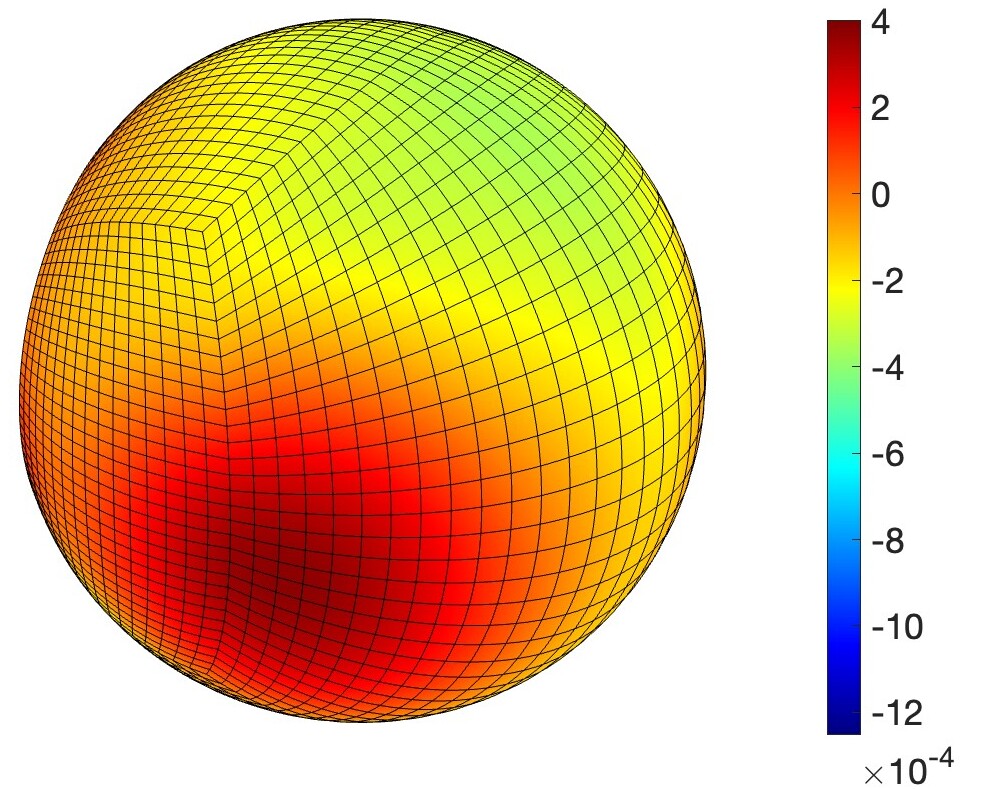}}
\put(3.7,3.25){\includegraphics[height=35mm]{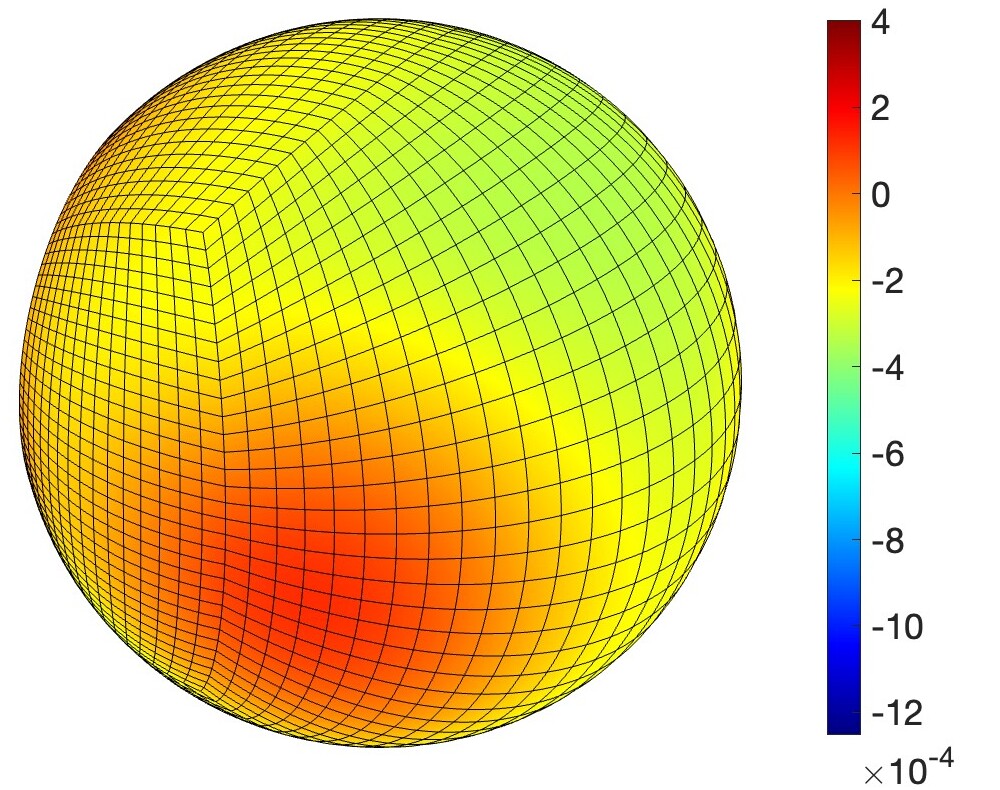}}
\put(-7.95,-.3){\includegraphics[height=35mm]{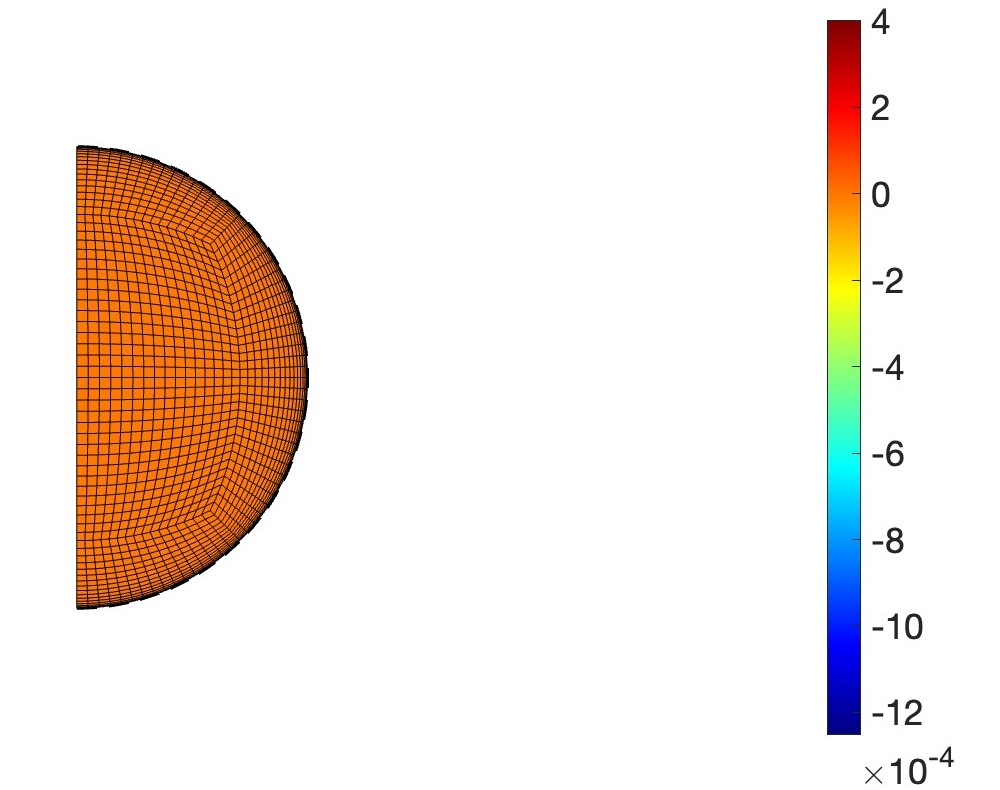}}
\put(-5.85,-.3){\includegraphics[height=35mm]{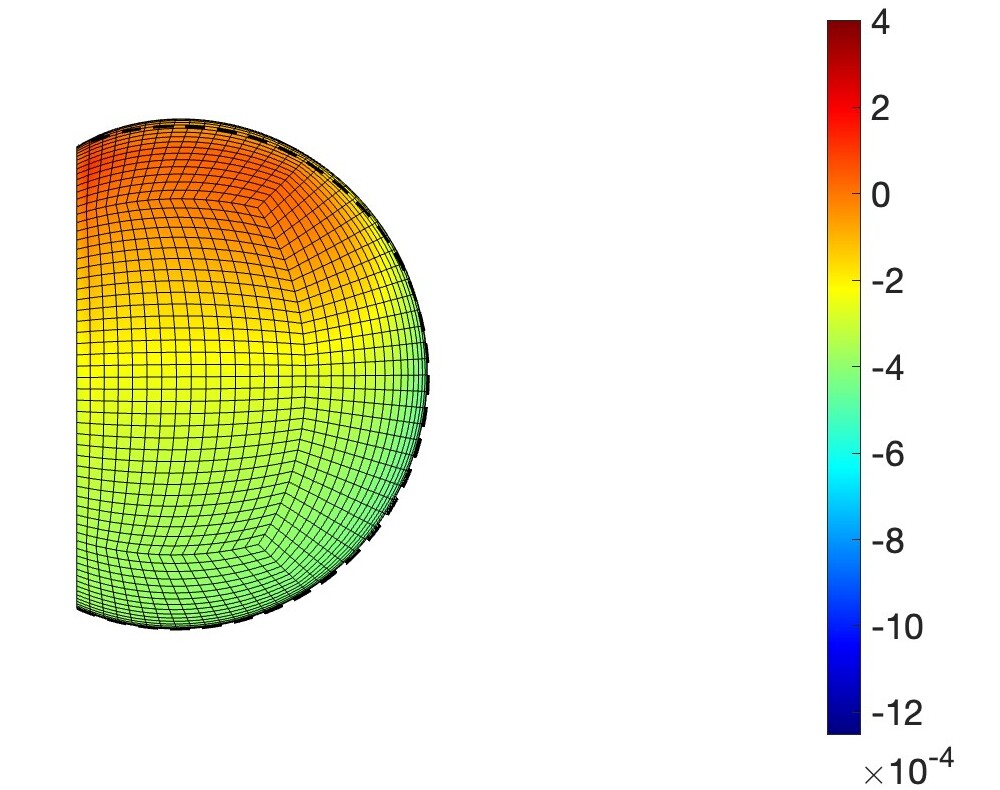}}
\put(-3.25,-.3){\includegraphics[height=35mm]{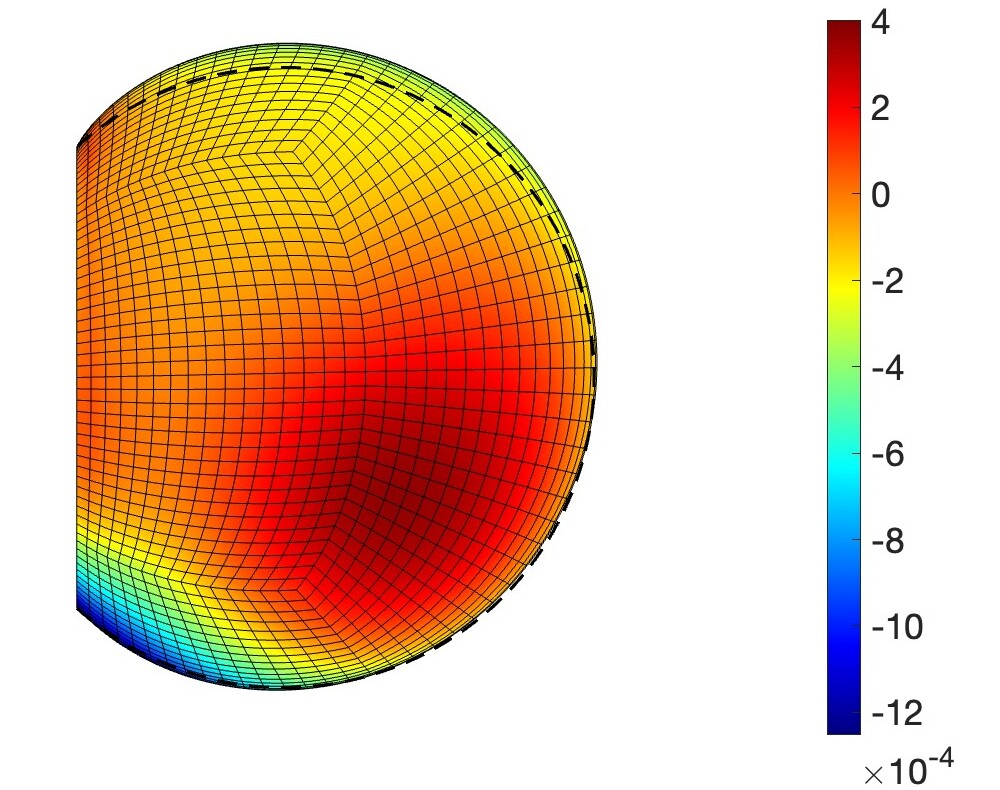}}
\put(0.05,-.3){\includegraphics[height=35mm]{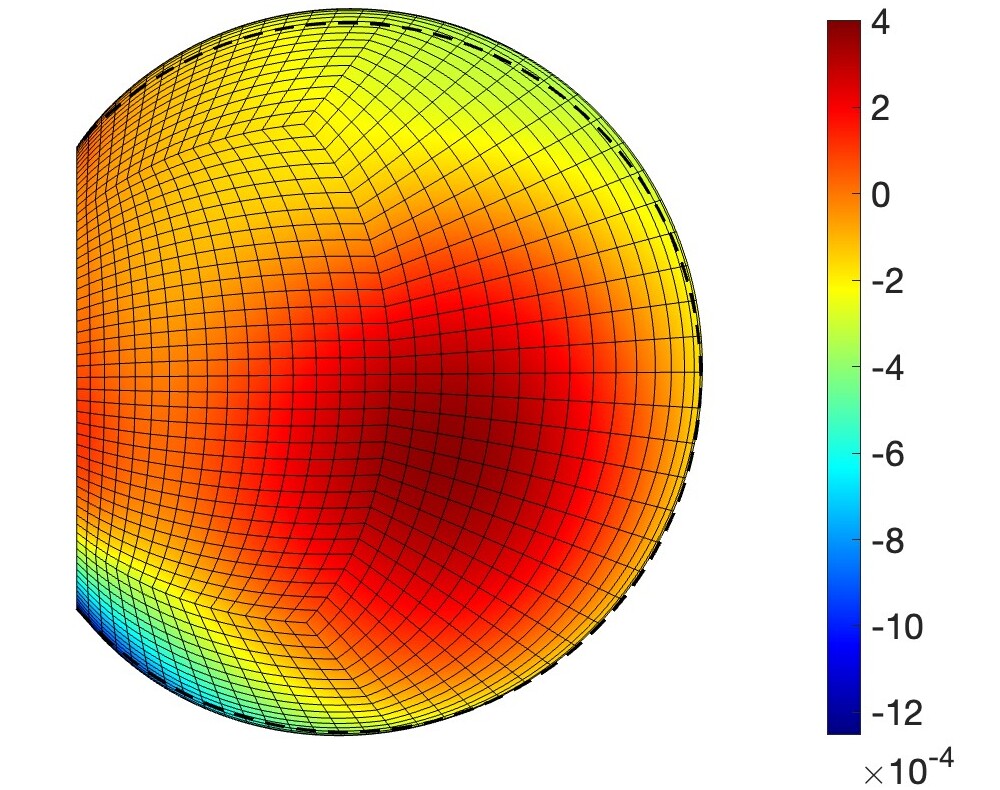}}
\put(3.7,-.3){\includegraphics[height=35mm]{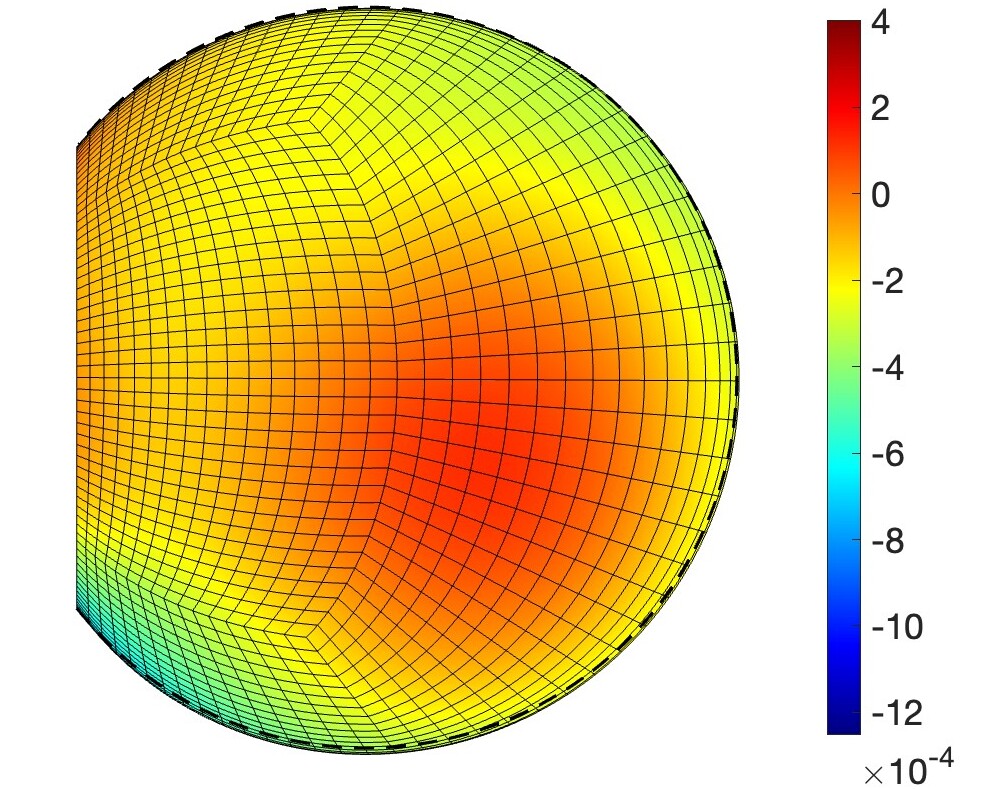}}
\end{picture}
\caption{
Non-uniform inflation of a soap bubble with low surface tension: % (Case 2 = case 7q)
Left to right: Bubble shape at $t = \{0,\,8,\,16,\,24,\,32\}$s.
Top row: Velocity field $\bv$ [m/s]. 
Middle \& bottom row: Surface tension change $\Delta q$ [$\mu$N/mm].
The dashed line below marks the shape of a perfect sphere with equal volume.
This shows that the bubble shape deviates significantly from a sphere, especially at the top, where the fluid velocity is high.
The results are from elastic mesh motion with $m=16$, $\Delta t = 16$ms, $R=100$mm, $q_0=0.01\mu$N/mm, $p_0=0.2$mPa, $\tmu = 10^{-4}\mu$N/mm and
$\alpha_\mathrm{DB}=10^4$ using $\eta$ and $\rho$ from Table~\ref{t:soapbubble}.
}
\label{f:SoapBublTurb3}
\end{center}
\end{figure}

Fig.~\ref{f:SoapBublTurb1} also shows that the bubble shape remains near spherical during inflation.
This is because the flow does not alter the high static surface tension $q_0$ significantly ($\Delta q$ is more than 1000 times less than $q_0$ as Fig.~\ref{f:SoapBublTurb1} shows).
This changes when $q_0$ is reduced drastically, as Fig.~\ref{f:SoapBublTurb3} shows.
%-----------------------------------------------------------------
% run sSymBubbleTurb3/*3 pSymBubbleTurb7
%-----------------------------------------------------------------
Here now $q_0 = 0.01 \mu $N/mm is used together with $R=100$mm (hence $p_0=0.2$mPa), $\alpha_\mathrm{DB} = (m/32)^2\cdot 10^6$ and $\tmu = 10^{-4}\mu$N/mm.
Parameters $\eta$ and $\rho$ are the same as before, see Table~\ref{t:soapbubble}.
The smaller $q_0$ necessitates a larger $\alpha_\mathrm{DB}$ value to capture the surface flow accurately.
Additionally, small out-of-plane viscosity has been added now according to Remark~\ref{r:p_visc}, using $\eta_\mrn = 0.001\mu$Nms/mm$^3$ in order to stabilize the out-of-plane bubble motion.
This value has been found to successfully bound out-of-plane surface oscillations while having negligible influence on the fluid flow.
The peak velocity now is about 0.22m/s leading to a Reynolds number of~$86.4$. 

As Fig.~\ref{f:SoapBublTurb3} shows, the bubble shape deviates significantly from a sphere during inflation.
This deviation is related to the surface velocity.
This was also seen in the example of Fig.~\ref{f:octoevolve2}, but in that case there was no tangential mesh motion and no inflow into the surface.
The proposed formulation thus also works for non-spherical shapes.
Although it has to be noted that not all parameter combinations work equally well.
In particular, the numerical parameters $\alpha_\mathrm{DB}$ and $\mu_\mrm$ have important influences on the accuracy and stability of the simulations.
This should be studied further in future work.

The example of Fig.~\ref{f:SoapBublTurb3} has the highest Reynolds number.
Hence the local P\'eclet number of Eq.~\eqref{e:Pe} will also be highest, and the question arises whether the computational meshes are sufficient to guarantee convective stability as noted in Remark~\ref{r:SUPG}.
%-----------------------------------------------------------------
\begin{figure}[h!]
\begin{center} \unitlength1cm
\begin{picture}(0,5.8)
\put(-8,-.15){\includegraphics[height=58mm]{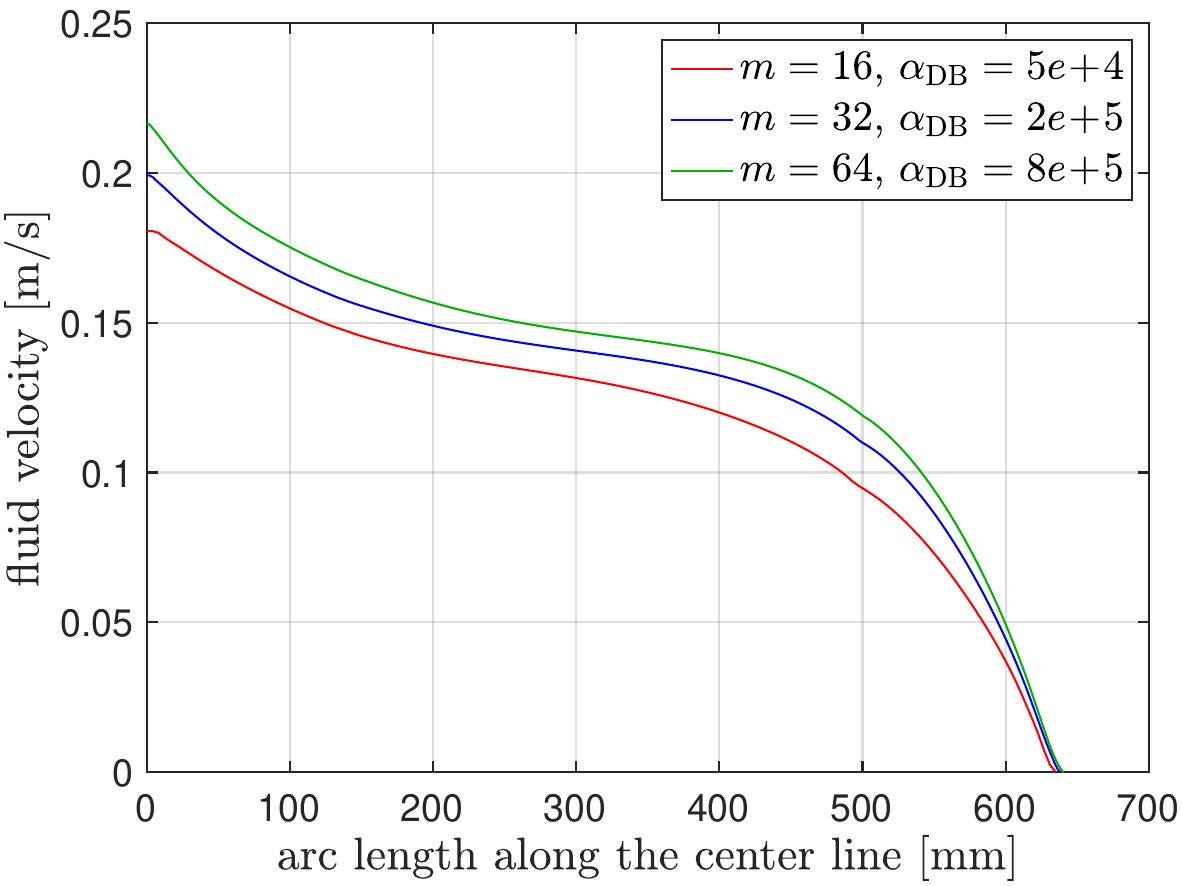}}
\put(0.2,-.2){\includegraphics[height=58mm]{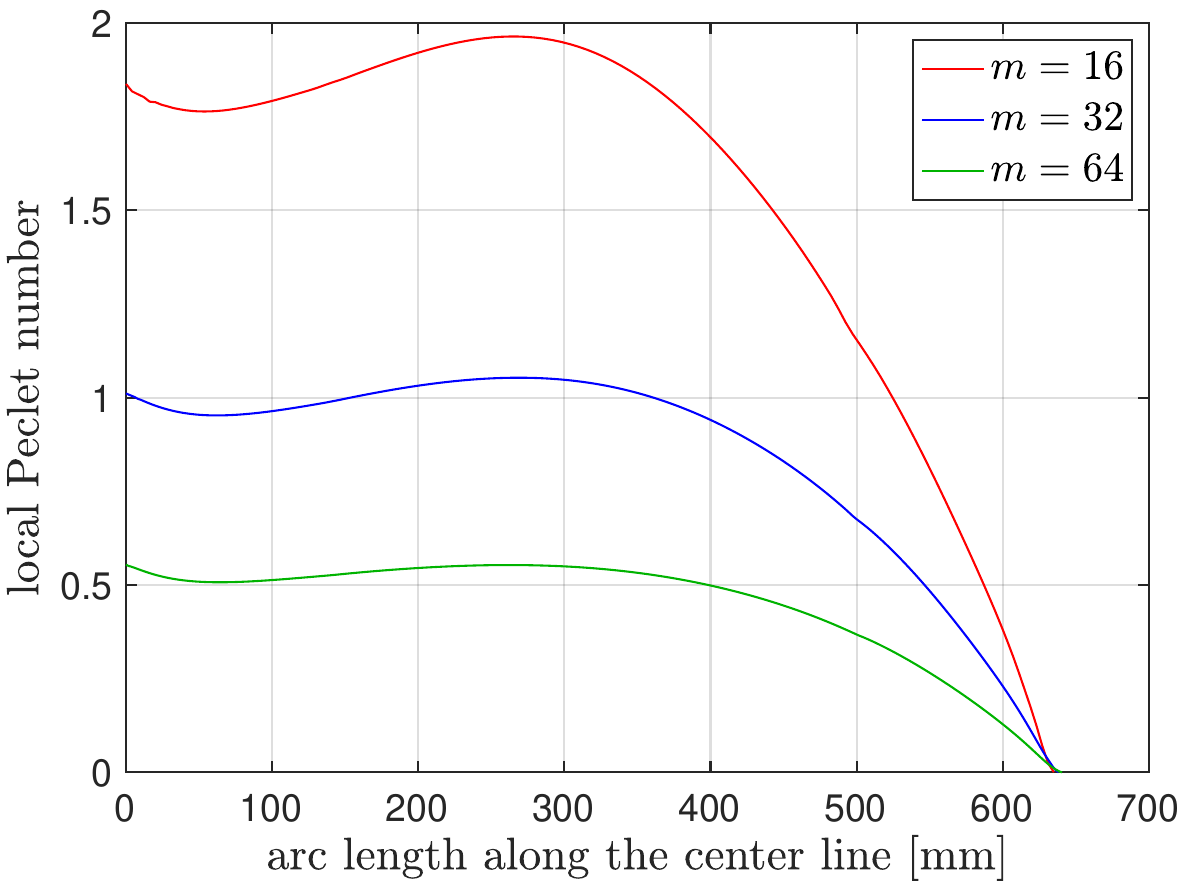}}
\put(-7.95,-.05){\footnotesize (a)}
\put(0.25,-.05){\footnotesize (b)}
\end{picture}
\caption{Non-uniform inflation of a soap bubble with low surface tension: % (Case 2 = case 7q))
(a) Fluid velocity $\norm{\bv}$ and (b) local P\'eclet number Pe along the symmetry line for time $t=16$s.}
\label{f:SoapBublTurb4}
\end{center}
\end{figure}
% run pSymBubbleTurb
%-----------------------------------------------------------------
Therefore, Fig.~\ref{f:SoapBublTurb4} shows the local P\'eclet number together with fluid velocity $\norm{\bv}$, which in this example is largest along the symmetry line and approximately equal to the convective velocity $\norm{\bv-\bv_\mrm}$ as $\bv_\mrm$ is quite small here. 
It is seen that for mesh $m=64$, Pe is well below one and hence sufficiently fine to guarantee stability, while for meshes $m = 32$, Pe is around one, and for mesh $m = 16$ Pe exceeds one.
Mesh $m = 16$ should thus be treated with care, even if no problems where detected in the present results.
All other examples above have P\'eclet numbers below one even for the coarse meshes.  
On the other hand, the example in Fig.~7a of \citet{ALEtheo} has Pe up to 6.4 for Eulerian mesh motion and shows severe velocity oscillations, but the proposed elastic mesh motion does not show any oscillations, even though Pe goes up to 8.2 for that example.
Thus, mesh elasticity may have also a stabilizing influence on the system.

\section{Conclusion}\label{s:concl}

This work presented a general surface ALE implementation for area-incompressible, transient Navier-Stokes flow on self-evolving manifolds.
The unknowns of the formulation are the material velocity, surface tension (negative in-plane pressure) and mesh motion.
The latter is determined from in-plane membrane elasticity in such a way that the material flow is not affected.
The formulation achieves optimal convergence rates, as is demonstrated through several examples.

Those include fixed surfaces, free surfaces that remain spherical through balancing surface pressures, and evolving surfaces with large surface changes.
If surface stretches are large, it is more accurate to integrate the Dohrmann-Bochev stabilization terms over the current surface instead of the initial surface.
Further, the proposed mesh elasticity formulation tends to be much more accurate than an Eulerian formulation, especially for $C^1$-continuous surface discretizations.

Future extensions of this work include the detailed study of $C^1$-continuous surface discretizations, bending resistance, and residual-based stabilization, such as streamline upwind Petrov-Galerkin (SUPG) stabilization. 
Adapting the latter to evolving surface flows would allow to use coarser meshes and thus increase efficiency.
Also the pressure stabilizing Petrov-Galerkin (PSPG) scheme could be explored as an alternative to the Dohrmann-Bochev scheme.
These extensions would then allow for more detailed studies of evolving soap bubbles, rolling droplets \citep{memFSI} and budding of lipid membranes \citep{liquidshell}.\\[-1mm]

\bigskip

{\Large{\bf Acknowledgements}}

The author is grateful to Thang X.~Duong for his comments.

\appendix

\section{Finite element linearization}\label{s:lin}

This section provides the linearization of finite element force vectors $\mf$, $\mbf$ and $\mtf$.
The notation $\mk:=\partial\mf/\partial\mx$, $\mcc:=\partial\mf/\partial\mv$, $\mm:=\partial\mf/\partial\mv'$, $\md:=\partial\mf/\partial\mq$ (and likewise for $\mbf$ and $\mtf$) is used to denote the tangent matrices corresponding to stiffness, viscosity, mass and the Lagrange multiplier. 
If the mesh motion is known (e.g.~for fixed surfaces) the linearization of the mesh ``forces" $\mtf$ is not needed.  
This is also the case for Lagrangian mesh motion, where $\bu :=  \bv - \bv_\mrm = \mathbf{0}$.

\subsection{Linearization of kinematic variables}

Since the shape functions do not change with deformation, the increments $\Delta\bx^h$, $\Delta\bv^h$, $\Delta\ba^h_\alpha$, $\Delta\bv_{\!,\alpha}^h$ and $\Delta q$ have the same structure as $\bx^h$, $\bv^h$, $\ba^h_\alpha$, $\bv^h_{\!,\alpha}$ and $q^h$, i.e.
\eqb{lll}
\Delta\bx^h \is \mN_e\,\Delta\mx_e\,, \\[1mm]
\Delta\bv^h \is \mN_e\,\Delta\mv_e\,, \\[1mm]
\Delta\ba_\alpha^h \is \mN_{e,\alpha}\,\Delta\mx_e\,, \\[1mm]
\Delta\bv_{\!,\alpha}^h \is \mN_{e,\alpha}\,\Delta\mv_e\,, \\[1mm]
\Delta q^h \is \mL_e\,\Delta\mq_e\,.
\label{e:Deltax}\eqe
The linearization of $\ba^\alpha$ yields
\eqb{l}
\Delta\ba^\alpha = \ba^{\alpha\beta}_\mrn\,\Delta\ba_\beta\,,
\label{e:Deltaa}\eqe
with
\eqb{l}
\ba^{\alpha\beta}_\mrn := a^{\alpha\beta}\,\bn\otimes\bn - \ba^\beta\otimes\ba^\alpha
\label{e:ban}\eqe
\citep{shelltheo}.
Note that Eqs.~\eqref{e:bas} and \eqref{e:ban} lead to $\ba^{\alpha\beta}_\mrs + \ba^{\alpha\beta}_\mrn = \bone\,a^{\alpha\beta}$, which offers an alternative way of calculating $\ba^{\alpha\beta}_\mrn$ -- one that does not require $\bn$.
The linearization of $J_\mrm$ is needed for the integration (since $\dif a = J_\mrm\,\dif A$).
The increment $\Delta J_\mrm$ follows analogous to $\dot J_\mrm$ \citep{ALEtheo} and hence is
\eqb{l}
\Delta J_\mrm = J_\mrm\,\ba^\beta\cdot\Delta\ba_\beta = J_\mrm\,\mD_e\,\Delta\mx_e\,,
\label{e:DeltaJm}\eqe
due to (\ref{e:Deltax}.3) and  \eqref{e:De}.
Therefore
\eqb{l}
\Delta(J_\mrm\ba^\alpha) = J_\mrm\,\ba^{\alpha\beta}_\mrJ\,\Delta\ba_\beta\,,
\label{e:DeltaaJm}\eqe
with
\eqb{l}
\ba^{\alpha\beta}_\mrJ := \ba^{\alpha\beta}_\mrn +  \ba^\alpha\otimes\ba^\beta\,. 
\eqe

\subsection{Linearization of the equation of motion}

The FE force vector $\mf$ representing the equation of motion consists of the parts derived in Sec.~\ref{s:f}.
Its linearization results in the increments
\eqb{l}
\Delta\mf = \Delta\mf_\mathrm{in} + \Delta\mf_\mathrm{intq} + \Delta\mf_\mathrm{intv} - \Delta\mf_\mathrm{ext}\,.
\label{e:Dmf}\eqe
They are discussed in the following.

\subsubsection{Linearization of $\mf^e_\mathrm{in}$}\label{s:Lfin}

Linearizing the elemental inertia vector $\mf^e_\mathrm{in}$ specified by \eqref{e:mf_trans} and \eqref{e:mf_conv} leads to
\eqb{l}
\Delta\mf^e_\mathrm{in} = \mm_e\,\Delta\mv'_{e} + \mcc^e_\mathrm{in}\,\Delta\mv_e + \mcc^e_\mrm\,\Delta\mx'_e  + \mk^e_\mathrm{in}\,\Delta\mx_e\,,
\label{e:Deltafin}\eqe
with
\eqb{lll}
\mm_e \dis \ds\int_{\Omega^e} \rho\,\mN_e^\mrT\,\mN_e \,\dif a\,, \\[3.5mm]
\mcc_\mrm^e \dis -\ds\int_{\Omega^e} \rho\,\mN_e^\mrT\, \nablas\bv\,\mN_e \,\dif a\,, \\[3.5mm]
\mcc_\mathrm{in}^e \dis \ds\int_{\Omega^e} \rho\,\mN_e^\mrT\,u^\alpha\,\mN_{e,\alpha} \,\dif a - \mcc^e_\mrm\,, \\[3.5mm]
\mk^e_\mathrm{in} \dis \ds\int_{\Omega^e} \rho\,\mN_e^\mrT\,\bu^\beta_\mrx\,\mN_{e,\beta} \,\dif a
+ \ds\int_{\Omega^e} \rho\,\mN_e^\mrT\,\bv'\otimes\ba^\beta\,\mN_{e,\beta} \,\dif a \,,
\label{e:kin}
\eqe
where $\nablas\bv$ and $u^\alpha = \dot\zeta^\alpha$ follow from \eqref{e:sgrad}, \eqref{e:badha} and \eqref{e:zetadoth}, and
\eqb{l}
\bu^\beta_\mrx := \bv_{\!,\alpha}\otimes\big(a^{\alpha\beta}\,u\,\bn - \ba^\alpha\,u^\beta + u^\alpha\,\ba^\beta \big)
\eqe
comes from linearizing $u^\alpha\,\dif a$.
The first term in $\bu^\beta_\mrx$ vanishes in theory, as $u = 0$.
But this may not be the case in the discretized system, especially not during Newton-Raphson iteration.
The last term in $\bu^\beta_\mrx$ comes from linearizing $\dif a$.
Similarly, the second term in $\mk^e_\mathrm{in}$ comes from linearizing $\dif a$ in mass matrix $\mm_e$. 

\subsubsection{Linearization of $\mf^e_\mathrm{int}$}\label{s:Lfint}

Applying \eqref{e:Deltax} and \eqref{e:DeltaaJm} to \eqref{e:felvisc} directly leads to
\eqb{llllll}
\Delta\mf^e_{\mathrm{int}q} \is \md_\mathrm{int}^e\,\Delta\mq_e \plus \mk^e_{\mathrm{int}q}\,\Delta\mx_e\,, \\[1mm]
\Delta\mf^e_{\mathrm{int}v} \is \mcc_\mathrm{int}^e\,\Delta\mv_e \plus \mk^e_{\mathrm{int}v}\,\Delta\mx_e\,,
\label{e:Deltafint}\eqe
with
\eqb{lll}
\md_\mathrm{int}^e \dis \ds\int_{\Omega^e} \mD_e^\mrT\,\mL_e \,\dif a\,, \\[3.5mm]
\mcc_\mathrm{int}^e \dis \ds\int_{\Omega^e} \eta\,\mN_{e,\alpha}^\mrT\,\ba^{\alpha\beta}_\mrs\,\mN_{e,\beta} \,\dif a\,,\\[3.5mm]
\mk^e_{\mathrm{int}q} \dis \ds\int_{\Omega^e} q\,\mN_{e,\alpha}^\mrT\,\ba^{\alpha\beta}_\mrJ\,\mN_{e,\beta} \,\dif a\,, \\[3.5mm]
\mk^e_{\mathrm{int}v} \dis \ds\int_{\Omega^e} 2\eta\,\mN_{e,\alpha}^\mrT\,\bd_\mrx^{\alpha\beta}\,\mN_{e,\beta} \,\dif a\,, 
\label{e:kint}
\eqe
where $\bd_\mrx^{\alpha\beta}$ is defined by 
\eqb{l}
\Delta\big(J_\mrm\,\ba_\mrs^{\alpha\beta}\big) \bv_{\!,\beta} = 2J_\mrm\,\bd_\mrx^{\alpha\beta}\Delta\ba_\beta
\eqe
and results in
\eqb{l}
2\bd_\mrx^{\alpha\beta} = \ba_\mrs^{\alpha\beta}\big( \nablas\bv^\mrT(\bn\otimes\bn) - \nablas\bv  \big) + 2\ba_\mrn^{\alpha\gamma}d^\beta_\gamma
+ 2d^{\alpha\gamma}\ba_\gamma\otimes\ba^\beta\,.
\eqe
The last part in $\bd_\mrx^{\alpha\beta}$ comes from the linearization of $J_\mrm$.

\subsubsection{Linearization of $\mf^e_\mathrm{ext}$}\label{s:Lfext}

In this work, the external force vector $\mf^e_\mathrm{ext}$ contains the two solution-dependent contributions $\mf^e_p$ and $ \mf^e_T$, i.e.~$\mf^e_\mathrm{ext} = \mf^e_p + \mf^e_T$.
The linearization of $\mf^e_\mathrm{ext}$ produces the three tangent contributions 
$\mk^e_\mathrm{ext} = \mk^e_p + \mk^e_T$,
$\mcc^e_\mathrm{ext} = \mcc^e_p$ and 
$\md^e_\mathrm{ext} = \md^e_T$ derived in the following.\\[-3mm]

\textit{A.2.3.1~~Surface pressure}

The first solution-dependent external force that is present in all examples is surface pressure.
It is captured by the FE force vector
\eqb{l}
\mf^e_p := \ds\int_{\Omega^e} \mN_e^\mrT\,p\,\bn\,\dif a\,.
\label{e:mfp}\eqe
Here, $p= \bar p + p_\mathrm{visc}$ contains the prescribed constant pressure $\bar p$ and the viscous pressure from Eq.~\eqref{e:pvisc}.
The linearization of the two parts is given in \citet{membrane} and \citet{dropslide}.
Together, they give
\eqb{l}
\Delta\mf^e_p = \mcc^e_p\,\Delta\mv_e + \mk^e_p \,\Delta\mx_e\,,
\label{e:Deltafp}\eqe
with the tangent matrices
\eqb{lll}
\mcc^e_p \dis
- \ds\int_{\Omega^e}\eta_\mrn\,\mN_e^\mrT(\bn\otimes\bn)\,\mN_e\,\dif a\,, \\[3.5mm]
\mk^e_p \dis \ds\int_{\Omega^e} \mN_e^\mrT\,\Big[p\,\big(\bn\otimes\ba^\alpha - \ba^\alpha \otimes\bn\big)
+ \eta_\mrn\,\bn\otimes\bn\,v^\alpha
\Big]\,\mN_{e,\alpha}\,\dif a\,,
\label{e:mckp}\eqe
due to \eqref{e:DeltaJm} and
\eqb{l}
\Delta\bn = -(\ba^\alpha\otimes\bn)\,\Delta\ba_\alpha
\label{e:dn}\eqe
\citep{membrane}.
Note, that these $\mcc^e_p$ and $\mk^e_p$ need to be subtracted from the other $\mcc$ and $\mk$ due to the minus sign in \eqref{e:Dmf}.\\[-3mm]

\textit{A.2.3.2~~Boundary forces in the soap bubble example}\label{s:Lft}

The soap bubble example of Sec.~\ref{s:SB} contains a free boundary, where the two solution-dependent boundary conditions $v_\mrn := \bv\cdot\bn = 0$ and $\bT = \bar\bT := q\,\bnu$ are prescribed

The first BC is enforced by the penalty method.
Thus the surface pressure $p_\epsilon = - \epsilon\,\bv\cdot\bn$ is applied along the boundary.
This is done as described in Eqs.~\eqref{e:mfp}-\eqref{e:mckp}, 
replacing pressure $p$ by $p_\epsilon$ and integration domain $\Omega^e$ by $\Gamma^e$, as the boundary condition is only enforced along the line elements along the boundary.  
However, the whole surface element $\Omega^e$ needs to be sampled in order to evaluate the normal vector $\bn$.

The second BC leads to the external FE force vector
\eqb{l}
\mf^e_T := \ds\int_{\Gamma^e} \mN^\mrT_e\,q\,\bnu\,\dif s
\label{e:ft}\eqe
according to \eqref{e:fext}.
Here $q = \mL_e\,\mq_e$ according to Eq.~\eqref{e:qha}.
Since the bubble is pinned, $\Gamma^e$ and $\dif s$ are fixed.
Thus the linearization of \eqref{e:ft} leads to
\eqb{l}
\Delta\mf^e_T = \md^e_T\,\Delta\mq_e + \mk^e_T\,\Delta\mx_e\,,
\label{e:dft}\eqe
with
\eqb{lll}
\md^e_T \dis \ds\int_{\Gamma^e} \mN^\mrT_e\,\bnu\,\mL_e\,\dif s\,, \\[3.5mm]
\mk^e_T \dis -\ds\int_{\Gamma^e} \mN^\mrT_e\,q\,(\btau\times\ba^\alpha)\,\bn\cdot\mN_{e,\alpha}\,\dif s\,.
\eqe
The latter expression follows from
\eqb{l}
\Delta\bnu = -(\btau\times\ba^\alpha)(\bn\cdot\Delta\ba_\alpha)\,, 
\eqe
which in turn follows from $\bnu = \btau\times\bn$ and \eqref{e:dn} for fixed $\btau$.
It is noted that along the boundary one of the $\ba^\alpha$ ($\alpha = 1,2$) is typically parallel to $\btau$, such that only the other $\ba^\alpha$ appears within~$\mk^e_T$.

\subsection{Linearization of $\mbf^e$}\label{s:Lfq}

According to \eqref{e:mbf1},
\eqb{l}
\Delta\mbf^e = \Delta\mbf^e_\mathrm{div} - \Delta\mbf^e_\mathrm{DB}\,.
\label{e:Deltambf}\eqe
From \eqref{e:sdivh}, \eqref{e:De} and \eqref{e:DeltaaJm} follows
\eqb{l}
\Delta\big(J_\mrm\,\divs^{\!\!\!\!h}\,\bv^h\big) = J_\mrm\,\mD_e\,\Delta\mv_e + J_\mrm\,\bv_{\!,\alpha}\cdot\ba^{\alpha\beta}_\mrJ\,\mN_{e,\beta}\,\Delta\mx_e\,.
\eqe
Thus (\ref{e:mbf2}.1) gives
\eqb{l}
\Delta\mbf^e_\mathrm{div} = \mbc_e\,\Delta\mv_e + \mbk_e\,\Delta\mx_e\,,
\label{e:Deltambf1}\eqe
with
\eqb{lll}
\mbc_e \dis \ds\int_{\Omega^e} \mL_e^\mrT\,\mD_e \,\dif a\,, \\[3.5mm]
\mbk_e \dis \ds\int_{\Omega^e} \mL_e^\mrT\,\bv_{\!,\alpha}\cdot\ba^{\alpha\beta}_\mrJ\,\mN_{e,\beta}\,\dif a\,,
\eqe
confirming the usual symmetry $\mbc_e^\mrT = \md_\mathrm{int}^e$. 
Also $\mbk_e$ and $\mk^e_{\mathrm{int}q}$ exhibit a certain symmetry.
Using \eqref{e:DeltaJm} and $\Delta(\mh_e^{-1}) = -\mh_e^{-1}\Delta\mh_e\,\mh_e^{-1}$, the linearization of \eqref{e:mbf3} leads to
\eqb{l}
\Delta\mbf^e_\mathrm{DB} = \mbd_e\,\Delta\mq_e +  \mbk^e_\mathrm{DB}\,\Delta\mx_e\,,
\label{e:Deltambf2}\eqe
with 
\eqb{lll}
\mbk^e_\mathrm{DB} \dis \ds\frac{\alpha_\mathrm{DB}}{\eta}\bigg[ \int_{\Omega^e} q\,\mL_e^\mrT\,\mD_e\,\dif a 
-\mg_e^\mrT\,\mh_e^{-1}\int_{\Omega^e} q\,\mathbf{\check L}_e^\mrT\,\mD_e\,\dif a\\[3mm]
\mi\ds\sum_{I=1}^{\check m_e}\bigg(\int_{\Omega^e} \mL_e^\mrT\,\mD_e\,\check L_I\,\dif a 
-\mg_e^\mrT\,\mh_e^{-1}\int_{\Omega^e} \mathbf{\check L}_e^\mrT\,\mD_e\,\check L_I\,\dif a \bigg) %v_\mathrm{hgq}^I\bigg]\,,
\,\check q_I \bigg]\,, 
\eqe
where $\check L_I$ and 
%$v_\mathrm{hgq}^I$ are the $I$-th components of vectors $\mathbf{\check L}_e$ and $\mv_\mathrm{hgq} := \mh_e^{-1}\,\mg_e\,\mq_e$.
$\check q_I$ are the $I$-th components of vectors $\mathbf{\check L}_e$ and $\mathbf{\check q}_e = \mh_e^{-1}\,\mg_e\,\mq_e$.

\subsection{Linearization of $\mtf^e$}\label{s:Lfh}

\subsubsection{Eulerian mesh motion}

From \eqref{e:mesh1} and \eqref{e:dn} follows
\eqb{l}
\Delta\mtf^e := \mtm_e\,\Delta\mx'_e - \mtn_e\,\Delta\mv_e + \mtn_e^v\,\Delta\mx_e\,,
\label{e:Deltamhf1}\eqe
with
\eqb{l}
\mtn_e^v := \ds\int_{\Omega^e_0}\,\mN_e^\mrT\,\big(v\,\ba^\alpha\otimes\bn + v^\alpha\,\bn\otimes\bn\big)\,\mN_{e,\alpha}\,\dif A\,.
\label{e:mhnv}\eqe

\subsubsection{Elastic mesh motion}

From \eqref{e:mesh2} and \eqref{e:fio} follows
\eqb{l}
\Delta\mtf^e := \Delta\mtf^e_\mri + \Delta\mtf^e_\mro\,,
\label{e:Deltamhf2}\eqe
with
\eqb{lll}
\Delta\mtf^e_\mri \dis \mtk^e_\mri\,\Delta\mx_e \\[1mm]
\Delta\mtf^e_\mro \dis \mtn_e\,\big(\Delta\mx_e'-\Delta\mv_e\big) + \mtn_e^u\,\Delta\mx_e\,.
\eqe
Here,
\eqb{lll}
\mtk^e_\mri \dis \ds\int_{\Omega_0^e}\mN^\mrT_{e,\alpha}\,\big(\bone\,\tau_\mrm^{\alpha\beta}+\tmu\,\ba_\mrs^{\alpha\beta}\big)\,\mN_{e,\beta}\,\dif A 
+ \ds\int_{\Omega_0^e}\mN^\mrT_e\,(\bn\otimes\bn)\,\tau_\mrm^{\alpha\beta}\,\mN_{e;\alpha\beta}\,\dif A \\[4mm]
\plus \ds\int_{\Omega_0^e}\mN^\mrT_e\,\big(2\tmu\,b^\gamma_\alpha\,\bn\otimes\ba^\alpha - \tau_\mrm^{\alpha\beta}b_{\alpha\beta}\,\ba^\gamma\otimes\bn\big)\,\mN_{e,\gamma}\,\dif A
\eqe
follows from \eqref{e:dn},
\eqb{l}
\Delta b_{\alpha\beta} = \bn\cdot\mN_{;\alpha\beta}\,\Delta\mx_e\,, 
\quad \mN_{;\alpha\beta} := \mN_{,\alpha\beta} - \Gamma^\gamma_{\alpha\beta}\,\mN_{e,\gamma}\,,
\eqe
\citep{solidshell} and
\eqb{l}
\Delta\tau_\mrm^{\alpha\beta} = c_\mrm^{\alpha\beta\gamma\delta}\,\ba_\gamma\cdot\Delta\ba_\delta\,,
\quad c_\mrm^{\alpha\beta\gamma\delta} := \tmu\big(a^{\alpha\gamma}\,a^{\beta\delta} + a^{\alpha\delta}\,a^{\beta\gamma}\big)\,,
\eqe
\citep{shelltheo}, while $\mtn_e^u$ is analogous to $\mtn_e^v$ from \eqref{e:mhnv} -- simply $\bv$ is replaced by $\bu := \bv-\bv_\mrm$ in \eqref{e:mhnv}, i.e.
\eqb{l}
\mtn_e^u:= \ds\int_{\Omega^e_0}\,\mN_e^\mrT\,\big(u\,\ba^\alpha\otimes\bn + u^\alpha\,\bn\otimes\bn\big)\,\mN_{e,\alpha}\,\dif A\,.
\eqe
As noted above, $u^\alpha=\dot\zeta^\alpha$ due to \eqref{e:bv}, but even though $u=0$ in theory, $u$ can become non-zero in the discrete system. 
It is noted that $\mtk^e_\mri$ is equal to the more lengthy tangent expression given in Appendix A.1 of \citet{droplet}.

\section{Tangent matrices of the time discretized system}\label{s:tang}

This section provides the FE tangent matrices for the monolithic solution of the coupled system in Sec.~\ref{s:NR}.
This is done in the framework of the generalized-$\alpha$ method \citep{chung93,jansen99}, even though all numerical examples use $\alpha_\mrf = \alpha_\mrm = 1$.
The generalized-$\alpha$ method simply evaluates the current residual $\mr_{n+1}$ at the asynchronous arguments
\eqb{llrlr}
\mx_{n+\alpha_\mrf} \is \alpha_\mrf\,\mx_{n+1} \plus (1-\alpha_\mrf)\,\mx_n\,, \\[1mm]
\mv_{n+\alpha_\mrf} \is \alpha_\mrf\,\mv_{n+1} \plus (1-\alpha_\mrf)\,\mv_n\,, \\[1mm]
\ma_{n+\alpha_\mrm} \is \alpha_\mrm\,\ma_{n+1} \plus (1-\alpha_\mrm)\,\ma_n\,,
\eqe
and so simply an extra factor of $\alpha$ is picked up in the corresponding linearization. 

The linearization of \eqref{e:vxnp1} leads to (reducing $n$ by 1)
\eqb{l}
\Delta\ma_n = \Delta\mv'_n = \ds\frac{1}{\gamma\,\Delta t_n}\,\Delta\mv_n\,,\quad
\Delta\mv_\mrm^n =  \ds\frac{1}{\gamma\,\Delta t_n}\,\Delta\mx_n\,.
\label{e:Deltavxnp1}\eqe
With this and Appendix \ref{s:lin}, one can obtain the elemental blocks of $\mK$ in \eqref{e:K}.
It is noted that indices $e$ and $n$ are placed wherever convenient.
The first three blocks of $\mK$ in \eqref{e:K} follow from Eqs.~\eqref{e:Deltafin}, \eqref{e:Deltafint}, \eqref{e:Deltafp} and \eqref{e:Deltavxnp1}, as
\eqb{l}
{\mk^e_{\mrv\mrv}}^{\back n} := \ds\pa{\mf^e_n}{\mv_n} 
	= \ds\frac{\alpha_\mrm}{\gamma\,\Delta t_n}\mm^n_e + \alpha_\mrf\big(\mcc^{e\,n}_\mathrm{in} + \mcc^{e\,n}_\mathrm{int} - \mcc^{e\,n}_\mathrm{ext} \big)\,,
\eqe
\eqb{l}
{\mk^e_{\mrv\mrx}}^{\back n} := \ds\pa{\mf^e_n}{\mx_n} = \ds\frac{\alpha_\mrf}{\gamma\,\Delta t_n}\mcc^{e\,n}_\mrm 
	+ \alpha_\mrf\big(\mk^{e\,n}_\mathrm{in} + \mk^{e\,n}_\mathrm{intq} + \mk^{e\,n}_\mathrm{intv} - \mk^{e\,n}_\mathrm{ext}\big)\,,	
\eqe
and
\eqb{l}
{\mk^e_{\mrv\mrq}}^{\back n} := \ds\pa{\mf^e_n}{\mq_n} = \md_\mathrm{int}^{e\,n}-\md_\mathrm{ext}^{e\,n}\,.
\eqe
The next three blocks follow from Eqs.~\eqref{e:Deltamhf1} and \eqref{e:Deltamhf2} as
\eqb{l}
{\mk^e_{\mrx\mrv}}^{\back n} := \ds\pa{\mtf^e_n}{\mv_n} = -\alpha_\mrf\,\mtn^n_e\,,
\eqe
\eqb{l}
{\mk^e_{\mrx\mrx}}^{\back n} := \ds\pa{\mtf^e_n}{\mx_n} 
	=  \ds\frac{\alpha_\mrf}{\gamma\,\Delta t_n}\begin{Bmatrix} \mtm_e \\
	\mtn_e \end{Bmatrix}_{\!n} + \alpha_\mrf\begin{Bmatrix}\mtn_e^v \\
	\mtk^e_\mri + \mtn_e^u \end{Bmatrix}_{\!n} ,
\eqe
and 
\eqb{l}
{\mk^e_{\mrx\mrq}}^{\back n} := \ds\pa{\mtf^e_n}{\mq_n} = \mathbf{0}\,.
\eqe
Here the top entries in the curly brackets are for Eulerian mesh motion (option I), while the lower entries are for elastic mesh motion (option II).
The last three blocks follow from Eq.~\eqref{e:Deltambf} as
\eqb{l}
{\mk^e_{\mrq\mrv}}^{\back n} := \ds\pa{\mbf^e_n}{\mv_n} = \alpha_\mrf\,\mbc_e^n = {{\mk^e_{\mrv\mrq}}^{\back n}}^\mrT \,,
\eqe
\eqb{l}
{\mk^e_{\mrq\mrx}}^{\back n} := \ds\pa{\mbf^e_n}{\mx_n} 	= \alpha_\mrf\big(\mbk_e^n - \mbk_\mathrm{DB}^{e\,n}\big)\,,
\eqe
and 
\eqb{l}
{\mk^e_{\mrq\mrq}}^{\back n} := \ds\pa{\mbf^e_n}{\mq_n} = -\mbd_e^n\,.
\eqe

\begin{remark}
If the mesh motion is known, e.g.~by prescribing either $\mx$ or $\mv_\mrm$, the corresponding blocks in $\mK$ are simply eliminated in the global system.
\end{remark}

\begin{remark}
Using \eqref{e:Deltavxnp1}, the unknowns can be easily exchanged. 
E.g.~replacing $\mx$ by $\mv_\mrm$ simply changes 
\eqb{l}
\begin{bmatrix}
\mK_{\mrv\mrv} & \mK_{\mrv\mrx} & \mK_{\mrv\mrq}\, \\[.5mm]
\mK_{\mrx\mrv} & \mK_{\mrx\mrx} & \mK_{\mrx\mrq}\, \\[.5mm]
\mK_{\mrq\mrv} & \mK_{\mrq\mrx} & \mK_{\mrq\mrq}\,
\end{bmatrix}_{\!n} \begin{bmatrix}
\Delta\mv\, \\
\Delta\mx\, \\
\Delta\mq\,
\end{bmatrix}_{\!n+1} \! = -\begin{bmatrix}
\mf \\
\mtf \\
\mbf
\end{bmatrix}_{\!n}
\eqe
into
\eqb{l}
\begin{bmatrix}
\mK_{\mrv\mrv} & \gamma\,\Delta t_n\,\mK_{\mrv\mrx} & \mK_{\mrv\mrq}\, \\[.5mm]
\mK_{\mrx\mrv} & \gamma\,\Delta t_n\,\mK_{\mrx\mrx} & \mK_{\mrx\mrq}\, \\[.5mm]
\mK_{\mrq\mrv} & \gamma\,\Delta t_n\,\mK_{\mrq\mrx} & \mK_{\mrq\mrq}\,
\end{bmatrix}_{\!n} \begin{bmatrix}
\Delta\mv \\
\Delta\mv_\mrm \\
\Delta\mq
\end{bmatrix}_{\!n+1} \! = -\begin{bmatrix}
\mf \\
\mtf \\
\mbf
\end{bmatrix}_{\!n}.
\eqe
\end{remark}

\section{Additional numerical results for fixed surfaces}\label{s:add}

\begin{figure}[h]
\begin{center} \unitlength1cm
\begin{picture}(0,11.6)
\put(-8,5.8){\includegraphics[height=58mm]{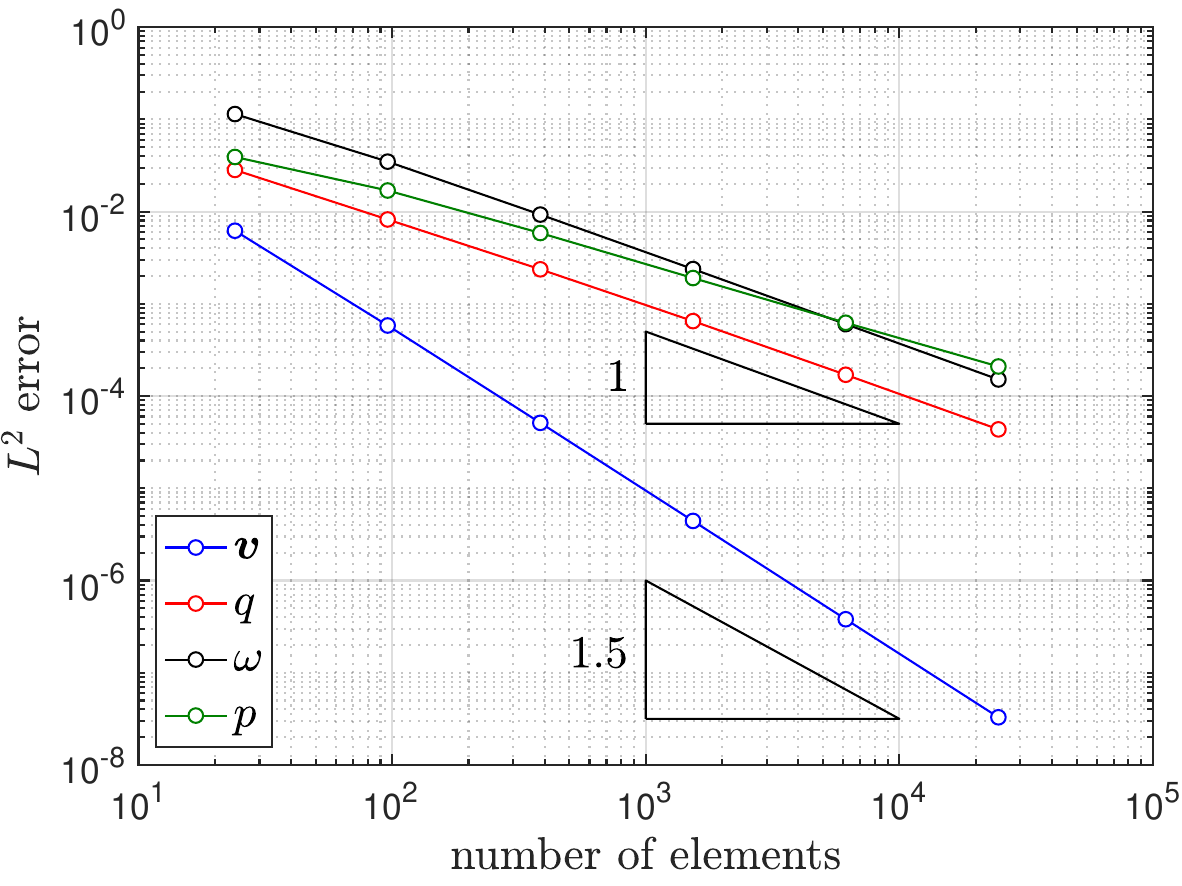}}
\put(0.2,5.8){\includegraphics[height=58mm]{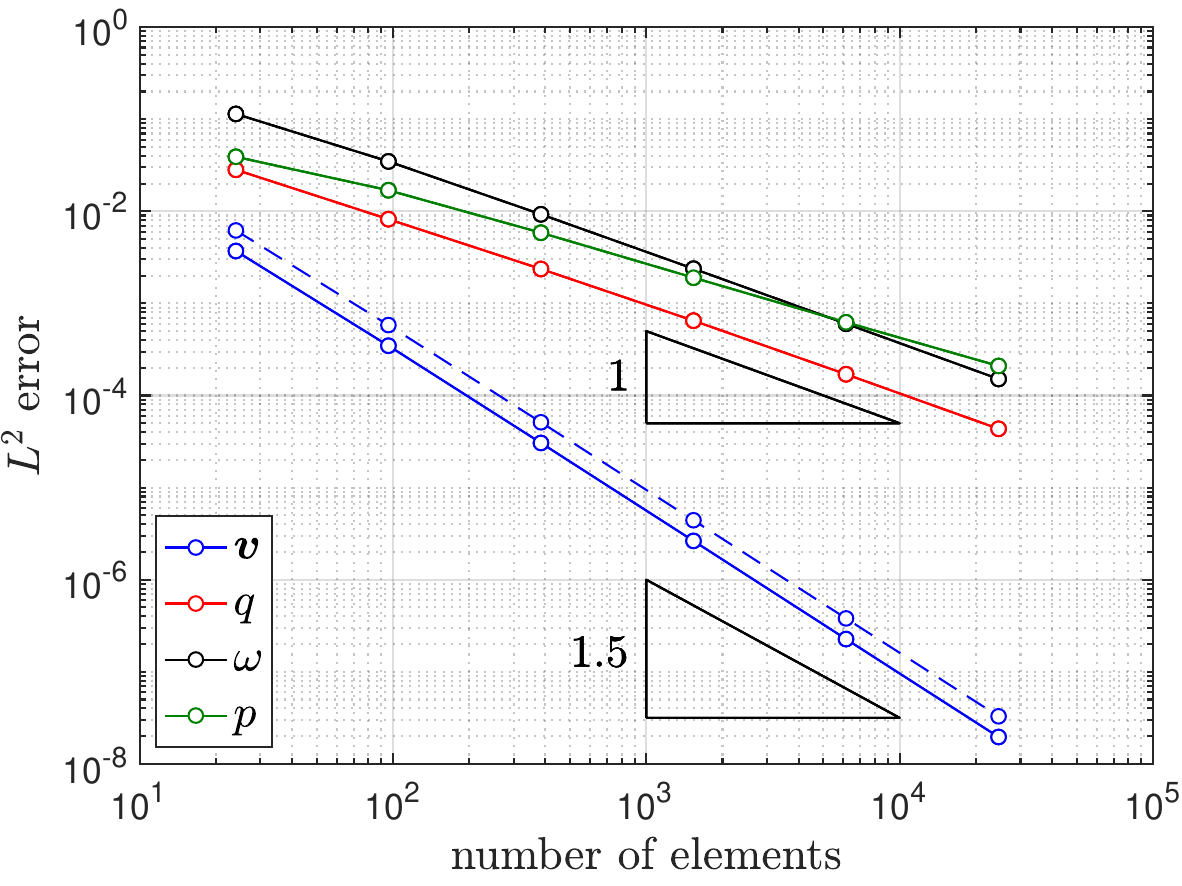}}
\put(-8,-.12){\includegraphics[height=58mm]{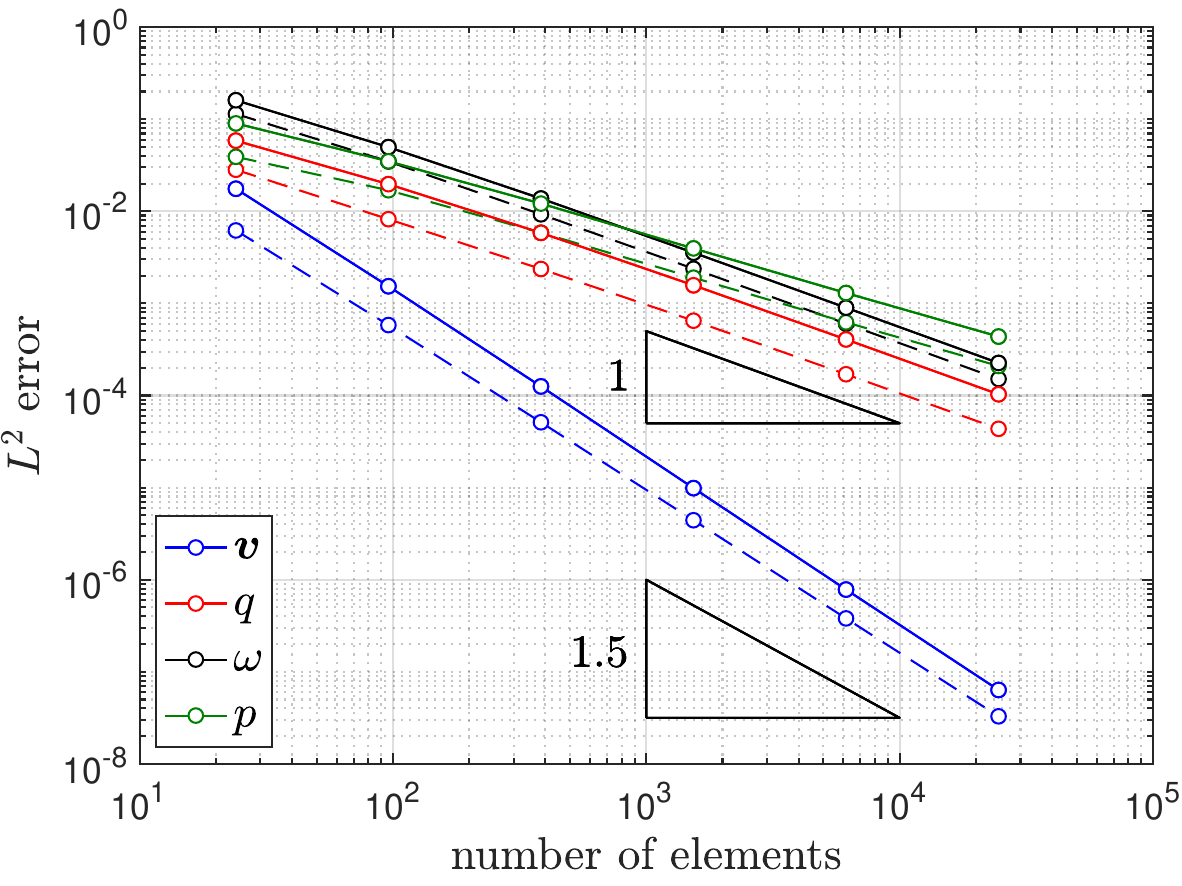}}
\put(0.2,-.2){\includegraphics[height=58mm]{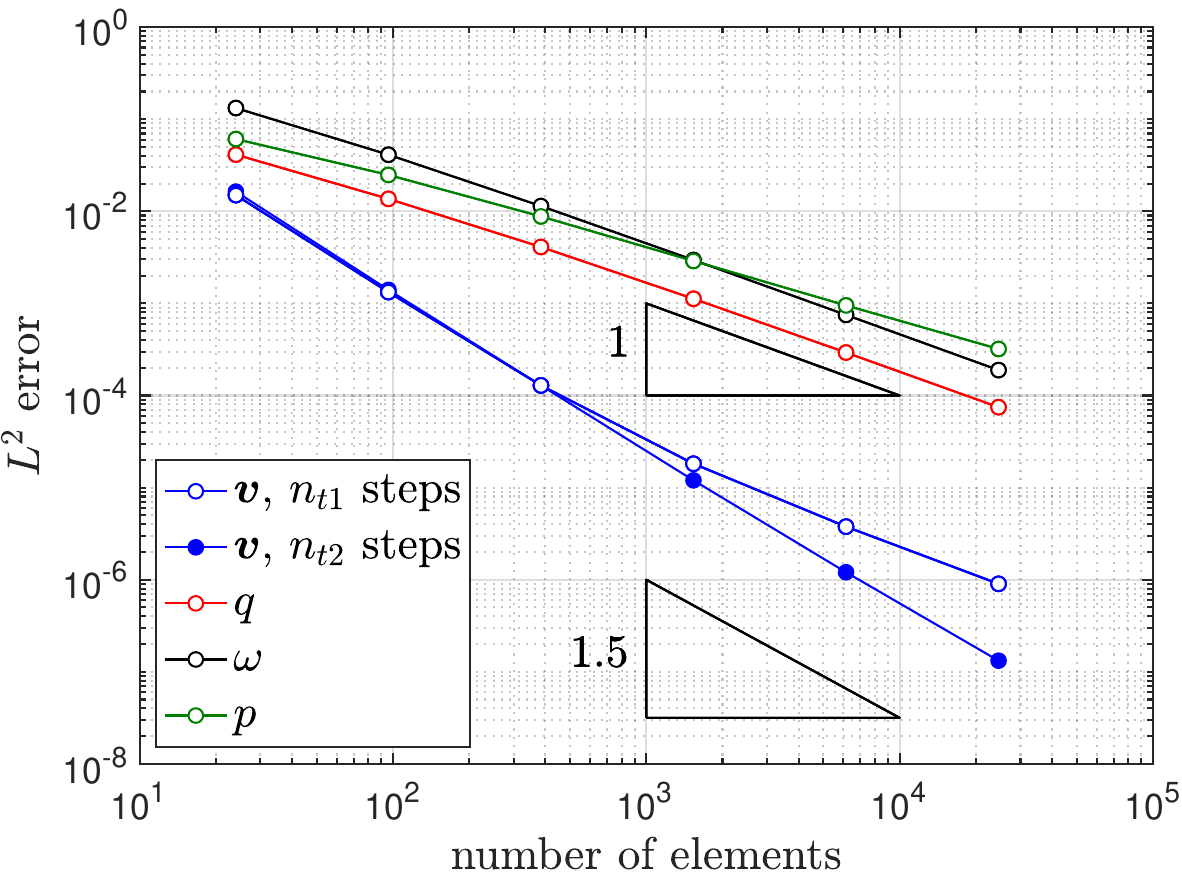}}
\put(-7.95,5.85){\footnotesize (a)}
\put(0.2,5.85){\footnotesize (b)}
\put(-7.95,-.15){\footnotesize (c)}
\put(0.2,-.15){\footnotesize (d)}
\end{picture}
\caption{Simple shear flow on a rigid sphere for load case 2: 
Convergence of velocity, surface tension and vorticity for the four ALE cases (a-d) in the $L^2$ error norm.
Optimal convergence rates are obtained for all fields.
The dashed lines in (b) \& (c) are the results from (a).}
\label{f:shearflowA1}
\end{center}
\end{figure}
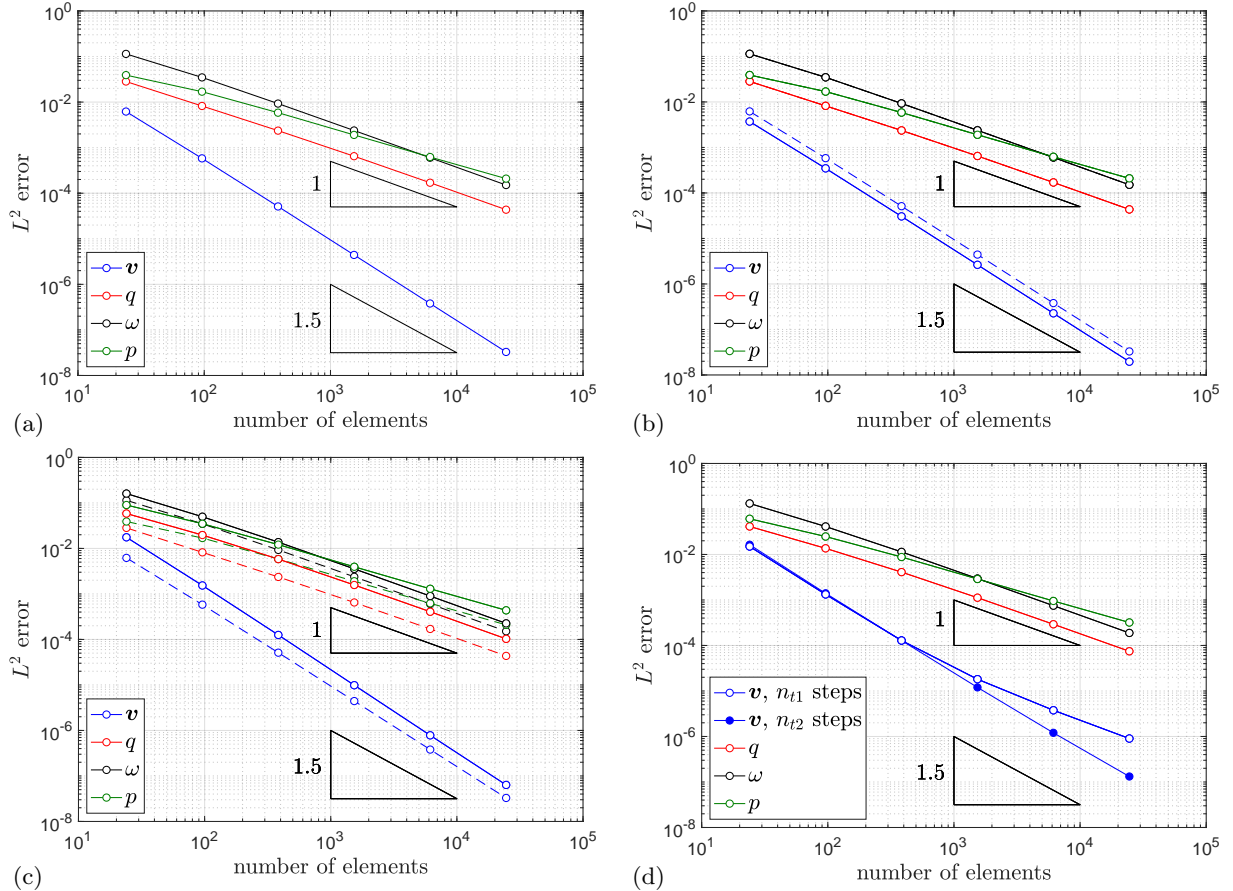

\begin{figure}[h]
\begin{center} \unitlength1cm
\begin{picture}(0,11.6)
\put(-8,5.8){\includegraphics[height=58mm]{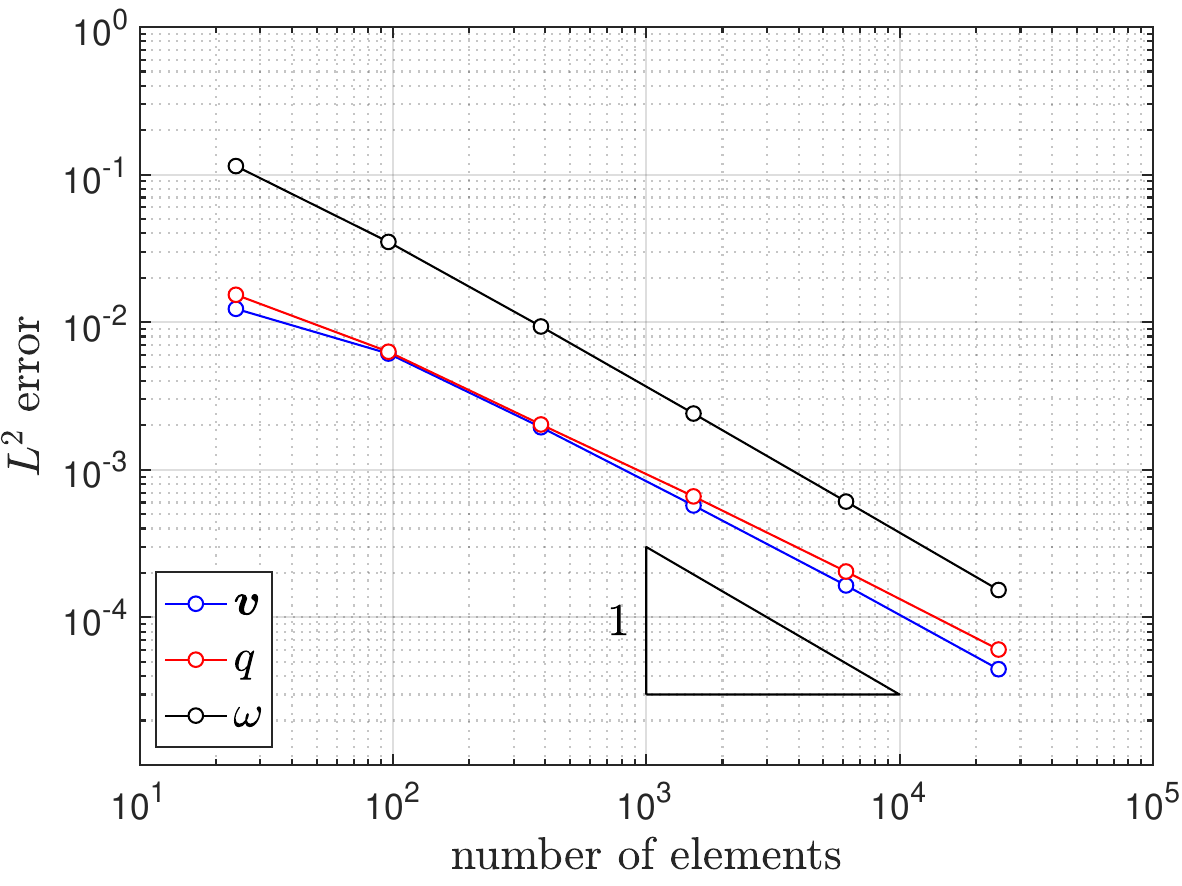}}
\put(0.2,5.8){\includegraphics[height=58mm]{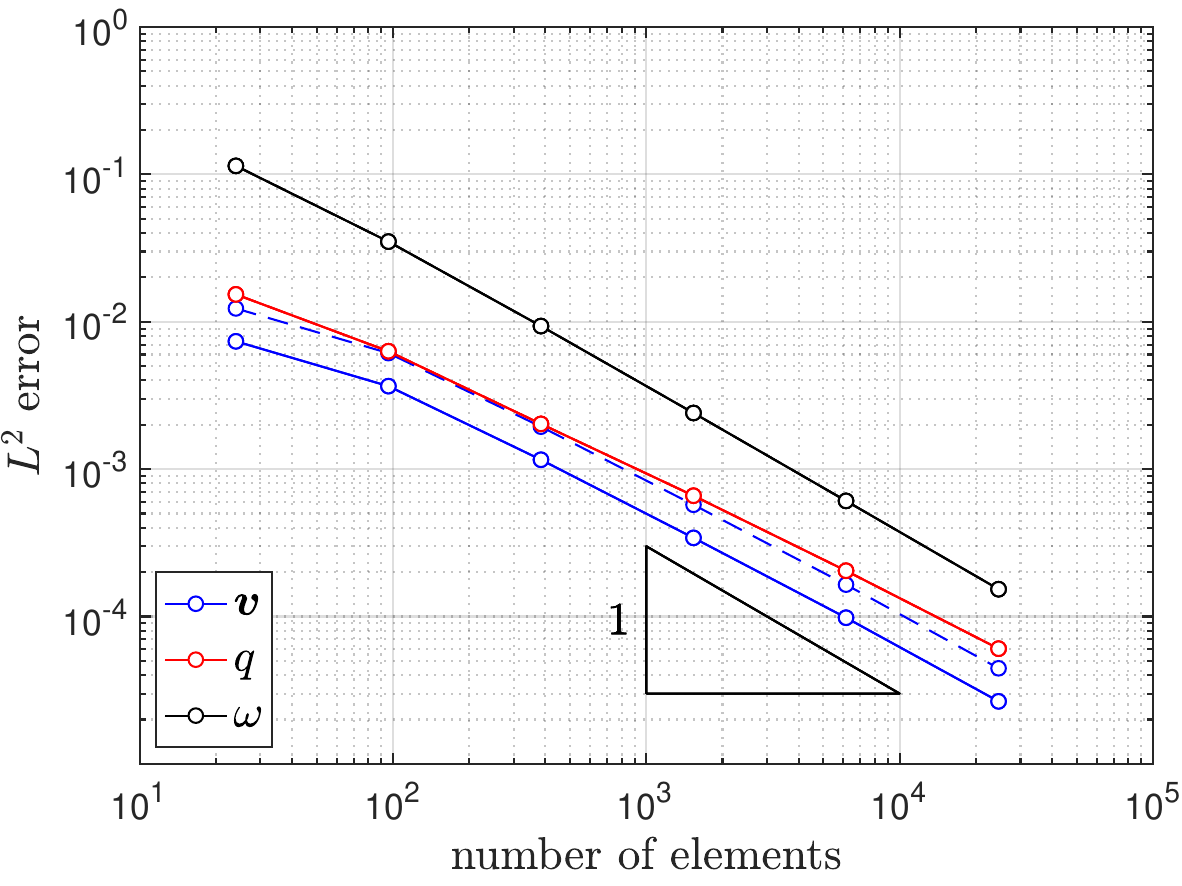}}
\put(-8,-.12){\includegraphics[height=58mm]{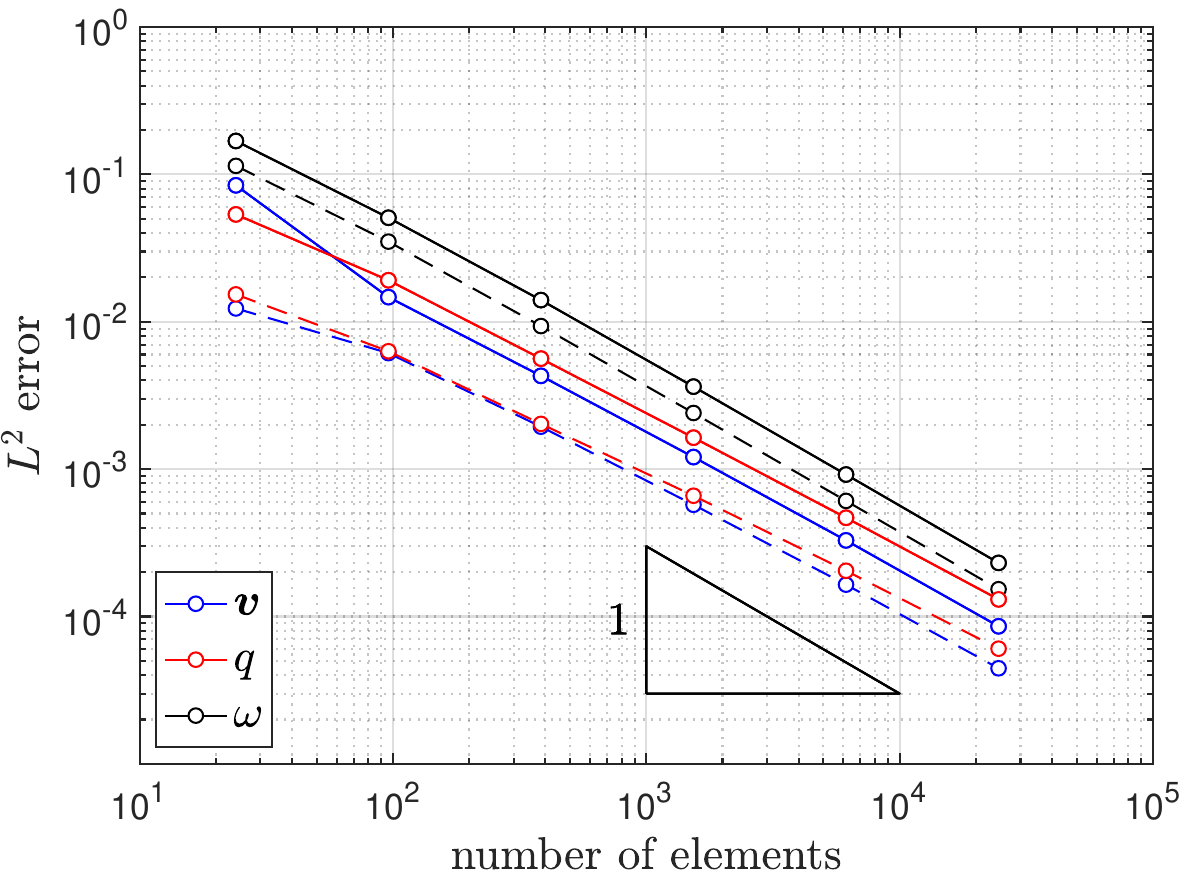}}
\put(0.2,-.2){\includegraphics[height=58mm]{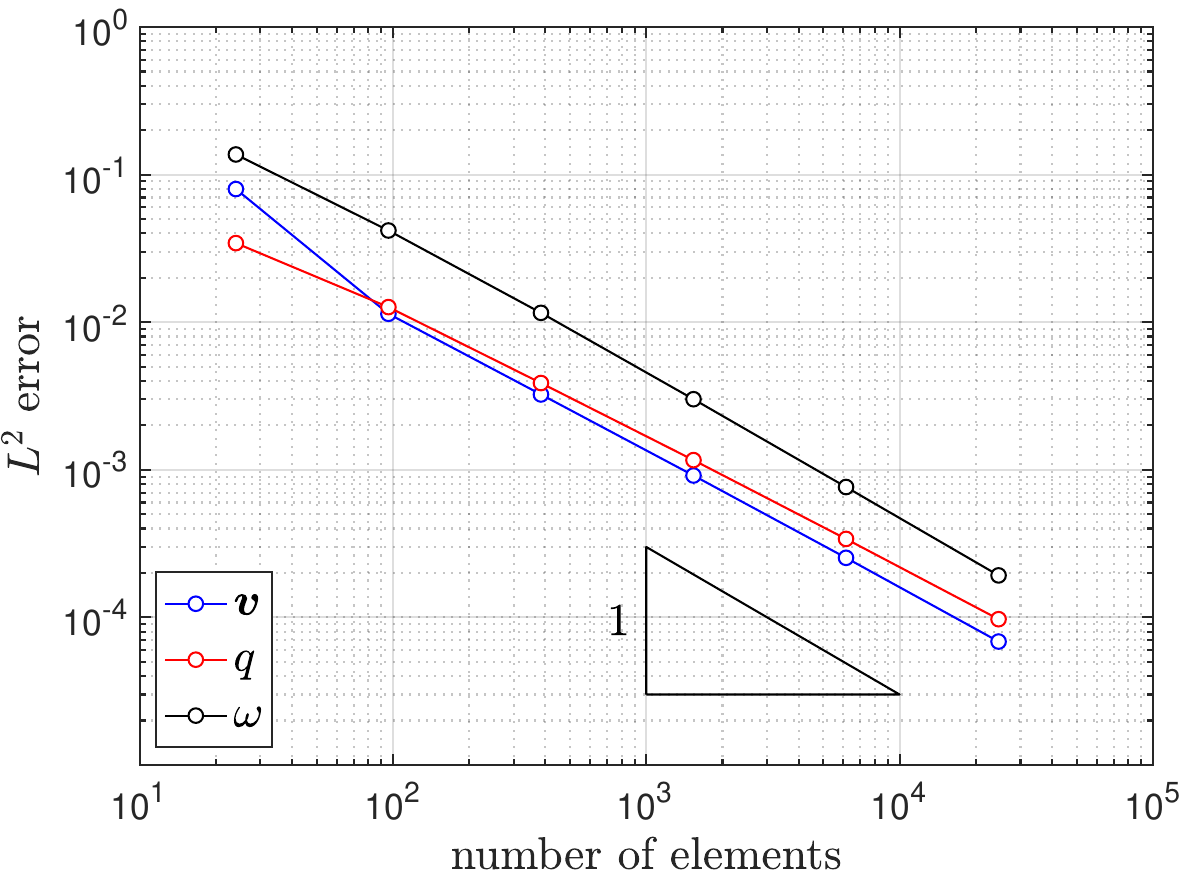}}
\put(-7.95,5.85){\footnotesize (a)}
\put(0.2,5.85){\footnotesize (b)}
\put(-7.95,-.15){\footnotesize (c)}
\put(0.2,-.15){\footnotesize (d)}
\end{picture}
\caption{Simple shear flow on a rigid sphere for load case 4: 
Convergence of velocity, surface tension and vorticity for the four ALE cases (a-d) in the $L^2$ error norm.
Nearly optimal convergence rates are obtained for all fields.}
\label{f:shearflowA2}
\end{center}
\end{figure}

This appendix shows the results for load cases 2 and 4 from the study of Sec.~\ref{s:shearflow}, see Table~\ref{t:shearflowcases}.
Load case 2, shown in Fig.~\ref{f:shearflowA1}, is the case where the normal surface velocity $v_\mrn$ is prescribed by a Dirichlet boundary condition.
%----------------------------------------------------------------------------------------------------------------------------------
% run pShearFlow.m & pShearFlowALEc.m with corresponding jALE & jcase
% data from sShearFlow/ vShearFlow* mat{1}.case = 2
%----------------------------------------------------------------------------------------------------------------------------------
Load case 4, shown in Fig.~\ref{f:shearflowA2}, is the case where the normal surface velocity $v_\mrn$ adjusts to the Neuman boundary condition on the surface pressure $p$.
%----------------------------------------------------------------------------------------------------------------------------------
% run pShearFlow.m & pShearFlowALEc.m
% data from sShearFlow/ vShearFlow*p0 mat{1}.case = 4
%----------------------------------------------------------------------------------------------------------------------------------
The convergence rate for $q$ does not look quite so good in this load case, which could be an effect of too low $\alpha_\mathrm{DB}$.

\bibliographystyle{apalike}
\bibliography{../../Tex/bibliography}
%\bibliography{bibliography}
%

\end{document}

%% file: neco.tex
\newcommand{\bitm}{\begin{itemize}}
\newcommand{\eitm}{\end{itemize}}
\newcommand{\bnumr}{\begin{enumerate}}
\newcommand{\enumr}{\end{enumerate}}

\newcommand {\eqb}[1]{\begin{equation}\begin{array}{#1}}
\newcommand {\eqe}{\end{array}\end{equation}}

\newcommand {\esb}[1]{\begin{equation*}\begin{array}{#1}}
\newcommand {\ese}{\end{array}\end{equation*}}
\newcommand {\ds}{\displaystyle}

\newcommand {\pa}[2]{\frac{\partial{#1}}{\partial{#2}}}

\newcommand {\back}{\! \! \!}
\newcommand {\is}{\back &=& \back}
\newcommand {\dis}{\back &:=& \back}

\newcommand {\plus}{\back &+& \back}
\newcommand {\mi}{\back &-& \back}

\newcommand {\norm}[1]{\|#1\|}

\newcommand {\dif}{\mathrm{d}}

\newcommand {\II}{{I\kern-.3em I}}
\newcommand {\III}{{I\kern-.3em I\kern-.3em I}}

\newcommand {\mra}{\mathrm{a}}
\newcommand {\mrb}{\mathrm{b}}
\newcommand {\mrc}{\mathrm{c}}

\newcommand {\mre}{\mathrm{e}}
\newcommand {\mrf}{\mathrm{f}}

\newcommand {\mri}{\mathrm{i}}

\newcommand {\mrm}{\mathrm{m}}
\newcommand {\mrn}{\mathrm{n}}
\newcommand {\mro}{\mathrm{o}}
\newcommand {\mrp}{\mathrm{p}}
\newcommand {\mrq}{\mathrm{q}}

\newcommand {\mrs}{\mathrm{s}}
\newcommand {\mrt}{\mathrm{t}}

\newcommand {\mrv}{\mathrm{v}}

\newcommand {\mrx}{\mathrm{x}}

\newcommand{\mrJ}{\mathrm{J}}

\newcommand{\mrT}{\mathrm{T}}

\newcommand {\ma}{\mathbf{a}}

\newcommand {\mcc}{\mathbf{c}}
\newcommand {\md}{\mathbf{d}}

\newcommand {\mf}{\mathbf{f}}
\newcommand {\mg}{\mathbf{g}}
\newcommand {\mh}{\mathbf{h}}

\newcommand {\mk}{\mathbf{k}}

\newcommand {\mm}{\mathbf{m}}

\newcommand {\mq}{\mathbf{q}}
\newcommand {\mr}{\mathbf{r}}

\newcommand {\muu}{\mathbf{u}}
\newcommand {\mv}{\mathbf{v}}
\newcommand {\mw}{\mathbf{w}}
\newcommand {\mx}{\mathbf{x}}

\newcommand {\mbc}{\mathbf{\bar c}}
\newcommand {\mbd}{\mathbf{\bar d}}

\newcommand {\mbf}{\mathbf{\bar f}}

\newcommand {\mbk}{\mathbf{\bar k}}

\newcommand {\mhm}{\mathbf{\hat m}}
\newcommand {\mhn}{\mathbf{\hat n}}

\newcommand {\mtf}{\mathbf{\tilde f}}
\newcommand {\mtk}{\mathbf{\tilde k}}
\newcommand {\mtm}{\mathbf{\tilde m}}
\newcommand {\mtn}{\mathbf{\tilde n}}

\newcommand {\ba}{\boldsymbol{a}}

\newcommand {\bc}{\boldsymbol{c}}
\newcommand {\bd}{\boldsymbol{d}}
\newcommand {\be}{\boldsymbol{e}}
\newcommand {\bff}{\boldsymbol{f}}

\newcommand {\bi}{\boldsymbol{i}}

\newcommand {\bn}{\boldsymbol{n}}

\newcommand {\bu}{\boldsymbol{u}}
\newcommand {\bv}{\boldsymbol{v}}
\newcommand {\bw}{\boldsymbol{w}}
\newcommand {\bx}{\boldsymbol{x}}

\newcommand {\bnu}{\mbox{\boldmath$\nu$}}

\newcommand {\mD}{\mathbf{D}}

\newcommand {\mK}{\mathbf{K}}
\newcommand {\mL}{\mathbf{L}}

\newcommand {\mN}{\mathbf{N}}

\newcommand {\mX}{\mathbf{X}}

\newcommand {\bA}{\boldsymbol{A}}

\newcommand {\bN}{\boldsymbol{N}}

\newcommand {\bT}{\boldsymbol{T}}

\newcommand {\bX}{\boldsymbol{X}}

\newcommand {\sig}{\sigma}

\newcommand {\bsig}{\mbox{\boldmath$\sigma$}}

\newcommand {\btau}{\mbox{\boldmath$\tau$}}

\newcommand {\bone}{\mathbf{1}}

\newcommand {\bbR}{\mathbb{R}}

\newcommand {\IR}{{\rm\kern.24em
   \vrule width.02em height1.53ex depth-.05ex
   \kern-.3em R}}
\newcommand {\ic}{{\rm\kern.20em
   \vrule width.02em height1.0ex depth-.05ex
   \kern-.22em c}}
\newcommand {\ia}{{\rm\kern.20em
   \vrule width.02em height1.05ex depth-.0ex
   \kern-.25em a}}
\newcommand {\IC}{{\rm\kern.24em
   \vrule width.02em height1.4ex depth-.05ex
   \kern-.26em C}}
\newcommand {\ID}{{\rm\kern.34em
   \vrule width.02em height1.5ex depth-.05ex
   \kern-.36em D}}
\newcommand {\IS}{{\rm\kern.24em
   \vrule width.02em height1.6ex depth.05ex
   \kern-.26em S}}
\newcommand {\IT}{{\rm\kern.50em
   \vrule width.02em height1.55ex depth-.05ex
   \kern-.52em T}}

\newcommand {\IE}{{\rm\kern.24em
   \vrule width.02em height1.55ex depth-.05ex
   \kern-.33em E}}
\newcommand {\IEa}{{\rm\kern.24em
   \vrule width.02em height1.55ex depth-.05ex
   \kern-.33em E}^{1}_{ijkl}}
\newcommand {\IEb}{{\rm\kern.24em
   \vrule width.02em height1.55ex depth-.05ex
   \kern-.33em E}^{2}_{ijkl}}

\newcommand {\sC}{\mathcal{C}}

\newcommand {\sP}{\mathcal{P}}
\newcommand {\sQ}{\mathcal{Q}}

\newcommand {\sS}{\mathcal{S}}
\newcommand {\sT}{\mathcal{T}}

\newcommand {\sV}{\mathcal{V}}
\newcommand {\sW}{\mathcal{W}}

\newcommand {\vaub}{\ba_\beta}

\newcommand{\nablas}{\nabla_{\!\mrs}}
\newcommand{\divs}{\mathrm{div_s}\,}

\newcommand {\Ass}[2]{\kern 0.9ex \vrule width0.45em height0.2ex depth0ex \kern -2.1ex \bigwedge_{#1}^{#2}}
\newcommand {\ASS}[2]{\kern 1.45ex \vrule width0.5em height0.2ex depth0ex \kern -2.65ex \bigwedge_{#1}^{#2}}